%% file: mainCSF.tex
\documentclass[conference]{IEEEtran}

\newif\ifshep
\sheptrue

\ifshep
\usepackage[users=S]{uchanges}
\else
\usepackage[users=S, hide] {uchanges} 
\fi

\newif\ifcameraready
\camerareadyfalse

\ChangesSetUserColor{S}{blue}

\usepackage{cite}
\usepackage{amsmath,amssymb,amsfonts}
\usepackage{algorithmic}
\usepackage{graphicx}
\usepackage{textcomp}
\usepackage{xcolor}
\usepackage{paralist}
\usepackage{tikz}
\usepackage{mathpartir}
\usetikzlibrary{positioning}
\usetikzlibrary{fit}  
\def\BibTeX{{\rm B\kern-.05em{\sc i\kern-.025em b}\kern-.08em
    T\kern-.1667em\lower.7ex\hbox{E}\kern-.125emX}}

\usepackage{styles/bitmlx_syntax}
\usepackage{styles/bitmlx_semantics}
\usepackage{styles/bitmlx_compiler}
\usepackage{styles/intermediate_semantics}
\usepackage{styles/bitmlx_comments}
\usepackage{styles/bitmlx_base}
\usepackage{styles/coherence}
\usepackage{styles/strategies}
\usepackage{styles/correctness}

\newcommand{\someParContract}{\isActiveContract{\contract{C}}{\isContractState}{\someBlockchain}{\someContract}}
\newcommand{\someParPrChoice}{\isActiveContract{\contract{D} \prchoice \contract{C}}{\isContractState}{\someBlockchain}{\someContract}}

\usepackage{pifont}
\usepackage[capitalize]{cleveref}

\usepackage{xspace}

\renewcommand{\paragraph}[1]{\smallskip \noindent \textbf{#1.}\;}

\makeatletter
\newcounter{subparagraph}[paragraph]
\renewcommand\thesubparagraph{%
  \theparagraph.\@arabic\c@subparagraph}
\newcommand\subparagraph{%
  \@startsection{subparagraph}    
    {6}                              
    {\parindent}                     
    {3.25ex \@plus 1ex \@minus .2ex} 
    {-1em}                           
    {\normalfont\normalsize\bfseries}}
\makeatother

\newcommand{\contractExchange}{\contract{Swap^x}}
\newcommand{\contractExchangeBit}{\contract{Swap}^{\money{\bitcoins}}}
\newcommand{\contractExchangeDoge}{\contract{Swap}^{\money{\dogecoins}}}

\newcommand{\movetext}{move\xspace}
\newcommand{\movestext}{moves\xspace}
\newcommand{\fundstext}{funds\xspace}
\newcommand{\blockchaintext}{blockchain\xspace}
\newcommand{\blockchainstext}{blockchains\xspace}

\usepackage{standalone}

\makeatletter
\newcommand{\linebreakand}{%
  \end{@IEEEauthorhalign}
  \hfill\mbox{}\par
  \mbox{}\hfill\begin{@IEEEauthorhalign}
}
\makeatother

\newif\ifshowcommentsmainbody
\showcommentsmainbodyfalse

\begin{document}

\title{\bitmlx: Secure Cross-chain Smart Contracts For Bitcoin-style Cryptocurrencies}

\author{
    \IEEEauthorblockN{Federico Badaloni}
    \IEEEauthorblockA{\textit{MPI-SP} \\
    federico.badaloni@mpi-sp.org}%
    \and
    \IEEEauthorblockN{Sebastian Holler}
    \IEEEauthorblockA{\textit{MPI-SP} \\
    sebastian.holler@mpi-sp.org}
    \linebreakand
    \IEEEauthorblockN{Chrysoula Oikonomou}
    \IEEEauthorblockA{\textit{Aristotle University of Thessaloniki} \\
    linen@protonmail.com}
    \and
    \IEEEauthorblockN{Pedro Moreno-Sanchez}
    \IEEEauthorblockA{\textit{IMDEA Software Institute} \\
    \textit{VISA Research}\\
    \textit{MPI-SP}\\
    pedro.moreno@imdea.org}
    \and
    \IEEEauthorblockN{Clara Schneidewind}
    \IEEEauthorblockA{\textit{MPI-SP} \\
    clara.schneidewind@mpi-sp.org}
}

\maketitle

\input{abstract}


\input{intro}    

\input{background}  

\input{key-ideas-new}  

\input{bitmlx_language} 

\input{compilation} 

\input{compiler-correctness} 

\input{discussion} 

\input{applications} 

\input{implementation_evaluation} 

\input{related_work} 

\input{conclusion} 

\bibliographystyle{IEEEtran}
\bibliography{bib}

\ifcameraready{}
\else{
    \newpage
    \onecolumn
    \appendices
    
    \section{\bitmlx Syntax}
    \label{sec:bitmlx-syntax}

\input{theory_docs/syntax}

    \section{\bitmlx Semantics}
    \label{sec:bitmlx-semantics}

\input{theory_docs/bitmlx_semantics.tex}

    \section{Compiling \bitmlx Contracts}
    \label{sec:app-compiler}

\input{theory_docs/compiler.tex}

    \section{Intermediate Semantics}
    \label{sec:app-intermediate-semantics}

\input{theory_docs/intermediate_semantics.tex}

    \section{Compiling \bitmlx Strategies}
    \label{sec:app-strategies}

\input{theory_docs/xStrategy.tex}

    \section{Frontiers}

\input{theory_docs/frontiers}

    \section{Round-Based Execution}

\input{theory_docs/rounds}

    \section{\bitmlx Soundness}
    \label{sec:app-bitmlx-soundness}

\input{theory_docs/xSoundness.tex}

    \section{Intermediate Semantics to \bitml coherence}

\input{theory_docs/low_level_proof}

    \section{Money Preservation}

\input{theory_docs/money_preservation}

    \section{Liquidity}

\input{theory_docs/final-statement}

    \section{Auxiliary Functions}

\input{theory_docs/auxiliary}

}
\fi

\end{document}

%% file: abstract.tex
\begin{abstract}

    A smart contract is an interactive program that governs funds in the realm of a single cryptocurrency. 
    Yet, the many existing cryptocurrencies have spurred the design of cross-chain applications that require  interactions with multiple cryptocurrencies simultaneously. 
    Currently, cross-chain applications are implemented as use-case-specific cryptographic protocols that serve as overlay to synchronize smart contract executions in the different cryptocurrencies. 
    Hence, their design 
    requires substantial expertise, as well as a security analysis in complex cryptographic frameworks.

    In this work, we present \bitmlx, the first domain-specific language for cross-chain smart contracts, enabling interactions with several users that hold \fundstext across multiple Bitcoin-like cryptocurrencies. 
    We contribute a compiler to automatically translate a \bitmlx contract into one contract per involved cryptocurrency and a user strategy that synchronizes the execution of these contracts. 
    We prove that an honest user, who follows the prescribed strategy when interacting with the several contracts, ends up with at least as many funds as in the corresponding execution of the \bitmlx contract. 
    Last, but not least, we implement the \bitmlx compiler and demonstrate its utility in the design of illustrative examples of cross-chain applications such as multi-chain donations or loans across different cryptocurrencies. 
\end{abstract}

%% file: intro.tex
\section{Introduction}
\label{sec:intro}
\newcommand{\blockchain}{blockchain\xspace}
\newcommand{\blockchains}{blockchains\xspace}

A smart contract is an interactive program that controls cryptocurrency \fundstext.
Smart contracts are registered on a \blockchain, a jointly maintained, append-only data structure that captures the system state of the cryptocurrency.
The smart contract logic is then enforced as part of the consensus protocol that is executed by the cryptocurrency users to 
advance the system state by appending transactions (that represent user actions) to the \blockchain.

While cryptocurrencies like Ethereum genuinely support complex stateful smart contracts that can be written in expressive (high-level) languages, the native smart contract support of other cryptocurrencies, foremost Bitcoin, is limited to basic payment conditions (e.g., digital signature-based authorization) expressed in a simple scripting language. 
This simple scripting functionality still gives rise to many interesting applications (e.g., escrows and lotteries) when composing multiple of the abovementioned payment conditions and integrating them with a cryptographic protocol.

To facilitate the development of such applications, Bartoletti and Zunino contribute \bitml~\cite{bitml}, a high-level specification format for smart contracts in Bitcoin-style cryptocurrencies. 
\bitml allows for defining smart contracts in terms of a simple process calculus that describes how the contract state evolves when interacting with a set of predefined contract users. 
For execution on a Bitcoin-style \blockchain, the contract users can then compile a \bitml smart contract into a cryptographic protocol over Bitcoin transactions that will realize the behavior of the high-level contract execution.

\paragraph{Problem} The execution of a smart contract is confined to the realm of a single \blockchain and its underlying consensus protocol. 
This applies both to natively-supported and \bitml smart contracts. 
However,  there exist many different \blockchains, which cater different use cases and many practical applications demand to govern \fundstext at several \blockchains \emph{simultaneously}. 
As an illustrative example,  a two-party swap~\cite{thyagarajan2022universal}  involves two users Alice and Bob who want to exchange \fundstext that they own at different \blockchains. 
This exchange must be atomic, meaning that Alice transfers their \fundstext to Bob in the first \blockchain if and only if Bob also transfers their \fundstext to Alice in the second \blockchain. 
Otherwise, both users should get their original \fundstext refunded. 
More complex applications involve a larger number of users, each of which hold \fundstext at potentially many \blockchains (e.g., $n$-party swaps~\cite{atomic-swaps} is a generalization of the two-party swap where $n$ parties swap \fundstext held at $k$ blockchains among each other). 

\paragraph{State-of-the-art and challenges} 
Existing works in the literature tackle the abovementioned problem of cross-chain smart contracts~\cite{zamyatin2021sokcrosschain} by contributing  cryptographic protocols used as a synchronization overlay among smart contracts executed otherwise independently at different \blockchains. 
These cryptographic protocols are application-specific meaning that they need to be adapted for each use case. 
The security of each of these protocols, hence, is proven from scratch, requiring involved proofs in complex cryptographic proof frameworks~\cite{TairiMS23}, a task that is cumbersome and error-prone as demonstrated by the security flaws found so far in protocols proposed by both academia and industry~\cite{multi-hop-locks,foundations-coin-mixing,foundations-adaptor-signatures,flood-loot}. 

\paragraph{Goal} 
In this work we want to answer the following question: 
\emph{Can we compile secure-by-design protocols over Bitcoin-style \blockchains from a domain-specific language for cross-chain smart contracts?}

\paragraph{Our approach}
To achieve this goal, we define \bitmlx, the first domain-specific language for cross-chain smart contracts. 
A smart contract expressed in \bitmlx is compiled into (i) several \bitml smart contracts, one per \blockchain; and 
(ii) a strategy per honest user that defines how they must interact with the different \bitml contracts to synchronize their execution.   
The smart contracts resulting from our compilation can be executed in Bitcoin-style \blockchains. 
Therefore, our approach imposes minimal requirements to the underlying \blockchains in that it only requires the support for basic scripting capabilities, and, hence, 
is compatible with a wide range of different \blockchains. 
Building upon \bitml, \bitmlx smart contracts can be easily executed on UTXO-based cryptocurrencies with support for basic payment conditions that match the limited expressiveness of Bitcoin's scripting language (e.g., Dogecoin, Bitcoin Cash, Litecoin, or Cardano).
Account-based cryptocurrencies with a (quasi) Turing-complete scripting language (such as Ethereum) could support \bitmlx contracts by emulating the execution of \bitml contracts.

Using \bitmlx, cross-chain applications (e.g., atomic swaps)
can be designed as a cross-chain smart contract abstracting away the cryptographic details to be handled when realizing applications with custom cross-currency protocols. 
In our approach, every honest user following the prescribed strategy is guaranteed that the execution of the several compiled \bitml contracts ends up with them obtaining at least as many \fundstext as in the corresponding execution of the \bitmlx contract. 
While we find that \bitmlx can express many interesting cross-chain applications, \bitmlx is not a complete language for writing such applications. 
Its expressiveness is inherently limited by the underlying \bitml language.
Further, to enable secure cross-chain execution, \bitmlx smart contracts, as opposed to \bitml contracts, need to follow a slightly more restrictive structure. 
We discuss those limitations in detail in~\Cref{sec:discussion}.

\paragraph{Our contributions} We make the following contributions:
\begin{asparaitem}
    \item We introduce \bitmlx, the first domain-specific language for cross-chain smart contracts (\Cref{sec:bitml-language}). \bitmlx permits expressing smart contracts among $n$ users whose functionality requires interacting with $k$ different \blockchains. 
    \item We contribute a compiler (\cref{sec:compilation}) to (i) translate \bitmlx contracts into  $k$ \bitml contracts; and (ii) translate strategies to execute the \bitmlx contract into strategies to interact with the compiled \bitml contracts. 
    The crucial technical challenge lies in the safe replication of the smart contract steps in the $k$ different blockchains.
        This must hold also in the presence of adversaries that can decide on the execution order of scheduled actions and may even insert new actions that may conflict with the honest user's attempts to enforce synchronous contract executions. 
        \item We prove the correctness of the \bitmlx compilation (\cref{sec:proof}). Our correctness result establishes that an honest user following the strategy prescribed by our compiler to execute the $k$ \bitml contracts will end up with at least as much \fundstext as in the corresponding strategy in the \bitmlx contract. 
    \item To showcase the practicality of our approach, we have implemented our compiler~\cite{bitmlx-code} and used it to implement and evaluate illustrative cross-chain applications (\Cref{sec:evaluation}) such as (i) multi\blockchain donations, where a non-profit organization accepts donations in different denominations; (ii) multi\blockchain payments with an exchange service, where a customer can decide to pay a service provider for its service in different denominations; and (iii) multi-\blockchain loan with mediator, where a user borrows a loan in a certain denomination whereas holding \fundstext in a different denomination as collateral. 
    We discuss the scope and limitations of the \bitmlx language for developing cross-currency applications (\Cref{sec:discussion}).
    \end{asparaitem}

%% file: background.tex
\section{Background}
\label{sec:relwork}

\paragraph{\bitml in a nutshell} 
In~\cite{bitml}, Bartoletti and Zunino introduce \bitml, a domain-specific language for smart contracts in Bitcoin and other Bitcoin-style cryptocurrencies. 
As opposed to cryptocurrencies like Ethereum, Bitcoin does not allow for locking funds into interactive stateful smart contracts. 
Instead, Bitcoin transactions transfer the ownership of coins based on static spending conditions written in a simple scripting language. 
Integrating this basic mechanism with cryptographic protocols still allows for implementing a wide range of interesting applications on Bitcoin, such as lotteries or escrows. 

\bitml serves as a high-level specification format to express such applications while completely abstracting from Bitcoin transactions and cryptographic details. 
The \bitml syntax permits defining contracts of the form $\{\preconditions{G}\}\contract{C}$ where $\contract{G}$ denotes the \emph{preconditions} for the contract whereas $\contract{C}$ denotes the code encoding the contract logic according to which funds governed by the contract will be distributed. 
The preconditions $\preconditions{G}$ 
specify the funds that users need to provide to the contract ($\depositsPre{A}{\money{v\bitcoins}}{x}$ denotes the requirement of $\participant{A}$ to fund the contract with $v$ bitcoins ($\money{\bitcoins}$)) and the secrets to which contracts users need to commit ($\participant{A}: ~ \bitmlcode{secret}~ \secret{s}$ denotes that $\participant{A}$ needs to commit to secret $\secret{s}$).

A contract $\contract{C} \Coloneqq \contract{D_1} + \contract{D_2} + \dots + \contract{D_k}$ is defined as a list of mutually exclusive execution choices represented by subcontracts $\contract{D_i}$.
Each sub-contract $\contract{D_i}$ has one of the following forms: 
(i) $\bitmlxAuthOp{A}{\contract{D}}$, denoting that $\participant{A}$ must authorize the execution of $\contract{D}$; 
(ii) $\bitmlcode{reveal}~\secret{s}~\bitmlcode{then}~\contract{C}$, denoting that secret $\secret{s}$ must be revealed to continue with the execution of contract $\contract{C}$;
(iii) $\bitmlcode{after}~t : \contract{D}$, denoting time $t$ must be reached before executing $\contract{D}$; 
(iv) $\bitmlcode{split}~\money{v\vec{\bitcoins}}~\rightarrow~\vec{\contract{C}}$, denoting that the contract funds are split according to the vector $\money{v\vec{\bitcoins}}$ and henceforth will be governed by the contracts as given by the vector $\vec{\contract{C}}$; 
and (v) $\bitmlcode{withdraw}~\participant{A}$, denoting that $\participant{A}$ can obtain the \fundstext controlled by the contract. 

For example, consider the following \bitml contract that implements a so-called \emph{timed commitment} where a user $\participant{A}$ commits to paying funds to another user $\participant{B}$ if $\participant{B}$ reveals a pre-agreed secret $\secret{s}$:

{\small 
\[
\begin{matrix}
    \{ \depositsPre{A}{\money{1\bitcoins{}}}{x}
    ~ \vert ~ \participant{B}: ~ \bitmlcode{secret} ~ \secret{s} \}  \contract{TC} \\
    \contract{TC} = \contract{Exchange} + \contract{Refund}\\
    \contract{Exchange} = \bitmlcode{reveal}~\secret{s}~.~\bitmlcode{withdraw}~\participant{B}\\
    \contract{Refund} = \bitmlcode{after}~t~.~\bitmlcode{withdraw}~\participant{A}
\end{matrix}
\]
}

The contract preconditions determine that $\participant{A}$ needs to fund the contract with $\money{1\bitcoins}$ (here $x$ serves as the identifier of the concrete fund) and that $\participant{B}$ needs to commit to a secret named $\secret{s}$.
Once this happened (in which case we say that the contract was successfully \emph{stipulated}), 
the \fundstext can be withdrawn by $\participant{B}$ if $\secret{s}$ is revealed (i.e., $\contract{Exchange}$ clause). Alternatively, starting from time $t$, $\participant{A}$ can withdraw the \fundstext (i.e., $\contract{Refund}$ clause). 
Such $\contract{Refund}$ clause is usually added to ensure that the funds of $\participant{A}$ cannot be locked indefinitely if $\participant{B}$ fails to collaborate.

While \bitml smart contracts can be considered interactive programs whose faithful execution is ensured by the blockchain consensus, we are usually interested in the guarantees that an honest user can obtain when interacting with these programs.
E.g., in the contract $\contract{TC}$, an honest user $\participant{B}$ is guaranteed to obtain the funds when revealing secret $\secret{s}$ and withdrawing the contract funds before time $t$. 
However, when $\participant{B}$ reveals $\secret{s}$ late, no such guarantees can be given. 
For this reason, we consider \bitml smart contracts in conjunction with honest user protocols (henceforth called honest user \emph{strategies}) that characterize the interactions of an honest user with the \bitml contract.

In~\cite{bitml}, Bartoletti and Zunino define a computational model of Bitcoin execution, as well as a symbolic model formally describing the execution of \bitml contracts.  
They contribute a compiler that translates \bitml contracts into a set of standard Bitcoin transactions and each symbolic honest user strategy into a computational strategy leveraging these transactions. 
Based on that, they provide a computational soundness result, showing that every protocol run in the computational model can be faithfully captured by a run in the symbolic model. 
This ensures that \bitml contract guarantees proven in the symbolic model carry over the executions of the compiled protocol in the computational world. 

%% file: key-ideas-new.tex
\section{Key Ideas}
\label{sec:keyideas}

Natively supported smart contracts can only control \fundstext from a single \blockchaintext because their execution is enforced by the underlying blockchain-based consensus protocol.
As a result, smart contracts on different \blockchainstext (and therefore subject to different consensus protocols) execute independently.
In practice, however, users are confronted with a heterogeneous environment of \blockchainstext and wish to manage their funds in these systems in a unified and synchronized manner. 
Still, there exists no systematic way for developing cross-chain smart contracts that simultaneously govern \fundstext on different \blockchainstext. 

In this section, we overview our approach to \bitmlx, the first domain-specific language for cross-chain smart contracts. 
As a running example, we use a two-party atomic swap, a cross-chain application that has been largely studied in the literature and is widely used in practice~\cite{zamyatin2021sokcrosschain}.
In a two-party atomic swap, Alice and Bob want to exchange \fundstext that they own on different blockchains (e.g., Alice owns  $\money{1\bitcoins}$ in Bitcoin whereas Bob owns $\money{1\dogecoins}$ in Dogecoin). 
The outcome of the two-party atomic swap must be that Alice sends their bitcoins to Bob if and only if Bob sends Alice their dogecoins. 

\paragraph{Using \bitml to express two-party atomic swaps} 
A naive approach towards realizing a two-party atomic swap would be to make Alice and Bob lock their coins into timed commitment contracts on the chains where they own their funds and to condition the release of \fundstext in both contracts to Bob revealing a secret $\secret{s}$:

{\footnotesize
\[
\begin{matrix}
    \{ \depositsPre{A}{\money{1\bitcoins{}}}{x_A}
    ~ \vert ~ \participant{B}: ~ \bitmlcode{secret} ~ \secret{s} \}  \contract{TC^{\bitcoins}} \\
    \contract{TC^{\bitcoins}} = \contract{Exchange^{\bitcoins}} + \contract{Refund^{\bitcoins}}\\
    \contract{Exchange^{\bitcoins}} = \bitmlcode{reveal}~\secret{s}~.~\bitmlcode{withdraw}~\participant{B}\\
    \contract{Refund^{\bitcoins}} = \bitmlcode{after}~t~.~\bitmlcode{withdraw}~\participant{A}\\ \\
    \{ \depositsPre{B}{\money{1\dogecoins{}}}{x_B}
    ~ \vert ~ \participant{B}: ~ \bitmlcode{secret} ~ \secret{s} \}  \contract{TC^{\dogecoins}} \\
    \contract{TC^{\dogecoins}} = \contract{Exchange^{\dogecoins}} + \contract{Refund^{\dogecoins}}\\
    \contract{Exchange^{\dogecoins}} = \bitmlcode{reveal}~\secret{s}~.~\bitmlcode{withdraw}~\participant{A}\\
    \contract{Refund^{\dogecoins}} = \bitmlcode{after}~t~.~\bitmlcode{withdraw}~\participant{B}
\end{matrix}
\]
}

If both users are honest, the execution of these two contracts does actually lead to a successful swap: 
Imagine that Alice and Bob agree to swap the coins. 
In such case, Bob can reveal the secret $\secret{s}$ to execute the subcontract $\contract{Exchange^{\bitcoins}}$, letting Alice learn $\secret{s}$ and consequently execute the subcontract $\contract{Exchange^{\dogecoins}}$. 
Similarly, this approach allows honest users to safely abort the swap:
Imagine that Bob does not wish to swap the coins. 
In such case, Bob does not reveal the secret $\secret{s}$, and after time $t$ expires, Alice (correspondingly Bob) can get refunded by executing the contract $\contract{Refund^{\bitcoins}}$ (correspondingly $\contract{Refund^{\dogecoins}}$). 

\paragraph{Issues with malicious users}
The challenges appear when any of two users does not obey the prescribed strategy to execute the \bitml contracts. 
Note that the contracts $\contract{TC^{\bitcoins}}$ and $\contract{TC^{\dogecoins}}$ reside in two blockchains and therefore execute independently. 
The coordination of actions in different contracts relies on the users observing and replicating smart contract interactions.
However, the blockchain-based consensus does not easily allow for atomic replication of actions observed on another blockchain:
Blockchain (and so, in particular smart contract) interactions conducted by (honest) users do not immediately come into effect but get scheduled for execution. 
A potentially malicious scheduler (e.g., a Bitcoin miner) then decides on the execution order of scheduled actions and may even insert new actions.
Attempts to replicate an action observed on another blockchain, hence, may be preempted by the execution of other, conflicting actions.

For instance, imagine that a malicious user Bob collaborating with the scheduler waits until time $t$ and only then reveals $\secret{s}$ and executes $\contract{Exchange^{\bitcoins}}$. 
Even if Alice observes this action and tries to replicate it by scheduling $\contract{Exchange^{\dogecoins}}$ for execution, it is not guaranteed that $\contract{Exchange^{\dogecoins}}$ will get successfully executed. 
Instead, Bob may schedule $\contract{Refund^{\dogecoins}}$ and the malicious scheduler may execute it before $\contract{Exchange^{\dogecoins}}$, hence making Bob receive the \fundstext of both contracts.

\subsection{Our Approach} 
In this work, we develop a general solution to the described atomic replication problem.
To this end, we design \bitmlx, a novel domain-specific language for expressing cross-currency smart contracts and show how to compile \bitmlx contracts into independently executing \bitml contracts, which support safe replication by design.

Using \bitmlx, we can express the two-party atomic swap as follows:

{\footnotesize 
\[
    \begin{matrix}
        \contractExchange &= \{
            \participant{A}: \money{1\bitcoins}
            | \participant{B}: \money{1\dogecoins}
        \}
        {\contract{Pay^x} \prchoice \contract{Abort^x}}\\
        \contract{Pay^x} ~ &= \bitmlcode{withdraw}(
            \money{1\dogecoins} \rightarrow \participant{A},
            ~ \money{1\bitcoins} \rightarrow \participant{B}
        )\\
        \contract{Abort^x} ~ &= \bitmlcode{withdraw}(
            \money{1\bitcoins} \rightarrow \participant{A},
            ~\money{1\dogecoins} \rightarrow \participant{B}
        )
    \end{matrix}
\]
}

Departing from \bitml and crucial to securely support cross-chain applications (as we explain later), a contract in \bitmlx is defined as a \emph{priority choice} between two clauses. In the running example, every contract user has the opportunity to execute the \movetext $\contract{Pay^x}$ (i.e., \emph{high priority \movetext}), and only if all users decide to skip this option, the alternative clause $\contract{Abort^x}$ (i.e., \emph{low priority \movetext}) will be available. 
The code $\contract{Pay^x}$ \emph{simultaneously} enables $\participant{A}$ to withdraw  $\money{1\dogecoins}$ and $\participant{B}$ to withdraw  $\money{1\bitcoins}$. 
The clause $\contract{Abort^x}$ enables $\participant{A}$ and $\participant{B}$ to reclaim their originally deposited funds. 
Effectively, the contract implements an atomic swap between users $\participant{A}$ (holding $\money{1\bitcoins}$) and $\participant{B}$ (holding $\money{1\dogecoins}$): Either both \fundstext are transferred to the respective counterparty or both \fundstext are refunded. 

\paragraph{Compilation into two \bitml contracts}
We compile the \bitmlx contract into two \bitml contracts, each of which can be executed on one of the involved blockchains, in our running example Bitcoin and Dogecoin. 
Moreover, we compile honest \bitmlx user strategies into a strategies describing what steps to perform at each of the \bitml contracts. 
Next, we overview the challenges and design rationale of our compilation.

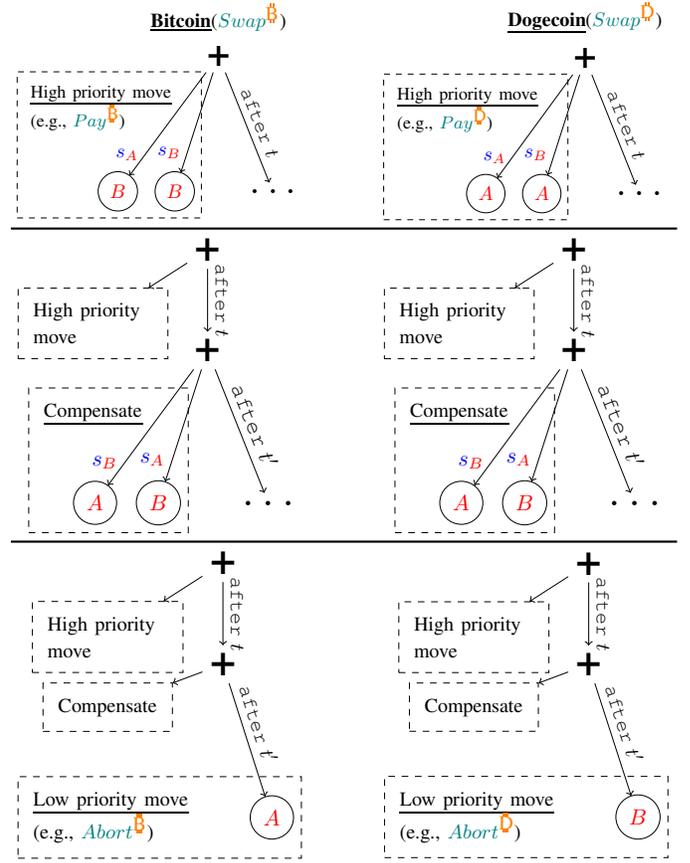
\begin{figure}[tb]
    \begin{minipage}{0.45\columnwidth}
        \resizebox{\textwidth}{!}{
            \begin{tikzpicture}
                \node (btctext) {\underline{\textbf{Bitcoin}}($\contractExchangeBit$)};
                \node [below=0.01cm of btctext] (topchoice)  {\huge \textbf{+}}; 
                \node [below left=0.15cm of topchoice, text width=2.8cm] (hpm) {{\small \underline{High priority move} (e.g., $\contract{Pay^{\money{\bitcoins}}}$)}};
                \node [below=2cm of topchoice] (levelone) {};
                \node [circle, draw, left=0.3cm of levelone] (withdB) {$\participant{B}$};
                \node [circle, draw, left=0.3cm of withdB] (withdA) {$\participant{B}$};
                \node[draw, dashed, fit=(hpm) (withdA) (withdB)] {};
                \node [right=0.3cm of levelone] (nextone) {\huge $\ldots$};
                \draw [->] (topchoice) -- (withdB) node[pos=0.8, left] {$\secret{s}_{\participant{B}}$};
                \draw [->] (topchoice) -- (withdA) node[pos=0.8, left] {$\secret{s}_{\participant{A}}$};
                \draw [->] (topchoice) -- (nextone) node[midway, above,sloped] {$\bitmlcode{after}~t$};
            \end{tikzpicture}
        }
    \end{minipage}
    \hfill
    \begin{minipage}{0.45\columnwidth}
        \resizebox{\textwidth}{!}{
            \begin{tikzpicture}
                \node (btctext) {\underline{\textbf{Dogecoin}}($\contractExchangeDoge$)};
                \node [below=0.01cm of btctext] (topchoice)  {\huge \textbf{+}}; 
                \node [below left=0.15cm of topchoice, text width=2.8cm] (hpm) {{\small \underline{High priority move} (e.g., $\contract{Pay^{\money{\dogecoins}}}$)}};
                \node [below=2cm of topchoice] (levelone) {};
                \node [circle, draw, left=0.3cm of levelone] (withdB) {$\participant{A}$};
                \node [circle, draw, left=0.3cm of withdB] (withdA) {$\participant{A}$};
                \node[draw, dashed, fit=(hpm) (withdA) (withdB)] {};
                \node [right=0.3cm of levelone] (nextone) {\huge $\ldots$};
                \draw [->] (topchoice) -- (withdB) node[pos=0.8, left] {$\secret{s}_{\participant{B}}$};
                \draw [->] (topchoice) -- (withdA) node[pos=0.8, left] {$\secret{s}_{\participant{A}}$};
                \draw [->] (topchoice) -- (nextone) node[midway, above,sloped] {$\bitmlcode{after}~t$};
            \end{tikzpicture}
        }
    \end{minipage}
    \begin{minipage}{\columnwidth}
        \vspace{0.1cm}
        \begin{tikzpicture}
            \draw[thick] (0,0) -- (\columnwidth,0);
        \end{tikzpicture}
    \end{minipage}
    \begin{minipage}{0.45\columnwidth}
        \resizebox{\textwidth}{!}{
            \begin{tikzpicture}
                \node [] (topchoice)  {\huge \textbf{+}}; 
                \node [below left=0.6cm of topchoice, text width=1.9cm] (hpm) {{\small High priority move}};
                \node[draw, dashed, fit=(hpm)] (hpmsquare) {};
                \node [below=1cm of topchoice] (levelone) {\huge \textbf{+}};
                \node [below=2cm of levelone] (leveltwo) {};
                \node [below=0.7cm of hpm] (comp) {{\small \underline{Compensate}}};
                \node [circle, draw, left=0.3cm of leveltwo] (withdB) {$\participant{B}$};
                \node [circle, draw, left=0.3cm of withdB] (withdA) {$\participant{A}$};
                \node[draw, dashed, fit=(comp) (withdB) (withdA)] (compsquare) {};
                \node [right=0.3cm of leveltwo] (nexttwo) {\huge $\ldots$};
                \draw [->] (topchoice) -- (hpmsquare) {};
                \draw [->] (topchoice) -- (levelone) node[midway, above,sloped] {{\small $\bitmlcode{after}~t$}};
                \draw [->] (levelone) -- (withdB) node[pos=0.8, left] {$\secret{s}_{\participant{A}}$};
                \draw [->] (levelone) -- (withdA) node[pos=0.8, left] {$\secret{s}_{\participant{B}}$};
                \draw [->] (levelone) -- (nexttwo) node[midway, above,sloped] {$\bitmlcode{after}~t'$};
            \end{tikzpicture}
        }
    \end{minipage}
    \hfill
    \begin{minipage}{0.45\columnwidth}
        \resizebox{\textwidth}{!}{
            \begin{tikzpicture}
                \node [] (topchoice)  {\huge \textbf{+}}; 
                \node [below left=0.6cm of topchoice, text width=1.9cm] (hpm) {{\small High priority move}};
                \node[draw, dashed, fit=(hpm)] (hpmsquare) {};
                \node [below=1cm of topchoice] (levelone) {\huge \textbf{+}};
                \node [below=2cm of levelone] (leveltwo) {};
                \node [below=0.7cm of hpm] (comp) {{\small \underline{Compensate}}};
                \node [circle, draw, left=0.3cm of leveltwo] (withdB) {$\participant{B}$};
                \node [circle, draw, left=0.3cm of withdB] (withdA) {$\participant{A}$};
                \node[draw, dashed, fit=(comp) (withdB) (withdA)] (compsquare) {};
                \node [right=0.3cm of leveltwo] (nexttwo) {\huge $\ldots$};
                \draw [->] (topchoice) -- (hpmsquare) {};
                \draw [->] (topchoice) -- (levelone) node[midway, above,sloped] {{\small $\bitmlcode{after}~t$}};
                \draw [->] (levelone) -- (withdB) node[pos=0.8, left] {$\secret{s}_{\participant{A}}$};
                \draw [->] (levelone) -- (withdA) node[pos=0.8, left] {$\secret{s}_{\participant{B}}$};
                \draw [->] (levelone) -- (nexttwo) node[midway, above,sloped] {$\bitmlcode{after}~t'$};
            \end{tikzpicture}
        }
    \end{minipage}
    \begin{minipage}{\columnwidth}
        \vspace{0.1cm}
        \begin{tikzpicture}
            \draw[thick] (0,0) -- (\columnwidth,0);
        \end{tikzpicture}
    \end{minipage}
    \begin{minipage}{0.45\columnwidth}
        \resizebox{\textwidth}{!}{
            \begin{tikzpicture}
                \node [] (topchoice)  {\huge \textbf{+}}; 
                \node [below left=0.6cm of topchoice, text width=1.9cm] (hpm) {{\small High priority move}};
                \node[draw, dashed, fit=(hpm)] (hpmsquare) {};
                \node [below=1cm of topchoice] (levelone) {\huge \textbf{+}};
                \node [below=2cm of levelone] (leveltwo) {};
                \node [below=0.4cm of hpm] (comp) {{\small Compensate}};
                \node[draw, dashed, fit=(comp)] (compsquare) {};
                \node [circle, draw, right=0.3cm of leveltwo] (nexttwo) {$\participant{A}$};
                \node [left=of nexttwo, text width=2.3cm] (lpm) {{\small \underline{Low priority move} (e.g., $\contract{Abort^{\money{\bitcoins}}}$)}};
                \node[draw, dashed, fit=(lpm) (nexttwo)] (lpmsquare) {};
                \draw [->] (topchoice) -- (hpmsquare) {};
                \draw [->] (topchoice) -- (levelone) node[midway, above,sloped] {{\small $\bitmlcode{after}~t$}};
                \draw [->] (levelone) -- (compsquare) {};
                \draw [->] (levelone) -- (nexttwo) node[pos=0.4, above,sloped] {{\small $\bitmlcode{after}~t'$}};
            \end{tikzpicture}
        }
    \end{minipage}
    \hfill
    \begin{minipage}{0.45\columnwidth}
        \resizebox{\textwidth}{!}{
            \begin{tikzpicture}
                \node [] (topchoice)  {\huge \textbf{+}}; 
                \node [below left=0.6cm of topchoice, text width=1.9cm] (hpm) {{\small High priority move}};
                \node[draw, dashed, fit=(hpm)] (hpmsquare) {};
                \node [below=1cm of topchoice] (levelone) {\huge \textbf{+}};
                \node [below=2cm of levelone] (leveltwo) {};
                \node [below=0.4cm of hpm] (comp) {{\small Compensate}};
                \node[draw, dashed, fit=(comp)] (compsquare) {};
                \node [circle, draw, right=0.3cm of leveltwo] (nexttwo) {$\participant{B}$};
                \node [left=of nexttwo, text width=2.3cm] (lpm) {{\small \underline{Low priority move} (e.g., $\contract{Abort^{\money{\dogecoins}}}$)}};
                \node[draw, dashed, fit=(lpm) (nexttwo)] (lpmsquare) {};
                \draw [->] (topchoice) -- (hpmsquare) {};
                \draw [->] (topchoice) -- (levelone) node[midway, above,sloped] {{\small $\bitmlcode{after}~t$}};
                \draw [->] (levelone) -- (compsquare) {};
                \draw [->] (levelone) -- (nexttwo) node[pos=0.4, above,sloped] {{\small $\bitmlcode{after}~t'$}};
            \end{tikzpicture}
        }
    \end{minipage}
    \caption{Illustrative example of the compilation. Here, elipses denote withdraw \fundstext by the corresponding participant; dashed boxes represent logical steps in our compilation. The underlined name indicates the active phase. We assume $t' > t$ so that the honest user has enough time to execute \emph{Compensate}. \label{fig:compilation}}
\end{figure}

Intuitively, one can think of 
a priority choice in \bitmlx as a temporal priority between the two \movestext in the sense that there is a time window where only the high priority \movetext can be taken, and after which only the low priority \movetext is viable to the detriment of the high priority \movetext. To realize such a behavior, our compilation first offers the high priority \movetext on both \blockchainstext. 
To execute such a high priority \movetext, an honest user is always required to move the contracts at both \blockchainstext simultaneously.
For instance, in~\cref{fig:compilation}, any of the users can take both, the choice of $\bitmlcode{withdraw}~\participant{B}$ in Bitcoin and the choice of $\bitmlcode{withdraw}~\participant{A}$ in Dogecoin. If one of the users does both \movestext simultaneously, then the contract $\contractExchange$ is faithfully executed in both \blockchainstext and the execution terminates. 
Otherwise, we require a mechanism to handle the situation where the high priority \movetext is done only at one of the chains, as we describe next.  

Instead of enabling an honest user to replicate partial moves of a malicious user to ensure synchronous execution of the two \bitml contracts, we design a compilation that allows honest users to get compensated when observing non-synchronous \movestext. 
For this, it is crucial that users can be held accountable for the \movestext that they make. 
We ensure this in our compilation by preparing individualized \movetext options for each user that require the user to reveal a personalized secret (called step secret) when taking a move. 
A step secret, hence identifies both the user and the \movetext they took. 
For instance, in our running example, $\participant{A}$ must reveal the step-secret $\secret{s}_{\participant{A}}$ to move $\contractExchangeBit$ into $\bitmlcode{withdraw}~\participant{B}$ and $\contractExchangeDoge$ into $\bitmlcode{withdraw}~\participant{A}$. 
Similarly, $\participant{B}$ must reveal the step-secret $\secret{s}_{\participant{B}}$ to do these \movestext instead.

With the step-secrets in place, after the high priority move times out, 
the contract can be moved into a compensation phase 
where an honest user can claim misbehavior of other users. 
Such claim is done by providing the step secret revealed by the misbehaving user that proves that they made the high priority \movetext on the other chain, while they did not on the current chain (since otherwise they would not have entered the compensation phase there). 
For instance,  in~\cref{fig:compilation} assume now that $\participant{B}$ makes the move of $\bitmlcode{withdraw}~\participant{B}$  in $\contractExchangeBit$ but does not make the move of 
$\bitmlcode{withdraw}~\participant{A}$ in $\contractExchangeDoge$ (i.e., a malicious $\participant{B}$ wants to obtain $\money{1\bitcoins}$ without paying $\money{1\dogecoins}$ to $\participant{A}$). 
In such case, after the high priority choice in the contract $\contractExchangeDoge$ times out, 
the honest participant $\participant{A}$ can get compensated by revealing the secret $\secret{s}_{\participant{B}}$ learned from the other chain. 

After the compensation phase, the contract execution moves to a phase where only the lower priority choice is available. 
In our running example, assume now that none of the users decided to synchronously move the contract in both chains to the high priority \movetext, nor needed to get compensated after a partial move from the other party. 
Then, after enough time to execute the compensation has passed, the honest user can synchronously move both contracts to the low priority choice.

Note that, to ensure a safe execution, an honest user needs to obey certain rules, such as only performing synchronous \movestext themselves, and always perform compensations when observing non-synchronous moves. 
To see this, as before consider a malicious Bob that makes the \movetext of $\bitmlcode{withdraw}~\participant{B}$  in $\contractExchangeBit$ but does not make the move of 
$\bitmlcode{withdraw}~\participant{A}$ in $\contractExchangeDoge$. 
The strategy of \participant{A} must be such that it executes the compensation as soon as it gets available. Otherwise, after enough time to execute compensation has passed, Bob successfully gets away with $\money{1\bitcoins}$ without paying $\money{1\dogecoins}$ to $\participant{A}$.

\subsection{Discussion}

\paragraph{Generalization to other contracts} 
To ease the explanation, the running example is limited to a contract for a single priority choice. 
However, the described procedure can be used to compile more complex smart contracts that consists of multiple execution steps (c.f.~\cref{sec:applications}). 
This is possible because, after executing each of the priority choices, the execution of the contract enters into a safe state from the point of view of the honest user. 
To see this, observe that in the simplest case we have that the priority choice was synchronously executed, either with the high priority move or the low priority one. 
For the case where the priority choice was asynchronously executed, assume w.l.o.g. that the high priority choice was taken on chain $\someBlockchain_1$ but not taken on chain $\someBlockchain_2$ (the other case is symmetric). 
In such case, the honest user can enter the compensation phase in $\someBlockchain_2$ and obtain the complete contract balance. Therefore, the honest user obtained the best possible outcome on chain $\someBlockchain_2$. 
In chain $\someBlockchain_1$, the original contract keeps being executed, according to the original contract's logic. 
This will result in an honest user obtaining a valid outcome (according to the original contract) on $\someBlockchain_1$ and the maximal outcome on $\someBlockchain_2$, which leaves them at least as good as in a synchronous execution in both chains. 

\paragraph{Collateral}
In the 2-party case, the compensation ensures that the non-cheating party has a beneficial outcome because 
they obtain all funds locked in the compensated \blockchainstext in the case of cheating.
To generalize this idea to $n$-party contracts, users will initially lock additional collateral that can be used for compensating honest users in case of cheating and released back after honest execution.
Assuming that the \bitmlx contract balance is $b_i$ for the $i$-th chain, each user needs to additionally lock $(n - 2) \cdot b_i$ coins for the $i$-th chain as collateral. 
This suffices to ensure that every honest user can always be compensated in case of misbehavior at each of the chains: the misbehaving user's collateral can be used to compensate $n-2$ users. 
The amount already held by the contract can be used to compensate the remaining honest user. 

\paragraph{Generalization to $k$ chains}
Our approach seamlessly generalizes to smart contracts that control funds in $k > 2$ different chains by allowing honest users to compensate whenever entering a compensation phase in a chain $\someBlockchain_i$ while obtaining the step secret of another user for the respective step. 
The step secret indicates that at least one chain moved to the high priority choice while $\someBlockchain_i$ (having entered the compensation phase) did not. 
Compensating on all chains $\someBlockchain_i$ that entered the compensation phase, safely aborts the execution on these chains (leaving honest users with the most beneficial outcome for them). 
All remaining chains are ensured to have taken the high priority choice synchronously and can resume execution. 

\paragraph{Correctness}
We formally define the semantics of \bitmlx and specify
(i) a contract compiler that translates a \bitmlx contract into $k$ \bitml contracts (running on $k$ different Bitcoin-style blockchains)
and 
(ii) a strategy compiler that transforms a strategy of an honest user interacting with a \bitmlx contract into a strategy for that user to interact with the $k$ \bitml contracts resulting from the contract compilation. 
We provide a generic correctness result for the compilation that shows that an honest user, when executing the compiled \bitml strategy, will always end up in a state that allows them to cash out at least as many funds as they could have obtained from execution of the corresponding \bitmlx strategy (executing the \bitmlx contract).

%% file: bitmlx_language.tex
\section{The \bitmlx Language}
\label{sec:bitml-language}

\newcommand{\setparticipants}{\ensuremath{\mathsf{Part}}}
\newcommand{\sethonest}{\ensuremath{\mathsf{Hon}}}
\newcommand{\depositletter}{\ensuremath{x}}
\newcommand{\depositlettervec}{\ensuremath{\vec{\depositletter}}}
\newcommand{\secretletter}{\ensuremath{s}}

In the following, we assume a set $\setparticipants$ of \emph{participants}, ranged over by $\participant{A}$, $\participant{B}$, $\ldots$. 
Further, we denote by $\sethonest \subseteq \setparticipants$ a non-empty set of \emph{honest} participants. 
We will use the following naming conventions: 
(i) \emph{chains} denoted by  $\someBlockchain[]$; 
(ii) \emph{deposits} denoted by $\depositletter$. We denote by $\depositlettervec$ a finite sequence of deposit names, and similar notation is adopted for sequences of other kinds;
(iii) \emph{secrets} denoted by $\secret{\secretletter}$.   

\subsection{Syntax}

\begin{figure}[tb]
    {\small
    \[
    \begin{array}{l l}
        \preconditions{G} \Coloneqq \\
        \quad \vert ~ \depositsPre{A}{\balance}{\vec{x}} & \text{deposit of $\balance$, expected from $\participant{A}$} \\
        \quad \vert ~ \participant{A}: \bitmlcode{secret} ~ \secret{s} & \text{committed secret by $\participant{A}$}\\
        \quad \vert ~ \preconditions{G} \| \preconditions{G'} & \text{composition} \\
        \preconditions{G} ~ \vert ~ \contractSettings{t_0}{\someStipulation}{} & \text{start time $t_0$ and unique identifier $\uniqueId_0$} \\
        \balance = [v_1\someBlockchain[1],\dots,v_k\someBlockchain[k]] & \text{Balance}
    \end{array}
    \]
    } 
    \caption{Syntax of preconditions of \bitmlx contracts.\label{fig:syntax-preconditions}}
\end{figure}

\paragraph{Contract preconditions} The syntax of contract preconditions is shown in~\cref{fig:syntax-preconditions}. 
Similar to \bitml, $\participant{A}: \bitmlcode{secret} ~ \secret{s}$ requires $\participant{A}$ to commit to a (randomly chosen) secret with name $\secret{s}$.
During the execution of the contract $\contract{C}$, $\participant{A}$ can decide whether to reveal $\secret{s}$ to other participants. 
Additionally, the preconditions of a \bitmlx contract specify the deposits required to stipulate the contract and the secrets to which contract users should commit prior to execution. 
As opposed to \bitml, in \bitmlx the contract deposits of users stem from multiple blockchains, and hence are provided as a vector:  
The precondition $\depositsPre{A}{\balance}{\vec{x}}$ requires $\participant{A}$ to own $v_i = \balance[][\someBlockchain_i]$ coins  in a deposit $x_i$ at chain $\someBlockchain[i]$.
Finally, \bitmlx contracts feature a single precondition of type $\contractSettings{t_0}{\someStipulation}{}$ that specifies the contract execution parameters $t_0$, denoting the contract's starting time and $\someStipulation$, denoting a unique contract identifier. 
Intuitively, \bitmlx contracts require a starting time to safely synchronize the cross-blockchain stipulation of the different \bitml contracts.

\begin{figure}[tb]
    {\small
    \[
    \begin{array}{l l}    
        \contract{C} \Coloneqq \contract{D} \prchoice \contract{C} & \text{top-level contracts}\\
        \quad \vert ~ \bitmlxWithdraw{\balance}{A} & \text{transfer the balances $\vec{\balance}$ to  $\vec{\participant{A}}$}\\
        \contract{D} \Coloneqq \bitmlxWithdraw{\balance}{A} & \text{guarded contracts} \\
        \quad \vert ~ \bitmlxSplit{\balance}{C} & \text{split the balance ($|\vec{\balance}| = |\vec{\contract{C}}|$)}\\
        \quad \vert ~ \bitmlxAuthOp{A}{\contract{D}} & \text{wait for authorization from $\vec{\participant{A}}$}\\
        \quad \vert~ \bitmlxReveal{\secret{s}}{p}{C} & \text{collect secrets $\vec{\secret{s}}$} 
    \end{array}
    \]
    }
    \caption{Syntax of \bitmlx contracts.\label{fig:syntax-bitmlx}}
\end{figure}

\begin{figure}[tb]
    {\small
  \begin{minipage}{0.24\textwidth}
      \begin{align*}
      		\predP 
      		\::= \; &
      		      		\textit{true}
      		&& \text{truth}
      		\\
      		\;\;|\;
      		& \predP \land \predP
      		&& \text{conjunction}
      		\\
      		\;\;|\;
      		& \neg \predP
      		&& \text{negation}
      		\\
      		\;\;|\;
      		& \expE = \expE 
      		&& \text{equality} 
      		\\
      		\;\;|\;
      		& \expE < \expE 
      		&& \text{less than}
      	\end{align*}
    \end{minipage}
    \begin{minipage}{0.24\textwidth}
      \begin{align*}
      		\expE
      		&::= \; 
      		&& \text{contract expression} \\
      		&\quad N
      		&& \text{32-bit constant}
      		\\
      		\;\;|\;
      		& | s |
      		&& \text{length of a secret}
      		\\
      		\;\;|\;
      		& \expE + \expE
      		&& \text{addition} 
      		\\
      		\;\;|\;
      		& \expE - \expE
      		&& \text{subtraction}
      	\end{align*}
  \end{minipage}
   \caption{Syntax of \bitmlx reveal conditions (as defined in~\cite{bitml}).\label{fig:syntax-reveal-conditions}}
    }
\end{figure}

\paragraph{Contracts} The syntax for \bitmlx contracts is given in~\cref{fig:syntax-bitmlx}. 
A contract $\contract{C}$ is given either by a contract $\bitmlxWithdraw{\balance}{A}$  
or a \emph{priority choice} between a guarded contract $\contract{D}$ and the continuation contract $\contract{C'}$. 
A contract $\bitmlxWithdraw{\balance}{A}$ represents the default case of a priority choice that distributes the contract funds to (possibly only part of) the contract users. 
It denotes that each user $\participant{A_i} \in \vec{\participant{A}}$ obtains a balance from the involved blockchains as defined by $\balance_i = [v_{i,1}\someBlockchain[1],\dots,v_{i,k}\someBlockchain[k]]$. 

The left (high priority) branch of a priority choice must be a guarded contract $\contract{D}$.
A guarded contract can be of one of the following forms:
(i) $\bitmlxWithdraw{\balance}{A}$ (same as the default case) distributes the contract funds on the different chains to the contract users
(ii) $\bitmlxSplit{\balance}{C}$ splits the contract balance according to the vector $\vec{\balance} = (\balance_1, \dots, \balance_n)$ and continues executing contracts $\contract{C}_1$, \dots  $\contract{C}_n$ as given by $\vec{\contract{C}}$ with the respective balances $\balance_1$, \dots, $\balance_n$.
(iii) $\bitmlxAuthOp{A}{\contract{D}}$ waits for the authorization of users $\vec{\participant{A}}$
and (iv) $\bitmlxReveal{\secret{s}}{p}{C}$ given the secrets in $\vec{\secret{s}}$ being revealed, evaluates the condition $p$ (potentially referencing those secrets) and continues to execute as $\contract{C}$ if $p$ evaluates to true. 
The syntax of conditions is shown in~\cref{fig:syntax-reveal-conditions}. 

\subsection{Semantics}

\begin{figure}[tb]
    {
        \small
        \begin{align*}
        & \Gamma \Coloneqq 
                                     \activeContract{\contract{C}}{\balance}{\someContract}
            ~\vert ~ \assignedContract{A}{\balance}{\someContract}
                        ~\vert ~  \secretCommitment{A}{s}{N}
            ~\vert ~ \secretReveal{A}{s}{N}
            ~\vert ~ \participant{A}:[\someContract\triangleright \contract{D}]
                                \end{align*}
    }
\caption{\bitmlx configurations for contact execution}
\label{fig:config}
\end{figure}

A configuration $\Gamma$  (cf. \cref{fig:config}) records the state of currently executing \bitmlx contracts as well as contract-relevant (user) interactions.
We define the semantics of \bitmlx as a relation $\Gamma \xrightarrow{\alpha} \Gamma'$ that describes how a configuration $\Gamma$ changes when executing (user) action $\alpha$. 

\begin{figure*}
    \footnotesize
    \begin{mathpar}
        \infer[D-Withdraw]
{
    \reductionRule
        {\activeContract{\contract{D} \prchoice \contract{C}}{\balance}{\someContract} ~ \vert ~ \Gamma}
        {dwithdraw(\someContract)}
        {
            \parallelComposition_{i=1}^{n}
                \assignedContract{A_i}{\balance[i]}{\someContract_i}
            ~|~
            \Gamma
        }
}
{
        \contract{D} = \bitmlxWithdraw{\balance}{A} \\
        \vec{\balance} = \balance[1], \dots, \balance[n] \\
        \vec{\participant{A}} = \participant{A_1}, \dots, \participant{A_n} \\
}
\\
\infer[D-Split]
{
    \reductionRule
        {\activeContract{\contract{D} \prchoice \contract{C}}{\balance}{\someContract} ~ \vert ~ \Gamma}
        {split(\someContract)}
        {
            \parallelComposition_{i=1}^k
                \activeContract{\contract{C_i}}{\balance[i]}{\someContract_i}
            ~|~ \Gamma
        }
}
{
        \contract{D} = \bitmlxSplit{\balance}{C} \\
        \vec{\balance} = \balance[1], \dots, \balance[n] \\
        \vec{\contract{C}} = \contract{C_1}, \dots, \contract{C_k} \\
        \someContract_1, \dots, \someContract_k ~ \text{fresh} \\
                }
\\
\infer[D-Reveal]
{
    \reductionRule
        {\activeContract{\contract{D} \prchoice \contract{C}}{\balance}{\someContract} ~ \vert ~ \Delta ~ \vert ~ \Gamma}
        {reveal(\someContract)}
        {
            \activeContract{\contract{C'}}{\balance}{\someContract'}
            ~ \vert ~ \Gamma
        }
}
{
        \contract{D} = \bitmlxReveal{s}{p}{C'} \\
        \vec{\secret{s}} = \secret{s_1}, \dots, \secret{s_k} \\
        \Delta = \parallelComposition_{i=1}^k \secretReveal{A}{s_i}{N_i} \\
        \llbracket p \rrbracket_\Delta = true \\
        \someContract' ~ \text{fresh} \\
}
\\
\infer[D-Auth]
{
    \reductionRule
        {\activeContract{\bitmlxAuthOp{A}{\contract{D}} \prchoice \contract{C}}{\balance}{\someContract} ~ \vert ~ \Delta ~ \vert ~ \Gamma}
        {\alpha}
        {\Gamma'}
}
{
        \Delta = \parallelComposition_{\participant{A_i} \in \vec{\participant{A}}}
            \participant{A_i} [\someContract\triangleright \contract{D}] \\
        \reductionRule
            {\activeContract{\contract{D} \prchoice \contract{C}}{\balance}{\someContract} \vert \Gamma}
            {\alpha}
            {\Gamma'} \\
}
\\
\infer[C-Withdraw]
{
    \reductionRule
        {\activeContract{\contract{C}}{\balance}{\someContract} ~ \vert ~ \Gamma}
        {cwithdraw(\someContract)}
        {
            \parallelComposition_{i=1}^{n}
                \assignedContract{A_i}{\balance[i]}{\someContract_i}
            ~|~
            \Gamma
        }
}
{
        \contract{C} = \bitmlxWithdraw{\balance}{A} \\
        \vec{\balance} = \balance[1], \dots, \balance[n] \\
        \vec{\participant{A}} = \participant{A_1}, \dots, \participant{A_n} \\
}
\\
\infer[C-Skip]
{
    \reductionRule
        {\activeContract{\contract{D} \prchoice \contract{C}}{\balance}{\someContract} ~ \vert ~ \Gamma}
        {skip(\someContract)}
        {\activeContract{\contract{C}}{\balance}{\someContract'} ~ \vert ~ \Gamma}
}
{
        \someContract' ~ \text{fresh} \\
}
\\
\infer[A-Signature]
{
    \reductionRule
        {\activeContract{\bitmlxAuthOp{A}{\contract{D}} \prchoice \contract{C}}{\balance}{\someContract} ~ \vert ~ \Gamma}
        {\participant{A}: \uniqueId}
        {
            \activeContract{\bitmlxAuthOp{A}{\contract{D}} \prchoice \contract{C}}{\balance}{\someContract}
            ~|~ \participant{A} [\someContract \triangleright \contract{D}]
            ~|~ \Gamma
        }
}
{
        \participant{A} \in \vec{\participant{A}} \\
}
\\
\infer[A-RevealSecret]
{
    \reductionRule
        {\secretCommitment{A}{a}{N} ~ \vert ~ \Gamma}
        {\participant A:\secret a}
        {\secretReveal{A}{a}{N} ~ \vert ~ \Gamma}
}
{
        N \neq \bot
}
    \end{mathpar}
    \caption{\bitmlx semantics for active contracts}
    \label{fig:semantics}
\end{figure*}

\newcommand{\xAdvStrategy}{\xStrategy[\textsf{Adv}]}
\newcommand{\mempool}{A}
To initiate the execution of a $\bitmlx$ contract, users can advertise a contract, commit to the secrets prescribed in $\preconditions{G}$ and authorize its stipulation. 
A stipulation results in an active contract $\activeContract{\contract{C}}{\balance}{\someContract}$, which evolves according to the semantic rules provided in~\cref{fig:semantics}.
Contracts of the form $\contract{D} \prchoice \contract{C}$ either execute  according to the high-priority choice $\contract{D}$ 
(rules \textit{D-Withdraw}, \textit{D-Split}, \textit{D-Reveal}, and \textit{D-Auth}) 
or discard the high-priority choice executing the \textit{skip} action (rule \textit{C-Skip}).
The \textit{D-Withdraw} rule splits the balance $\balance$ of a contract
 $\activeContract{\bitmlxWithdraw{\balance}{A} \prchoice \contract{C}}{\balance}{\someContract}$ 
 according to the vector 
$\vec{\balance}$ and assigns these balances to the users in $\vec{\participant{A}}$ resulting in terminal contracts of the form $\assignedContract{A_i}{\balance[i]}{\someContract_i}$ that assign the balances to the corresponding users $\participant{A_i}$. 
The \textit{D-Split} rule proceeds similarly but transfers the balances to subcontracts $\activeContract{\contract{C_i}}{\balance[i]}{\someContract_i}$ as specified by a vector $\vec{\contract{C}}$. 
The \textit{D-Reveal} rule allows for evolving contract $\activeContract{\bitmlxReveal{s}{p}{C'} \prchoice \contract{C}}{\balance}{\someContract}$ into a contract $\activeContract{\contract{C'}}{\balance}{\someContract'}$ provided that the secrets in $\vec{\secret{s}}$ have been revealed (indicated by elements $\secretReveal{A_i}{s_i}{N_i}$ in the configuration) and that the condition $p$ evaluates to true ($\llbracket p \rrbracket_\Delta = \textit{true}$). 
Users can reveal secrets that they committed to previously any time (cf. rule \textit{RevealSecret}). 
Revealing a secret, in particular, reveals the secret's length $N_i$ which can be referenced by the condition $p$ (cg. \cref{fig:syntax-reveal-conditions}). 
Correspondingly, the condition $p$ is evaluated in an environment that resolves this reference for revealed secrets.
Finally, rule \textit{D-Auth} allows for evolving contracts $\activeContract{\bitmlxAuthOp{A}{\contract{D}} \prchoice \contract{C}}{\balance}{\someContract}$ whose high priority branch requires authorization from the users in $\vec{\participant{A}}$. 
Such authorizations are indicated by configuration elements of the form $\participant{A_i}:[\someContract\triangleright \contract{D}]$, which can be added by the corresponding user (cf. rule \textit{A-Signature}).
If such authorizations are provided by all users, $\activeContract{\bitmlxAuthOp{A}{\contract{D}} \prchoice \contract{C}}{\balance}{\someContract}$ evolves the same as $\activeContract{\contract{D} \prchoice \contract{C}}{\balance}{\someContract}$.

The only contract that can occur standalone (outside a priority choice) is of the form $\activeContract{\bitmlxWithdraw{\balance}{A}}{\balance}{\someContract}$ and evolves exactly as it would do when appearing as the high-priority branch of a priority choice (cf. rule \textit{C-Withdraw}).

\paragraph{Strategies and execution model}
The \bitmlx semantics specifies possible execution steps of \bitmlx smart contracts. 
However, many of these execution steps are restricted to be performed by specific users $\participant{A}$ (as indicated by labels of the form $\participant{A}: \alpha$).
To characterize valid \bitmlx smart contract interactions, we hence consider the \bitmlx semantics in conjunction with the strategies of the different contract users (as it is also done for \bitml). 

Formally, we represent a \bitmlx strategy of user $\participant{A}$ as a function $\xStrategy[\participant{A}]$ that takes as input a run $\xRun$ of the form $\Gamma_0 \xrightarrow{\alpha_0} \dots \xrightarrow{\alpha_{n-1}} \Gamma_n$ and outputs a set of actions $\mempool = \xStrategy[\participant{A}](\xRun)$.
For $\xStrategy[\participant{A}]$ to be valid it needs to hold that 
(i) if $\alpha \in \xStrategy[\participant{A}](\xRun)$ then there exists $\Gamma$ such that $\Gamma_n \xrightarrow{\alpha} \Gamma$ (so $\alpha$ is a valid move w.r.t the semantics)
(ii) if $\participant{B}: \alpha \in \xStrategy[\participant{A}](\xRun)$ then $\participant{B} = \participant{A}$ (so $\xStrategy[\participant{A}]$ only outputs actions that $\participant{A}$ is entitled to perform)
(iii) if $\alpha \in \xStrategy[\participant{A}](\xRun)$ and $\hat{\xRun} = \xRun \xrightarrow{\hat{\alpha}} \hat{\Gamma}$ 
is a valid extension of $\xRun$ then also $\alpha \in \xStrategy[\participant{A}](\hat{\xRun})$.

The last restriction captures that $\xStrategy[\participant{A}]$ is \emph{persistent}, meaning that users cannot take back actions that they wanted to execute at some point. 
This requirement becomes relevant when considering that actions output by $\xStrategy[\participant{A}]$ are not executed instantaneously but are handed to a malicious scheduler (e.g., a miner) who decides on the execution order.

More precisely, we assume the existence of an adversarial strategy $\xAdvStrategy$ that models the behavior of malicious protocol participants and the malicious scheduler. 
Such a strategy $\xAdvStrategy$, in addition to the run $\xRun$, gets as input the output $\mempool = \xStrategy[\participant{A}](\xRun)$ of the honest user strategy and based on that outputs the next action $\alpha$ to append to $\xRun$. 
$\xAdvStrategy$, similar to honest user strategies, is limited to only output actions $\alpha = \xAdvStrategy(\xRun, \mempool)$ that are valid extensions of $\xRun$ according to the \bitmlx semantics.
Further, $\xAdvStrategy$ may not schedule any actions $\participant{A}:\alpha$ unless those are included in $\xStrategy[\participant{A}](\xRun)$.
The same holds for actions $\alpha = skip(\someContract)$, reflecting that the honest user needs to agree to skipping a high priority branch of a priority choice contract.
We say that a run $\xRun = \Gamma_0 \xrightarrow{\alpha_0} \Gamma_1 \xrightarrow{\alpha_1} \dots \xrightarrow{\alpha_{n-1}} \Gamma_n$
conforms to $\xStrategy[\participant{A}]$ (written $\xStrategy[\participant{A}] \vdash \xRun$)
if there exists some adversarial strategy $\xAdvStrategy$ such that $\xRun$ results from the interactions of $\xStrategy[\participant{A}]$ with $\xAdvStrategy$, meaning that 
 $\alpha_i = \xAdvStrategy(\xRun[i], {\mempool_i})$ 
 for $\xRun[i] = \Gamma_0 \xrightarrow{\alpha_0} \dots \xrightarrow{\alpha_{i-1}} \Gamma_i$ and $\mempool_i =  \xStrategy[\participant{A}](\xRun[i])$ for all $i = 0, \dots, n-1$.

A full specification of the \bitmlx semantics can be found in Appendix~\ref{sec:bitmlx-semantics}.

%% file: compilation.tex
\section{Compilaton}
\label{sec:compilation}

In this section, we describe how to compile a \bitmlx contract $\contract{C}$ operating across $k$ blockchains into $k$ individual \bitml contracts $\contract{C}_1$, \dots $\contract{C}_k$ running on these chains. 
Further, we show how to translate an honest user strategy $\xStrategy[\participant{A}]$ for a participant $\participant{A}$ of $\contract{C}$ into a \bitml strategy enforcing the synchronous execution of $\contract{C}_1$, \dots $\contract{C}_k$ that corresponds to an execution of $\contract{C}$ conforming to $\xStrategy[\participant{A}]$.

\subsection{Compilation of contracts}
To facilitate compilation, we define a compiler state $\settings{\Omega}$ that carries relevant compilation information. 
The compiler state $\settings{\Omega}$ gets instantiated based on the information from the contract advertisement $\contractAdv{G}{C}$ (e.g., with the contract's starting time $t_0$).
During the compilation, $\settings{\Omega}$ gets updated reflecting that certain compilation parameters depend on the position of a compiled subcontract within the original contract's syntax tree. 
This applies in particular to the timelocks used to implement the priority choices, which increase with each tree level.

\paragraph{Priority choices} Remember that a \bitmlx priority choice is of the form $\contract{D} \prchoice \contract{C'}$. 
The compiler translates such a contract into two clauses: 
(i) one clause corresponding to the high priority contract $\contract{D}$ and 
(ii) one clause implementing the compensation phase followed by the execution of low-priority contract $\contract{C'}$. 

To ensure that an honest user can not be hindered from taking clause (i), clause (ii) is guarded by a timelock $t$. 
The timelock makes clause (ii) only available starting from time $t$, giving the honest user sufficient time to claim clause (i) before. 
The time $t$ is accessed from the compiler state $\settings{\Omega}$ where it increases with each level that the compilation descends into the syntax tree, ensuring that the timelocks increase with increasing depth. 

Following the same idea, to ensure a that an honest user has sufficient time to compensate a malicious user, clause (ii) consists of a choice between a compensation contract and a continuation contract guarded with timelock $t + \secDelay$. The parameter $\secDelay > 0$ is a system parameter, intuitively denoting the maximal time that a malicious scheduler can delay the execution of an honest user action.

E.g., our running example 
$\contract{C} \Coloneqq \contract{Pay^x} \prchoice \contract{Refund^x}$
is translated into 

{\footnotesize 
\[
    \begin{split}
 \guardedCompiler{\someBlockchain}(\contract{Pay^x}, \settings{\Omega})& ~ + ~\bitmlcode{after} ~t : (\\
 &\punishCompiler{\someBlockchain}(\settings{\Omega})~+~ \bitmlcode{after} ~(t + \secDelay): \topLevelCompiler{}(\contract{Refund^x}, \settings{\Omega}) )
    \end{split}
\]
}

Here, $\guardedCompiler{\someBlockchain}$ denotes the compilation of a \bitmlx guarded contract into a \bitml contract for chain $\someBlockchain$; $\punishCompiler{\someBlockchain}$ denotes the compilation of the compensation mechanism into \bitml for chain $\someBlockchain$; and $\topLevelCompiler{}$ denotes the compilation of the continuation contract into a \bitml contract for chain $\someBlockchain$.

\paragraph{Compiling \bitmlx guarded contracts} 
The compilation of a \bitmlx guarded contract $\contract{D}$ into a \bitml contract depends on the concrete form of $\contract{D}$.
We illustrate the compilation using the example of a \bitmlcode{withdraw} contract and refer to Appendix~\ref{sec:app-compiler} for a detailed description of the other cases.

A \bitmlx withdraw contract is compiled into a \bitml contract for target chain $\someBlockchain$ according to the following rule: 

{\footnotesize 
\[
\infer[]{
    \guardedCompiler{\someBlockchain}(\contract{D}, \settings{\Omega}) =\sum_{\stepSecret{A}{\someContract} \in S}
            \bitmlcode{reveal} ~ \stepSecret{A}{\someContract} ~. ~
                \bitmlcode{split}
                ~ \vec{\mu}
                \rightarrow \vec{\contract{C^{\someBlockchain}}}
}
{
    \begin{gathered}
        \contract{D} = \bitmlxWithdraw{\balance}{A} \\
        S = \{ \stepSecret{A}{\someContract} \in \settings{\Omega}.stepSecrets : \someContract = \settings{\Omega}.currentLabel\} \\
        \vec{\mu} = \outCompiler{\someBlockchain}(\vec{\participant{A}}, \vec{\balance}, \settings{\Omega}) \\
        \vec{\contract{C^{\someBlockchain}}} = [
            \bitmlcode{withdraw} ~ \participant{A_i}
            ~ | ~ \participant{A_i} \leftarrow \settings{\Omega}.participants
        ]\\
    \end{gathered}
}
\]
}

On the high level, the compilation consists of a choice guarded by a step secret $\stepSecret{A}{\someContract}$ for each contract user $\participant{A}$.
To this end, at the beginning of the compilation,  $\settings{\Omega}$ got initialized with a set of step secrets ($\settings{\Omega}.stepSecrets$) for each contract user $\participant{A}$ and position $\someContract$ in the original's contract's syntax tree.
The current position $\someContract$ of the compiled priority choice can be accessed from $\settings{\Omega}$ (by $\settings{\Omega}.currentLabel$) and is updated as the compiler traverses the syntax tree. 
The $\bitmlcode{split}$ distributes the balance of the contract as computed by $\outCompiler{\someBlockchain}(\vec{\participant{A}}, \vec{\balance}, \settings{\Omega})$, which namely (1) assigns the collateral back to each participant (accessed by $\settings{\Omega}.participants$); 
and (2) additionally assigns part of the contract's balance to each user as encoded in $\vec{\balance}$ of the \bitmlx $\bitmlcode{withdraw}$ statement.  

As an illustrative example, the output of $\guardedCompiler{\someBlockchain}(\contract{Pay^x}, \settings{\Omega})$ for the chain $\bitcoinBlockchain$ is as shown below. Here, $s^{\participant{A}}_{\someContract}$ is the step secret for $\participant{A}$ for choosing the high-priority choice where $\someContract = [\someStipulation, L]$, encoding the first left ($L$) choice of a contract with identifier $\someStipulation$. 
Similarly, $s^{\participant{B}}_{\someContract}$ is the step secret for $\participant{B}$ and label $\someContract$. 

{\footnotesize
\begin{equation*}
    \begin{split}
        \bitmlcode{reveal}~s^{\participant{A}}_{\someContract} . &\bitmlcode{split} [1, \bitmlcode{withdraw}~\participant{B} ]\\
        &+ \\
        \bitmlcode{reveal}~s^{\participant{B}}_{\someContract} . &\bitmlcode{split} [1, \bitmlcode{withdraw}~\participant{B} ]
    \end{split}
\end{equation*}
}

The compilation of $\bitmlcode{split}$, $\bitmlcode{reveal}$, and authorization contracts follows the same pattern.

\paragraph{Compiling the compensation mechanism}
The compensation mechanism is designed to compensate honest users $\vec{\participant{A_i}}$ when a dishonest participant $\participant{A_j} \not 
\in \vec{\participant{A_i}}$ did not do the high-priority move in all chains. 
In such case, the misbehaving user must have revealed their corresponding step secret $s^{\participant{A_j}}_\someContract$ to execute the move in some chains whereas in the other chains, honest users can enter the compensation phase at time $t + \secDelay$. 
At this point, the compilation ensures the existence of the following compensation contract: 

{\footnotesize
\[ 
    \sum_{i} \bitmlcode{reveal}~s^{\participant{A_j}}_\someContract ~.~(\bitmlcode{split}~\vec{v_i} \rightarrow \vec{\contract{C_i}})
\]
}

Each clause in the sum requires to reveal the step secret $s^{\participant{A_j}}_\someContract$ corresponding to the misbehaving user $\participant{A_j}$. Then, the total balance of the contract (i.e., $\vec{v_i}$) is equally split among the honest participants. This is encoded in $\contract{C_i} \Coloneqq [\bitmlcode{withdraw}~\vec{\participant{A_i}}]$. 
Remember that, at the stipulation of the contract, every user must not only include their deposits 
but also a collateral of $(n-2) \cdot b_0$ (where $b_0$ denotes the initial contract balance). 
Therefore, a contract, at any point of execution, is ensured to effectively hold balance $n * (n-2) * b + b$ (where $b$ is the contract's logical balance).
This ensures, in particular, that when entering the compensation contract, $(n-1)$ users (all users but the misbehaving one) can retrieve $(n-2) *b + b$ corresponding to a refund of the provided collateral for that subcontract as well as the maximal payout $b$ for the subcontract.

As an illustrative example, the output of $\punishCompiler{\someBlockchain}(\settings{\Omega})$ in our running example for target chain $\bitcoinBlockchain$ is as shown below. 
$\participant{A}$ gets the contract balance if  $\participant{B}$'s step secret was revealed in another chain. 
The other choice is symmetric.

{\footnotesize
\begin{equation*}
    \begin{split}
        \bitmlcode{reveal}~s^{\participant{A}}_{\someContract} . &\bitmlcode{split} [1, \bitmlcode{withdraw}~\participant{B} ]\\
        &+ \\
        \bitmlcode{reveal}~s^{\participant{B}}_{\someContract} . &\bitmlcode{split} [1, \bitmlcode{withdraw}~\participant{A} ]
    \end{split}
\end{equation*}
}
\subsection{Compilation of strategies}
To mimic the behavior of a \bitmlx strategy $\xStrategy[\participant{A}]$, a corresponding low-level strategy 
$\orStrategy[\participant{A}]$, operating on the compiled \bitml contract on different chains, needs to fulfill two functions:
(1) (synchronously) perform moves scheduled by $\xStrategy[\participant{A}]$
(2) maintain a synchronous execution state by detecting non-synchronous moves of other users and safely aborting execution on chains that deviate from the synchronous execution.

For $\orStrategy[\participant{A}]$ to provide the first function, we need to map a \bitml run $R$ to its corresponding \bitmlx run $\xRun$. 
To this end, we will introduce a coherence relation $\xCoherence[\participant{A}]$, relating a \bitml run $R$ and a \bitmlx run $\xRun$, such that intuitively, 
$\xRun \xCoherence[\participant{A}] R$ denotes that from the perspective of user $\participant{A}$, $R$ represents an execution corresponding to \bitmlx run $\xRun$. 

\paragraph{Intermediate semantics} 
As a first step towards defining coherence, 
we establish a direct link between \bitmlx and \bitml executions
by defining an intermediate semantics $\isConfig[] \xrightarrow{\alpha} \hat{\isConfig[]}$ that describes the concurrent execution of compiled contracts $\contract{C}_1$, \dots $\contract{C}_k$ executing on chains $\someBlockchain_1, \dots \someBlockchain_k$ in terms of the original \bitmlx contract $\contract{C}$. 
The intermediate semantics has a direct correspondence to the \bitml semantics, more precisely, there are functions $\isconfigcompile[]{}$ and $\isactioncompile{}$ such that 
\begin{align*}
    \small
    \isConfig[0] \xrightarrow{\alpha_0} \dots  \xrightarrow{\alpha_{n-1}} \isConfig[n]
     \Leftrightarrow
     \isconfigcompile[]{\orConfig[0]}  \xrightarrow{\isactioncompile{\alpha_0}} \dots \xrightarrow{\isactioncompile{\alpha_{n-1}}} \isconfigcompile[]{\orConfig[n]}
\end{align*}
Elements in an intermediate configuration $\isConfig$ are of the form $\isActiveContract{\contract{C}}{\isContractState}{\someBlockchain}{\someContract, x}$, indicating that they represent the execution of a \bitml contract corresponding to the compilation of \bitmlx contract $\contract{C}$ (with \bitmlx identifier $\someContract$) in blockchain $\someBlockchain$ with state $\isContractState$. 
The identifier $x$ establishes a direct link to the active \bitml contract matching $\isActiveContract{\contract{C}}{\isContractState}{\someBlockchain}{\someContract, x}$ on the \bitml level. 
The state $\isContractState$ indicates the contract's execution phase (e.g., whether $\contract{C}$ is idle, its compensation phase has been entered, or it got compensated).
Due to the direct correspondence between the \bitml and the intermediate semantics, we will, henceforth, state all results in terms of the intermediate semantics, using $\isRun$ to denote runs in the intermediate semantics and $\isStrategy[\participant{A}]$ to denote the honest user strategy based on the intermediate semantics. 
A full specification of the intermediate semantics can be found in Appendix~\ref{sec:app-intermediate-semantics}.  

\paragraph{Coherence}
Defining relation $\xCoherence[\participant{A}]$ is challenging because there exist many possible runs $\isRun$ that represent (potentially incomplete) executions of a run $\xRun$: 
Users may take the high-priority branches of priority choices in a some chain $\someBlockchain_i$ and in this way, run ahead with the execution of the \bitml contract on $\someBlockchain_i$ without (immediately) matching these moves on the other blockchains.

In such cases, it is still ensured that, eventually, the moves taken by the contract in the most advanced chain $\someBlockchain_i$ can be matched because, either the execution on all other chains $\someBlockchain_j$ will advance to the same state, or the execution on $\someBlockchain_j$ will enter the compensation phase and hence the execution on $\someBlockchain_j$ can be safely aborted. 

For this reason, $\xCoherence[\participant{A}]$ will associate a run $\isRun$ with a \bitmlx run $\xRun$ reflecting the "most-advanced" executions of compiled contracts $\contract{C}_1$, \dots $\contract{C}_k$ in $\isRun$.
Since those contracts, during execution, can be split in independently executing subcontracts, it is not guaranteed that there is a single blockchain $\someBlockchain_i$ featuring the most advanced executions of all those subcontracts, but different chains could feature the "most-advanced" executions of different subcontracts. 

Taking this into account, we define the coherence relation $\xCoherence[\participant{A}]$ as follows: 

\begin{definition}[Coherence, simplified]
    \footnotesize
\begin{align*}
& \xRun \xCoherence \isRun ::= \\
& \quad \maxFrontierFun{\xRun} = \bigjoin_{\someBlockchain \in \activeBlockchains}{\isMaxFrontierFun{\isRun}{\someBlockchain}} \\
&\quad\land \forall \someContract \in \maxFrontierFun{\xRun}: 
\contractAuths{\someContract}{\xRun} = \finishedContractAuths{A}{\someContract}{\isRun} \\ 
& \qquad \qquad \qquad \qquad \qquad \qquad \qquad \qquad \qquad \cup (\contractAuths{\someContract}{\isRun} \cap \{ \participant{A }\})
\end{align*}
\end{definition}
Here, $\maxFrontierFun{\xRun}$ denotes the \bitmlx contracts (represented by their identifiers $\kappa$) in the last configuration of run $\xRun$. 
Similarly, $\isMaxFrontierFun{\isRun}{\someBlockchain}$ denotes the \bitmlx contracts, for which configuration elements $\isActiveContract{\contract{C}}{\isContractState}{\someBlockchain}{\someContract, x}$ are available in the last configuration of the intermediate run $\isRun$.
So, intuitively, $\isMaxFrontierFun{\isRun}{\someBlockchain}$ denotes how far the \bitml execution of the compiled contract has advanced on chain $\someBlockchain$. 
Since $\isMaxFrontierFun{\isRun}{\someBlockchain_i}$ may differ across different chains $\someBlockchain_i$, we take the supremum over $\isMaxFrontierFun{\isRun}{\someBlockchain_i}$ for all chains to obtain a representation of the most advanced state of all subcontracts across all chains.
For $\xRun$ to be coherent with $\isRun$, it should match this most advanced state.
The second requirement reflects that the authorizations (for guarded contracts $\someContract$) provided on runs $\xRun$ and $\isRun$ should match in that 
the users that authorized $\someContract$ in $\xRun$ (denoted by $\contractAuths{\someContract}{\xRun}$) coincide with the users that authorized $\someContract$ on all (not yet compensated) blockchains in $\isRun$ (denoted by $\finishedContractAuths{A}{\someContract}{\isRun}$). 
Here, an exception is made for authorizations of the honest user $\participant{A}$ who may have only authorized $\someContract$ in a single chain in $\isRun$ (in this case $\contractAuths{\someContract}{\isRun} \cap \{ \participant{A }\} \neq \emptyset$).
The reasoning behind that is that honest users can safely replicate their authorizations on all chains, and hence already an authorization on a single chain serves as an evidence that the honest user will synchronously perform such authorizations in the future.

\paragraph{Intermediate strategies}
Using coherence, we can formally define how to obtain an intermediate strategy $\isStrategy[\participant{A}] = \xCompiledStrategy$ from a corresponding \bitmlx strategy $\xStrategy[\participant{A}]$.
We illustrate the main idea behind the strategy compilation using some exemplary compilation rules: 
\begin{mathpar}
    \small
\inferrule[]
{
        \xRun \xCoherence \isRun | t \\
    \alpha(\someContract) \in \xStrategy(\xRun) \\
    \alpha \in \{dwithdraw, split, reveal\} \\
    \isActiveContract{\contract{C}}{\isContractState}{\someBlockchain}{\someContract} \in \lastConfigOf{\isRun} \\
    \isContractState.status = \isStatusChoice \\
    \someUser \notin \userStepSecrets{\someContract}{\isRun} \\
    t < \isContractState.time
}
{
    (\participant{A}: \stepSecret{A}{\someContract}) \in \isStrategy(\isRun) 
}
\end{mathpar}

This rule captures the case where strategy $\xStrategy$ schedules a (high-priority) move $\alpha(\someContract)$ on a contract $\someContract$ in $\xRun$. 
Given a coherent intermediate run $\isRun$ ($\xRun \xCoherence \isRun$), $\isStrategy$ will mimic the move by revealing step secret $\stepSecret{A}{\someContract}$. 
Revealing the step secret initiates the high-priority move of $\participant{A}$ for contract $\someContract$. 
Once the step secret is revealed, the strategy $\isStrategy$ will continue scheduling the corresponding $\someContract$ moves synchronously on all blockchains. 
By requiring that the current execution time (denoted by $t$) of the intermediate run $\isRun$ is smaller than the timeout of contract $\someContract$ (as captured in the $\isContractState.time$ field), \participant{A} can be sure to have sufficient time to perform the consecutive $\someContract$ moves on all chains. 
Note that the requirement $\isContractState.status = \isStatusChoice$ ensures that $\contract{C}$ is in an idle state (e.g., has not yet been moved by another user).

Most other rules defining $\isStrategy$ are independent of $\xStrategy$, since they are concerned with maintaining a synchronized execution state. 
E.g., the following rule mandates $\isStrategy$ to schedule a \textit{ileft} move (corresponding to perform the top-level \bitmlcode{reveal} step in the compiled contract with the step secret) on all blockchains $\someBlockchain$ whenever $\participant{A}$'s step secret has been revealed (indicated by $\someUser \in \userStepSecrets{\someContract}{\isRun}$ ):
\begin{mathpar}
    \small
\inferrule
{
        \someUser \in \userStepSecrets{\someContract}{\isRun} \\
    \isRun \xrightarrow{ileft(\someContract, \someBlockchain)} \\
}
{
    ileft(\someContract, \someBlockchain) \in \isStrategy(\isRun)
}
\end{mathpar}

Similarly, $\isStrategy$ will eagerly perfom \textit{right} moves (corresponding to entering a contract $\someContract$'s compensation phase) whenever possible:
\begin{mathpar}
    \small
    \inferrule
    {
                                        \isRun \xrightarrow{right(\someContract, \someBlockchain)} \\
            }
    {
        right(\someContract, \someBlockchain)  \in \isStrategy(\isRun)
    }
\end{mathpar}
A full specification of $\isStrategy$ can be found in Appendix~\ref{sec:app-strategies}.

%% file: compiler-correctness.tex
\section{Compiler Correctness}
\label{sec:proof}

In this section, we sketch the final compiler correctness result and its proof.  

\paragraph{Soundness}
The core of our compiler correctness proof consists of establishing a general soundness result, essentially stating that for every intermediate run $\isRun$ resulting from an execution with $\isStrategy = \xCompiledStrategy$, we can find a coherent run $\xRun$ resulting from an execution with $\xStrategy$. 

This result ensures that the contracts in $\isRun$ execute synchronously, in the sense that they do not \emph{diverge} (taking the high-priority choice in one chain $\someBlockchain_i$, while taking the low-level branch in another chain $\someBlockchain_j$). 
Such diverging contract executions could never be matched by a single coherent $\xRun$.

For stating soundness, we need to consider that given that $\xStrategy$ does not aim to schedule a high-priority move, $\isStrategy = \xCompiledStrategy$ eagerly proceeds to the compensation phase and, if no compensation is possible, to the low-priority move. 
We reflect this behavior as a requirement on the \bitmlx strategy to be \emph{eager}, meaning that whenever a move on contract $\someContract$ in a run $\xRun$ is possible, then $\xStrategy$ should schedule a move related to $\someContract$ (e.g., performing an authorization, revealing a secret required by the contract or perform a move $\alpha(\kappa)$). 

\begin{lemma}[Soundness, simplified]
    Let $\theHonestUser$ be an honest user with an eager $\bitmlx$ strategy $\xStrategy$ and an intermediate semantics strategy $\isStrategy = \xCompiledStrategy$. Then
    \[
        \begin{aligned}        
        &\forall \isRun ~s.t.~ \isStrategy \models \isRun, \, \exists \xRun:
            \xStrategy \xConforms \xRun 
            \land \xRun \xCoherence[\theHonestUser] \isRun \\
                                                                    \end{aligned}
    \]
\end{lemma}
Proving soundness requires establishing a multitude of invariants on the configurations of $\isRun$. 
In particular, we need to show that as time progresses, the execution of subcontracts whose timeouts are reached are finalized, meaning that they must have advanced synchronously on all chains $\someBlockchain_i$, or they got compensated on all blockchains $\someBlockchain_j$ that did not match eventual high-priority moves before reaching the subcontract-specific timeout. 
The full proof can be found in Appendix~\ref{sec:app-bitmlx-soundness}.

\paragraph{Liquidity}
Soundness ensures the existence of a coherent run $\xRun$ for every intermediate run $\isRun$. 
However, such run $\xRun$ with $\xRun \xCoherence \isRun$ intuitively only constitutes an execution state, which $\isRun$ should be guaranteed to match \emph{eventually}. 
Further, for our final correctness result, we also want to show that the compensation mechanism indeed makes sure that an honest user will always receive adequate compensation.
To this end, we want to compare the amount of money that an honest user $\participant{A}$ obtained (on all blockchains) from running a compiled contract using $\isStrategy = \xCompiledStrategy$ with the amount of money $\participant{A}$ would have obtained from running $\xStrategy$. 

We make use of the fact that we can show $\xStrategy$ and $\isStrategy$ to be \emph{liquid}, meaning that these strategies will be able to cash out any \bitmlx or compiled contract, respectively. 
Intuitively, this is because the structure of \bitmlx contracts ensures that the contract execution cannot get stuck (given an eager honest user strategy): If a malicious user blocks a high-priority choice, the contract can always (unconditionally) advance to the low-priority branch until reaching the default case, which is an unconditional \bitmlcode{withdraw} of funds. 
This structure is reflected in the compiled contracts. 

More formally, a \bitmlx strategy is considered \emph{liquid}, if it can complete any run $\xRun$ with $\xStrategy \xConforms \xRun$ such that its final configuration does not contain any active contracts anymore. 
We call $\xRun$ \emph{liquidated} in this case. 
A similar notion can be established for intermediate strategies $\isStrategy$. 
To show the liquidity of a \bitmlx strategy $\xStrategy$ (or an intermediate strategy, respectively), we need to ensure that $\xStrategy$ does not keep on advertising an infinite amount of new contracts, hence hindering the termination of runs. 
We formally capture this requirement by requiring the contracts advertised by $\xStrategy$ to be bounded by a finite set of contract identifiers $\contractbound$.

With this, we can prove the following result:

\begin{lemma}[Intermediate Security]
    Let $\participant{A}$ be an honest participant and $\contractbound$ be a set of contract identifiers. 
    Let $\xStrategy[\participant{A}]$ be an eager strategy of $\participant{A}$ bounded by $\contractbound$ 
    and $\isStrategy[\participant{A}] = \xCompiledStrategy$ its corresponding intermediate strategy. 
    Then for every run $\isRun$ with $\isStrategy[\participant{A}] \vDash \isRun$
    there exists an extension $\hat{\isRun} = \isRun \xrightarrow[]{\alpha_1} \cdots \xrightarrow[]{\alpha_n}$ such that 
    \begin{enumerate}
        \item $\forall i \in 1 \dots n : ~ \alpha_i \in \isStrategy[\participant{A}](\isRun \xrightarrow[]{\alpha_1} \cdots \xrightarrow[]{\alpha_{i-1}})$
        \item $\hat\isRun$ is liquidated 
    \end{enumerate}
    and there exists $\hat{\xRun}$ such that $\xStrategy[\participant{A}] \vdash \hat{\xRun}$, $\hat{\xRun}$ is liquidated and 
    $$ \footnotesize 
        \forall \someBlockchain \in \activeBlockchains: \isUserPayout{A}{\someBlockchain}{\hat{\isRun}}{} -\isUserInputs{\participant{A}}{\someBlockchain}{\hat{\isRun}} 
        \geq 
        \xUserPayout{A}{\someBlockchain}{\hat{\xRun}}{} - \xUserInputs{\participant{A}}{\someBlockchain}{\hat{\xRun}} 
    $$
    \end{lemma}

This lemma states that the compiled user strategy $\isStrategy[\participant{A}] = \xCompiledStrategy$ can complete any run $\isRun$ with $\isStrategy[\participant{A}] \vDash \isRun$ to a liquidated run $\hat{\isRun}$ such that user $\participant{A}$ on all blockchains $\someBlockchain$ earns at least as much as they would earn in a corresponding (liquidated) run $\hat{\xRun}$  with $\xStrategy[\participant{A}] \vdash \hat{\xRun}$.
Here, $\isUserPayout{A}{\someBlockchain}{\hat{\isRun}}{}$ denotes the amount of money that gets assigned to $\participant{A}$ on chain $\someBlockchain$ during the run $\hat{\isRun}$ (due to regular payouts or compensations) and $\isUserInputs{\participant{A}}{\someBlockchain}{\hat{\isRun}}$ denotes the inputs of $\participant{A}$ on $\someBlockchain$ locked into contracts during $\hat{\isRun}$.
Similarly, $\isUserPayout{A}{\someBlockchain}{\isRun}{} -\isUserInputs{\participant{A}}{\someBlockchain}{\hat{\isRun}}$ denotes the money assigned to $\participant{A}$ on $\someBlockchain$ during $\hat{\xRun}$ and $\xUserInputs{\participant{A}}{\someBlockchain}{\hat{\xRun}}$ the inputs of $\participant{A}$ on $\someBlockchain$ locked into contracts during $\hat{\xRun}$.

%% file: discussion.tex
\section{Discussion}
\label{sec:discussion}

In this section, we discuss the rationale for some of the design decisions of \bitmlx and the limitations of our approach, hence identifying venues for future work. 

\paragraph{Syntax: \bitmlx vs \bitml} 
We have made two simplifications in the syntax of \bitmlx when compared to that of \bitml:
(1) We merged \bitml's choice operator ($+$) and time lock mechanism ($\bitmlcode{after}$) into the priority choice operator ($\prchoice$); and 
(2) we removed \bitml's support for volatile deposits.
In the following, we discuss the motivation behind these design choices and their implications on the expressiveness of \bitmlx.

    \emph{Priority Choices.} \bitmlx syntax supports neither the simultaneous choice $+$ nor the user-defined timelock $\bitmlcode{after}$ operators from \bitml.
    Instead, \bitmlx syntax supports the operator $\prchoice$, indicating the precedence of moves to the left over moves to the right side of the operator.
        So, a \bitmlx contract is of the form 
    $\contract{C} \coloneqq \contract{D_1} \prchoice (\contract{D_2} \prchoice (\ldots \prchoice \contract{D_k}) \ldots )$ where the lowest priority option $\contract{D_k}$ needs to be a $\bitmlcode{withdraw}$ statement serving as the contract's fallback case that will be executed if none of the higher priority moves were taken. 
    Intuitively, this design ensures that every \bitmlx contract is \emph{liquid}, meaning that funds can never be indefinitely locked in a contract.

    Introducing priority choices (as opposed to synchronous choices) in \bitmlx is both a necessity and a feature:
        
    Explicit precedence is what allows our compiler to synchronize the contracts in many chains by executing choices in rounds. 
    Further, as outlined in the previous section, liquidity is crucial for proving compiler correctness.
        
    On the other side, liquidity is a desirable contract security feature~\cite{bartoletti2019verifying}, which \bitmlx (as opposed to \bitml) enforces by design.
    Further, we argue that priority choices suitably abstract the combined usage of explicit timelocks and synchronous choices in a way that facilitates easier and more secure smart contract design.

    Due to \bitmlx's (as well as \bitml's) strong adversarial execution model, contracts $\contract{D_1} + \contract{D_2}$ that enable synchronous choices can easily lead to situations where both moves $\contract{D_1}$ and $\contract{D_2}$ are enabled, and hence the attacker can freely choose which move to take.  
    
        To avoid such situations, in many practical use cases, \bitml's time lock mechanism ($\bitmlcode{after}$) serves exactly for implementing precedence between two possible moves: In a \bitml contract of the form $\contract{D_1} + \bitmlcode{after}~t:~\contract{D_2}$ the clause $\bitmlcode{after}~t:~\contract{D_2}$ effectively lowers the priority of $\contract{D_2}$ with respect to $\contract{D_1}$ since before time $t$, contract $\contract{D_1}$ can be executed exclusively, enabling the honest user to make this move without interference by the attacker.
        For implementing precedence, the concrete choice of $t$ does not matter as long as it gives users sufficient time to execute $\contract{D_1}$ if desired. 
    The $\prchoice$ operator, hence, suitably abstracts the usage of $\bitmlcode{after}$ in those scenarios, relieving the developer from its manual implementation (including the choice of safe values for $t$).

    One drawback of replacing explicit time locks (as supported by $\bitmlcode{after}$) with priority choices is that the contract developer can only establish a relative execution order but not schedule executions for concrete time points.
    For specific use cases, contract developers may like to set $t$ to a concrete point in time (e.g., to denote a concrete expiration date). 
    Such a feature is currently not supported by \bitmlx.
        In principle, it would be possible to integrate a corresponding language feature into the \bitmlx language, 
    e.g., one could consider adding a timed priority choice operator, where the user explicitly sets the time lock that will be used for compiling this priority choice. 
    However, such a language feature would need to be subject to well-formedness checks ensuring that the user-defined timelock does not interfere with the timelocks automatically inferred by the compiler.
        Adding such a language feature would consequently make it necessary to expose the user to the implementation details of the compiler. 
    Even worse, for reflecting user-defined timelocks, the \bitmlx semantics would need to be extended with an explicit model of time. 
    This model of time would, in turn, make the time-specific execution details of (non-timed) priority choices explicit since otherwise it would not be possible to faithfully model the interference between timed and non-timed priority choices.

\emph{Volatile Deposits.} \bitml supports volatile deposits, denoting deposits from users that can be funded only at the time where the contract needs them. 
Instead, \bitmlx in its current form does not support volatile deposits. Every deposit in \bitmlx requires that every other user includes a collateral equal to the amount of such deposit, so that the honest user can be compensated if needed. 

    If \bitmlx were to support volatile deposits, this would require that every user also includes additional collateral for those deposits. 
    But at the point in time when the volatile deposit is consumed by the contract, it cannot be guaranteed that all contract users will have sufficient funds to provide the required collateral. 
    As a consequence, it is hard to enforce the synchronicity of such an execution step on all chains.
    This issue could be circumvented by requiring all contract users to provide the collaterals for all eventual volatile deposits at the contract stipulation time.
    However, such an implementation would defeat the original purpose of volatile deposits, which is to avoid binding too many financial resources in a contract for the whole duration of the contract execution.

\paragraph{System assumptions and supported compilation targets}
\bitmlx compiles to \bitml smart contracts, which in turn can be translated to transactions in Bitcoin-like cryptocurrencies.
We consider cryptocurrencies Bitcoin-like if they are UTXO-based and support a scripting language that subsumes the one of Bitcoin. 
More precisely, such scripting languages need to be able to restrict the spending of \emph{unspent transaction outputs (UTXOs)} based on conditions that include 
(i) signature checks (providing signatures that correctly verify for pre-specified public keys);
(ii) hash locks (providing preimages for predefined hash values);
(iii) time locks (requiring a minimal system time has been reached);
(iv) the evaluation of conditions (on input values) as defined in~\cref{fig:syntax-reveal-conditions}.
Examples of such cryptocurrencies are sharing the Bitcoin scripting language (such as Dogecoin~\cite{dogecoin-whitepaper}, Bitcoin Cash~\cite{bitcoin-cash-website} or Litecoin~\cite{litecoin-website}), or more expressive scripting languages (e.g., Cardano~\cite{extended-utxo} supports an extended UTXO model featuring a Turing-complete scripting language).
Account-based cryptocurrencies (such as Ethereum~\cite{ethereum-whitepaper}, Solana~\cite{solana-whitepaper}, Ripple~\cite{ripple-paper}, or Stellar~\cite{stellar-paper}), which explicitly track user and smart contract balances, or currencies with more limited scripting support (such as Monero~\cite{monero-paper}, Zerocoin~\cite{zerocoin-paper}, or Zerocash~\cite{zerocash-paper}) are currently not in scope.

More practically, for compiling a \bitmlx contract to $k$ blockchains $\someBlockchain[1], \dots, \someBlockchain[k]$, it is sufficient to ensure that the \bitml compiler supports the chains $\someBlockchain[1], \dots, \someBlockchain[k]$ as compilation targets. 
Fortunately, the existing implementation of the \bitml compiler outputs an abstract transaction format (called Balzac~\cite{formal-model-bitcoin,balzac-language}). 
Hence, it suffices to provide serializations from Balzac to transactions in the desired target blockchains $\someBlockchain[1], \dots, \someBlockchain[k]$. 
In particular, since our compilation targets \bitml, this means that our compiler correctness result is independent of the choice of targeted blockchains.

Expanding the scope of \bitmlx beyond Bitcoin-like cryptocurrencies could be done by encoding Balzac transactions and emulating their execution within the targeted system.
This could be a viable path towards extending \bitmlx to account-based cryptocurrencies with (quasi) Turing-complete contract support (such as Ethereum or Solana).

\paragraph{Assumptions on honest users} 
The execution of a compiled \bitmlx contract assumes that honest users are online during the execution of the contract, monitoring the potentially $k$ blockchains to attempt to include the transactions corresponding the next move when needed. 
In practice, the honest user would also need to pay a fee for each of such moves, since the inclusion of the corresponding transaction in the $i$-th blockchain requires a transaction fee. 
Relaxing these practical assumptions on honest users is an interesting future research direction, for instance, by compiling \bitmlx contracts into off-chain solutions such as payment channels or rollups~\cite{GudgeonSokOffChain2020}. 

\paragraph{Required collateral}
For simplicity, our solution uses an overly conservative formula for calculating the collateral needed in a \bitmlx contract. Through static analysis of contract code, upper bounds for the best-case withdraw output for each user can be determined, and the needed collateral could be reduced to the sum of those upper bounds for all users. An additional optimization in this direction could include allowing users to withdraw part of their collateral as the contract progresses and these upper bounds are lowered.

\paragraph{Supported applications}
Since \bitmlx relies on \bitml as its compilation target, \bitmlx inherits known limitations of the \bitml language. 
In particular, \bitmlx (as \bitml) only supports smart contracts involving a fixed set of users, and its computational expressiveness is limited (e.g., both \bitml and \bitmlx do not support recursion).
This excludes applications that are open-ended (such as payment channels) or that rely on dynamically changing sets of users (such as general decentralized exchanges or crowdfunding).
Recent extensions of the \bitml language~\cite{bitml-recursion} allow for expressing a limited form of recursion by enabling contract users to renegotiate parts of the contract during execution.
It remains interesting future work to investigate whether \bitmlx{} can be extended to support this feature as well.

The simplifications of \bitmlx with respect to \bitml (omission of explicit time locks and volatile deposits) exclude applications that either
(i) rely on payments that shall be scheduled for a specific time in the future (e.g., modeling loans that shall be returned at a specific date); or
(ii) enable users to top up the contract funds during execution (e.g., to pay a mediator to resolve conflicts on demand).

Despite these limitations, \bitmlx can still express many interesting cross-chain applications, as we will illustrate in the following section.

%% file: applications.tex
\newcommand{\borrower}{\ensuremath{\participant{B}}\xspace}
\newcommand{\lender}{\ensuremath{\participant{L}}\xspace}
\newcommand{\mediator}{\ensuremath{\participant{M}}\xspace}
\newcommand{\customer}{$\participant{C}$\xspace}
\newcommand{\restaurant}{$\participant{R}$\xspace}
\newcommand{\exchange}{$\participant{X}$\xspace}

\section{Applications}
\label{sec:applications}
Here, we overview cross-chain applications and express them with \bitmlx language. 
We then evaluate the resources required to execute them  in the underlying cryptocurrencies.

\paragraph{Multichain donations}
Several organizations such as non-profit, charities, universities, faith-based or mission-driven organizations accept cryptocurrency donations in different cryptocurrencies to support their cause~\cite{donate-crypto}. 
For instance, organizations in~\cite{donate-crypto} accept donations in over 80 different cryptocurrencies. 
Users might have a varied portfolio of cryptocurrencies, and they can decide to donate in different denominations. 
In the following, we show how to use \bitmlx to implement several instances of such use case, depending on how the denomination is chosen.

First, the receiver can select the denomination of the donation, with the \bitmlx contract shown in~\cref{fig:donation}. Here, the donor $\participant{A}$ offers to donate either $\money{1\bitcoins}$ or $\money{1\dogecoins}$ to $\participant{B}$, who can then decide the denomination  by exercising the priority choice between  $\contract{PayBTC}$ and $\contract{PayDGC}$. 
Second, both participants can jointly decide the denomination of the donation by means of a coin flip, as shown in~\cref{fig:donation-random}. Here, $\participant{A}$ commits to a secret bit $x$ and $\participant{B}$ commits to a secret bit $y$. If both are heads or  tails, the donor pays $\money{1\bitcoins}$. Otherwise, the donor pays $\money{1\dogecoins}$.

\begin{figure}[tb]
\centering
\fbox{
\begin{minipage}{0.95\columnwidth}
{\footnotesize
\[
\begin{array}{l l}
    \multicolumn{2}{l}{\{
            \participant{A}: \money{1\bitcoins}
            | \participant{A}: \money{1\dogecoins}
    \} \contract{Donate}}\\ 
    \contract{Donate} &= \contract{PayBTC} \prchoice \contract{PayDGC} \prchoice \contract{Refund}\\ 
    \contract{PayBTC} &= \participant{B}: [\bitmlcode{withdraw} (\money{1\bitcoins}, \money{0\dogecoins}) \rightarrow \participant{B},\\
    &\quad \quad \quad \bitmlcode{withdraw} (\money{0\bitcoins}, \money{1\dogecoins}) \rightarrow \participant{A}]\\
    \contract{PayDGC}  & =  [\bitmlcode{withdraw} (\money{0\bitcoins}, \money{1\dogecoins}) \rightarrow \participant{B},\\
    &\quad \bitmlcode{withdraw} (\money{1\bitcoins}, \money{0\dogecoins}) \rightarrow \participant{A}]\\
    \contract{Refund}  &=  \bitmlcode{withdraw}(\money{1\bitcoins}, \money{1\dogecoins}) \rightarrow \participant{A}
\end{array}
\]
}
\end{minipage}
}
\fbox{
\begin{minipage}{0.95\columnwidth}
\footnotesize{
\[
    \begin{array}{l l}
        \multicolumn{2}{l}{\{
            \participant{A}: \money{1\bitcoins}
            | \participant{A}: \money{1\dogecoins}
        \}
        {\contract{DonateAgreed}}}\\ 
        \contract{DonateAgreed} ~ = & \contract{CoinToss} \prchoice \contract{Refund}\\
        \multicolumn{2}{l}{\contract{CoinToss} ~ = \bitmlcode{reveal}^*~  x  \bitmlcode{.}(\contract{WinBTC} \prchoice \contract{WinDGC})} \\
        \multicolumn{2}{l}{\contract{WinBTC} ~ = \bitmlcode{reveal}^*~ y~ \land |x| = |y| \bitmlcode{.}(\contract{PayBTC} \prchoice \contract{PayDGC})} \\
        \multicolumn{2}{l}{\contract{WinDGC} ~ = \bitmlcode{reveal}^*~ y~\land |x| \neq |y| \bitmlcode{.}(\contract{PayDGC} \prchoice \contract{PayBTC})}\\
        \contract{PayBTC} ~ = &[\bitmlcode{withdraw} (\money{1\bitcoins}, \money{0\dogecoins}) \rightarrow \participant{B},\\
        &\bitmlcode{withdraw} (\money{0\bitcoins}, \money{1\dogecoins}) \rightarrow \participant{A}]\\
        \contract{PayDGC} ~ =  &[\bitmlcode{withdraw} (\money{0\bitcoins}, \money{1\dogecoins}) \rightarrow \participant{B},\\
        &\bitmlcode{withdraw} (\money{1\bitcoins}, \money{0\dogecoins}) \rightarrow \participant{A}]\\
        \contract{Refund} ~ = ~&\bitmlcode{withdraw}(\money{1\bitcoins}, \money{1\dogecoins}) \rightarrow \participant{A}
    \end{array}  
\]
}
\end{minipage}
}
\fbox{
\begin{minipage}{0.95\columnwidth}
{\footnotesize 
\[
    \begin{array}{l l}
        \multicolumn{2}{l}{\{
            \participant{C}: \money{10\bitcoins}
            | \participant{X}: \money{100\dogecoins}
        \}
        {\contract{ExchangeService}}}\\
        \contract{ExchangeService} &= \contract{Payment} \prchoice \contract{RefundAll} \\
        \multicolumn{2}{l}{\contract{Payment} ~ = \contract{DirectPay} \prchoice \contract{ExchangePay}}\\ 
        \contract{DirectPay} ~ &= \participant{C}: [\bitmlcode{withdraw} (\money{10\bitcoins}, \money{0\dogecoins}) \rightarrow \participant{R},\\
        &\quad \quad \quad \bitmlcode{withdraw} (\money{0\bitcoins}, \money{100\dogecoins}) \rightarrow \participant{X}]\\
        \contract{ExchangePay} ~ &= [\bitmlcode{withdraw} (\money{10\bitcoins}, \money{0\dogecoins}) \rightarrow \participant{X},\\
        &\quad \bitmlcode{withdraw} (\money{0\bitcoins}, \money{100\dogecoins}) \rightarrow \participant{R}]\\
        \contract{RefundAll} ~ &= [\bitmlcode{withdraw}(\money{10\bitcoins}, \money{0\dogecoins}) \rightarrow \participant{C}, \\
        &\quad \bitmlcode{withdraw}(\money{0\bitcoins}, 
        \money{100\dogecoins}) \rightarrow \participant{X}]
    \end{array}  
\]
}
\end{minipage}
}
\fbox{
\begin{minipage}{0.95\columnwidth}
{\footnotesize 
    \[
      \begin{array}{l l}
        \multicolumn{2}{l}{\{
                \borrower: \money{3\bitcoins}
                | \lender: \money{30\dogecoins}
            \}\contract{Loan}}\\
            \multicolumn{2}{l}{\contract{Loan} =  {\contract{ExecLoan} \prchoice \contract{RefundLoan}}} \\
            \contract{ExecLoan} ~ &= \bitmlcode{split}~ [\bitmlcode{withdraw}~ (\money{0\bitcoins}, \money{30\dogecoins}) \rightarrow \borrower , \\ &(\money{3\bitcoins}, \money{0\dogecoins}) \rightarrow \contract{Installment[3]}]\\
            \contract{Installment}[i] ~ &= \mediator : \bitmlcode{split}~ [\bitmlcode{withdraw} (\money{1\bitcoins}, \money{0\dogecoins}) \rightarrow \borrower, \\
            &(\money{i-1\bitcoins}, \money{0\dogecoins}) \rightarrow \contract{Installment[i-1]}] \\ 
            &\prchoice \bitmlcode{withdraw}(\money{i~\bitcoins}, \money{0\dogecoins}) \rightarrow \lender\\
            \contract{RefundLoan} ~ &= [\bitmlcode{withdraw}(
                \money{3\bitcoins}, \money{0\dogecoins}) \rightarrow \borrower, \\
                &\bitmlcode{withdraw}(\money{0\bitcoins}, \money{30\dogecoins} \rightarrow \lender]
      \end{array}
    \]
    }
\end{minipage}
}
\caption{\bitmlx contract for applications (from top to bottom): multichain donation (with receiver chosen denomination), multichain donation (with jointly agreed denomiination), multichain payment with exchange service, multichain loan with mediator. For readability, we use $\bitmlcode{reveal}^*~s$ as a shortcut for $\bitmlcode{reveal}~s~\bitmlcode{if}~0 \leq |s| \leq 1$.  \label{fig:donation} \label{fig:donation-random} \label{fig:payment-exchange-rate} \label{fig:multichain-loan}}
\end{figure}

\paragraph{Multichain payments with exchange service}
As  with fiat currencies, cross-cryptocurrency payments often require an intermediate exchange service that performs the exchange between different cryptocurrencies. 
Next, we illustrate how \bitmlx can be used to implement such scenarios 
(c.f.~\cref{fig:payment-exchange-rate}).  

Assume that a customer \customer only holds bitcoins and goes  to a restaurant (i.e., \restaurant). 
Further assume that \restaurant's portfolio consist of both bitcoins and dogecoins. 
In this setting, after the dinner is finished,  the restaurant offers the customer \customer two options for the payment:
(i) pay in bitcoins directly to \restaurant (i.e., $\contract{DirectPay}$); or 
(ii) pay in dogecoins using a predefined exchange rate -- $\money{1\bitcoins} : \money{10\dogecoins}$ in this example-- offered by an exchange service \exchange (i.e., $\contract{ExchangePay}$). 
In the former case, the restaurant \restaurant gets paid $\money{10\bitcoins}$ while the exchange \exchange gets back $\money{100\dogecoins}$. 
In the latter case, the exchange \exchange receives bitcoins whereas the restaurant \restaurant gets the  payment in dogecoins.

\paragraph{Multichain loan with mediator}
Assume a borrower \borrower and a lender \lender. The borrower has $\money{3\bitcoins}$ and wants to take a loan of $\money{30\dogecoins}$. 
We assume that the exchange rate between bitcoins and dogecoins is $\money{1\bitcoins}$:$\money{10\dogecoins}$. 
Having a surplus of dogecoins, the lender is willing to grant such a loan of $\money{30\dogecoins}$. 
However, since lender and borrower are mutually distrusted, the lender wants to hold the $\money{3\bitcoins}$ from the borrower as a collateral for the duration of the loan (or until one of the installment's payment of the loan is missed by the borrower).

In such a setting, a lending protocol works in three stages as follows. 
In the first stage, the lender and borrower input the loan coins (i.e., $\money{30\dogecoins}$) 
and the collateral coins (i.e., $\money{3\bitcoins}$) into the contract (i.e., $\contract{Loan}$). 
In such a contract, (i) borrower receives the   $\money{30\dogecoins}$ of the loan (i.e., $\contract{ExecLoan}$) and the $\money{3\bitcoins}$ of the collateral are locked into a second contract (i.e., $\contract{Installment}$), triggering the second stage.

The contract $\contract{Installments}$ works as follows.
At each step $i$, if the borrower has paid $\money{10\dogecoins}$ to the lender, the mediator \mediator, trusted to monitor when an installment payment of $\money{10\dogecoins}$ from the borrower to lender occurs,  authorizes the release of  $\money{1\bitcoins}$ in favor of the borrower. 
This second stage of the loan protocol ends when either 
(i) the last step is finished; 
or (ii) the borrower does not pay the installment at the $i$-th step. 
In the former case, the loan protocol is finished since the borrower returned the complete loan to the lender, getting the complete collateral back in exchange. 
In the latter case, the lender can claim the amount of bitcoins that remain as collateral as a payment for the default of the borrower.

The application described so far can be implemented with the \bitmlx contract shown in~\cref{fig:multichain-loan}. 
Note that we have made several simplifications of the protocol for ease of exposition. 
It is possible to extend the \bitmlx contract to model other practical features such as 
(i) different amounts for the loan and the collateral to account for exchanges rates different to $\money{1\bitcoins}$:$\money{10\dogecoins}$; 
(ii) different amounts for each of the installment payments;
(iii) consider a third blockchain where the mediator \mediator gets paid by the borrower \borrower and the lender \lender if the mediator's task in the loan is successfully carried out.

%% file: implementation_evaluation.tex
\newcommand{\ftx}{\ensuremath{\textit{Txs}}\xspace}
\newcommand{\nparticipants}{\ensuremath{\textit{U}}\xspace}
\paragraph{Evaluation}
\label{sec:evaluation}
Now, we discuss the blockchain resources required to execute the \bitmlx contracts presented so far in this section.  
For that, we have compiled the \bitmlx contracts into \bitml contracts using our compiler (\cref{sec:compilation}), whose source code is available at~\cite{bitmlx-code}. Then, using the \bitml compiler in~\cite{bitml}, we have transformed the \bitml contracts into Bitcoin transactions. 
As a result, we have obtained the total number of transactions required in Bitcoin to encode the complete logic of each application, as shown in~\cref{table:resources}.
However, not all transactions need to be included in the blockchain, only those required by the user strategies.  
Even in the worst case, this corresponds to the depth of the execution tree, which is considerably smaller (see~\cref{table:resources}).

\newcommand{\fdepth}{\ensuremath{\textit{Depth}}\xspace}
\newcommand{\fmax}{\ensuremath{\textit{Max}}\xspace}

 \newcommand{\ftime}{\ensuremath{\textit{Time}}\xspace}

\begin{table}
    \caption{Summary of resources per blockchain required by \bitmlx contracts described in~\cref{sec:applications}. \label{table:resources}}
    \centering
\begin{tabular}{c | c c }
     & Generated $\ftx$ & Executed $\ftx$   \\
     \hline
    $\contract{Donate}$~(\cref{fig:donation}) & $51$ & $6$    \\
    $\contract{DonateAgreed}$~(\cref{fig:donation-random}) & $51$  & $6$  \\
    $\contract{ExchangeService}$~(\cref{fig:payment-exchange-rate}) & $207$ & $6$  \\
    $\contract{Loan}$~(\cref{fig:multichain-loan}) & $4098$ & $7$ 
\end{tabular}

\end{table}

%% file: related_work.tex
\section{Related Work}
\label{sec:relatedwork}

Existing approaches in the literature~\cite{zamyatin2021sokcrosschain} for cross-chain applications, which all focus on custom designs tailored to the specific application, can be grouped into (i) migration protocols or bridges; and (ii) cryptographic protocols. 
Next, we overview  and compare them with our work.

\paragraph{Migration protocols or bridges} 
The cornerstone of this protocol class is to rely on a coordinator to (i) lock funds in the different cryptocurrencies as required by the cross-chain smart contract; 
(ii) release these funds into a (possibly different) cryptocurrency with support for expressive smart contracts (e.g., Ethereum) that it is in then used to execute the smart contract; 
and (iii) redistribute the final distribution of funds back to the individual cryptocurrencies, as indicated by the final state of the cross-chain smart contract. 
The role of the coordinator is implemented by trusted hardware module~\cite{tesseract2019} or a committee of users that jointly authorize the back and forth transfer of funds across cryptocurrencies~\cite{zkbridge2022}.  

The security of this approach is built upon an additional trust assumption on the coordinator itself, either in the form of honest majority across the committee or the trust on the hardware module itself. 
Moreover, this approach requires a cryptocurrency with support for stateful smart contracts, limiting its applicability to cryptocurrencies like Bitcoin. 

\paragraph{Custom cross-chain cryptographic protocols} 
Alternatively to migration protocols or bridges, 
this group of works contributes an application-dependent cryptographic protocol as a synchronization overlay among the cryptocurrencies involved in the cross-chain application. 
Examples of these protocols exist for payment channels~\cite{AumayrChannels2021}, payment channel networks~\cite{AMHL2019}, coin mixing~\cite{foundations-coin-mixing}, oracle-based contracts~\cite{OracleContracts}, and beyond.

While this approach departs from the additional trust assumption as in migration protocols, the design of these custom cryptographic protocols requires substantial (cryptographic) expertise.
Moreover, the security of these protocols is proven from scratch, requiring involved proofs in complex cryptographic proof frameworks, an error-prone task as demonstrated so far in the literature~\cite{TairiMS23}.

Different to existing approaches, in our work we provide a domain-specific language that can be used to describe the functionality of the cross-chain smart contract, abstracting away the cryptographic details of the protocol realizing the contract execution. 
The \bitmlx smart contract is then compiled into several contracts (i.e., one per involved cryptocurrency) that are compatible with Bitcoin-like cryptocurrencies and a strategy for each honest user on how to interact with them. 
Our approach comes with formally proven correctness guarantees: 
If an honest user follows the prescribed strategy when interacting with the compiled smart contract at each cryptocurrency, they are guaranteed to end up with at least as many funds as described by the execution of the \bitmlx cross-chain contract.

%% file: conclusion.tex
\section{Conclusion}
\label{sec:conclusion}

We present \bitmlx, the first domain-specific language for cross-chain smart contracts. 
We show a compiler to automatically translate a \bitmlx contract into one contract per involved cryptocurrency and a user strategy to interact with them, 
and prove that a user following such strategy will end up with as many funds as 
in the corresponding execution of the \bitmlx contract. 
We have implemented the \bitmlx compiler and showcased its utility in the design of illustrative examples of cross-chain applications, e.g., multichain loans.

\section*{Acknowledgments}

We would like to thank the reviewers for their helpful feedback. 
Furthermore, we would like to thank Minhua Liu for assisting in the preparation of the paper artifacts and helping improve the paper presentation.
This work has been supported by the Heinz Nixdorf Foundation through a Heinz Nixdorf Research Group (HN-RG) and funded by the Deutsche Forschungsgemeinschaft (DFG, German Research Foundation) under Germany’s Excellence Strategy—EXC 2092 CASA—390781972.
Further, this work has been partially supported by the ESPADA project (grant PID2022-142290OB-I00), MCIN/AEI/10.13039/501100011033/ FEDER, UE; 
and by the PRODIGY project (grant ED2021-132464B-I00), funded by MCIN/AEI/10.13039/501100011033/ and the European Union NextGenerationEU/ PRTR.

%% file: theory_docs/syntax.tex
\maketitle

\subsection{Preconditions}
The \bitmlx syntax tries to remain as close to it as possible to the single chain version of BitML. Notable changes are:
\begin{itemize}
    \item Persistent deposits take balance, which is a list of currencies.
    \item No support for volatile deposits     \item We add a new precondition for establishing contract start time unique initial id. This is limited syntactically to a single precondition of this kind.
\end{itemize}

\begin{align*}
    & \preconditions{G} \Coloneqq \\
    & \quad \vert ~ \depositsPre{A}{\balance}{\vec{x}} \\
    & \quad \vert ~ \participant{A}: \bitmlcode{secret} ~ \secret{s} \\
    & \quad \vert ~ \preconditions{G} \| \preconditions{G'} \\
    & \preconditions{G} ~ \vert ~ \contractSettings{t_0}{\someStipulation} \\
    & \balance = [v_1\someBlockchain[1],\dots,v_k\someBlockchain[k]] & \quad \text{Balance}\\
\end{align*}

Preconditions are commutative.

\subsection{Contracts}
\begin{itemize}
    \item The contract sum has been replaced with priority choices.
    \item Guarded contracts are only allowed on the left side of a priority choice.
    \item \bitmlcode{withdraw} can also be used as a top-level contract and acts as a terminal constructor. It has also been generalized to map users to the corresponding funds they withdraw on each blockchain.
    \item We remove the \bitmlcode{put} statement because again, we don't support volatile deposits.
    \item We also remove the \bitmlcode{after} statement because our timed semantics will be coupled to our choice operator.
    \item We write signature authorizations with lists of signing users. Technically, concatenating multiple single-user authorizations is still allowed by the syntax.
    \item \bitmlcode{split} now takes a vector of balances to split for each contract. \bitmlcode{withdraw} also has a list of balances for each user.
\end{itemize}

\begin{align*}    
    & \contract{C} \Coloneqq \contract{D} \prchoice \contract{C} & \text{Top-level contracts}\\
    & \quad \vert ~ \bitmlxWithdraw{\balance}{A} \\
    & \contract{D} \Coloneqq \bitmlxWithdraw{\balance}{A} & \text{Guarded contracts} \\
    & \quad \vert ~ \bitmlxSplit{\balance}{C} \\
    & \quad \vert ~ \bitmlxAuthOp{A}{\contract{D}} \\
    & \quad \vert~ \bitmlxReveal{\secret{s}}{p}{C} \\
\end{align*}

\subsection{WellFormdness Check}

The well-formedness check further constraints what kind of \bitmlx contract advertisements are actually valid. 
It rules out things like inconsistent funds distribution on splits and withdraws or references to non-existing participants.

The auxiliary functions we use throughout the rules are defined at the end of the appendix.

\[
    \textit{wellFormed}(\preconditions{G}, \contract{D}) := \wf{\preconditions{G}, \contract{D} \prchoice \contract{C}, \balance(balance(\contract{G}})
\]

\subsubsection{Top-Level-Contracts}

A top-level contract is well-formed if it is a priority choice between well-formed contracts or if it is a well-formed withdraw (which we explain on the guarded version).

\[
\infer[WF-PChoice]
{
    \wf{\preconditions{G}, \contract{D} \prchoice \contract{C}, \balance}
}
{
    \wf{\preconditions{G}, \contract{C}, \balance} &
    \wfg{\preconditions{G}, \contract{D}, \balance} 
} 
\]

\[
\infer[WF-Withdraw]
{
        \wf{\preconditions{G}, \bitmlxWithdraw{\balance}{A}, \balance}
}
{
    \vec{\balance} = \balance_1, \dots, \balance_k  &
        \sum_1^k \balance_i = \balance &
    \forall \participant{A_i} \in \vec{\participant{A}}.~ \participant{A_i} \in \participants{G}
}
\]

\subsubsection{Guarded Contracts}

\[
\infer[WFG-Split]
{
    \wfg{\preconditions{G}, \bitmlxSplit{\balance}{C}, \balance}
}
{
    \begin{gathered}
            \vec{\balance} = \balance_1, \dots, \balance_k  \\
    \vec{C} = C_1, \dots, C_k  
    \end{gathered} &
    \begin{gathered}
                \sum_1^k \balance_i = \balance 
    \end{gathered} &
    \begin{gathered}
        \forall \participant{A_i} \in \vec{\participant{A}}.~ \participant{A_i} \in \participants{G} \\
        \forall i\in [1,\dots,k].\wf{\preconditions{G}, \contract{C_i},\balance_i} 
    \end{gathered} 
}
\]

\[
\infer[WFG-Reveal]
{
    \wfg{\preconditions{G}, \bitmlxReveal{s}{p}{C}, \balance}
}
{
    \forall \secret{s'} \in \vec{\secret{s}}.~ \secret{s'} \in secrets(\preconditions{G}) &
    \wf{\preconditions{G}, \contract{C}, \balance} 
}
\]

\[
\infer[WFG-AuthOp]
{
    \wfg{\preconditions{G}, \bitmlxAuthOp{A}{\contract{D}}, \balance}
}
{
    \wfg{\preconditions{G}, \contract{D}, \balance}  &
    \forall \participant{A_i} \in \vec{\participant{A}}.~ \participant{A_i} \in \participants{G} &
    \contract{D} \neq \bitmlxAuthOp{A'}{\contract{D'}}
}
\]

\[
\infer[WFG-Withdraw]
{
    \wfg{\preconditions{G}, \bitmlxWithdraw{v}{A}, \balance}
}
{
    \vec{\balance} = \balance_1, \dots, \balance_k  &
    \sum_1^k \balance_i = \balance &
    \forall \participant{A_i} \in \vec{\participant{A}}.~ \participant{A_i} \in \participants{G}
}
\]

%% file: theory_docs/bitmlx_semantics.tex
\maketitle

\subsection{Configurations}

A configuration $\Gamma$ encapsulates the complete current state including contract advertisements, active contracts, assigned contracts, commited and revealed (user) secrets and authorizations.
\begin{itemize}
    \item Active contracts $\activeContract{\contract{C}}{\balance}{\someContract}$ have a generelized balance, expressed as a vector of coins.
    \item User deposits are explicit about the blockchain they live in.
\end{itemize}

\begin{align*}
    & \Gamma \Coloneqq \contractAdv{G}{C} & \text{Configurations} \\
        & \quad \vert ~ \activeContract{\contract{C}}{\balance}{\someContract} \\
    & \quad \vert ~ \assignedContract{A}{\balance}{\someContract} \\
    & \quad \vert ~ \participant{A}: [\phi] \\
    & \quad \vert ~  \secretCommitment{A}{s}{N} \\
    & \quad \vert ~ \secretReveal{A}{s}{N} \\
    & \quad \vert ~ \Gamma \| \Gamma' \\
    & \balance = [v_1\someBlockchain[1],\dots,v_k\someBlockchain[k]] & \text{Contract balance}\\
    & \phi ::= \someContract \triangleright \contract{D} ~\vert ~ \# \triangleright \contract{D} ~|~\contractAdv{G}{C} \\
\end{align*}

\subsection{Guarded contracts}

Rules in this section describe how guarded contracts transition. They are always assumed to be the left choice in the context of a priority choice with another contract.

\subsubsection{Withdraw}

The $D-Withdraw$ rule says that if:
\begin{itemize}
    \item We have a \bitmlcode{withdraw} guarded contract called $\contract{D}$, where each user in $\vec{ \participant A}$ is given a corresponding balance in $\vec{\balance}$.
    \item The values on the assigned balances add up to the contract balance, on each blockchain. Notice that this is expressed in vector arithmetic.
    \item We have fresh identifiers for the all new user deposits.
\end{itemize}

Then, when $\contract{D}$ is on the left side of a priority choice, we can transition by consuming the active contract from the configuration and including new assigned contracts for each participant $\participant{A_i}$ with balance $\balance[i]$ and identifier $\someContract_i$. If there's extra context $\Gamma$, then it's conserved.

\[
\infer[D-Withdraw]
{
    \reductionRule
        {\activeContract{\contract{D} \prchoice \contract{C}}{\balance}{\someContract} ~ \vert ~ \Gamma}
        {dwithdraw(\someContract)}
        {
            \parallelComposition_{i=1}^{n}
                \assignedContract{A_i}{\balance[i]}{\someContract_i}
            ~|~
            \Gamma
        }
}
{
    \begin{gathered}
        \contract{D} = \bitmlxWithdraw{\balance}{A} \\
        \vec{\balance} = \balance[1], \dots, \balance[n] \\
        \vec{\participant{A}} = \participant{A_1}, \dots, \participant{A_n} \\
    \end{gathered}
    \hspace{10pt}
    \begin{gathered}
    \end{gathered}
}
\]

\subsubsection{Split}

The $D-Split$ rule says that if:
\begin{itemize}
    \item We have a \bitmlcode{split} guarded contract called $\contract{D}$, where the contract funds are divided among new contracts $\vec{ \contract C}$ with corresponding balances $\vec{\balance}$.
    \item The values on the funds mapping add up to the contract balance, on each blockchain.
    \item We have fresh identifiers for the new user active contracts.
\end{itemize}

Then, when $\contract{D}$ is on the left side of a priority choice, we can transition by removing the original active contract from the configuration and including a new active contract for each $\contract{C_i}$ with balance $\balance[i]$ and identifiers $\someContract_i$. If there's extra context $\Gamma$, then it's conserved.

\[
\infer[D-Split]
{
    \reductionRule
        {\activeContract{\contract{D} \prchoice \contract{C}}{\balance}{\someContract} ~ \vert ~ \Gamma}
        {split(\someContract)}
        {
            \parallelComposition_{i=1}^k
                \activeContract{\contract{C_i}}{\balance[i]}{\someContract_i}
            ~|~ \Gamma
        }
}
{
    \begin{gathered}
        \contract{D} = \bitmlxSplit{\balance}{C} \\
        \vec{\balance} = \balance[1], \dots, \balance[n] \\
        \vec{\contract{C}} = \contract{C_1}, \dots, \contract{C_k} \\
        \someContract_1, \dots, \someContract_k ~ \text{fresh} \\
                    \end{gathered}
    \hspace{10pt}
    \begin{gathered}
                                            \end{gathered}
}
\]

\subsubsection{Reveal}

The $D-Reveal$ rule says that if:
\begin{itemize}
    \item We have a \bitmlcode{reveal} guarded contract called $D$ with secrets $\vec{\secret{s}}$, predicate $p$ and follow-up contract $\contract{C'}$. 
    \item All of the secrets are revealed in $\Delta$.
    \item The semantics on $p$ when substituting it's free variables by the values revealed in $\Delta$ evaluate to true.
\end{itemize}

Then, when $\contract{D}$ is on the left side of a priority choice, we can transition to a configuration where we replace the active contract with a new one containing $\contract{C'}$.

\[
\infer[D-Reveal]
{
    \reductionRule
        {\activeContract{\contract{D} \prchoice \contract{C}}{\balance}{\someContract} ~ \vert ~ \Delta ~ \vert ~ \Gamma}
        {reveal(\someContract)}
        {
            \activeContract{\contract{C'}}{\balance}{\someContract'}
            ~\vert~ \Delta ~ \vert ~ \Gamma
        }
}
{
    \begin{gathered}
        \contract{D} = \bitmlxReveal{s}{p}{C'} \\
        \vec{\secret{s}} = \secret{s_1}, \dots, \secret{s_k} \\
        \Delta = \parallelComposition_{i=1}^k \secretReveal{A}{s_i}{N_i} \\
    \end{gathered}
    \hspace{10pt}
    \begin{gathered}
        \llbracket p \rrbracket_\Delta = true \\
        \someContract' ~ \text{fresh} \\
    \end{gathered}
    \hspace{10pt}
}
\]

The $D-Auth$ rule says that if:
\begin{itemize}
    \item We have a contract $\contract{D}$ guarded by the signatures of participants $\vec{\participant{A}}$.
    \item All of the required signatures are provided in $\Delta$.
    \item An active contract $\someContract$ including a priority choice with $\contract{D}$ on the left side in a context $\Gamma$ can transition to a configuration $\Gamma'$.
\end{itemize}

Then, $\bitmlxAuthOp{A}{\contract{D}}$ can also do the same transition when including $\Delta$ in the context.

\[
\infer[D-Auth]
{
    \reductionRule
        {\activeContract{\bitmlxAuthOp{A}{\contract{D}} \prchoice \contract{C}}{\balance}{\someContract} ~ \vert ~ \Delta ~ \vert ~ \Gamma}
        {\alpha}
        {\Gamma'}
}
{
    \begin{gathered}
        \Delta = \parallelComposition_{\participant{A_i} \in \vec{\participant{A}}}
            \participant{A_i} [\someContract\triangleright \contract{D}] \\
        \reductionRule
            {\activeContract{\contract{D} \prchoice \contract{C}}{\balance}{\someContract} \vert \Gamma}
            {\alpha}
            {\Gamma'} \\
    \end{gathered}
    \hspace{10pt}
}
\]

\subsection{Top-Level Contracts}

Rules here describe the transitions to take a right choice, either by skipping to another contract, or finishing with a Withdraw.

\subsubsection{Withdraw}

The transition rule for a top-level \bitmlcode{withdraw} has the same final effect as a guarded \bitmlcode{withdraw}contract.

\[
\infer[C-Withdraw]
{
    \reductionRule
        {\activeContract{\contract{C}}{\balance}{\someContract} ~ \vert ~ \Gamma}
        {cwithdraw(\someContract)}
        {
            \parallelComposition_{i=1}^{n}
                \assignedContract{A_i}{\balance[i]}{\someContract_i}
            ~|~
            \Gamma
        }
}
{
    \begin{gathered}
        \contract{C} = \bitmlxWithdraw{\balance}{A} \\
        \vec{\balance} = \balance[1], \dots, \balance[n] \\
        \vec{\participant{A}} = \participant{A_1}, \dots, \participant{A_n} \\
    \end{gathered}
    \hspace{10pt}
}
\]
\subsubsection{Skip}

The $C-Skip$ rule says that a priority choice can transition to the right side. Restrictions in the the strategies will only allow the attacker to skip when there's consensus for it.

\[
\infer[C-Skip]
{
    \reductionRule
        {\activeContract{\contract{D} \prchoice \contract{C}}{\balance}{\someContract} ~ \vert ~ \Gamma}
        {skip(\someContract)}
        {\activeContract{\contract{C}}{\balance}{\someContract'} ~ \vert ~ \Gamma}
}
{
    \begin{gathered}
        \someContract' ~ \text{fresh} \\
    \end{gathered}
}
\]

\subsection{Authorizations}

These rules don't transform active contracts but instead only add some authorization by some participant in the context, as preconditions for other rules.

\subsubsection{Signatures}

The $A-Auth$ rule says that if:
\begin{itemize}
    \item We have an active contract where the left side is guarded with authorizations by $\vec{\participant{A}}$.
    \item $\participant{A}$'s signature is not present in the configuration.
\end{itemize}

Then we can transition by adding the authorization in the configuration.

\[
\infer[A-Signature]
{
    \reductionRule
        {\activeContract{\bitmlxAuthOp{A}{\contract{D}} \prchoice \contract{C}}{\balance}{\someContract} ~ \vert ~ \Gamma}
        {\participant{A}: \uniqueId}
        {
            \activeContract{\bitmlxAuthOp{A}{\contract{D}} \prchoice \contract{C}}{\balance}{\someContract}
            ~|~ \participant{A} [\someContract \triangleright \contract{D}]
            ~|~ \Gamma
        }
}
{
    \begin{gathered}
        \participant{A} \in \vec{\participant{A}} \\
    \end{gathered}
}
\]

\subsubsection{Secrets}

The $A-RevealSecret$ rule says that if $\participant{A}$ really has the value for a secret they commited, then they can always reveal it. Remember that dishonest users can commit to any secret, even if they don't actually know the value, which is represented by the $\bot$ length.

\[
\infer[A-RevealSecret]
{
    \reductionRule
        {\secretCommitment{A}{a}{N} ~ \vert ~ \Gamma}
        {\participant A:\secret a}
        {\secretReveal{A}{a}{N} ~ \vert ~ \Gamma}
}
{
    \begin{gathered}
        N \neq \bot
    \end{gathered}
}
\]

\subsection{Stipulation Protocol}\label{stipulation-protocol}

BitMLx stipulation protocol looks very similar to BitML's, but has the option to abort a contract, which destroys the advertisement, commitment and signatures. This design is much simpler than our previous one, which required an extra intermediate step. My reasoning for it is that, just like we can map a compensated contract as a left move, we can map all combinations of not publishing, refunding or compensating to an abort.

\subsubsection{Advertisement}

A contract advertisement $\contractAdv{G}{C} ~ \vert ~ \Gamma$ is valid if
\begin{itemize}
    \item All secrets in the contract are fresh.
    \item At least one user is honest.
\end{itemize}

\[
\infer[S-Advertisement]
{
    \reductionRule
        {\Gamma}
        {advertise(\preconditions{G}, \contract{C})}
        {\contractAdv{G}{C} ~ \vert ~ \Gamma}
}
{
    \begin{gathered}
        \participants{G} \cap \Hon \neq \emptyset \\
        wellFormed(\contractAdv{G}{C}) \\
        \forall \secret{s} \in secrets(\preconditions{G}), ~\secret{s} ~\fresh \\
    \end{gathered}
    \hspace{10pt}
}
\]

Where:
\begin{itemize}
    \item $secrets(\preconditions{G})$ is the set of all secrets that appear in the preconditions $\preconditions{G}$.
    \item $\participants{G}$ is the set of all participants that have a deposit in $\preconditions{G}$. 
    \item $wellFormed(\preconditions{G}, \contract{C})$ checks some basic correctness conditions on the contract, such as all secrets used in \bitmlcode{reveal} statements being commited to in the preconditions and all branches on \bitmlcode{split} and \bitmlcode{withdraw} statements add up to the original balance. 
\end{itemize}

\subsubsection{Secret Commitment}

This is taken from BitML almost verbatim. Each participant $\participant A$ can commit to the contract advertisement $\contractAdv{G}{C}$ if
\begin{itemize}
    \item The contract is advertised in the current configuration.
    \item If they are honest, all of their secrets have some length $N \in \mathbb N$. If not, the length can also be $\bot$, which represents a secret for which they don't know the value.
    \item They have not commited any of their secrets in $\preconditions{G}$ in the past.
    \item They now commit to all of their secrets in $\preconditions{G}$.
\end{itemize}

\[\infer[S-AuthCommit]
{
    \reductionRule
        {\contractAdv{G}{C} ~ \vert ~ \Gamma}
        {commit(\participant A, \contractAdv{G}{C})}
        {
            \contractAdv{G}{C}
            ~ \vert ~ \Delta
            ~ \vert ~ \userAuthIn{A}{\#}{\contractAdv{G}{C}}
            ~ \vert ~ \Gamma
        }
}
{
    \begin{gathered}
                \participant{A} \in users(\preconditions{G}) \\
        \forall \secret{s} \in secrets_{\participant{A}}(\preconditions{G}), N \in
            \begin{cases}
                \mathbb{N},& \text{if } \participant{A} \in \Hon \\
                \mathbb{N} \cup \{\bot\},              & \text{otherwise}
            \end{cases}
        \\ 
        \forall \secret{s} \in secrets_{\participant{A}}(\preconditions{G}),
            \nexists N:
                \secretCommitment{A}{s}{N} \in \Gamma \\
        \Delta = \parallelComposition_{\secret{s} \in secrets_{\participant{A}}(\preconditions{G})}
            \secretCommitment{A}{s}{N_s}
    \end{gathered}
    \hspace{10pt}
}
\]

Where:
\begin{itemize}
    \item $secrets_{\participant{A}}(\preconditions{G})$ is the set of all secrets in $\preconditions{G}$ that $\participant{A}$ commits to.
\end{itemize}

\subsubsection{Stipulation Authorization}

Each participant $\participant{A}$ can authorize an advertisement if everyone has commited to their secrets. This corresponds in the low levels to authorizing their deposits to be spent in favor of the contract.

\[\infer[S-AuthInit]
{
    \reductionRule
        {\contractAdv{G}{C} ~ \vert ~ \Gamma}
        {sign(\participant A, \contractAdv{G}{C})}
        {
            \contractAdv{G}{C}
            ~|~ \participant{A}[\contractAdv{G}{C}]
            ~|~ \Gamma
        }
}
{
    \begin{gathered}
    \forall \participant{A_i} \in \participants{G}. ~
        \userAuthIn{A_i}{\#}{\contractAdv{G}{C}} \in \Gamma \\
    \end{gathered}
}
\]

\subsubsection{Initialize}

Initialization consumes the user authorizations to transition from a contract advertisement to an active contract.

\[
\infer[S-Initialize]{
    \reductionRule
        {
            \contractAdv{G}{C}
                        ~|~ secrets
            ~|~ signatures
            ~|~ \Gamma
        }
        {initialize(\someContract, \contractAdv{G}{C})}
        {
            \activeContract{\contract{C}}{\balance}{\someContract}
            ~|~ \Gamma
        }
}
{
    \begin{gathered}
        \participants{G} = \participant{A_1}, \dots, \participant{A_n} \\
        \totalBalance{\preconditions{G}} = \balance \\
        \someContract ~ \text{fresh} \\
    \end{gathered}
    \hspace{10pt}
    \begin{gathered}
        secrets = \parallelComposition_{i=1}^{n}
            \userAuthIn{A_i}{\#}{\contractAdv{G}{C}} \\
        signatures = \parallelComposition_{i=1}^{n}
            (\participant{A_i}[\contractAdv{G}{C}]) \\
    \end{gathered}
    \hspace{10pt}
}
\]

\subsubsection{Abort}

Abort stops the initialization of a contract, refunding all users their deposit.

\[
\infer[S-Abort]{
    \reductionRule
        {
            \contractAdv{G}{C}
                        ~|~ \Gamma
        }
        {abort(\contractAdv{G}{C})}
        {
            \parallelComposition_{i=1}^{n}
                    \assignedContract{A_i}{\balance[i]}{\someContract}
            ~|~ \Gamma
        }
}
{
    \begin{gathered}
                                                                        \deposits{G} = \parallelComposition_{i=1}^{n} \depositsPre{A_i}{\balance[i]}{\vec{x^i}} \\
    \end{gathered}
}
\]

%% file: theory_docs/compiler.tex
\maketitle

\subsection{Introduction and Notation}

\subsubsection{The compiler functions}

The main contract compiler functions are $\mathcal{F}_C$ and $\mathcal{F}_D$ for top-level contracts and guarded contracts respectively. They include a superscript $\someBlockchain$ indicating the target blockchain. They also take as extra input a record of settings (defined in section \ref{subsec:settings-record}) that give context to the compilation.

Other auxiliary functions used during compilation are defined in the ending section.

\subsubsection{Node labels and step secrets}

Our implementation enforces virtually synchronous execution with a mechanism of step secrets and compensations. Every resolution of a priority choice in one blockchain will be either replicated on the other blockchains or the protocol will punish the participant responsible for the asynchronicity and compensate the rest for their loses.

To do this, we first need to tag every node in our contract's syntax tree with a unique label $\someContract$. An auxiliary function that traverses the syntax tree generating the set containing all these labels is defined in section \ref{sec:auxiliary-functions}.

Then, every participant $\participant{A}$ generates for every label $\someContract$ a step secret $\stepSecret{A}{\someContract}$. The reveal of this secret serves as irrefutable of proof of $\participant{A}$'s intent to execute the subcontract labeled as $\someContract$.

The compilation of every guarded contract will require revealing a step secret from any participant but with the current node label. On section \ref{subsec:compile-priority-choice}, we will use this same step secret to punish asynchronous behaviour.

\subsubsection{Compiler Settings}
\label{subsec:settings-record}
BitMLx contracts compilation needs to keep track of many variables relating to funds, step secrets and time. However, having functions with 10 arguments would result in very poor readability. For that reason, we encapsulate everything other than the contract we are compiling in a compiler settings record:

\begin{align*}
    &\settings{\Omega} ::= \{ \\
    &\quad participants, & \text{A list containing all the participants in the contract.}\\
    &\quad balance, & \text{Logical balance at the BitMLx level.}\\
    &\quad collateral, & \text{How much each participant deposits as collateral.}\\
    &\quad currentTime, & \text{A time counter for priority choices.}\\
    &\quad timeElapse, & \text{Safety time period needed for an action.}\\
    &\quad currentLabel, & \text{Label of the node we are compiling.}\\
    &\quad stepSecrets, & \text{A set of all step secrets for each participant and node label in the contract.}\\
    &\quad initSecrets, & \text{A set of all init secrets for each participant.}\\
    &\}
\end{align*}

We refer to an attribute in the record using dot notation. For example, $\settings{\Omega}.participants$ is the $participants$ attribute of the settings record $\settings{\Omega}$. We will use replace syntax to create updated settings records. For example, $\settings{\Omega'} = \settings{\Omega}[balance \mapsto b]$ means that $\settings{\Omega'}$ is the record with same values as $\settings{\Omega}$, except for $balance$ where the new value is $b$.

Initializing the settings record is discussed in section \ref{sec:initial_settings}

\subsubsection{Lists and sets by comprehension}
We write lists and sets by comprehension when convenient. For example $\{f(x) ~|~ x \leftarrow X\}$ is the set of elements  $f(x)$ for every $x \in X$. Similarly, $[f(x_ii) ~|~ x_i \leftarrow \vec{x}]$ is the list whose elements are $f(x_i)$ for every element $x_i$ of $\vec{x}$, in the same order.

\subsection{Guarded Contracts}

\subsubsection{Withdraw}

A withdraw contract is compiled as a BitML split between withdraw contract, where all participants withdraw their collateral and whatever funds they are assigned in the parameter mappings, if any.

The auxiliary function $\outCompiler{\someBlockchain}$ that assigns the exact amount each participant withdraws, is defined below in section \ref{sec:auxiliary-functions}. The result of this auxiliary function is a list of values in the cryptocurrency of $\someBlockchain$.

In this guarded contract version, we also require the reveal of any step secret corresponding to this execution step.

\[
\infer[D-Withdraw]{
    \guardedCompiler{\someBlockchain}(\contract{D}, \settings{\Omega}) =\sum_{\stepSecret{A}{\someContract} \in S}
            \bitmlcode{reveal} ~ \stepSecret{A}{\someContract} ~. ~
                \bitmlcode{split}
                ~ \vec{\mu}
                \rightarrow \vec{\contract{C^{\someBlockchain}}}
}
{
    \begin{gathered}
        \contract{D} = \bitmlxWithdraw{\balance}{A} \\
        S = \{ \stepSecret{A}{\someContract} \in \settings{\Omega}.stepSecrets : \someContract = \settings{\Omega}.currentLabel\} \\
        \vec{\mu} = \outCompiler{\someBlockchain}(\vec{\participant{A}}, \vec{\balance}, \settings{\Omega}) \\
        \vec{\contract{C^{\someBlockchain}}} = [
            \bitmlcode{withdraw} ~ \participant{A_i}
            ~ | ~ \participant{A_i} \leftarrow \settings{\Omega}.participants
        ]\\
    \end{gathered}
}
\]

\subsubsection{Split}

To compile a \bitmlcode{split} guarded contract among subcontracts $\vec{\contract{C}}$, we choose the appropriate funds allocation list, according to the target blockchain, compile the subcontracts and return the BitML \bitmlcode{split} with both, conditioned by the reveal of any step secret.

The settings for the subcontracts compilation update the balance and collateral apropriately and add the branch index to the new label. Notice that the label used to trigger the execution (and on the other side, the punishment) does not include the branch index.

\[
\infer[D-Split]{
    \guardedCompiler{\someBlockchain}(\contract{D}, \settings{\Omega}) =
        \sum_{\stepSecret{A}{\someContract} \in S}
            \bitmlcode{reveal} ~ \stepSecret{A}{\someContract} ~. ~
                \bitmlcode{split}
                    ~ \vec{\mu}
                    \rightarrow
                    \vec{\contract{C^{\someBlockchain}}}
}
{
    \begin{gathered}
        \contract{D} = \bitmlxSplit{\balance}{C} \\
        k = |\vec{\balance}| = |\vec{\contract{C}}| \\
        n = |\settings{\Omega}.participants| \\
        b = \settings{\Omega}.balance \\
        \someContract = \settings{\Omega}.currentLabel \\
        t = \settings{\Omega}.currentTime \\
        \secDelay = \settings{\Omega}.timeElapse \\
        S = \{ \stepSecret{A}{\someContract} \in \settings{\Omega}.stepSecrets : \someContract = \settings{\Omega}.currentLabel\} \\
    \end{gathered} &
    \begin{gathered}
                \forall i \in 1,\dots,k: ~b_i = \balance[i][\someBlockchain] \\
        \forall i \in 1,\dots,k: ~c_i = (n - 2) ~b_i \\
        \forall i \in 1,\dots,k: ~\mu_i = b_i + n ~ c_i \\
        \begin{aligned}
                &\forall i \in 1,\dots,k: ~\settings{\Omega_i} = \settings{\Omega}[\\
        &\quad balance \mapsto b_i, \\
        &\quad collateral \mapsto c_i, \\
        &\quad currentLabel \mapsto \someContract ~|~ i, \\
        &]\\ 
        \end{aligned} \\
        \vec{\contract{C^{\someBlockchain}}} = [
            \topLevelCompiler{\someBlockchain}(\contract{C_i}, \settings{\Omega_i})
            ~ | ~ \contract{C_i} \leftarrow \vec{\contract{C}}
        ]\\
    \end{gathered}
}
\]

\subsubsection{Authorization}

An authorization is with the list of participants $\participant{A}$ is compiled to the sequential authorizations by all $\participant{A_i}$s with the reveal of the step secret.

We are assuming that the $wellFormed$ check prohibits $\contract{D'}$ from also being an authorization.

\[
\infer[D-Auth]{
    \guardedCompiler{\someBlockchain}(\contract{D}, ~ \settings{\Omega}) =
        \sum_{i=1}^{k}
            \big( \participant{A_1}: ~\dots ~:\participant{A_n}:
                \contract{D^{'\someBlockchain}_i} \big)
}
{
    \begin{gathered}
        \contract{D} = \vec{\participant{A}}: \contract{D'} \\
        \vec{\participant{A}} = \participant{A_1}, ~\dots~ ,\participant{A_n}\\
        \sum_{i=1}^{k} \contract{D^{'\someBlockchain}_i} = \guardedCompiler{\someBlockchain}(\contract{D'}, \settings{\Omega}) \\
    \end{gathered}
}
\]

\subsubsection{Reveal}

To compile a \bitmlcode{reveal} statement, we compile the subcontract and wrap it with a BitML \bitmlcode{reveal} conditioned by both the original secrets and the step secrets, keeping the predicate $p$.

\[
\infer[D-Reveal]{
    \guardedCompiler{\someBlockchain}(\contract{D}, ~ \settings{\Omega}) =
        \sum_{\stepSecret{A}{\someContract} \in S} \bitmlcode{reveal} ~ \vec{\secret{s}}|\stepSecret{A}{\someContract} ~
            \bitmlcode{if} ~ p ~
            . ~ \contract{C}^{\someBlockchain{}}
}
{
    \begin{gathered}
        \contract{D} = \bitmlxReveal{s}{p}{C} \\
        S = \{ \stepSecret{A}{\someContract} \in \settings{\Omega}.stepSecrets : \someContract = \settings{\Omega}.currentLabel\} \\
        \contract{C^{\someBlockchain}} = \topLevelCompiler{\someBlockchain}(\contract{C}, \settings{\Omega}) \\
    \end{gathered}
}
\]

\subsection{Top-level Contracts}

\subsubsection{Priority Choice}
\label{subsec:compile-priority-choice}

We want to compile a priority choice $\contract{C}$ between a guarded contract $\contract{D}$ and a contract $\contract{C'}$ with label $l$ at time $t$ and with time elapse $\secDelay$, for the target blockchain $\someBlockchain$.

We first compile $\contract{D}$ to $\contract{D^{\someBlockchain}}$, using the current label appended with an $'L'$ and increasing the time by one elapse. Similarly, we compile $\contract{C'}$ to $\contract{C'^{\someBlockchain}}$ appending $'R'$ to the label and increasing the time by two elapses.

The compilation result then, is the choice between executing $\contract{D^{\someBlockchain}}$ or, after $t + \secDelay$, skipping. When skipping, we can either punish (defined further below in section \ref{sec:auxiliary-functions}) or, after after $t + \secDelay$, move to $\contract{C'^{\someBlockchain}}$.

\[
\infer[C-PriorityChoice]{
    \topLevelCompiler{\someBlockchain}(\contract{C}, \settings{\Omega}) = 
            \contract{D^{\someBlockchain}}
            ~ + ~ \bitmlcode{after} ~ (t + \secDelay): ~
                \tau(\punishCompiler{\someBlockchain}(\settings{\Omega_{D}})
                ~ + ~ \bitmlcode{after} ~ (t + 2\secDelay): ~ \tau(\contract{C'^{\someBlockchain}}))
}
{
    \begin{gathered}
        \contract{C} = \contract{D} \prchoice \contract{C'} \\
        t = \settings{\Omega}.currentTime \\
        \secDelay = \settings{\Omega}.timeElapse \\
        \someContract = \settings{\Omega}.currentLabel \\
        \settings{\Omega_{D}} = \settings{\Omega}[
            currentTime \mapsto t + 2\secDelay,
            ~ currentLabel \mapsto \someContract|'L'
        ] \\
        \settings{\Omega_{C'}} = \settings{\Omega}[
            currentTime \mapsto t + 2\secDelay,
            ~ currentLabel \mapsto \someContract|'R'
        ] \\
        \contract{D^{\someBlockchain}} = \guardedCompiler{\someBlockchain}(\contract{D}, \settings{S_{D}}) \\
        \contract{C'^{\someBlockchain}} = \topLevelCompiler{\someBlockchain}(\contract{C'}, \settings{S_{C'}}) \\
    \end{gathered}
}
\]
The auxiliary function $\tau$ is an empty \bitmlcode{reveal} statement and it works as a small hack to turn a BitML $C$ (a sum of choices) into a $D$ (a guarded contract). This is needed here because the \bitmlcode{after} clause only takes $D$ as inputs. The trade-off here is that doing this reveal step, is compiled as publishing a new transaction.

\[
    \tau(\contract C) = reveal [] ~.~ \contract C 
\]

\subsubsection{Withdraw}

A top-level BitMLx withdraw contract is compiled in a very similar way to its guarded counterpart, only removing the requirement for revealing a step secret.

\[
\infer[C-Withdraw]{
    \topLevelCompiler{\someBlockchain}(\contract{C}, \settings{\Omega}) = 
            \bitmlcode{split}
            ~ \vec{\mu}
            \rightarrow \vec{\contract{C^{\someBlockchain}}}
}
{
    \begin{gathered}
        \contract{C} = \bitmlxWithdraw{v}{A} \\
        \vec{\mu} = \outCompiler{\someBlockchain}(\vec{\participant{A}}, ~\vec{v}, ~\settings{\Omega}) \\
        \vec{\contract{C^{\someBlockchain}}} = [
            \bitmlcode{withdraw} ~ \participant{A_i}
            ~ | ~ \participant{A_i} \leftarrow \settings{\Omega}.participants
        ]\\
    \end{gathered}
}
\]

\subsection{Stipulation Protocol}

\subsubsection{Initial Settings}\label{sec:initial_settings}

The following function generates an initial settings record for a contract advertisement. We basically need to
\begin{itemize}
    \item Define the set of participants as the ones with deposits.
    \item Add up the balance from the deposits on this target blockchain.
    \item Determine the collaterals according to our formula.
    \item Initialize the label with the empty string $\epsilon$.
    \item Compute the set of step secrets using the auxiliary functions $\someContract_C$ and $\someContract_D$ defined below.
    \item Extract the starting time $t_0$ and time elapse $\secDelay$ from the timing precondition.
\end{itemize}

\[
\infer[S-Settings]{
    \settingsCompiler{\someBlockchain{}}(\contractAdv{G}{C}) = \settings{\Omega}
}
{
            \begin{gathered}
                                                                                                                b = \partialBalance{\someBlockchain}{\preconditions{G}} \\
                P = \participants{G} \\
                (t_0, \someStipulation) = initSettings({\preconditions{G}}) \\
                \Lambda = \nodes{\contract{C}}{[\someContract_0]} \\
            \end{gathered} &
            \begin{gathered}
                \begin{aligned}
                    &\settings{\Omega} = \{ \\
                    &\quad participants = P \\
                    &\quad balance = b \\
                    &\quad collateral = (|P|-2)b\\
                    &\quad currentLabel = \someContract | `L` \\
                    &\quad stepSecrets =
                        \bigcup\limits_{\substack{\participant{A} \in P \\ \someContract \in \Lambda}}
                        \{\stepSecret{A}{\someContract}\} \cup \bigcup\limits_{\substack{\participant{A} \in P}} \{ \stepSecret{A}{\someContract_0} \}\\
                                                                                &\quad currentTime = t_0 \\
                    &\quad timeElapse = \secDelay \\
                    &\}
                \end{aligned}
            \end{gathered}
}
\]

\subsubsection{Preconditions}

When compiling preconditions, we need to:

\begin{itemize}
    \item Pick the right deposits for the target blockchain and add the collaterals.
    \item Keep all regular secrets and add step secrets.
\end{itemize}

We will divide this function into two different rules, one for each target blockchain.

\[
\infer[S-Preconditions]{
    \preconditionsCompiler{\someBlockchain{}}(\preconditions{G}, \settings{\Omega}) = \preconditions{G}^{\someBlockchain}
}
{
    \begin{gathered}
        \begin{aligned}
        & \preconditions{G} = \parallelComposition_{i=1}^n \big(\depositsPre{A_i}{\balance[i]}{\vec{x}} \big) \\
        &\quad | \parallelComposition_{j=1}^m \big(\participant{A_j}: \texttt{secret } \secret{s_{j}} \big ) \\
        &\quad | ~(t_0, \someStipulation) \\    
        \end{aligned}
    \end{gathered} &
                                                \begin{gathered}
        c = \settings{\Omega}.collateral \\
        \begin{aligned}
                & \preconditions{G} = \parallelComposition_{i=1}^n \big(\depositsPre{A_i}{(\balance[i][\someBlockchain]+c)}{(\vec{x}[\someBlockchain])} \big) \\
                                        &\quad | \parallelComposition_{j=1}^m \big(\participant{A_j}: \texttt{secret } \secret{s_{j}} \big ) \\
                                                        \end{aligned} \\
    \end{gathered}
}
\]

\[
\infer[S-GlobalSecrets]{
    \secretsCompiler(\preconditions{G}) = \secret{S}
}
{
    \begin{gathered}
        P = \participants{G} \\
        (t_0, \someStipulation) = initSettings({\preconditions{G}}) \\
        \Lambda = \nodes{\contract{C}}{\someContract_0} \\
                                                            \end{gathered} &
    \begin{gathered}
        \begin{aligned}
        & \secret{S} = \parallelComposition\limits
            _{\substack{\participant{A} \in P \\
                \someContract \in \Lambda}}
            \big(\participant{A}: \texttt{secret } \stepSecret{A}{\someContract} \big ) \\
        &\quad~ | \parallelComposition\limits_{\substack{\participant{A} \in P}}
            \big(\participant{A}: \texttt{secret } \initSecret{A}{\someContract_0} \big ) \\
        \end{aligned}
    \end{gathered}
}
\]

\subsubsection{Refund}

The refund contract will be executed on a cooperative abort and will just return every participant it's deposit. 

\[
\infer[S-Refund]{
    \refundCompiler(\preconditions{G}, \settings{\Omega}) = \bitmlcode{split}
            ~ \vec{\mu}
            \rightarrow \vec{\contract{C^{\someBlockchain}}}
}
{
    \begin{gathered}
       \contract{C} = \bitmlxWithdraw{\balance}{A} \\
        \vec{\mu} = \outCompiler{\someBlockchain}(\vec{\participant{A}}, ~\vec{\balance}, ~\settings{\Omega}) \\
        \vec{\contract{C^{\someBlockchain}}} = [
            \bitmlcode{withdraw} ~ \participant{A_i}
            ~ | ~ \participant{A_i} \leftarrow \settings{\Omega}.participants
        ]\\ 
    \end{gathered}
    \begin{gathered}
        \begin{aligned}
        & \preconditions{G} = \parallelComposition_{i=1}^n \big(\depositsPre{A_i}{\balance[i]}{\vec{x^i}} \big) \\
        &\quad | \parallelComposition_{i=1}^m \big(\participant{A_j}: \texttt{secret } \secret{s_{j}} \big ) \\
        &\quad |(t_0, \someStipulation) \\    
        \end{aligned} \\
    \end{gathered} &
    \begin{gathered}
        \vec{\balance} = \balance[1], ~\dots~ , \balance[n] \\
        \vec{\participant{A}} = [\participant{A_1}, ~\dots~ ,\participant{A_n}] \\
    \end{gathered} \\
}
\]

\subsubsection{Advertisement}

This is the entry point of our compiler. To compile a contract advertisement $\contractAdv{G}{C}$, we need to:
\begin{itemize}
    \item Check that the advertisement is well-formed.
    \item Generate the initial compilation settings.
    \item Compile the preconditions.
    \item Create a refund contract.
    \item Compile the contract in a priority choice with our refund contract.
    \item Additionally, we guard the contract with the requirement that all participants reveal their init secret. This condition means that all participants will explicitly agree to starting.
\end{itemize}

\[
\infer[S-Stipulation]{
    \stipulationCompiler^{\someBlockchain}(\contractAdv{G}{C}, \settings{\Omega}) = 
            \contract{D^{\someBlockchain}}
            ~ + ~ \bitmlcode{after} ~ (t + \secDelay): ~
                \tau(\punishCompiler{\someBlockchain}(\settings{\Omega_{D}})
                ~ + ~ \bitmlcode{after} ~ (t + 2\secDelay): ~ \contract{C'^{\someBlockchain}})
}
{
    \begin{gathered}
                t = \settings{\Omega}.currentTime \\
        \secDelay = \settings{\Omega}.timeElapse \\
        \someContract = \settings{\Omega}.currentLabel \\
        \settings{\Omega_{D}} = \settings{\Omega}[
            currentTime \mapsto t + 2\secDelay,
            ~ currentLabel \mapsto \someContract|'L'
        ] \\
        \settings{\Omega_{C'}} = \settings{\Omega}[
            currentTime \mapsto t + 2\secDelay,
            ~ currentLabel \mapsto \someContract|'R'
        ] \\
        P = \participants{G} \\
        (t_0, \someStipulation) = initSettings({\preconditions{G}}) \\
        IS = \bigcup_{\participant{A} \in P} \{\initSecret{A}{\someContract_0}\} \\
        \contract{D^{\someBlockchain}} =\guardedCompiler{\someBlockchain}(\big( \bitmlcode{reveal} ~ IS ~ \bitmlcode{then} ~ \contract{C} \big) , \settings{\Omega_{D}}) \\
        \contract{C'^{\someBlockchain}} = \topLevelCompiler{\someBlockchain}(\refundCompiler(\preconditions{G}), \settings{\Omega_{C'}}) \\
    \end{gathered}
}
\]
\[
\infer[S-Advertise]{
    \advCompiler(\contractAdv{G}{C}) =
                \batchAdvertisement{S}{G}{C}
}
{
    \begin{gathered}
        wellFormed(\contractAdv{G}{C}) \\
                                                \forall \someBlockchain \in \chains{G}:
            \settings{\Omega^{\someBlockchain}} =
                \settingsCompiler{\someBlockchain}(\contractAdv{G}{\contract{C_x}}) \\
        \forall \someBlockchain \in \chains{G}:
            \preconditions{G}^{\someBlockchain} =
                \preconditionsCompiler{\someBlockchain}(
                    \preconditions{G},
                    ~\settings{\Omega^{\someBlockchain}}
                ) \\
        \forall \someBlockchain \in \chains{G}:
            \contract{C}^{\someBlockchain} =  \stipulationCompiler^{\someBlockchain}(
                \contractAdv{G}{C},
                ~\settings{\Omega^{\someBlockchain}}
            ) \\
        \overrightarrow{\contractAdv{G}{C}} = [\contractAdv{\preconditions{G}^{\someBlockchain}}{\contract{C}^{\someBlockchain}}: \forall \someBlockchain \in \chains{G}]\\
        \secret{S} = \secretsCompiler(\preconditions{G}) \\
    \end{gathered}
}
\]

\subsection{Auxiliary Functions}\label{sec:auxiliary-functions}

\subsubsection{Step secret generation}

\[
\begin{aligned}
    &nodes(\contract{D} \prchoice \contract{C}, \someContract) =
        nodes_D(\contract{D}, \someContract|'L')
        ~\cup~
        nodes(\contract{C}, \someContract|'R') \\
    &nodes(\bitmlxWithdraw{\balance}{A}, \someContract) = \{\} \\
\end{aligned}
\]

\[
\begin{aligned}
    &nodes_D(\bitmlxSplit{\balance}{C}, \someContract) =
        \{\someContract\} ~\cup~
        \bigcup_{\contract{C_i} \in \vec{\contract{C}}}
            nodes(\contract{C_i}, \someContract|i) \\
    &nodes_D(\bitmlxReveal{s}{p}{C}, \someContract) = \{\someContract\} ~\cup~ nodes(\contract{C}, \someContract) \\
    &nodes_D(\vec{\participant{A}}: \contract{D}, \someContract) = nodes_D(\contract{D}, \someContract) \\
    &nodes_D(\bitmlxWithdraw{\balance}{A}, \someContract) = \{\someContract\} \\
\end{aligned}
\]

\subsubsection{Compensate}

The compensation contract is a sum of reveals where for each step secret at the current label, we give every participant, except the owner of the revealed step secret, a balance $b$ worth of the cryptocurrency on $\someBlockchain$.

\[
\infer[Compensate]{
    \punishCompiler{\someBlockchain}(\settings{\Omega}) =
        \sum_{i = 1}^{n}
            \bitmlcode{reveal} ~ \stepSecret{A_i}{\someContract} ~ . ~
                (\bitmlcode{split} ~ \vec{v^i} \rightarrow ~ \vec{\contract{C^i}})
}
{
    \begin{gathered}
        \participant{A_1}, ~\dots, ~\participant{A_n} = \settings{\Omega}.participants \\
        b = \settings{\Omega}.balance \\
        \someContract = \settings{\Omega}.currentLabel \\
        \forall i \in [1, \dots, n]: ~ \stepSecret{A_i}{\someContract} \in \settings{\Omega}.stepSecrets \\
        \forall i \in [1, \dots, n]: ~ \vec{v^i} = (b)^{n-1} \\
        \forall i \in [1, \dots, n]: ~ \vec{\contract{C^i}} = [
            \bitmlcode{withdraw} ~ \participant{A_j}  ~ 
            | ~ j \leftarrow [1, ~\dots, ~n] \backslash \{i\}
        ] \\
    \end{gathered}
}
\]

\subsubsection{Withdraw outputs}

The following functions compute how many bitcoins and how many dogecoins should withdraw each participant, by adding their collateral, and whatever was stateted in the BitMLx $withdraw$ statement.

\[
\infer[Out-\someBlockchain]{
    \outCompiler{\someBlockchain}(\vec{\participant{A}}, ~\vec{\balance}, ~\settings{\Omega}) =
        [v_{\participant{A}} ~ | ~ \participant{A} \leftarrow P]
}
{
    \begin{gathered}
        \vec{\participant{A}} = \participant{A_1}, ~\dots~ ,\participant{A_n} \\
        \vec{\balance} = \balance[1], ~\dots~ ,\balance[n] \\
        c = \settings{\Omega}.collateral \\
        P = \settings{\Omega}.participants \\
        \forall \participant{A} \in P:
            ~ v_{\participant{A}} =
            \begin{cases}
                \balance[i][\someBlockchain] + c & \text{if} ~ \exists i: \participant{A} = \participant{A_i} \\
                c & \text{otherwise}
            \end{cases} 
    \end{gathered}
}
\]

%% file: theory_docs/intermediate_semantics.tex
\maketitle

\subsection{Configurations}

Intermediate semantics configurations look very similar to \bitmlx configurations, but deposits and contracts reside now on a single blockchain, indicated by a superscript. They also have a state record $\isContractState$, with a lot more context on the contract, instead of just the balance. One of them, is the "contract status", which models an internal state machine for the contracts. The status can be:

\begin{itemize}
    \item A $\isStatusChoice$ contract is in it's neutral state, waiting to de between the left or right move.
    \item A $\isStatusLeft$ contract has executed the \bitmlcode{reveal} statement corresponding to the step secret, but didn't execute the inner statement yet.
    \item An $\isStatusAssigned{A}$ contract has finished executing and can be withdrawn by A.
    \item A $\isStatusRight$ contract has executed the $\tau$ step of disabling tthe left move and is waiting for possible compensations.
    \item A $\isStatusSlashed{\participant{A}}$ contract has executed the \bitmlcode{reveal} statement corresponding to compensating for a late step secret reveal from user $\participant{A}$, but has not executed the inner \bitmlcode{split} yet.
    \item A $\isStatusCompensated{\participant{A}}$ contract has executed the \bitmlcode{split} statement corresponding to compensating for a late step secret reveal from user $\participant{A}$.
    \item A $\isStatusStipChoice$ contract is going through the stipulation protocol. It is just published and needs to either initialize or abort.
    \item A $\isStatusStipRight$ contract is going through the stipulation protocol. It has taken the sright $\tau$ step and is waiting for possible compensations before refunding everyone's deposit.
    \item A $\isStatusStipSlashed{A}$ contract is going through the stipulation protocol. It has executed the \bitmlcode{reveal} statement corresponding to compensating for a late initial step secret reveal from user $\participant{A}$, but has not executed the inner \bitmlcode{split} yet.
    \item A $\isStatusStipCompensation{A}$ contract is going through the stipulation protocol. It has executed the \bitmlcode{split} statement corresponding to compensating for a late initial step secret reveal from user $\participant{A}$.
    \item A $\isStatusStipRefunded{\participant{A}}$ contract is going through the stipulation protocol. It has executed the \bitmlcode{split} statement corresponding to refunding their deposits and is now assigned to user $\participant{A}$.
\end{itemize}

Finally, there are new secret objects that explicitly differentiate step secrets and init secrets from logical secrets (secrets that are part of the \bitmlx contract logic).

\begin{align*}
    & \isConfiguration \Coloneqq \isContractAdv{G}{C}{\someBlockchain} \\
    & \quad \vert ~ \isActiveContract{\contract{C}}{\isContractState}{\someBlockchain}{\someContract, x} \\
    & \quad \vert ~ \isParticipantDeposit{\participant{A}}{v}{\someBlockchain}{x} \\
    & \quad \vert ~ \participant{A}: [\phi] \\
    & \quad \vert ~ \secretCommitment{A}{s}{N} \\
    & \quad \vert ~ \secretReveal{A}{s}{N} \\
    & \quad | ~\isCommitedStepSecret{A}{\someContract} \\
    & \quad | ~\isRevealedStepSecret{A}{\someContract} \\
    & \quad | ~\isCommitedInitSecret{A}{\someContract} \\
    & \quad | ~\isRevealedInitSecret{A}{\someContract} \\
    & \quad \vert ~ \isConfiguration \| \isConfiguration[\secDelay] \\
    & \isConfiguration ~ \vert ~ t \\
\end{align*}

Active contracts need to store a lot of additional information, so we hide all of this into a state record $\isContractState$.

\begin{align*}
    & \isContractState \Coloneqq \{ \\
    & \quad balance: v \\
    & \quad status: s \\
    & \quad time: ~t \\
    & \quad participants: P \\
    & \quad deposits: D \\
        & \}
\end{align*}

\subsection{Secrets}

The $|-RevealSecret$ rule says that if $\participant{A}$ really has the value for a secret they commited, then they can always reveal it. Remember that dishonest users can commit to any secret, even if they don't actually know the value, which is represented by the $\bot$ length.

\[
\infer[||-RevealSecret]
{
    \reductionRule
        {
            \secretCommitment{A}{a}{N} ~ \vert ~ \isConfiguration
        }
        {\participant A:\secret a}
        {
            \secretReveal{A}{a}{N} ~ \vert ~ \isConfiguration
        }
}
{
    \begin{gathered}
        N \neq \bot
    \end{gathered}
}
\]

\subsection{Priority choice internals}

These rules handle the tracking and use of step secrets to transition among the different internal states we use to implement priority choices.

\subsubsection{Authorize a left}

Moving a contract to an intermediate $left$ state requires a revealed step secret. Any commited step secret can be revealed at any time. Honest users will only do so when they want to execute a left side choice, but they need to be able to parse and keep track of malicious users revealing theirs at any time.

\[
\infer[||-StepSecretRev]
{
    \reductionRule
        {
            \isCommitedStepSecret{A}{\someContract}
            ~\vert~ \isConfiguration
        }
        {\isRevealedStepSecret{A}{\someContract}}
        {
            \isRevealedStepSecret{A}{\someContract}
            ~\vert~ \isConfiguration
        }
}
{}
\]

\subsubsection{Introduction of left}

This step corresponds to executing the \bitmlcode{reveal} statement guarded by the step secret and transitioning to the actual $D$ contract, when $D$ is not a \bitmlcode{reveal} itself. The $left$ modifier will then be eliminated by the actual execution of $D$ to the rules on the next section. The step secret is not consumed in the transition and it can be used in the other chain for either replication or compensation.

We additionally require any needed signatures for auth guards.

\[
\infer[||-ILeft]
{
    \reductionRule
        {
            \isActiveContract{
                \vec{\participant{A}}:
                    \contract{D}
                \prchoice
                \contract{C}}{\isContractState}{\someBlockchain}{\someContract, x}
            ~\vert~ ~\isRevealedStepSecret{A}{\someContract}
            ~\vert~ sigs
            ~\vert~ \isConfiguration
        }
        {ileft(\someContract, \someBlockchain, x)}
        {
            \isActiveContract{\contract{D} \prchoice \contract{C}}{\isContractState}{\someBlockchain}{\someContract, x'}
            ~\vert~ ~\isRevealedStepSecret{A}{\someContract}
            ~\vert~ \isConfiguration
        }
}
{
    \begin{gathered}
    \contract{D} \neq \bitmlcode{reveal} ~\vec{s} ~\bitmlcode{if} ~p ~\bitmlcode{then} ~\contract{C'} \\
    \contract{D} \neq \vec{\participant{A'}}: \contract{D'} \\
    \isContractState = [status=\isStatusChoice, ~\dots] \\
    sigs = \parallelComposition_{\participant{A_i} \in \vec{\participant{A}}}
                    \participant{A_i} [(\someContract, \someBlockchain, x) ~ \triangleright ~ \contract{D}]\\
    \isContractState' = \isContractState[
        status \mapsto \isStatusLeft
    ] \\
    x' ~ \text{fresh} \\
    \end{gathered}
}
\]

The \bitmlcode{reveal} statement needs special treatment. The reveal of \bitmlx secrets is merged with that of step secrets and there is no intermediate state before transitioning to the follow-up contract.

\[
\infer[||-Reveal]
{
    \reductionRule
        {
                \isActiveContract{
                    \vec{\participant{A}}:
                        \contract{D}
            \prchoice
                \contract{C}}{\isContractState}{\someBlockchain}{\someContract, x}
            ~\vert~ ~\isRevealedStepSecret{A}{\someContract}
            ~\vert~ secrets
            ~\vert~ sigs
            ~\vert~ \isConfiguration
        }
        {reveal(\someContract, \someBlockchain, x)}
        {
            \isActiveContract{\contract{C'}}{\isContractState'}{\someBlockchain}{\someContract', x'}
            ~\vert~ ~\isRevealedStepSecret{A}{\someContract}
            ~\vert~ secrets
            ~\vert~ \isConfiguration
        }
}
{
    \begin{gathered}
    \contract{D} = \bitmlcode{reveal} ~\vec{s} ~\bitmlcode{if} ~p ~\bitmlcode{then} ~\contract{C'} \\
    \isContractState = [time = t, status=\isStatusChoice, ~\dots] \\
    sigs = \parallelComposition_{\participant{A_i} \in \vec{\participant{A}}}
                    \participant{A_i} [(\someContract, \someBlockchain, x) ~ \triangleright ~ \contract{D}]\\
    secrets = \parallelComposition_{\secret{s_j} \in \vec{s}}
            \secretReveal{A_j}{s_j}{N_j} \\
    \llbracket p \rrbracket_{secrets} = true \\
    \isContractState' = \isContractState[
        time \mapsto t + 2\secDelay
    ] \\
        \someContract' = \someContract | L
    \end{gathered}
}
\]

The signatures can be added with $||-AuthControl$ rule.

\[
\infer[||-AuthControl]
{
    \reductionRule
        {
            \isActiveContract{\vec{\participant{A}}: \contract{D} \prchoice \contract{C}}{\isContractState}{\someBlockchain}{\someContract, x}
            ~\vert~ \isConfiguration
        }
        {\participant{A}: (\someContract, \someBlockchain, x)}
        {   
            \isActiveContract{\vec{\participant{A}}: \contract{D} \prchoice \contract{C}}{\isContractState}{\someBlockchain}{\someContract, x}
            ~|~ \participant{A_i} [(\someContract, \someBlockchain, x) ~ \triangleright ~ \contract{D}]
            ~\vert~ \isConfiguration
        }
}
{
    \begin{gathered}
        \isContractState = [status=\isStatusChoice, ~\dots] \\
        \participant{A_i} \in \vec{\participant{A}} \\
    \end{gathered}
}
\]

\subsubsection{Introduction of right}

$ISkip$ moves the contract to an intermediate state where we can either move right or compensate. This corresponds in the BitML level to executing the $\tau$ (empty reveal) statement to discard $D$.

\[
\infer[||-IRight]
{
    \reductionRule
        {
            \isActiveContract{\contract{D} \prchoice \contract{C}}{\isContractState}{\someBlockchain}{\someContract, x}
            ~\vert~ \isConfiguration
        }
        {from ~ t: ~ skip(\someContract, \someBlockchain, x)}
        {
            \isActiveContract{\contract{D} \prchoice \contract{C}}{\isContractState'}{\someBlockchain}{\someContract, x'}
            ~\vert~ \isConfiguration
        }
}
{
    \begin{gathered}
    \isContractState = [time = t, status=\isStatusChoice, ~\dots] \\
    \isContractState' = \isContractState [
        time \mapsto t + \secDelay,
        ~status \mapsto \isStatusRight
    ] \\
        \end{gathered}
}
\]

\subsubsection{Elimination of right}

A skip contract can finish skipping and move to the right side of the priority choice after some time condition is met. The label gets updated accordingly.

\[
\infer[||-ERight]
{
    \reductionRule
        {
            \isActiveContract{\contract{D} \prchoice \contract{C'}}{\isContractState}{\someBlockchain}{\someContract, x}
            ~\vert~ \isConfiguration
        }
        {from ~ t: ~ right(\someContract, \someBlockchain, x)}
        {
            \isActiveContract{\contract{C'}}{\isContractState'}{\someBlockchain}{\someContract', x'}
            ~\vert~ \isConfiguration
        }
}
{
    \begin{gathered}
    \isContractState = [time = t, status=\isStatusRight, ~\dots] \\
    \isContractState' = \isContractState [
        time \mapsto t+\secDelay,
        ~status \mapsto \isStatusChoice
    ] \\
    x' ~ \text{fresh} \\
    \someContract' = \someContract | R
    \end{gathered}
}
\]

\subsubsection{Introduction of Compensation}

A skipping contract can also transition to a compensation when there's a step secret for the corresponding left choice in the context. Again, the step secret is not consumed because it might be needed for the other blockchain.

Notice that we transition to an intermediate state and not directly to the withdraw states, because this is implemented as a reveal step.

\[
\infer[||-ICompensation]
{
    \reductionRule
        {
            \isActiveContract{\contract{D} \prchoice \contract{C}}{\isContractState}{\someBlockchain}{\someContract, x}
            ~\vert~ \isRevealedStepSecret{A}{\someContract}
            ~\vert~ \isConfiguration
        }
        {slash(\someContract, \someBlockchain, x, \participant{A})}
        {
            \isActiveContract{\contract{D} \prchoice \contract{C}}{\isContractState'}{\someBlockchain}{\someContract, x'}
            ~\vert~ \isRevealedStepSecret{A}{\someContract}
            ~\vert~ \isConfiguration
        }
}
{
    \begin{gathered}
    \isContractState = [status=\isStatusRight, ~\dots] \\
        \isContractState' = \isContractState [
        status \mapsto \isStatusSlashed{\participant{A}}
    ] \\
    x' ~ \text{fresh} \\
    \end{gathered}
}
\]

\subsubsection{Elimination of Compensation}

A compensation is just a withdraw with a special name that will help us later on. Because we move to withdrawals, we add the collaterals.

\[
\infer[||-ECompensation]
{
    \reductionRule
        {
            \isActiveContract{\contract{C}}{\isContractState}{\someBlockchain}{\someContract, x}
            ~\vert~ \isConfiguration
        }
        {compensate(\someContract, \someBlockchain, x, \participant{A})}
        {
            \isActiveContract{\contract{C}}{\isContractState'}{\someBlockchain}{\someContract, \vec x}
            ~\vert~ \isConfiguration
        }
}
{
    \begin{gathered}
        \isContractState = [status=\isStatusSlashed{\participant{A}}, ~\dots] \\
        \isContractState' = \isContractState [
            status \mapsto \isStatusCompensated{\participant{A}}
        ] \\
     \vec x ~\fresh \\
    \end{gathered}
}
\]

\subsection{Eliminations of left}

After the $left$ state has been introduced, guarded contracts basically transition just like they would in \bitmlx. Most importantly, they update the $\someContract$ when needed.

\subsubsection{Guarded Withdraw}

A \bitmlcode{withdraw} contract in intermediate left state transitions to assigned funds for each specified participant on the corresponding blockchain.

\[
\infer[||-GuardedWithdraw]
{
    \reductionRule
        {
            \isActiveContract{\contract{D} \prchoice \contract{C}}{\isContractState}{\someBlockchain}{\someContract, x}
            ~\vert~ \isConfiguration
        }
        {dwithdraw(\someContract, \someBlockchain, x)}
        {
            \parallelComposition_{i=1}^{n}
                \isActiveContract{\contract{D} \prchoice \contract{C}}{\isContractState'}{\someBlockchain}{\someContract_i, x_i}
            ~\vert~ \isConfiguration
        }
}
{
    \begin{gathered}
    \contract{D} = \bitmlxWithdraw{\balance}{A} \\
    \isContractState = [status=\isStatusLeft, ~\dots] \\
    \vec{\balance} = \balance[1], \dots, \balance[n] \\
    \vec{\participant{A}} = \participant{A_1}, \dots, \participant{A_n} \\
    \forall i \in 1,\dots,n: \isContractState_i = \isContractState [
        balance \mapsto \balance[i][\someBlockchain],
        ~status \mapsto \isStatusAssigned{A_i}
    ] \\
    \forall i \in 1 \dots n: ~ \someContract_i = \someContract|L_i \\
    x_1,\dots,x_n ~\fresh \\
    \end{gathered}
}
\]

\subsubsection{Split}

A \bitmlcode{split} left contract transitions to the active subcontracts, each with their corresponding balance and label.

\[
\infer[||-Split]
{
    \reductionRule
        {
            \isActiveContract{\contract{D} \prchoice \contract{C}}{\isContractState}{\someBlockchain}{\someContract, x}
            ~\vert~ \isConfiguration
        }
        {split(\someContract, \someBlockchain, x)}
        {
            \parallelComposition_{i=1}^n \isActiveContract{\contract{C_i}}{\isContractState_i}{\someBlockchain}{\someContract_i, x_i}
            ~\vert~ \isConfiguration
        }
}
{
    \begin{gathered}
    \contract{D} = \bitmlxSplit{\balance}{C} \\
    \vec{\contract{C}} = \contract{C_1}, \dots, \contract{C_k} \\
    \vec{\balance} = \balance[1], \dots, \balance[k] \\
    \isContractState = [time = t, ~status=\isStatusLeft, ~\dots] \\
        \forall i \in 1 \dots k: ~
        \isContractState_i = \isContractState[
            balance \mapsto \balance[i][\someBlockchain],
            ~time \mapsto t + 2\secDelay,
            ~status=\isStatusChoice, 
        ] \\
    \forall i \in 1 \dots n: ~ \someContract_i = \someContract|L_i \\
    x_1,\dots,x_n ~\fresh \\
    \end{gathered}
}
\]

\subsection{Top-level Withdraw}

A right-most withdraw needs no step secret and can always be executed.

\[
\infer[||-Withdraw]
{
    \reductionRule
        {
            \isActiveContract{\contract{C}}{\isContractState}{\someBlockchain}{\someContract, x}
            ~\vert~ \isConfiguration
        }
        {cwithdraw(\someContract, \someBlockchain, x)}
        {
            \parallelComposition_{i=1}^{n}
                \isActiveContract{\contract{C}}{\isContractState}{\someBlockchain}{\someContract_i, x_i}
            ~\vert~ \isConfiguration
        }
}
{
    \begin{gathered}
    \contract{C} = \bitmlxWithdraw{\balance}{A} \\
    \vec{\participant{A}} = \participant{A_1}, \dots, \participant{A_n} \\
    \vec{\balance} = \balance[1], \dots, \balance[n] \\
    \isContractState = [status=\isStatusChoice, ~\dots] \\
        \forall i \in 1,\dots,n: \isContractState_i = \isContractState [
        balance \mapsto \balance[i][\someBlockchain],
        ~status \mapsto \isStatusAssigned{A_i}
    ] \\
    \forall i \in 1 \dots n: ~ \someContract_i = \someContract|L_i \\
    x_1,\dots,x_n ~\fresh \\
    \end{gathered}
}
\]

\subsection{Stipulation Protocol}

We advertise all contracts simultaneously, corresponding to advertising a batch. We require users to have enough funds to cover for both the balance and their collateral.

Then of course, we have regular advertisement conditions such as using fresh secrets, having at least one honest user and for the advertisement to be well-formed.

\[
\infer[||-Advertise]
{\reductionRule
    {\isConfiguration}
    {advertise(\contractAdv{G}{C})}
    {
        \parallelComposition_{\someBlockchain \in \activeBlockchains}
            \isContractAdv{G}{C}{\someBlockchain}
        ~\vert~ \isConfiguration
    }
}
{
    \begin{gathered}
    wellFormed(\contractAdv{G}{C}) \\
    (t_0, \someStipulation) = initSettings({\preconditions{G}}) \\
    \someStipulation ~\fresh \\
    \participants{G} \cap \Hon \neq \emptyset \\
    \forall \secret{s} \in \secrets{G}. ~\secret{s} ~\fresh \\
    \forall (\depositsPre{A}{\balance}{\vec{x}}) \in deposits(\preconditions{G}).
        \forall \someBlockchain \in \activeBlockchains.
            ~\isParticipantDeposit
                {\participant{A}}
                {(\balance[][\someBlockchain] + \partialBalance{\someBlockchain}{\preconditions{G}} \times (|\participants{G}|-2))}{\someBlockchain}{x_i}  
            \in \isConfiguration \\
    \end{gathered}
}
\]

In the commit phase, users commit to their logical secrets, their step secrets and their init secrets. These are represented as global secrets in the batch and that's why this is a single action. The $nodes$ function is defined in the compiler theory document.

We don't require freshness conditions on the step and init secrets because we are already using a fresh $\someContract$.

\[
\infer[||-AuthCommit]
{
    \reductionRule
        {
            \parallelComposition_{\someBlockchain \in \activeBlockchains}
                \isContractAdv{G}{C}{\someBlockchain}
            ~|~ \isConfiguration
        }
        {commit(\participant A, \contractAdv{G}{C}, \Delta)}
        {
            \parallelComposition_{\someBlockchain \in \activeBlockchains}
                \isContractAdv{G}{C}{\someBlockchain}
            ~|~ \Delta
            ~|~ \isConfiguration
        }
}
{
    \begin{gathered}
    (t_0, \someStipulation) = initSettings({\preconditions{G}}) \\
    \participant{A} \in \participants{G} \\
    logical = \parallelComposition_{\secret{s} \in secrets_{\participant{A}}(\preconditions{G})}
        \secretCommitment{A}{s}{N} \\
    step = \parallelComposition_{\someContract \in \nodes{C}{\someContract}}
        \isCommitedStepSecret{A}{\someContract} \\
    init = \isCommitedInitSecret{A}{\someContract} ~|~ \isCommitedStepSecret{A}{\someStipulation}\\
    \forall \secretCommitment{A}{s}{N} \in logical.
        ~N \in \begin{cases}
            \mathbb{N} & \text{if} ~\participant{A} \in \Hon \\
            \bot & \text{otherwise} \\
        \end{cases}\\
    \Delta = logical ~|~ step ~|~ init
            ~|~ \parallelComposition_{\someBlockchain \in \activeBlockchains}
                \participant{A}: [\# \triangleright \isContractAdv{G}{C}{\someBlockchain}]
    \end{gathered}
}
\]

Init authorizations are by contract, because each contract requires it's own signatures.

\[
\infer[||-AuthInit]
{
    \reductionRule
        {\isContractAdv{G}{C}{\someBlockchain} ~|~ \isConfiguration}
        {authInit(\participant{A}, \isContractAdv{G}{C}{\someBlockchain})}
        {
            \isContractAdv{G}{C}{\someBlockchain}
            ~|~ \participant{A}: [x \triangleright \isContractAdv{G}{C}{\someBlockchain}]
            ~|~ \isConfiguration }
}
{
    \begin{gathered}
        \forall \participant{A_i} \in \participants{G}.
            ~\participant{A_i}: [\# \triangleright \isContractAdv{G}{C}{\someBlockchain}] \in \isConfiguration \\
        \depositsPre{A}{\balance}{\vec{x}} \in \preconditions{G} \\
        x \in \vec{x} \\
    \end{gathered}
}
\]

An authorized contract can be published on it's blockchain, consuming the deposits and authorizations on it. The result is a starting contract, that will go through the synchronization protocol. 

\[
\infer[||-Publish]
{
    \reductionRule
        {
            \isContractAdv{G}{C}{\someBlockchain}
            ~|~ \Delta
            ~|~ \isConfiguration}
        {publish(\isContractAdv{G}{C}{\someBlockchain})}
        {
            \isActiveContract{\contract{C}}{\isContractState}{\someBlockchain}{\someStipulation, x}
            ~|~ \isConfiguration
        }
}
{
    \begin{gathered}
    P = \participants{G} \\
    v = \partialBalance{\someBlockchain}{\preconditions{G}} \\
    D = \deposits{G} \\
    (t_0, \someContract_0) = initSettings({\preconditions{G}}) \\
        \Delta = \parallelComposition_{\participant{A_i} \in P}
        (\isParticipantDeposit{\participant{A_i}}{v_i + c}{\someBlockchain}{x_i}
        ~|~ \participant{A_i}: [\# \triangleright \isContractAdv{G}{C}{\someBlockchain}]
        ~|~ \participant{A_i}: [x_i \triangleright \isContractAdv{G}{C}{\someBlockchain}]) \\
    x ~\fresh \\
    \end{gathered}
    \begin{gathered}
    \begin{aligned}
    &\isContractState = [ \\
    &\quad status = \isStatusStipChoice, \\
    &\quad balance = v, \\
    &\quad participants = P, \\
    &\quad time = t_0, \\
    &\quad deposits = D, \\
    &] \\        
    \end{aligned}
    \end{gathered}
}\]

Alternatively, a user can double-spend the funds reserved as their deposit in the advertisement, rendering it useless. We are interested in distinguishing this case, because we can use it to emit a specific transition label.

\[
\infer[||-DoubleSpendDeposit]
{
    \reductionRule
        {
            \isParticipantDeposit{\participant{A}}{v}{\someBlockchain}{ x}
            ~|~ \isContractAdv{G}{C}{\someBlockchain}
            ~|~ \isConfiguration
        }
        {doubleSpend(\isContractAdv{G}{C}{\someBlockchain}, \participant{A}, x)}
        {
                        \isContractAdv{G}{C}{\someBlockchain}
            | \isConfiguration
        }
}{
    \begin{gathered}
    \depositsPre{A}{\balance}{\vec x} \in \preconditions{G} \\
    x \in \vec{x} \\
        \end{gathered}
}
\]

Assuming all contract have been published, then honest users can start revealing their stipulation secrets. The rule is very similar to revealing a step secret.

\[
\infer[||-InitSecretRev]
{
    \reductionRule
        {
            \isCommitedInitSecret{A}{\someStipulation}
            ~\vert~ \isConfiguration
        }
        {\isRevealedInitSecret{A}{\someStipulation}}
        {
            \isRevealedInitSecret{A}{\someStipulation}
            ~\vert~ \isConfiguration
        }
}
{}
\]

If everyone revealed their stipulation secrets and someone also revealed a corresponding step secret, we can initialize the contract, resulting in an active contract, ready to start running.

\[
\infer[||-Init]
{
    \reductionRule
        {
            \isActiveContract{\contract{C}}{\isContractState}{\someBlockchain}{\someStipulation, x}
            ~|~ \secDelay
            ~|~ \isConfiguration
        }
        {init(\someStipulation, \someBlockchain, x)}
        {
            \isActiveContract{\contract{C}}{\isContractState'}{\someBlockchain}{[\someStipulation], x'}
            ~|~ \isConfiguration
        }
}
{
    \begin{gathered}
    \begin{aligned}
    &\isContractState = [ \\
    &\quad status = \isStatusStipChoice, \\
    &\quad time = t_0, \\
    &\quad participants=P, \\
            &\quad \dots \\
    &] \\
    \end{aligned} \\
    \secDelay = \isRevealedStepSecret{A}{\someStipulation} 
        ~|~ \parallelComposition_{\participant{A_i} \in P}
            \isRevealedInitSecret{A_i}{\someStipulation} \\
    \end{gathered}
    \begin{gathered}
    \begin{aligned}
    &\isContractState' = \isContractState[ \\
    &\quad status \mapsto \isStatusChoice, \\
    &\quad time \mapsto t_0 + 2\secDelay, \\
    &] \\
    \end{aligned} \\
    x' ~\fresh \\
    \end{gathered}
}
\]

After the timeout, we can abort the contract, transitioning it to an aborted contract.

\[
\infer[||-SSkip]
{
    \reductionRule
        {
            \isActiveContract{\contract{C}}{\isContractState}{\someBlockchain}{\someStipulation, x}
            ~|~ \isConfiguration
        }
        {from~t: ~sright(\someStipulation, \someBlockchain)}
        {
            \isActiveContract{\contract{C}}{\isContractState}{\someBlockchain}{\someStipulation, x'}
            ~|~ \isConfiguration
        }
}
{
    \begin{gathered}
    \isContractState = [time = t, ~status=\isStatusStipChoice, ~\dots] \\
    \isContractState' = \isContractState [
        time \mapsto t + \secDelay,
        ~status \mapsto \isStatusStipRight
    ] \\
    x' ~\fresh \\
    \end{gathered}
}
\]

After the timeout, an aborted contract can be refunded to all participants.

\[
\infer[||-Abort]
{
    \reductionRule
        {
            \isActiveContract{\contract{C}}{\isContractState'}{\someBlockchain}{\someStipulation, x}
            ~|~ \isConfiguration
        }
        {from~t:~abort(\someStipulation, \someBlockchain, x)}
        {
            \parallelComposition_{\participant{A_i} \in P}
                \isActiveContract{\contract{C}}{\isContractState_i}{\someBlockchain}{\someContract_i, x_i}
            ~\vert~ \isConfiguration
        }
}
{
    \begin{gathered}
    \isContractState = [status=\isStatusStipRight, ~time = t, ~participants=P, ~deposits=D, ~\dots] \\
    \forall \participant{A_i} \in P: (\participant{A_i}, v_{i}\someBlockchain) \in D \\
    \forall \participant{A_i} \in P: \isContractState_i = \isContractState [
        balance \mapsto v_i,
        ~status \mapsto \isStatusStipRefunded{A_i}
    ] \\
    \forall \participant{A_i} \in P: x_i ~\fresh \\
    \end{gathered}
}
\]

If we have a step secret, the aborted contract can transition to a compensation for all the participants, except the owner of the step secret.

\[
\infer[||-StipCompensation]
{
    \reductionRule
        {
            \isActiveContract{\contract{C}}{\isContractState}{\someBlockchain}{\someStipulation, x}
            ~\vert~ \isRevealedStepSecret{A}{\someStipulation}
            ~\vert~ \isConfiguration
        }
        {sslash(\someStipulation, \someBlockchain, \participant{A}, x)}
        {
            \isActiveContract{\contract{C}}{\isContractState'}{\someBlockchain}{\someStipulation, x'}
            ~\vert~ \isRevealedStepSecret{A}{\someStipulation}
            ~\vert~ \isConfiguration
        }
}
{
    \begin{gathered}
    \isContractState = [status=\isStatusStipRight, ~\dots] \\
        \isContractState' = \isContractState [
        status \mapsto \isStatusStipCompensation{A}
    ] \\
    x' ~\fresh \\
    \end{gathered}
}
\]

\[
\infer[||-EStipCompensation]
{
    \reductionRule
        {
            \isActiveContract{\contract{C}}{\isContractState}{\someBlockchain}{\someStipulation, x}
            ~\vert~ \isConfiguration
        }
        {scompensate(\someStipulation, \someBlockchain, \participant{A}, x)}
        {
            \isActiveContract{\contract{C}}{\isContractState'}{\someBlockchain}{\someStipulation, \vec x}
            ~\vert~ \isConfiguration
        }
}
{
    \begin{gathered}
        \isContractState = [status=\isStatusStipSlashed{A}, ~\dots] \\
        \isContractState' = \isContractState [
            status \mapsto \isStatusStipCompensation{\participant{A}}
        ] \\
    \vec x ~\fresh \\
    \end{gathered}
}
\]

\subsection{Timed Semantics}\label{timed-semantics}

Timed semantics are like a higher level to base semantics described so far. They describe how to transition in cases where labels have time conditions like $before$ and $from$. Note that our $from$ is called $after$ in BitML.

\subsubsection{Action}

Actions without time conditions can and enabled on an untimed context, can be performed with any time on a timed context.

\[
\infer[T-Action]
{
    \reductionRule
        {\isConfiguration_0 ~ \vert ~ t}
        {\alpha}
        {\isConfiguration ~ \vert ~ t}
}
{
    \begin{gathered}
        \reductionRule
            {\isConfiguration_0}
            {\alpha}
            {\isConfiguration} \\
    \end{gathered}
}
\]

\subsubsection{Delay}

Time can always be advanced.

\[
\infer[T-Delay]
{
    \reductionRule
        {\isConfiguration ~ \vert ~ t}
        {\secDelay}
        {\isConfiguration ~ \vert ~ t + \secDelay}
}
{
    \secDelay > 0
}
\]

\subsubsection{From}

This rule eliminates $from$ conditionals on transitions. Notice that we call it $from$ and not $after$ as in BitML, because the range is inclusive.

\[
\infer[T-From]
{
    \reductionRule
        {\isConfiguration_0 ~ \vert ~ t}
        {\alpha}
        {\isConfiguration ~ \vert ~ t}
}
{
    \begin{gathered}
        \isConfiguration_0 \xrightarrow{from ~ t' : ~ \alpha} \isConfiguration \\
        t \ge t'
    \end{gathered}
}
\]

%% file: theory_docs/xStrategy.tex
\maketitle

Given a \bitmlx user strategy $\xStrategy$ we define by the following rules a function $\isStrategy = \xCompiledStrategy$ that produces the compiled \istitle strategy.

\subsection{Active Contracts}

\subsubsection{Left Move}

\[
\infer[S-RevealStepSecret]
{
    (\participant{A}: \stepSecret{A}{\someContract}) \in \isUntimedStrategy(\isRun)
}
{
    \begin{gathered}
    \xRun \xCoherence \isRun | t \\
    \alpha(\someContract) \in \xStrategy(\xRun) \\
    \alpha \in \{dwithdraw, split, reveal\} \\
    \isActiveContract{\contract{C}}{\isContractState}{\someBlockchain}{\someContract} \in \lastConfigOf{\isRun} \\
    \isContractState.status = \isStatusChoice \\
    \someUser \notin \userStepSecrets{\someContract}{\isRun} \\
    t < \isContractState.time
    \end{gathered}
}
\]

\[
\infer[S-Reveal]
{
    reveal(\someContract, \someBlockchain) \in \isUntimedStrategy(\isRun)
}
{
    \begin{gathered}
        \someUser \in \userStepSecrets{\someContract}{\isRun} \\
    \isRun \xrightarrow{reveal(\someContract, \someBlockchain)} \\
    \end{gathered}
}
\]

\[
\infer[S-ILeft]
{
    ileft(\someContract, \someBlockchain) \in \isUntimedStrategy(\isRun)
}
{
    \begin{gathered}
        \someUser \in \userStepSecrets{\someContract}{\isRun} \\
    \isRun \xrightarrow{ileft(\someContract, \someBlockchain)} \\
    \end{gathered}
}
\]

\[
\infer[S-ELeft]
{
    \alpha(\someContract, \someBlockchain) \in \isUntimedStrategy(\isRun)
}
{
    \begin{gathered}
            \isRun \xrightarrow{\alpha(\someContract, \someBlockchain)} \\
    \alpha \in \{dwithdraw, split\} \\
    \end{gathered}
}
\]

\subsubsection{Right and Skip}

\[
\infer[S-IRight]
{
    right(\someContract, \someBlockchain)  \in \isUntimedStrategy(\isRun)
}
{
    \begin{gathered}
                    \isRun \xrightarrow{right(\someContract, \someBlockchain)} \\
        \end{gathered}
}
\]

\[
\infer[S-ERight]
{
    skip(\someContract, \someBlockchain) \in  \isUntimedStrategy(\isRun)
}
{
    \begin{gathered}
                    \isRun \xrightarrow{skip(\someContract, \someBlockchain)} \\
            \end{gathered}
}
\]
\subsubsection{Compensation}

\[
\infer[S-Slash]
{
    slash(\someContract, \someBlockchain, x, \participant{B})  \in \isUntimedStrategy(\isRun)
}
{
    \begin{gathered}
                            \isRun \xrightarrow{slash(\someContract, \someBlockchain, \participant{B})} \\
    \end{gathered}
}
\]

\[
\infer[S-Compensate]
{
    compensate(\someContract, \someBlockchain, \participant{B})  \in \isUntimedStrategy(\isRun)
}
{
    \begin{gathered}
                            \isRun \xrightarrow{compensate(\someContract, \someBlockchain, x, \participant{B}) } \\
    \end{gathered}
}
\]
\subsubsection{Authorizations}

\[
\infer[S-AuthControlStart]
{
    \participant{A}: (\someContract, \someBlockchain)  \in\isUntimedStrategy(\isRun)
}
{
    \begin{gathered}
    \xRun \xCoherence \isRun \\
        (\participant{A}: \someContract) \in \xStrategy(\xRun) \\
    \isRun \xrightarrow{\participant{A}: (\someContract, \someBlockchain)} \\
    \end{gathered}
}
\]

\[
\infer[S-AuthControlFinish]
{
    \participant{A}: (\someContract, \someBlockchain)  \in\isUntimedStrategy(\isRun)
}
{
    \begin{gathered}
    \someUser \in \contractAuths{\someContract}{\isRun} \\
    \isRun \xrightarrow{\participant{A}: (\someContract, \someBlockchain)} \\
    \end{gathered}
}
\]

\[
\infer[S-RevealSecret]
{
    (\participant A:\secret a)  \in  \isUntimedStrategy(\isRun)
}
{
    \begin{gathered}
    \xRun \xCoherence \isRun \\
    (\participant A:\secret a) \in \xStrategy(\xRun) \\
    \end{gathered}
}
\]

\subsection{Pre-Stipulation}

\[
\infer[S-Adv-Advertise]
{
      advertise(\contractAdv{G}{C}) \in \isUntimedStrategy(\isRun)
}
{
    \begin{gathered}
    \xRun \xCoherence \isRun \\
      advertise(\contractAdv{G}{C}) \in \xStrategy(\xRun) \\
    \isRun \xrightarrow{advertise(\contractAdv{G}{C})} \\
    \end{gathered}
}
\]

\[
\infer[S-Adv-AuthCommit]
{
     commit(\participant{A}, \contractAdv{G}{C})  \in \isUntimedStrategy(\isRun)
}
{
    \begin{gathered}
    \xRun \xCoherence \isRun \\
    commit(\participant A, \contractAdv{G}{C}) \in \xStrategy(\xRun) \\
    \isRun \xrightarrow{commit(\contractAdv{G}{C})} \\
    \end{gathered}
}
\]

\[
\infer[S-Adv-AuthInitStart]
{
    (\participant{A}: \isContractAdv{G}{C}{\someBlockchain})  \in \isUntimedStrategy(\isRun)
}
{
    \begin{gathered}
    \xRun \xCoherence \isRun \\
    (\someUser: \contractAdv{G}{C}) \in \xStrategy(\xRun) \\
        \isRun \xrightarrow{\someUser: \isContractAdv{G}{C}{\someBlockchain}} \\
    (t_0, \someStipulation) = initSettings(\preconditions{G}) \\
    t < t_0 \\
    \end{gathered}
}
\]

\[
\infer[S-AuthInitFinish]
{
    (\participant{A}: \isContractAdv{G}{C}{\someBlockchain})  \in \isUntimedStrategy(\isRun)
}
{
    \begin{gathered}
    \someUser \in \contractAuths{\someStipulation}{\isRun} \\
    \isRun \xrightarrow{\someUser: \isContractAdv{G}{C}{\someBlockchain}} \\
    \end{gathered}
}
\]

\subsection{x-Stipulation}

\[
\infer[S-Adv-Publish]
{
     publish(\isContractAdv{G}{C}{\someBlockchain}) \in \isUntimedStrategy(\isRun)
}
{
    \begin{gathered}
    \xRun \xCoherence \isRun \\
    (t_0, \someStipulation) = initSettings(\preconditions{G}) \\
    init(\someStipulation) \in \xStrategy(\xRun) \\
    \isRun \xrightarrow{publish(\isContractAdv{G}{C}{\someBlockchain})} \\
    t < t_0 \\
    \end{gathered}
}
\]

\[
\infer[S-Adv-DoubleSpend]
{
     doubleSpend(\isContractAdv{G}{C}{\someBlockchain}, \theHonestUser) \in \isUntimedStrategy(\isRun)
}
{
    \begin{gathered}
    \xRun \xCoherence \isRun \\
    (t_0, \someStipulation) = initSettings(\preconditions{G}) \\
    not ~\isRun \xrightarrow{publish(\isContractAdv{G}{C}{\someBlockchain})} \\
    t \geq t_0 \\
    \end{gathered}
}
\]

\[
\infer[S-Stip-InitSecret]
{
    \initSecret{A}{\someStipulation} \in \isUntimedStrategy(\isRun)
}
{
    \begin{gathered}
    \xRun \xCoherence \isRun|t \\
    (t_0, \someStipulation) = initSettings(\preconditions{G}) \\
    init(\someStipulation) \in \xStrategy(\xRun) \\
        \forall \someBlockchain \in \activeBlockchains:
        \exists \isActiveContract{\contract{C}}{\isContractState}{\someBlockchain}{\someStipulation}
        \in \lastConfigOf{\isRun} \\
    \someUser \not \in \userInitSecrets{\someStipulation}{\isRun} \\
    t < t_0 \\
    \end{gathered}
}
\]

\[
\infer[S-Stip-StepSecret]
{
     \stepSecret{A}{\someStipulation} \in \isUntimedStrategy(\isRun)
}
{
    \begin{gathered}
    \xRun \xCoherence \isRun|t \\
    (t_0, \someStipulation) = initSettings(\preconditions{G}) \\
    init(\someStipulation) \in \xStrategy(\xRun) \\
    \userInitSecrets{\someStipulation}{\isRun} = users(\someStipulation) \\
    \someUser \notin \userStepSecrets{\someStipulation}{\isRun} \\
    t < t_0  \\
    \end{gathered}
}
\]

\[
\infer[S-Stip-Initialize]
{
    init(\someStipulation, \someBlockchain)  \in \isUntimedStrategy(\isRun)
}
{
    \begin{gathered}
    \xRun \xCoherence \isRun \\
    \someUser \in \userStepSecrets{\someStipulation}{\isRun} \\
    \isRun \xrightarrow{init(\someStipulation, \someBlockchain)} \\
    \end{gathered}
}
\]

\[
\infer[S-Stip-Abort]
{
    doubleSpend(\someStipulation, \someBlockchain, \participant{A}) \in \isUntimedStrategy(\isRun)
}
{
    \begin{gathered}
    \xRun \xCoherence \isRun \\
    abort(\someStipulation) \in \xStrategy(\xRun) \\
    \isRun \xrightarrow{doubleSpend(\someStipulation, \someBlockchain, \participant{A})} \\
    \end{gathered}
}
\]

\[
\infer[S-Stip-Right]
{
    sright(\someStipulation, \someBlockchain) \in \isUntimedStrategy(\isRun)
}
{
    \begin{gathered}
    \isRun \xrightarrow{sright(\someStipulation, \someBlockchain)} \\
    \end{gathered}
}
\]

\[
\infer[S-Stip-Abort]
{
    abort(\someStipulation, \someBlockchain) \in \isUntimedStrategy(\isRun)
}
{
    \begin{gathered}
    \isRun \xrightarrow{abort(\someStipulation, \someBlockchain)} \\
    \end{gathered}
}
\]

\[
\infer[S-ICompensation]
{
    scompensate(\someContract, \someBlockchain, x, \participant{A'})  \in \isUntimedStrategy(\isRun)
}
{
    \begin{gathered}
    \isRun \xrightarrow{scompensate(\someContract, \someBlockchain, x, \participant{A'})} \\
    \end{gathered}
}
\]

\[
\infer[S-ECompensation]
{
    doCompensation(\someContract, \someBlockchain, x, \participant{A'}) \in  \isUntimedStrategy(\isRun)
}
{
    \begin{gathered}
    \isRun \xrightarrow{doCompensation(\someContract, \someBlockchain, x, \participant{A'})} \\
    \end{gathered}
}
\]

\subsection{Timed Strategy}

\[
\infer[S-Action]
{
    \isStrategy(\isRun) = \isUntimedStrategy(\isRun)
}
{
    \begin{gathered}
    \isUntimedStrategy(\isRun) \neq \emptyset \\
    \end{gathered}
}
\]

\[
    \nextDeadline{\isRun|t} = min \{
        \secDelay : \isActiveContract{\contract{C}}{\isContractState}{\someBlockchain}{\someContract} \in \xConfig(\isRun) \land t+\secDelay = \isContractState.timeout
    \}
\]

\[
\infer[S-Wait]
{
    \isStrategy(\isRun) = \{ \secDelay \}
}
{
    \begin{gathered}
    \isUntimedStrategy(\isRun) = \emptyset \\
    \secDelay = \nextDeadline{\isRun} \\
    \end{gathered}
}
\]

%% file: theory_docs/frontiers.tex
\maketitle

\subsection{Frontier Definitions}

We will introduce the concept of a \emph{contract frontier} in a run. 
Intuitively, a frontier is a set of node identifiers that covers all contract executions in the run.

More formally we define for \bitmlx runs $\xRun$:

\begin{definition}[\bitmlx frontiers]
Let $\xRun$ be a \bitmlx run. 
Then a set $\front$ of identifiers is a frontier of $\xRun$ (written $\frontier{\front}{\xRun}$) if the following conditions hold:
\begin{enumerate}
    \item $\forall \someContract \in \front.~ \someContract \in \xRun$ i.e. $\kappa$ is in some configuration of $\xRun$ 
    \item $\forall \someContract' \in \xRun.~ \exists \someContract \in \front.~ \someContract \in \ancestor(\someContract') \lor \someContract \in \descendant(\someContract')$
    \item $\forall \someContract, \someContract' \in \front.~ \someContract \neq \someContract' \Rightarrow \someContract \not \in \ancestor(\someContract')$
\end{enumerate}

\end{definition}

We can extend a similar definition to runs $\isRun$ in the intermediate semantics: 

\begin{definition}[Intermediate frontiers]
Let $\isRun$ be a run in the intermediate semantics. 
Then a set $\front$ of identifiers is a frontier of $\isRun$ on blockchain $\someBlockchain$ (written $\isFrontier{\front}{\isRun}{\someBlockchain}$) if the following conditions hold:
\begin{enumerate}
        \item $\front \subseteq \isRun$
        \item $\forall \someContract' \in \isRun.~ \exists \someContract \in \front.~ \someContract \in \ancestor(\someContract') \lor \someContract \in \descendant(\someContract')$
    \item $\forall \someContract \someContract' \in \front.~ \someContract \neq \someContract' \Rightarrow \someContract \not \in \ancestor(\someContract')$
\end{enumerate}

\end{definition}

We define an ordering among frontiers. 

\begin{definition}[Frontier ordering]
We say that a frontier $\front_1$ is at least as small as a frontier $\front_2$ (written $\front_1 \frontpreorder \front_2$) if
$$
    \forall \someContract_2 \in \front_2.~ \exists \someContract_1 \in \front_1.~ \someContract_1 \in \ancestor(\someContract_2)
$$
Similarly, we say that frontier $\front_1$ is smaller than frontier $\front_2$ (written $\front_1 \frontorder \front_2$) if
$\front_1 \frontpreorder \front_2$ but $\front_1 \neq \front_2$.
\end{definition}

Note that the minimal frontier of a run contains exactly the roots of all contracts appearing in a run.

\begin{definition}[Root Contracts]
We say that a set $\front$ of identifiers are the root contracts of a run $\xRun$ (written $\rootContracts(\xRun, \front)$) if the following conditions hold:
    \begin{enumerate}
        \item $\frontier{\front}{\xRun}$
        \item $\forall \front'.~ \frontier{\front'}{\xRun} \Rightarrow \front \frontpreorder \front'$
    \end{enumerate}
\end{definition}

\begin{lemma}[Existence of \bitmlx root contracts]
    Let $\xRun$ be a (valid) \bitmlx run. 
    Then there exists a unique set $\front$ of root contracts:
    \begin{align*}
        \exists \front: \rootContracts(\xRun, \front)
        \land \forall \front': \rootContracts(\front', \xRun) \Rightarrow \front = \front'
    \end{align*}
    We write $\front = \rootContracts(\xRun)$ in this case. 
\end{lemma}

Note that each frontier can be partitioned by their root contracts. 

\begin{definition}[Maximal Frontier, \bitmlx]
    A frontier $\front$ of a run $\xRun$ is maximal (written $\maxFrontier{\xRun}{\front}$) if the following conditions hold
        \begin{enumerate}
        \item $\frontier{\front}{\xRun}$
        \item $\forall \front'.~ \frontier{\front'}{\xRun} \Rightarrow \front' \frontpreorder \front$
    \end{enumerate}
\end{definition}

\begin{lemma}[Existence of maximal \bitmlx frontier]
    Let $\xRun$ be a (valid) \bitmlx run. 
    Then there exists a unique maximal frontier:
    \begin{align*}
        \exists \front: \maxFrontier{\front}{\xRun} 
        \land \forall \front': \maxFrontier{\front'}{\xRun} \Rightarrow \front = \front'
    \end{align*}
    We write $\front = \maxFrontierFun{\xRun}$ in this case. 
\end{lemma}

\begin{definition}[Maximal Frontier, intermediate]
    A frontier $\front$ of a run $\isRun$ is maximal (written $\isMaxFrontier{\xRun}{\front}{\someBlockchain}$) if the following conditions hold
        \begin{enumerate}
        \item $\isFrontier{\front}{\isRun}{\someBlockchain}$
        \item $\forall \front'.~ \isFrontier{\front'}{\isRun}{\someBlockchain} \Rightarrow \front' \frontpreorder \front$
    \end{enumerate}
\end{definition}

\begin{lemma}[Existence of maximal intermediate frontier]
    Let $\isRun$ be a (valid) intermediate run and $\someBlockchain \in \activeBlockchains$. 
    Then there exists a unique maximal frontier for $\someBlockchain$:
    \begin{align*}
        \exists \front: \isMaxFrontier{\front}{\isRun}{\someBlockchain} 
        \land \forall \front': \isMaxFrontier{\front'}{\isRun}{\someBlockchain} \Rightarrow \front = \front'
    \end{align*}
    We write $\front = \maxFrontierFun{\isRun}{\someBlockchain}$ in this case. 
\end{lemma}

\subsection{Lemmas on Frontiers}

\begin{lemma}[Frontier configurations]
    Let $\xRun$ be a \bitmlx run and $\front$ be the maximal frontier of $\xRun$ ($\front = \maxFrontierFun{\xRun}$). 
    Then if $\someContract \in \front$ there exist $\contract{C}$, $\balance$ such that $ \activeContract{\contract{C}}{\balance}{\someContract} \in \xConfig(\xRun)$.
\end{lemma}

\textit{Helper lemma H1:}
\[
        \someContract \in \xConfig(\xRun[0]) \land \someContract \in \xConfig(\xRun) \land \maxFrontier{\front_0}{\xRun[0]} \land \maxFrontier{\front}{\xRun} \implies (\someContract \in \ \front_0 \iff \someContract \in \front) 
\]

\begin{proof}[Proof: Lemma H1]
    Induction on $\xRun$ with proof by contradiction in both directions.
\end{proof}

\newcommand{\newset}{\textit{New}}
\newcommand{\delset}{\textit{Deleted}}
\newcommand{\perset}{\textit{Persistent}}

\begin{proof}[Proof: Frontier configurations]
    Proof by Induction on $\xRun$:
    \begin{itemize}
        \item Base case: $\front = \emptyset$
        \item Inductive case:
        For every \bitmlx transition $\reductionRule{\xRun[0]}{\alpha}{\xRun}$, we define sets $\newset = \{ \someContract \in \xConfig(\xRun) | \someContract \not \in \xConfig(\xRun[0])\}$, $\delset = \{ \someContract \not \in \xConfig(\xRun) | \someContract \in \xConfig(\xRun[0])\}$ and $\perset = \{ \someContract \in \xConfig(\xRun) | \someContract \in \xConfig(\xRun[0])\}$. \\

        Let $\front_0$ and $\front$ be the unique maximal frontiers of $\xRun[0]$ and $\xRun$ (i.e. $\maxFrontier{\front_0}{\xRun[0]}$ and $\maxFrontier{\front}{\xRun}$).

        We will show that $\front = (\front_0\setminus \delset) \cup \newset$ by 
        \begin{itemize}
                        \item $\delset \cap \front = \emptyset$. Every \bitmlx semantics rule that deletes a contract from a configuration, appends a set of descendants to the new configuration:
            \[
                \forall \someContract_0 \in \delset.~ \exists \someContract \in (\descendant(\someContract_0) \cap \xConfig(\xRun))
            \]
            By condition (3) of frontier, we know that for each such pair of $\someContract \in \descendant(\someContract_0)$ at most one of them can be part of $\front$. By definition of maximality, we conclude that $\someContract_0 \not \in \front$.
            \item $\newset \subseteq \front$. By definition of frontiers we know that $\someContract \in \newset \implies \exists \someContract' \in \front.~ \someContract' \in \ancestor(\someContract) \lor \someContract' \in \descendant(\someContract)$. However $\someContract'$ cannot be a strict ancestor since $\front$ is maximal. Additionally, by \bitmlx semantics we know that no descendant of $\someContract$ is part of run $\xRun$ since it was freshly added. 
            Hence, $\someContract \in \front$.
                        \item $\perset \subseteq \front_0 \cap \front$: By lemma (H1)
            \item $\not \exists \someContract.~ \someContract \not \in \xConfig(\xRun) \land \someContract\not\in\xConfig(\xRun[0]) \land \someContract \in \front \land \someContract \not \in \front_0$ \\ 
            Towards contradiction, we derive that $\someContract \in \xRun[0]$ i.e. there exists $\someContract' \in \descendant(\someContract)$ that is in the maximal frontier ($\someContract' \in \front_0$) and the configuration $\xConfig(\xRun[0])$. 
            Hence, either $\someContract'$ or its direct descendant is in $\xConfig(\xRun)$ by \bitmlx semantics.
            This contradict maximality.
        \end{itemize}

        To be shown $\forall \someContract \in \front.~ \exists \contract{C},\balance.~ \activeContract{\contract{C}}{\balance}{\someContract} \in \xConfig(\xRun)$. Case distinction on $\someContract$:
        \begin{itemize}
            \item $\someContract \in \perset$: By IH
            \item $\someContract \in \newset$: By definition of $\newset$
        \end{itemize}
    \end{itemize}
\end{proof}

\begin{lemma}
    Let $\xRun$ be a \bitmlx run and $\front$ be the maximal frontier of $\xRun$ ($\front = \maxFrontierFun{\xRun}$). 
    Then for every $\activeContract{\contract{C}}{\balance}{\someContract} \in \lastConfigOf{\xRun}$, $\someContract \in \front$.
\end{lemma}

\paragraph{On frontiers and coherence}
Intuitively, the maximal frontier of a \bitmlx run coincides with the frontier derived from the union over the maximal frontiers (per blockchain, by taking their maximum) of a coherent intermediate run on all non-compensated contracts.

Similarly, the maximal frontier of the common prefix coincides with the frontier derived from the union over the maximal frontiers (per blockchain, by taking their minimum) of a coherent intermediate run on all non-compensated contracts.
We spell this out more formally in the following.

\begin{definition}[Join of intermediate frontiers]
Let $\front_1$ and $\front_2$ be frontiers of a run $\isRun$. Then the join of $\front_1$ and $\front_2$ (written $\front_1 \join \front_2$) is a set satisfying the following conditions
\begin{enumerate}
    \item $\front_1 \join \front_2 \subseteq \front_1 \cup \front_2$
    \item $\forall \someContract' \in \front_1 \cup \front_2.~ \exists \someContract \in \front_1 \join \front_2.~ \someContract \in \descendant(\someContract')$
            \item $\forall \someContract \someContract' \in \front_1 \join \front_2.~ \someContract \neq \someContract' \Rightarrow \someContract \not \in \ancestor(\someContract')$
    \end{enumerate}
    \end{definition}
Intuitively, the join of two frontiers is the upper bound of the frontiers. 

Note that the join $\front_1 \join \front_2$ of two frontiers always contains the far most progressed contracts out of the chains. 
This accounts for both moves that ran ahead, as well as compensated chains.

\begin{lemma}[Single-blockchain frontier update]
    Let $\isRun = \isRun[0] \xrightarrow{\alpha(\someContract, \someBlockchain)}$ be an  intermediate semantics run with the move $\alpha$ resulting in successor contracts $\vec\someContract$. 
    Then it holds that:
    \[
         \isMaxFrontierFun{\isRun}{\someBlockchain} =
            \isMaxFrontierFun{\isRun[0]}{\someBlockchain} \setminus \{ \someContract \} \cup \{\someDescendant \in \vec\someContract \}
    \]
\end{lemma}

\begin{definition}
    We say that an intermediate semantics run $\isRun$ is "non-divergent" (we denote this $\nondivergent{\isRun}$) if for every pair of moves $\alpha(\someContract, \someBlockchain), \alpha'(\someContract, \someBlockchain') \in \isRun$, with $\alpha, \alpha' \in \{dwithdraw, cwithdraw, reveal, split, right\}$, it holds that $\alpha = \alpha'$.
\end{definition}

\begin{lemma}[Eventual Synchronicity Frontier Update]
    Let $\isRun = \isRun[0] \xrightarrow{\alpha(\someContract, \someBlockchain)}$ be a non-divergent run with the move $\alpha$ resulting in successor contracts $\vec\someContract$. 
    Then it holds that:
    \[
        \bigjoin_{\someBlockchain_i \in \activeBlockchains} \isMaxFrontierFun{\isRun}{\someBlockchain_i} = \begin{cases}
            \bigjoin_{\someBlockchain_i \in \activeBlockchains} \isMaxFrontierFun{\isRun[0]}{\someBlockchain_i} \setminus \{ \someContract \} \cup \{\someDescendant \in \vec\someContract \} & \text{if}~ \someContract \in \bigjoin_{\someBlockchain_i \in \activeBlockchains}\isMaxFrontierFun{\isRun[0]}{\someBlockchain_i} \\
            \bigjoin_{\someBlockchain_i \in \activeBlockchains} \isMaxFrontierFun{\isRun[0]}{\someBlockchain_i} & \text{otherwise}
        \end{cases} 
    \]
                            
\end{lemma}
\begin{proof}

We are doing a $\alpha(\someContract, \someBlockchain)$ move resulting in successors $\vec\someContract$. We then know that $\someContract \in \isMaxFrontierFun{\someBlockchain}{\isRun[0]}$ and the updated maximal frontier for blockchain $\someBlockchain$ will be $\isMaxFrontierFun{\someBlockchain}{\isRun} = \isMaxFrontierFun{\someBlockchain}{\isRun[0]} \setminus \{ \someContract \} \cup \{ \someContract' \in \vec{\someContract} \}$. But, because the $\alpha(\someContract, \someBlockchain)$ only affects blockchain $\someBlockchain$, for other blockchains $\someOtherBlockchain \neq \someBlockchain$, we will have $\isMaxFrontierFun{\someBlockchain'}{\isRun} = \isMaxFrontierFun{\someBlockchain'}{\isRun[0]}$.

Let's first consider the join of maximal frontiers for all blockchains.
\[
\begin{aligned}
    \isJoinFrontier{\isRun} &= \bigjoin_{\someBlockchain' \neq \someBlockchain}{\isMaxFrontierFun{\isRun}{\someBlockchain'}} \join \isMaxFrontierFun{\someBlockchain}{\isRun} \\
    &= \bigjoin_{\someBlockchain' \neq \someBlockchain}
        \isMaxFrontierFun{\isRun[0]}{\someBlockchain'}
        \join \big(
            \isMaxFrontierFun{\someBlockchain}{\isRun[0]}
            \setminus \{ \someContract \}
            \cup \{ \someContract' \in \vec{\someContract} \}
        \big) \\
\end{aligned}
\]

We split by cases on whether $\someContract$ is up to date in $\someBlockchain$.

\begin{itemize}
    \item If \underline{$\uptodate{\isRun[0]}{\someBlockchain}{\someContract}$}, then $\someContract \in \isJoinFrontier{\isRun[0]}$. We will prove, by applying the definition of the join of frontiers, that:  
    \[
    \begin{aligned}
        &\bigjoin_{\someBlockchain' \neq \someBlockchain}
                \isMaxFrontierFun{\isRun[0]}{\someBlockchain'}
                \join \big(
                    \isMaxFrontierFun{\someBlockchain}{\isRun[0]}
                    \setminus \{ \someContract \}
                    \cup \{ \someContract' \in \vec{\someContract} \}
                \big) \\
        &= \isJoinFrontier{\isRun[0]} \setminus \{ \someContract \} \cup \{ \someContract' \in \vec{\someContract} \}    
    \end{aligned}
    \]
    \begin{enumerate}
        \item For every $\someOtherContract \in \isJoinFrontier{\isRun[0]} \setminus \{ \someContract \} \cup \{ \someContract' \in \vec{\someContract} \}$, we can see that either $\someOtherContract \in \isJoinFrontier{\isRun[0]} \setminus \{ \someContract \}$ or $\someOtherContract \in \vec{\someContract} \subseteq \isMaxFrontierFun{\someBlockchain}{\isRun[0]} \setminus \{ \someContract \} \cup \{ \someContract' \in \vec{\someContract} \}$.
        \item Let $\someContract' \in \bigjoin_{\someBlockchain' \neq \someBlockchain} \isMaxFrontierFun{\isRun[0]}{\someBlockchain'} \cup \big( \isMaxFrontierFun{\someBlockchain}{\isRun[0]} \setminus \{ \someContract \} \cup \{ \someContract' \in \vec{\someContract} \} \big)$. We split by case on whether $\someContract'$ is an ancestor of $\someContract$ or not. If $\someContract' \notin \ancestor(\someContract)$, then there is a descendant $\someDescendant \in \descendant(\someContract')$ in $\bigjoin_{\someBlockchain' \neq \someBlockchain} \isMaxFrontierFun{\isRun[0]}{\someBlockchain'}$ and it cannot be $\someContract$, so $\someDescendant \in \isMaxFrontierFun{\someBlockchain}{\isRun[0]}\setminus \{ \someContract \} \cup \{ \someContract' \in \vec{\someContract} \} $. If instead $\someContract' \in \ancestor(\someContract)$, then all the the successors of $\someContract$ are also descendants of $\someContract'$.
        \item Let $\someContract_1, \someContract_2 \in \isJoinFrontier{\isRun[0]} \setminus \{ \someContract \} \cup \{ \someContract' \in \vec{\someContract} \}$. We will prove by contradiction that $\someContract_1 \notin \ancestor(\someContract_2)$. If $\someContract_1 \in \ancestor(\someContract_2)$ then that would imply that either:
        \begin{itemize}
            \item $\isJoinFrontier{\isRun[0]}$ is not a frontier, in the case where both $\someContract_1$ and $\someContract_2$ are in $\in \isJoinFrontier{\isRun[0]} \setminus \{ \someContract \}$. We know by inductive hypothesis that $\isJoinFrontier{\isRun[0]}$ is a frontier. 
            \item $\someContract$ had an ancestor in $\isJoinFrontier{\isRun[0]}$, in the case where $\someContract_2 \in \{ \someContract' \in \vec{\someContract} \}$. But we know that $\someContract \in \isJoinFrontier{\isRun[0]}$ and by the definition of frontiers, there cannot be another ancestor.
        \end{itemize}

            \end{enumerate}

    \item If \underline{not $\uptodate{\isRun[0]}{\someBlockchain}{\someContract}$}, because we know that $\someContract \in \isMaxFrontierFun{\isRun}{\someBlockchain}$ it must be the case that $\someContract \notin \isJoinFrontier{\isRun[0]}$. We will prove, by applying the definition of the join of frontiers, that:  
    
    \[
    \begin{aligned}
        &\underbrace{\bigjoin_{\someBlockchain' \neq \someBlockchain}
            \isMaxFrontierFun{\isRun[0]}{\someBlockchain'}}_{\textcircled{1}}
            \join \big(
                \underbrace{\overbrace{\isMaxFrontierFun{\someBlockchain}{\isRun[0]}
                \setminus \{ \someContract \}}^{\textcircled{2a}}
                \cup \overbrace{\{ \someContract' \in \vec{\someContract} \}}^{\textcircled{2b}}}_{\textcircled{2}}
            \big) \\
        &= \isJoinFrontier{\isRun[0]} \\
        &= \underbrace{\bigjoin_{\someBlockchain' \neq \someBlockchain}
            \isMaxFrontierFun{\isRun[0]}{\someBlockchain'}}_{\textcircled{1}}
            \join \underbrace{\isMaxFrontierFun{\someBlockchain}{\isRun[0]}}_{\textcircled{3}} \\
    \end{aligned}
    \]

    Using the definition of the join of frontiers:
    \begin{enumerate}
        \item Let $\someOtherContract \in \isJoinFrontier{\isRun[0]} \subseteq \textcircled{1} \cup \textcircled{3}$. Then, either $\someOtherContract \in \textcircled{1} \subseteq \textcircled{1} \cup \textcircled{2}$ or $\textcircled{3}$.  If $\someOtherContract \in \textcircled{3}$ then, if $\someOtherContract \neq \someContract$, we know that $\someOtherContract \in \textcircled{2}$. But know, because we are in the case where $\someBlockchain$ was not up to date, that $\someContract \notin \isJoinFrontier{\isRun[0]}$
        
        \item Let $\someOtherContract \in \textcircled{1} \cup \textcircled{2}$. If $\someOtherContract \in \textcircled{1} \cup \textcircled{2a} \subseteq \textcircled{1} \cup \textcircled{3}$, then by condition $(2)$ of definition 12, we know that there exists an ancestor $\someOtherDescendant \in \descendant(\someOtherContract)$ such that $\someOtherDescendant \in \textcircled{1} \join \textcircled{3}$.

        If $\someOtherContract \in \textcircled{2b}$, then $\someOtherContract$ is a successor of $\someContract$. We know that $\someContract \in \textcircled{3} \subseteq \bigcup_{\someBlockchain' \in \activeBlockchains}\isMaxFrontierFun{\isRun[0]}{\someBlockchain'}$. By condition $(2)$ of definition 12, this means that there exist a descendant $\someDescendant$ of $\someContract$ in $\isJoinFrontier{\isRun[0]}$. By the definition of frontier, this implies that there is, for some other blockchain $\someOtherBlockchain$ a move $\hat\alpha(\someContract, \someOtherBlockchain) \in \isRun$. But, because we know that $\isRun$ is non-divergent, we know that $\hat\alpha = \alpha$ and resulted in the same successors $\textcircled{2b}$. Then, $\someDescendant$ is also a descendant of $\someOtherContract$.

        \item Let $\someContract_1, \someContract_2 \in \isJoinFrontier{\isRun[0]}$, then we already know by condition $(3)$ of definition 12 that $\someContract_1 \notin \ancestor(\someContract_2)$.

            \end{enumerate}
            
\end{itemize}

\end{proof}

\subsection{Stipulation Status}

\begin{definition}[\bitmlx stipulation status]
Let $\xRun$ be a \bitmlx run and $\contractAdv{G}{C} \in \xRun$ a contract advertisement in the run. Then, the stipulation status of $\contractAdv{G}{C}$ in $\xRun$ is defined as:

\[
    \xStipStatus{\xRun}{\contractAdv{G}{C}} =
    \begin{cases}
        Initialized & \text{if}~ desc(\xRun, \someStipulation) \neq \emptyset \\
        Aborted & \text{if}~ desc(\xRun, \someStipulation) = \emptyset \land \contractAdv{G}{C} \notin \xConfig \\
        Advertised & \text{if}~ \contractAdv{G}{C} \in \lastConfigOf{\xRun}
    \end{cases}
\]
    
\end{definition}

\begin{definition}[Intermediate stipulation status]

Let $\isRun$ be a run in the intermediate semantics $\contractAdv{G}{C} \in \isRun$ a contract advertisement in the run such that $initSettings(\preconditions{G}) = (t_0, \someStipulation)$. Then, the intermediate stipulation status of $\contractAdv{G}{C}$ for blockchain $\someBlockchain$ in $\isRun$ is defined as:

\[
    \isStipStatus{\isRun}{\contractAdv{G}{C}}{\someBlockchain} =
    \begin{cases}
        \isContractState.status & \text{if}~ \exists\isActiveContract{\contract{C}}{\isContractState}{\someBlockchain}{\someStipulation} \in \lastConfigOf{\isRun} \\
        Initialized & \text{if}~ ldesc(\someStipulation, \someBlockchain) \neq \emptyset \\
        Refunded & \text{if}~ rdesc(\someStipulation, \someBlockchain) \neq \emptyset \\
        Advertised &\text{if}~\isContractAdv{G}{C}{\someBlockchain} \in \isRun \\
            &\quad\land ~\forall (\depositsPre{A}{\balance}{\vec{x}}) \in deposits(\preconditions{G}).
                \forall \someBlockchain \in \activeBlockchains.        
                    \exists\isParticipantDeposit{\participant{A}}{v}{\someBlockchain}{x_i}  
                    \in \isConfig \\
        DoubleSpent &\text{if}~\isContractAdv{G}{C}{\someBlockchain} \in \isRun \\
            &\quad\land ~\exists (\depositsPre{A}{\balance}{\vec{x}}) \in deposits(\preconditions{G}).
                \exists\someBlockchain \in \activeBlockchains.
                    \not\exists\isParticipantDeposit{\participant{A}}{v}{\someBlockchain}{x_i}
                    \in \isConfig \\
    \end{cases}
\]
    
\end{definition}

\begin{definition}[Eventual synchronicity stipulation status]

Let $\isRun$ be a run in the intermediate semantics $\contractAdv{G}{C} \in \isRun$ a contract advertisement in the run. Then, the eventual synchronicity stipulation status of $\contractAdv{G}{C}$ is defined as:

\[
    \esStipStatus{\isRun}{\contractAdv{G}{C}} =
    \begin{cases}
        Initialized & \text{if}~ \exists \someBlockchain \in \activeBlockchains:
            \isStipStatus{\isRun}{\contractAdv{G}{C}}{\someBlockchain} = Initialized \\
            &\quad \land ~\forall \someBlockchain \in \activeBlockchains:
                \isStipStatus{\isRun}{\contractAdv{G}{C}}{\someBlockchain} \notin \{
                    Refunded, DoubleSpent, Advertised
                \} \\
        Aborted & \text{if}~ (\exists \someBlockchain \in \activeBlockchains: 
            \isStipStatus{\isRun}{\contractAdv{G}{C}}{\someBlockchain} \in \{
                Refunded, DoubleSpent
            \} \\
            &\quad \lor ~\forall \someBlockchain \in \activeBlockchains:
                \isStipStatus{\isRun}{\contractAdv{G}{C}}{\someBlockchain} \in \{ Right, Slashed, Compensated \}) \\
            &\quad \land ~\forall \someBlockchain \in \activeBlockchains:
                \isStipStatus{\isRun}{\contractAdv{G}{C}}{\someBlockchain} \notin \{
                    Initialized 
                \} \\
        Advertised & \text{if}~ \forall \someBlockchain \in \activeBlockchains:
            \isStipStatus{\isRun}{\contractAdv{G}{C}}{\someBlockchain} \notin \{
                Initialized, Refunded, DoubleSpent
            \} \\
    \end{cases}
\]

    \end{definition}

%% file: theory_docs/rounds.tex
\maketitle

\subsection{Round-based execution}

The execution in the intermediate semantics proceeds in rounds based on the depth of the contract. 
To make this explicit and to derive guarantees based on the round structure we first define a function to access the time-based round status of a contract $\someContract$ in an intermediate run $\isRun$:

\begin{definition}[Time-based Round Status]
Let $\isRun$ be an intermediate run ending in $\isConfig | t$ and $\front_R$ be its set of root contracts ($\front_R = \rootContracts(\isRun)$). 
Let $\someContract \in \isRun$ and $\someStipulation \in \front_R \cap \ancestor(\someContract)$ be the root contract of $\someContract$ in $\isRun$. 
Let further $advertise(\contractAdv{G}{C}) \in \isRun$ with $(t_0, \someStipulation) = initSettings({\preconditions{G}})$.
Then we define the round status of $\someContract$ in $\isRun$ (written $\roundStatus{\someContract}{\isRun}$) to be the tuple $(r, p)$ where

\begin{align*}
    r &:= \left \lfloor \frac{t - t_0}{2\secDelay} \right \rfloor \\
    p &:= 
    \begin{cases}
    0 & \text{if}~ 2 r \secDelay \leq t - t_0 < (2r+1) \secDelay \\
    1 & \text{if}~ (2r+1) \secDelay \leq t - t_0 < (2r+2) \secDelay
    \end{cases}
\end{align*}
\end{definition}

Intuitively, $r$ represents the last execution round (where each round takes $2 \secDelay$) and $p$ describes the phase within the current execution round. 
The phases correspond to the different stages of skipping and compensation.

We make the implications of these phases more explicit with the following property, which we will prove as an invariant of our soundness statement:

\begin{property}[Round-based Execution]
    Let $\isRun$ be an intermediate run ending in $\isConfig | t$ and $\theHonestUser$ an honest user. We say that the round fulfils the Round-Based Execution property (we write this $roundBasedExecution(\theHonestUser, \isRun)$) if the following conditions hold:

    \begin{itemize}
        \item For every contract advertisement $\isContractAdv{G}{C}{\someBlockchain} \in \isConfig$ with stipulation time $t_0$, the following statements hold:
        \begin{enumerate}
            \item If $t < t_0 + \secDelay$:
                \begin{enumerate}
                    \item $\theHonestUser \notin \userStepSecrets{\someStipulation}{\isRun}$
                    \item $\theHonestUser \notin \userInitSecrets{\someStipulation}{\isRun}$
                    \item $desc(\someStipulation) = \emptyset$
                \end{enumerate}
                
            We say that $\contractAdv{G}{C}$ is in publishing phase.
                
            \item If $t \geq t_0 + \secDelay$, then $\isStipStatus{\isRun[0]}{\someStipulation}{\someBlockchain} = DoubleSpent$.
            
            We say that $\contractAdv{G}{C}$ is invalidated.
        \end{enumerate}

        \item For every stipulation contract $\isActiveContract{\contract{C}}{\isContractState}{\someBlockchain}{\someStipulation} \in \lastConfigOf{\isRun}$  such that $advertise(\contractAdv{G}{C}) \in \isRun$ with $(t_0, \someStipulation) = initSettings({\preconditions{G}})$ the following statements hold: 
        \begin{enumerate}
            \item If $t < t_0 $, then
            \begin{enumerate}
                \item $ \isContractState.status = \isStatusStipChoice$
                \item $\theHonestUser \in \userInitSecrets{\someStipulation}{\isRun} \implies \finishedContractAuths{A}{\contractAdv{G}{C}}{\isRun} = \contractUsers{\someStipulation}{\isRun}$
                                \item $desc(\someStipulation) \neq \emptyset \lor \theHonestUser \in \userStepSecrets{\someStipulation}{\isRun} \implies \userInitSecrets{\someStipulation}{\isRun} = \contractUsers{\someContract}{\isRun}$
            \end{enumerate}
    
            We say that $\someStipulation$ is in initialization phase.
            
            \item If $t_0 \leq t < t_0 + \secDelay $, then
            \begin{enumerate}
                \item $\isContractState.status \notin \{ \isStatusStipSlashed{\theHonestUser}, \isStatusStipCompensation{\theHonestUser}\}$ 
                \item $\theHonestUser \notin \userStepSecrets{\someStipulation}{\isRun}$
                \item $desc(\someStipulation) \neq \emptyset \implies \userInitSecrets{\someStipulation}{\isRun} = \contractUsers{\someContract}{\isRun} \land \userStepSecrets{\someStipulation}{\isRun} \setminus \{\theHonestUser\} \neq \emptyset$
            \end{enumerate}
    
            We say that $\someStipulation$ is in compensation phase.
            
            \item If $t_0 + \secDelay \leq t < t_0 + 2 \times \secDelay $, then
            \begin{enumerate}
                \item $
                                \exists \participant{B} \in \contractUsers{\someContract}{\isRun} \setminus \{\theHonestUser\}: \isContractState.status \in \{ \isStatusStipSlashed{\participant{B}}, \isStatusStipCompensation{\participant{B}} \}$
                                \item $\theHonestUser \notin \userStepSecrets{\someStipulation}{\isRun}$
                \item $desc(\someStipulation) = \emptyset$
            \end{enumerate}
    
            We say that $\someStipulation$ is in refund phase.
            
            \item If $t_0 + 2 \times \secDelay < t $, then
            \begin{enumerate}
                \item
                                $\exists \participant{B} \in \contractUsers{\someContract}{\isRun} \setminus \{\theHonestUser\}: \isContractState.status = \isStatusStipCompensation{\participant{B}}$
                            \end{enumerate}
    
            We say that $\someStipulation$ is finalized.
        \end{enumerate}

        \item for every active contract $\isActiveContract{\contract{C}}{\isContractState}{\someBlockchain}{\someStipulation} \in \lastConfigOf{\isRun}$ with round status $\roundStatus{\someContract}{\isRun} = (r, p)$, the following statements hold:    
        \begin{enumerate}
            \item If $\length{\someContract} > r$, then
            \begin{enumerate}
                \item $\isContractState.status \in \{\isStatusChoice, \isStatusLeft\} \lor \exists \participant{B} \in \contractUsers{\someContract}{\isRun}: \isContractState.status = \isStatusAssigned{\participant{B}} $
                \item $rdesc(\isRun, \someContract) = \emptyset$
            \end{enumerate}
    
            We say that $\someContract$ is ahead of its round.
            
            \item If $\length{\someContract} = r \land p = 0 $, then
            \begin{enumerate}
                \item $\isContractState.status \notin \{\isStatusSlashed{\theHonestUser}, \isStatusCompensated{\theHonestUser}\}$
                \item $(\forall \participant{B} \in \contractUsers{\someContract}{\isRun}: \isContractState.status \neq \isStatusAssigned{\participant{B}}) \implies \theHonestUser \notin \userStepSecrets{\someContract}{\isRun}$
                \item $rdesc(\isRun, \someContract) = \emptyset$
                \item $\isContractState.status = \isStatusLeft \lor desc(\someContract) \neq \emptyset \implies \userStepSecrets{\someStipulation}{\isRun} \setminus \{\theHonestUser\} \neq \emptyset$
            \end{enumerate}
    
            We say that $\someContract$ is in compensation phase.
            
            \item If $\length{\someContract} = r \land p = 1 $, then
            \begin{enumerate}
                \item $\isContractState.status \notin \{ \isStatusChoice, \isStatusLeft, \isStatusSlashed{\theHonestUser}, \isStatusCompensated{\theHonestUser}\}$
                \item $(\forall \participant{B} \in \contractUsers{\someContract}{\isRun}.\isContractState.status \neq \isStatusAssigned{\participant{B}}) \implies \theHonestUser \notin \userStepSecrets{\someContract}{\isRun}$
                \item $ldesc(\isRun, \someContract) \neq \emptyset \implies rdesc(\isRun, \someContract) = \emptyset \land \exists \participant{B} \in \contractUsers{\someContract}{\isRun} \setminus \{\theHonestUser\}: \isContractState.status = \isStatusCompensated{\participant{B}}$
            \end{enumerate}
    
            We say that $\someContract$ is in skipping phase.
            
            \item If $\length{\someContract} < r $, then
            \begin{enumerate}
                \item $\exists \participant{B} \in \contractUsers{\someContract}{\isRun}: \isContractState.status = \isStatusAssigned{\participant{B}}$ or $\exists \participant{B} \in \contractUsers{\someContract}{\isRun} \setminus \{\theHonestUser\}: \isContractState.status = \isStatusCompensated{\participant{B}}$
                \item $ldesc(\isRun, \someContract) \neq \emptyset \implies rdesc(\isRun, \someContract) = \emptyset$
            \end{enumerate}
    
            We say that $\someContract$ is finalized.
        \end{enumerate}
    \end{itemize}
\end{property}

\begin{corollary}[Non-divergence]
    Let $\isRun$ be an intermediate run and $\theHonestUser$ be an honest user such and $roundBasedExecution(\theHonestUser, \isRun)$. Let further $\someContract \in \isRun$ be a contract in the run. Then, for every pair of moves $\alpha(\someContract, \someBlockchain), \alpha'(\someContract, \someBlockchain') \isRun$ in different blockchains for the contract, $\alpha = \alpha'$.
\end{corollary}

\begin{corollary}[No honest compensations]
    Let $\isRun$ be an intermediate run and $\theHonestUser$ be an honest user such and $roundBasedExecution(\theHonestUser, \isRun)$. Then,
    \begin{itemize}
        \item For every stipulation contract $\isActiveContract{\contract{C}}{\isContractState}{\someBlockchain}{\someStipulation} \in \lastConfigOf{\isRun}$ , $\isContractState.status \notin \{ Stip-Slashed(\theHonestUser), Stip-CompensatedFrom(\theHonestUser)\}$
        \item For every active contract $\isActiveContract{\contract{C}}{\isContractState}{\someBlockchain}{\someContract} \in \lastConfigOf{\isRun}$ , $\isContractState.status \notin \{Slashed(\theHonestUser), CompensatedFrom(\theHonestUser)\}$
    \end{itemize}
\end{corollary}

%% file: theory_docs/xSoundness.tex
\maketitle

We will now prove the soundness of \bitmlx. We start by defining some auxiliary lemmas, and some properties that we will prove as invariants.

The first lemma is a property of intermediate runs and says that if we have stipulation contracts in all blockchains, then the advertisement is fully authorized by all participants and we no longer have contract advertisements in the configuration.

\begin{lemma}[Fully Published]
Let $\isRun$ be an intermediate ending in $\isConfig$ and $\someAdvertisement$ a contract advertisement with $initSettings(\someAdvertisement) = (t_0, \someStipulation)$ such that for all blockchains $\someBlockchain \in \activeBlockchains$, we have an intermediate stipulation contract $\isActiveContract{\contract{C}}{\isContractState}{\someBlockchain}{\someStipulation} \in \lastConfigOf{\isRun}$. Then, we know that:
\begin{itemize}
    \item $\finishedContractAuths{A}{\contractAdv{G}{C}}{\isRun} = \participants{G}$
    \item $\forall \someBlockchain \in \activeBlockchains: \isContractAdv{G}{C}{\someBlockchain} \notin \lastConfigOf{\isRun}$
\end{itemize}
\end{lemma}
\begin{proof}
    By standard induction over $\isRun$.
\end{proof}

The second lemma says that if there was an authorization move at some point in the run and we still have the contract for that authorization, then we still have the authorization object in the configuration.

\begin{lemma}[\bitmlx Persistence of Authorizations]
Let $\xRun$ be a \bitmlx run ending in $\xConfig$ and $\someContract$ a contract in the run. If $\activeContract{\bitmlxAuthOp{A}{\contract{D}} \prchoice \contract{C}}{\balance}{\someContract} \in \isConfig$ and $\participant{A} \in \contractAuths{\someContract}{\xRun}$ for some user $\participant{A} \in \vec{\participant{A}}$, then $\userAuthIn{\participant{A}}{\someContract}{\contract{D}} \in \isConfig$.
\end{lemma}

\begin{proof}
    By standard induction over $\xRun$.
\end{proof}

The Configuration Consistency invariant will keep track of the relation of the objects in the intermediate and \bitmlx runs.

\begin{property}[Configuration Consistency]
    Let $\isRun$ be an intermediate ending in $\isConfig$ and $\xRun$ a \bitmlx run ending in $\xConfig$. We say that the runs fulfil the Configuration Consistency condition (we write this $confConsistency(\xRun, \isRun)$) if
    \begin{enumerate}
        \item $\forall \isContractAdv{G}{C}{\someBlockchain} \in \isConfig: initSettings({\preconditions{G}}) = (\someStipulation, t_0) \land \esStipStatus{\isRun}{\contractAdv{G}{C}} = Advertised 
        \implies \contractAdv{G}{C} \in \xConfig$
        \item $\forall \isActiveContract{\contract{C}}{\isContractState}{\someBlockchain}{\someStipulation} \in \isConfig: \esStipStatus{\isRun}{\contractAdv{G}{C}} = Advertised 
        \implies  \contractAdv{G}{C} \in \xConfig: initSettings({\preconditions{G}}) = (\someStipulation, t_0) \land \partialBalance{\someBlockchain}{\preconditions{G}} = \isContractState.balance$
        \item $\forall \someParContract \in \isConfig: \someContract \in \maxFrontierFun{\xRun} \implies \activeContract{\contract{C}}{\balance}{\someContract} \in \xConfig \land \balance[][\someBlockchain] = \isContractState.balance$
        \item $\forall \secretCommitment{B}{s}{N} \in \isConfig: \secretCommitment{B}{s}{N} \in \xConfig$
        \item $\forall (\secretReveal{B}{s}{N}) \in \isConfig: (\secretReveal{B}{s}{N}) \in \xConfig$
        \item $\forall \userAuthIn{B}{\#}{\contractAdv{G}{C}} \in \isConfig: \userAuthIn{B}{\#}{\contractAdv{G}{C}} \in \xConfig$
    \end{enumerate}
\end{property}

The Timeout Consistency property relates the timeout for status transition of a contract to it's depth and current status.

\begin{property}[Timeout Consistency]
    Let $\isRun$ be an intermediate run and $\front_R$ be its set of root contracts ($\front_R = \rootContracts(\isRun)$). 

    Let $\isActiveContract{\contract{C}}{\isContractState}{\someBlockchain}{\someStipulation} \in \lastConfigOf{\isRun}$ be a stipulation contract. Let further $advertise(\contractAdv{G}{C}) \in \isRun$ with $(t_0, \someStipulation) = initSettings({\preconditions{G}})$.
    
    Then it holds that:
    $$
        \isContractState.time = t_0 + s \secDelay
    $$
    
    where $s=0$ if $\isContractState.status = \isStatusStipChoice$ or $1$ otherwise. \\
    
    Let $\isActiveContract{\contract{C}}{\isContractState}{\someBlockchain}{\someContract} \in \lastConfigOf{\isRun}$ be a non-terminal active contract and $\someStipulation \in \front_R \cap \ancestor(\someContract)$ be the root contract of $\someContract$ in $\isRun$. 
    Let further $advertise(\contractAdv{G}{C}) \in \isRun$ with $(t_0, \someContract^*) = initSettings({\preconditions{G}})$.
    Then it holds that:
    $$
        \isContractState.time = t_0 + (2|\someContract|+s) \secDelay
    $$
    
    where $s=0$ if $\isContractState.status \in \{ \isStatusChoice, \isStatusLeft \}$ or $1$ otherwise.
\end{property}

The Honest Skip invariant says that if a contract is in compensation phase, then the honest user wants to do a \bitmlx $skip$ move.

\begin{property}[Honest Skip]
Let $\isRun$ be an intermediate ending in $\isConfig$ and $\xRun$ a \bitmlx run. Further let $\theHonestUser$ be an honest user with strategy $\xStrategy$. Then, we say that the runs and the user fulfil the Honest Skip property (we write this $honestSkip(\theHonestUser, \xRun, \isRun)$) if for every $\isActiveContract{\contract{D} \prchoice \contract{C}}{\isContractState}{\someBlockchain}{\someContract} \in \isConfig$ with $(r, p) = \roundStatus{\someContract}{\isRun}$ such that $|\someContract| = r$ (that is, $\someContract$ is in compensation phase) it holds that  $skip(\someContract) \in \xStrategy(\xRun)$.

                                \end{property}

The Authorized Left invariant says that whenever we have an active contract in the intermediate state for a started left move and the contract requires the authorization of the honest user, then we have their authorization in the \bitmlx run.

\begin{property}[Authorized Left]
Let $\isRun$ be an intermediate ending in $\isConfig$, $\xRun$ a \bitmlx run ending in $\xConfig$ and $\theHonestUser$ and honest user. We say that the user and the runs fulfil the Authorized Left property (we write this $authorizedLeft(\theHonestUser, \xRun, \isRun)$) if for every $\isActiveContract{\bitmlxAuthOp{B}{\contract{D}} \prchoice \contract{C}}{\isContractState}{\someBlockchain}{\someContract} \in \isConfig$ with $\isContractState.status = \isStatusLeft$ such that $\theHonestUser \in \vec{\participant{B}}$ and $\activeContract{\bitmlxAuthOp{B}{\contract{D}} \prchoice \contract{C}}{\balance}{\someContract} \in \lastConfigOf{\xRun}$ it holds that $\userAuthIn{\theHonestUser}{\someContract}{\contract{D}} \in \lastConfigOf{\xRun}$.
                                        \end{property}

And finally, we prove our \bitmlx Soundness theorem.

\begin{lemma}[\bitmlx Soundness]

Let $\theHonestUser$ be an honest user with an eager $\bitmlx$ strategy $\xStrategy$ and an intermediate semantics strategy $\isStrategy = \xCompiledStrategy$.

\[
    \begin{aligned}        
    &\forall \isRun ~s.t.~ \isStrategy \vDash \isRun, \\
    &\quad \exists \xRun:
        \xStrategy \xConforms \xRun 
        \land \xRun \xCoherence[\theHonestUser] \isRun \\
        &\quad \land confConsistency(\xRun, \isRun) \\
        &\quad \land honestSkip(\theHonestUser, \xRun, \isRun) \\
        &\quad \land authorizedLeft(\theHonestUser, \xRun, \isRun) \\
        &\quad \land timeoutConsistency(\isRun) \\
        &\quad \land roundBasedExecution(\theHonestUser, \isRun) \\
    \end{aligned}
\]
\end{lemma}

\begin{proof}

We proceed by induction on $\isRun$.

\paragraph{\underline{Initial Configuration}}
$\isRun = \isConfig[0] = \isParticipantDeposit{\participant{A_0}}{v_0}{\someBlockchain[0]}{x^0} | \dots | \isParticipantDeposit{\participant{A_n}}{v_n}{\someBlockchain[n]}{x_n}$.

With no contracts or advertisements, timeout consistency and roun-based execution are trivial.

We will choose the initial run to be the empty configuration $\xRun = \xConfig[0] = \emptyset$. This run is trivially conforming to the honest user strategy, trivially coherent to $\isRun$ and trivially fulfilling the Configuration Consistency, Honest Skip and Authorized Left invariants. \\

The initial configuration is our only base case. We move now to proving the inductive cases of run transitions. For the Honest Skip invariant, we will prove only the case of the time delay $\delta$ as all other cases are trivial. Similar for the Authorized Left invariant, where we only prove the $ileft$ case.

\paragraph{\underline{Advertise}}
$\isRun = \isRun[0] \xrightarrow{advertise(\contractAdv{G}{C})} \isConfig$ and we know by inductive hypothesis that $\xStrategy[A] \xConforms \xRun[0] \land \xRun[0] \xCoherence \isRun[0]$. \\

We are advertising, simultaneously for every blockchain $\someBlockchain$, a new contract $\isContractAdv{G}{C}{\someBlockchain}$ with a fresh unique id $\someStipulation$ and stipulation time $t_0$. We know that the contract is well-formed and $\theHonestUser \in \participant{G}$.

If for the current time $t$ of the intermediate run $t < t_0$, then we will mirror this movement in the \bitmlx run and choose $\xRun = \xRun[0] \xrightarrow{advertise(\contractAdv{G}{C})} \xConfig$. Otherwise, we ignore the advertisement and keep $\xRun = \xRun[0]$. \\

\textbf{Run Conformance} \\

We know by the intermediate semantics that $\contractAdv{G}{C}$ is well formed, it's secrets are fresh and $\theHonestUser \in \participant{G}$, so this is a valid \bitmlx advertisement. \\

\textbf{Coherence} \\

The new advertisment $\contractAdv{G}{C}$ has no authorizations both in the \bitmlx and intermediate semantics level and $\xStipStatus{\xRun}{\contractAdv{G}{C}} = \esStipStatus{\isRun}{\contractAdv{G}{C}} = Advertised$. \\

\textbf{Configuration Consistency} \\

By mirroring the intermediate move $advertise(\contractAdv{G}{C})$ in the \bitmlx run, we add the advertisement in both levels, fulfilling the invariant condition. \\

\textbf{Round-Based Execution} \\

We know by the rule for advertisements that $\contractAdv{G}{C}$ is well-formed. This means in particular that for it's advertisement time $t_0$ and the current time $t$ of $\isRun[0]$, it holds that $t < t_0$  \\

\paragraph{\underline{Stipulation Double Spend}}
$\isRun = \isRun[0] \xrightarrow{doubleSpend(\isContractAdv{G}{C}{\someBlockchain}, \participant{B})} \isConfig$ and we know by inductive hypothesis that $\xStrategy[A] \xConforms \xRun[0] \land \xRun[0] \xCoherence \isRun[0]$. \\

User $\participant{B}$ chose to double-spend his deposit $\isParticipantDeposit{\participant{B}}{v}{\someBlockchain}{x} \in \lastConfigOf{\isRun[0]}$ which is required by the preconditions $\preconditions{G}$. If he is the first to do so, he is invalidating the contract advertisement, which is now aborted. \\

This move will change the intermediate stipulation status of $\someAdvertisement$ from $\isStipStatus{\isRun[0]}{\contractAdv{G}{C}}{\someBlockchain} = Advertised$ to $\isStipStatus{\isRun}{\contractAdv{G}{C}}{\someBlockchain} = DoubleSpent$.

We split by cases on the stipulation status for proving Run Conformance and Coherence.
\begin{itemize}
    \item We first prove by contradiction that \underline{$\esStipStatus{\isRun[0]}{\contractAdv{G}{C}} \neq Initialized$}. Suppose that the opposite is true and $\esStipStatus{\isRun[0]}{\contractAdv{G}{C}} = Initialized$. We split by cases on the round status of $\contractAdv{G}{C}$.
    \begin{itemize}
        \item If $t < t_0 + \secDelay$, then $desc(\someStipulation) = \emptyset$  for the unique id $\someStipulation$ of $\preconditions{G}$. But, if $\esStipStatus{\isRun[0]}{\contractAdv{G}{C}} = Initialized$ then $\exists \someOtherBlockchain \in \activeBlockchains: \isStipStatus{\isRun}{\contractAdv{G}{C}}{\someOtherBlockchain} = Initialized$, which in turn means that $ldesc(\someStipulation, \someBlockchain) \subseteq desc(\someStipulation) \neq \emptyset$
        \item If $t \geq t_0 + \secDelay$, then $\isStipStatus{\isRun[0]}{\contractAdv{G}{C}}{\someBlockchain} = DoubleSpent$ (that is, some other user double-spent already). But if $\esStipStatus{\isRun[0]}{\contractAdv{G}{C}} = Initialized$ then $\forall \someBlockchain \in \activeBlockchains: \isStipStatus{\isRun}{\contractAdv{G}{C}}{\someBlockchain} \neq DoubleSpent$.
    \end{itemize}

    \item If \underline{$\esStipStatus{\isRun[0]}{\contractAdv{G}{C}} = Advertised$}, then this is the first double-spent deposit. We will choose the \bitmlx run $\xRun = \xRun[0] \xrightarrow{abort(\someStipulation)} \isConfig$. \\
    
    \textbf{Run Conformance} \\

    Because we are doing a $doubleSpend(\isContractAdv{G}{C}{\someBlockchain}, \participant{B})$ move, we know that we have an intermediate contract advertisement $\isContractAdv{G}{C}{\someBlockchain}$ in $\lastConfigOf{\isRun[0]}$.
    We know by the Configuration Consistency invariant (because $\esStipStatus{\isRun[0]}{\contractAdv{G}{C}} = Advertised$), that $\contractAdv{G}{C} \in \lastConfigOf{\xRun[0]}$.
    
    Given that this is the only precondition for a \bitmlx $abort(\contractAdv{G}{C})$ move, $\xRun$ is conformant to $\theHonestUser$'s strategy. \\

    \textbf{Coherence} \\
    
    Given that $\isParticipantDeposit{\participant{B}}{v}{\someBlockchain}{x} \notin \isConfig$, 
    by definition we have $\isStipStatus{\isRun}{\contractAdv{G}{C}}{\someBlockchain} = DoubleSpent$ and $\esStipStatus{\isRun}{\contractAdv{G}{C}} = Aborted$.
    Meanwhile, in $\xRun$ the $abort$ move will consume the advertisement $\contractAdv{G}{C}$ and generate no descendants, so we also have $\xStipStatus{\xRun}{\contractAdv{G}{C}} = Aborted$.

    Other conditions are trivially held by inductive hypothesis, so we can claim that $\xRun \xCoherence[\theHonestUser] \isRun$. \\

    \textbf{Configuration Consistency} \\

    Holds for $\contractAdv{G}{C}$ because $\esStipStatus{\isRun}{\contractAdv{G}{C}} \neq Advertised$ and for the successor right-descendant contracts, the $abort(\contractAdv{G}{C})$ move will introduce the assigned contracts in the \bitmlx configuration too. \\
        
    \item If \underline{$\esStipStatus{\isRun[0]}{\contractAdv{G}{C}} = Aborted$}, then we keep $\xRun = \xRun[0]$, which is trivially conformant, coherent and configuration consistent by inductive hypothesis. \\
    
\end{itemize}

Round-based execution is trivial for this case.

\paragraph{\underline{AuthCommit}}
$\isRun = \isRun[0] \xrightarrow{commit(\participant B, \contractAdv{G}{C})} \isConfig$ and we know by inductive hypothesis that $\xStrategy[A] \xConforms \xRun[0] \land \xRun[0] \xCoherence \isRun[0]$. \\

A user $\participant{B}$ is committing, simultaneously for every blockchain $\someBlockchain$, to it's secrets in the contract advertisement $\isContractAdv{G}{C}{\someBlockchain}$. This includes the step secrets, the init secret and the stipulation step secret.

If for the current time $t$ of the intermediate run $t < t_0$, then we will mirror this movement in the \bitmlx run and choose $\xRun = \xRun[0] \xrightarrow{commit(\participant B, \contractAdv{G}{C})} \xConfig$. Otherwise, we ignore the move and keep $\xRun = \xRun[0]$. In both cases, Coherence, round-bases execution and timeout consistency are trivial. \\

\textbf{Run Conformance} \\

For the case where $\participant{B} = \theHonestUser$, we know by the honest user strategy that if $commit(\theHonestUser, \contractAdv{G}{C} \in \isStrategy(\isRun[0])$ then $commit(\theHonestUser, \contractAdv{G}{C} \xStrategy(\xRun[0])$. \\

\textbf{Configuration Consistency} \\

By mirroring the intermediate move $commit(\participant B, \contractAdv{G}{C})$ in the \bitmlx run, we add the secret commitments in both levels, fulfilling the invariant condition. \\

\paragraph{\underline{Stipulation Authorization}}
$\isRun = \isRun[0] \xrightarrow{authInit(\participant{A}, \isContractAdv{G}{C}{\someBlockchain})} \isConfig$ and we know by inductive hypothesis that $\xStrategy[A] \xConforms \xRun[0] \land \xRun[0] \xCoherence \isRun[0]$. \\

We need to first distinguish by cases on the stipulation status of $\contractAdv{G}{C}$. We know that $\esStipStatus{\isRun}{\contractAdv{G}{C}} = Initialized$ because that would imply that $\isStipStatus{\isRun}{\contractAdv{G}{C}}{\someBlockchain} \neq Advertised$ but $\isContractAdv{G}{C}{\someBlockchain}) \lastConfigOf{\isRun}$ as a precondition for the advertisement authorization move. If $\esStipStatus{\isRun}{\contractAdv{G}{C}} = Aborted$ then we will ignore this transition and keep $\xRun = \xRun[0]$, which is trivially conforming and (given that the stipulation status doesn't change) coherent.

If $\esStipStatus{\isRun}{\contractAdv{G}{C}} = Advertised$ then, similar to active contract authorizations, we will use different tactics if the authorization is from the honest user or the potentially dishonest ones.

\begin{itemize}
    \item If \underline{$\participant{B} = \theHonestUser$}, we need to further distinguish whether this is the first blockchain where the honest user is authorizing the move or not.
    \begin{itemize}
        \item If \underline{$\theHonestUser \notin \contractAuths{\contractAdv{G}{C}}{\isRun[0]}$} then we will mirror this authorization in the \bitmlx run. That is, we will choose $\xRun = \xRun[0] \xrightarrow{\theHonestUser \colon~ \contractAdv{G}{C}} \isConfig$. \\

        \textbf{Run Conformance}

        We first need to prove that $\theHonestUser \colon~ \contractAdv{G}{C}$ is a valid move in $\xRun[0]$. For this, the \bitmlx semantics requires that we have a contract advertisement $\contractAdv{G}{C} \in \lastConfigOf{\xRun[0]}$ and that we have all of the needed secret commitments.

        Because we are doing an $authInit(\participant{A}, \isContractAdv{G}{C}{\someBlockchain})$ move, we know we have a the secret commitments $\userAuthIn{\participant{C}}{\#}{\contractAdv{G}{C}} \in \lastConfigOf{\isRun[0]}$ for every participant $\participant{C} \in P$, and with our Configuration Consistency invariant we can also claim that $\userAuthIn{\participant{C}}{\#}{\contractAdv{G}{C}} \in \lastConfigOf{\xRun[0]}$.

        Finally, we know this authorization to be conforming to the \bitmlx strategy of $\theHonestUser$ by the definition of our strategy compilation that states that we only start an authorization in the intermediate semantics level when the user is doing so in the \bitmlx level. \\

        \textbf{Coherence}

        We need to prove that after the new authorization, $\contractAuths{\contractAdv{G}{C}}{\xRun} = \finishedContractAuths{A}{\contractAdv{G}{C}}{\isRun} \cup (\contractAuths{\contractAdv{G}{C}}{\isRun} \cap \{ \participant{A }\})$ still holds. We first observe that $\contractAuths{\contractAdv{G}{C}}{\isRun} = \contractAuths{\contractAdv{G}{C}}{\isRun[0]} \cup \{ \theHonestUser \}$. Then, we can do the following reasoning, which is the same as in the case for active contract authorizations:

        \[
        \begin{aligned}
            \contractAuths{\contractAdv{G}{C}}{\xRun} &= \contractAuths{\contractAdv{G}{C}}{\xRun[0]} \cup \{ \theHonestUser \} \\
            & \finishedContractAuths{A}{\contractAdv{G}{C}}{\isRun[0]} \cup (\contractAuths{\contractAdv{G}{C}}{\isRun[0]} \cap \{ \theHonestUser\}) \cup \{ \theHonestUser \} \\
            & \finishedContractAuths{A}{\contractAdv{G}{C}}{\isRun[0]} \cup (\contractAuths{\contractAdv{G}{C}}{\isRun[0]} \cup \{ \theHonestUser \}) \cap \{ (\theHonestUser\} \cup \theHonestUser) \\
            & \finishedContractAuths{A}{\contractAdv{G}{C}}{\isRun[0]} \cup (\contractAuths{\contractAdv{G}{C}}{\isRun} \cap \{ \theHonestUser \}) \\
        \end{aligned}
        \]
    
    \item If \underline{$\theHonestUser \in \contractAuths{\contractAdv{G}{C}}{\isRun[0]}$}, then we don't extend the run and keep $\xRun = \xRun[0]$, which is trivially conforming. To see that it's also coherent, we can do:
    
        \[
            \begin{aligned}        
            \contractAuths{\contractAdv{G}{C}}{\xRun} &= \contractAuths{\contractAdv{G}{C}}{\xRun[0]} \\
                &= \finishedContractAuths{A}{\contractAdv{G}{C}}{\isRun[0]}
                    \cup (\contractAuths{\contractAdv{G}{C}}{\isRun[0]} \cap \{\theHonestUser\}) \\
                &= \finishedContractAuths{A}{\contractAdv{G}{C}}{\isRun[0]}
                    \cup ((\contractAuths{\contractAdv{G}{C}}{\isRun[0]} \cup \{\theHonestUser\})
                    \cap \{\theHonestUser\}) \\
                &= \finishedContractAuths{A}{\contractAdv{G}{C}}{\isRun[0]}
                    \cup (\contractAuths{\contractAdv{G}{C}}{\isRun}
                    \cap \{\theHonestUser\}) \\
            \end{aligned}
        \]
    \end{itemize}

     \item If \underline{$\participant{B} \neq \theHonestUser$}, we need to further distinguish whether this is the last blockchain where the potentially malicious user is authorizing the contract or not, up to compensations for $\theHonestUser$. 

     \begin{itemize}
        \item If \underline{$\participant{B} \in \finishedContractAuths{A}{\contractAdv{G}{C}}{\isRun}$}, then we extend the \bitmlx run with the corresponding authorization. We will choose $\xRun = \xRun[0] \xrightarrow{\participant{B} \colon~ \contractAdv{G}{C}} \isConfig$. \\

        \textbf{Conformance} \\

        Idem for the case of $\theHonestUser$, except that this time we don't care about conforming to the strategy, because it's not an honest user. \\
        
        \textbf{Coherence} \\

        We need to prove that $\contractAuths{\contractAdv{G}{C}}{\xRun} = \finishedContractAuths{A}{\contractAdv{G}{C}}{\isRun} \cup 
        (\contractAuths{\contractAdv{G}{C}}{\isRun} \cap \{\theHonestUser\}$. We start by observing that $\participant{B}$ is finishing the authorization in this action, so $\finishedContractAuths{A}{\contractAdv{G}{C}}{\isRun} = \{\participant{B}\} \cup  \finishedContractAuths{A}{\contractAdv{G}{C}}{\isRun[0]}$. Then, we have:

        \[
            \begin{aligned}        
            \contractAuths{\contractAdv{G}{C}}{\xRun} &= \{\participant{B}\} \cup \contractAuths{\contractAdv{G}{C}}{\xRun[0]} \\
                &= \{\participant{B}\} \cup  \finishedContractAuths{A}{\contractAdv{G}{C}}{\isRun[0]}
                    \cup (\contractAuths{\contractAdv{G}{C}}{\isRun[0]} \cap \{\theHonestUser\}) \\
                &= \finishedContractAuths{A}{\contractAdv{G}{C}}{\isRun}
                    \cup (\contractAuths{\contractAdv{G}{C}}{\isRun}
                    \cap \{\theHonestUser\}) \\
            \end{aligned}
        \]

        \item If \underline{$\participant{B} \notin \finishedContractAuths{A}{\contractAdv{G}{C}}{\isRun}$}, then we don't extend the run and keep $\xRun = \xRun[0]$, which is trivially conforming. To see that it's also coherent, we can do: 
    
        \[
            \begin{aligned}        
            \contractAuths{\contractAdv{G}{C}}{\xRun} &=  \contractAuths{\contractAdv{G}{C}}{\xRun[0]} \\
                &= \finishedContractAuths{A}{\contractAdv{G}{C}}{\isRun[0]}
                    \cup (\contractAuths{\contractAdv{G}{C}}{\isRun[0]} \cap \{\theHonestUser\}) \\
                &= \finishedContractAuths{A}{\contractAdv{G}{C}}{\isRun}
                    \cup (\contractAuths{\contractAdv{G}{C}}{\isRun}
                    \cap \{\theHonestUser\}) \\
            \end{aligned}
        \]
        
     \end{itemize}
\end{itemize}

Other invariants are trivial in this case.

\paragraph{\underline{Publish}}
$\isRun = \isRun[0] \xrightarrow{publish(\isContractAdv{G}{C}{\someBlockchain})} \isConfig$ and we know by inductive hypothesis that $\xStrategy[A] \xConforms \xRun[0] \land \xRun[0] \xCoherence \isRun[0]$. \\

We are doing a $publish(\isContractAdv{G}{C}{\someBlockchain})$ move that will consume a contract advertisement $\isContractAdv{G}{C}{\someBlockchain}$ and introduce a stipulation contract $\isActiveContract{\contract{C}}{\isContractState}{\someBlockchain}{\someStipulation}$ in status $\isContractState.status = \isStatusStipChoice$.

We do not extend the \bitmlx and instead keep $\xRun = \xRun[0]$, so Run Conformance is trivial. For Coherence and configuration cosnsitency, we split by cases in the eventual synchronicity stipulation status of $\contractAdv{G}{C}$.

\begin{itemize}
    \item We first prove by contradiction that \underline{$\esStipStatus{\isRun[0]}{\contractAdv{G}{C}} \neq Initialized$}. The proof is analogous to the $doubleSpend$ case.
    
    \item If \underline{$\esStipStatus{\isRun[0]}{\contractAdv{G}{C}} in \{ Advertised, Aborted$\}}, then the move will change the intermediate stipulation status in $\someBlockchain$ from $\isStipStatus{\isRun[0]}{\contractAdv{G}{C}}{\someBlockchain} = Advertised$ to $\isStipStatus{\isRun}{\contractAdv{G}{C}}{\someBlockchain} = \isStatusStipChoice$ which will not change the eventual synchronicity stipulation status.
    
\end{itemize}

\textbf{Configuration Consistency} \\

Because we are introducing a stipulation contract $\isActiveContract{\contract{C}}{\isContractState}{\someBlockchain}{\someStipulation}$ in the intermediate configuration and its advertisement has stipulation status $\isStipStatus{\isRun[0]}{\contractAdv{G}{C}}{\someBlockchain} = Advertised$, we need to prove that $\contractAdv{G}{C} \in \lastConfigOf{\xRun}$. But because $\isContractAdv{G}{C}{\someBlockchain} \in \lastConfigOf{\isRun[0]}$, we know by inductive hypothesis that $\contractAdv{G}{C} \in \lastConfigOf{\xRun[0]} = \lastConfigOf{\xRun}$. \\

\textbf{Round-Based Execution} \\

We know by inductive hypothesis that $t < t_0 + \secDelay$. This means that the new stipulation contract $\someStipulation$ can either be in initialization phase or in compensation phase. In both cases, the status $\isContractState.status = \isStatusStipChoice$ is consistent. We also know by the inductive hypothesis that $\theHonestUser \notin \userStepSecrets{\someStipulation}{\isRun}$. If $t \geq t_0$, then we know that $desc(\someStipulation) = \emptyset$ because (again by inductive hypothesis) $\theHonestUser \notin \userInitSecrets{\someStipulation}{\isRun}$.

\textbf{Timeout Consistency} \\

Follows directly from the intermediate rule for init. \\

\paragraph{\underline{Init secret reveal}}
$\isRun = \isRun[0] \xrightarrow{\isRevealedInitSecret{\participant{B}}{\someStipulation}} \isConfig$ and we know by inductive hypothesis that $\xStrategy[A] \xConforms \xRun[0] \land \xRun[0] \xCoherence \isRun[0]$. \\

Some user $\participant{B}$ is revealing their init secret for stipulation contract $\someStipulation$. \\

We do not extend the \bitmlx and instead keep $\xRun = \xRun[0]$. Since this intermediate move doesn't change the frontiers or authorizations, Run Conformance and Coherence are trivial. The Honest Skip and Authorized Left invariants are also trivial in this case. \\

\textbf{Round-Based Execution} \\

Let $adv(\someStipulation) = \contractAdv{G}{C}$. If $\participant{B} = \theHonestUser$, we know by our strategy compilation that $\forall \someBlockchain \in \activeBlockchains: \exists\isActiveContract{\contract{C}}{\isContractState}{\someBlockchain}{\someStipulation}$. Then, by the fully published lemma, we also know that $\finishedContractAuths{A}{\contractAdv{G}{C}}{\isRun} = \participants{G}$ which fulfils the stipulation contract condition and $\forall \someBlockchain \in \activeBlockchains: \isContractAdv{G}{C}{\someBlockchain} \notin \lastConfigOf{\isRun}$ which makes the contract advertisement condition trivial.

\paragraph{\underline{Stipulation step secret reveal}}
$\isRun = \isRun[0] \xrightarrow{\isRevealedStepSecret{\participant{B}}{\someStipulation}} \isConfig$ and we know by inductive hypothesis that $\xStrategy[A] \xConforms \xRun[0] \land \xRun[0] \xCoherence \isRun[0]$. \\

Some user $\participant{B}$ is revealing their step secret for stipulation contract $\someStipulation$. \\ 

We do not extend the \bitmlx and instead keep $\xRun = \xRun[0]$. Since this intermediate move doesn't change the frontiers or authorizations, Run Conformance and Coherence are trivial. The Honest Skip and Authorized Left invariants are also trivial in this case. \\

\textbf{Round-Based Execution} \\

If $\participant{B} = \theHonestUser$, we know by our strategy compilation that $t < t_0$ where $t_0$ is the stipulation time of contract $\someStipulation$, so the contract is in initialization phase. But we also know by the strategy compilation that $\userInitSecrets{\someStipulation}{\isRun} = users(\someStipulation)$, that is, all users have revealed their init secrets. \\

\paragraph{\underline{Stipulation Init}}
$\isRun = \isRun[0] \xrightarrow{init(\contractAdv{G}{C}, \someBlockchain)} \isConfig$ and we know by inductive hypothesis that $\xStrategy[A] \xConforms \xRun[0] \land \xRun[0] \xCoherence \isRun[0]$. \\

We are doing an $init(\someStipulation, \someBlockchain)$ move in the intermediate semantics that will consume a stipulation contract $\isActiveContract{\contract{C}}{\isContractState}{\someBlockchain}{\someStipulation}$ with $\isContractState.status = \isStatusStipChoice$ and introduce a new active contract $\isActiveContract{\contract{C}}{\isContractState'}{\someBlockchain}{\someContract}$ with $\isContractState'.status = \isStatusChoice$,  $\isContractState'.time = \isContractState.time + 2 \secDelay$ and $\someContract = [\someStipulation]$.

Intuitively, we are initializing the contract in blockchain $\someBlockchain$. This means that all participants, (\theHonestUser{} included), have revealed their stipulation secrets. We can assume then that the contract has been authorized and published in all blockchains. \\

For proving $\bitmlx$ Run Conformance and Coherence, we will split by cases on the eventual synchronicity stipulation status of $\someStipulation$.

\begin{itemize}
    \item We prove first by contradiction that \underline{$\esStipStatus{\isRun[0]}{\contractAdv{G}{C}} \neq Aborted$}. Suppose the opposite is true and $\esStipStatus{\isRun[0]}{\contractAdv{G}{C}} = Aborted$. We split by cases according to the definition:
    \begin{itemize}
        \item If $\exists \someOtherBlockchain \in \activeBlockchains: \isStipStatus{\isRun}{\contractAdv{G}{C}}{\someOtherBlockchain} = Refunded$, then according to the round-based execution invariant, $t \geq t_0 + \secDelay$. But, because $\isContractState.status = \isStatusChoice$, we know that $t < t_0 + \secDelay$
        \item If $\exists \someOtherBlockchain \in \activeBlockchains: \isStipStatus{\isRun}{\contractAdv{G}{C}}{\someOtherBlockchain} = DoubleSpent$, then $\isContractAdv{G}{C}{\someOtherBlockchain} \in \lastConfigOf{\isRun[0]}$ and round-based execution says that $\theHonestUser \notin \userInitSecrets{\someStipulation}{\isRun[0]}$. But, according to the intermediate semantics rule for $init$, we require that $\theHonestUser \in \userInitSecrets{\contractAdv{G}{C}}{\isRun[0]}$.
        \item If $\forall \someOtherBlockchain \in \activeBlockchains: \isStipStatus{\isRun}{\contractAdv{G}{C}}{\someOtherBlockchain} \in \{ Right, Slashed, Compensated \}$, then this is also true in particular for our currently initializing blockchain $\someBlockchain$. But, according to the intermediate semantics rule for $init$, we require that $\isContractState.status \neq \isStatusStipChoice$.
    \end{itemize}

    \item If \underline{$\esStipStatus{\isRun[0]}{\contractAdv{G}{C}} = Advertised$}, then this is the first blockchain to initialize. We will choose $\xRun = \xRun[0] \xrightarrow{init(\contractAdv{G}{C})}$. We now prove conformance and Coherence of this new run. \\

    \textbf{Run Conformance}

    Because we are doing an $init(\someStipulation, \someBlockchain)$ move, we know by the intermediate semantics that we have a stipulation contract $\isActiveContract{\contract{C}}{\isContractState}{\someBlockchain}{\someStipulation} \in \lastConfigOf{\isRun[0]}$. Then, by the Configuration Consistency invariant, we know that $\contractAdv{G}{C} \in \lastConfigOf{\xRun[0]}$.

    Similarly, we know that for all users $\participant{B} \in \participants{G}: \userAuthIn{B}{\#}{\contractAdv{G}{C}} \in \lastConfigOf{\isRun[0]}$ so we also have $\userAuthIn{B}{\#}{\contractAdv{G}{C}} \in \lastConfigOf{\xRun[0]}$.

    Finally, we need to prove that we have all the needed authorizations. Again, because we are doing a $init(\someStipulation, \someBlockchain)$, we know that $\theHonestUser$ has revealed $\initSecret{A}{\someStipulation}$. According round-based execution, this means that in the intermediate semantics, all users have fully authorized, formally $\finishedContractAuths{A}{\contractAdv{G}{C}}{\isRun} = \contractUsers{\someStipulation}{\isRun}$. Then, from the Coherence invariant, we get that

    \[
        \contractAuths{\contractAdv{G}{C}}{\xRun} =
            \finishedContractAuths{A}{\contractAdv{G}{C}}{\isRun}
            \cup (\contractAuths{\contractAdv{G}{C}}{\isRun} \cap \{ \participant{A} \})
                                    = \contractUsers{\contractAdv{G}{C}}{\isRun}
    \]

    In conclusion, $init(\contractAdv{G}{C})$ is a valid move in the \bitmlx semantics and so $\xStrategy[A] \xConforms \xRun$. \\
    
    \textbf{Coherence}
    
    This $init(\someStipulation, \someBlockchain)$ move will change the intermediate stipulation status of $\someAdvertisement = adv(\isRun, \someStipulation)$ from $\isStipStatus{\isRun[0]}{\contractAdv{G}{C}}{\someBlockchain} = \isStatusStipChoice$ to $\isStipStatus{\isRun}{\contractAdv{G}{C}}{\someBlockchain} = Initialized$. We also know by inductive hypothesis, because $\esStipStatus{\isRun[0]}{\contractAdv{G}{C}} = Advertised$, that for all other blockchains $\someOtherBlockchain$, $\isStipStatus{\isRun[0]}{\contractAdv{G}{C}}{\someOtherBlockchain} \notin \{ Refunded, DoubleSpent \}$. By definition we then have that now $\esStipStatus{\isRun}{\contractAdv{G}{C}} = Initialized$. Similarly, because we are introducing an active contract $[\someStipulation] \in desc(\xRun, \someStipulation)$, we also have $\xStipStatus{\xRun}{\someStipulation} = Initialized$.

    Other conditions are trivially held by inductive hypothesis, so we can claim that $\xRun \xCoherence[\theHonestUser] \isRun$. \\
    
    \textbf{Configuration Consistency} \\

    Because we are adding a new active contract $\isActiveContract{\contract{C}}{\isContractState}{\someBlockchain}{[\someStipulation]}$ to the intermediate configuration and $[\someStipulation] \in \maxFrontierFun{\xRun}$, we need to prove that $\activeContract{\contract{C}}{\balance}{\someContract} \in \xConfig$ and $\balance[][\someBlockchain] = \isContractState.balance$.

    But, because $\isActiveContract{\contract{C}}{\isContractState}{\someBlockchain}{\someStipulation} \in \lastConfigOf{\isRun[0]}$ and $\esStipStatus{\isRun[0]}{\contractAdv{G}{C}} = Advertised$ we know by inductive hypothesis that $\contractAdv{G}{C} \in \xConfig$ and $\partialBalance{\someBlockchain}{\preconditions{G}} = \isContractState.balance$. And according to the \bitmlx semantics, the $init(\someAdvertisement)$ move, will introduce into the configuration $\activeContract{\contract{C}}{\balance}{\someContract}$ with $\totalBalance{\preconditions{G}} = \balance$. \\
    
    \item If \underline{$\esStipStatus{\isRun[0]}{\contractAdv{G}{C}} = Initialized$}, then this is not the first blockchain to initialize. We will not extend the \bitmlx run and instead keep $\xRun = \xRun[0]$, which is trivially conformant to the honest user strategy. \\

    \textbf{Coherence} \\

    We know by inductive hypothesis that $\xStipStatus{\xRun[0]}{\someStipulation} = \esStipStatus{\isRun[0]}{\contractAdv{G}{C}} = Initialized$. On the intermediate semantics side, we are doing an $init(\someStipulation, \someBlockchain)$ move, which will not change the status so $\esStipStatus{\isRun}{\contractAdv{G}{C}} = \esStipStatus{\isRun[0]}{\contractAdv{G}{C}}$. And we also defined $\xRun = \xRun[0]$, so we can conclude that $\xStipStatus{\xRun}{\contractAdv{G}{C}} = \esStipStatus{\isRun}{\contractAdv{G}{C}}$.

    Again, other conditions are trivially held by inductive hypothesis, so we can claim that $\xRun \xCoherence[\theHonestUser] \isRun$. \\
    
\end{itemize}

\textbf{Timeout Consistency} \\

Because $\isContractState.status = \isStatusStipChoice$, we know by inductive hypothesis that $\isContractState.time = t_0$ where $advertise(\contractAdv{G}{C}) \in \isRun$ and $(t_0, \someContract^*) = initSettings({\preconditions{G}})$. Then, for the new status $\isContractState'$ after the move, we can reason in the following way:

\[
    \isContractState'.time
    = \isContractState.time + 2 \secDelay
    = t_0 + 2|\someContract|\secDelay
\]

\textbf{Round-Based Execution} \\

Because $\isContractState.status = \isStatusChoice$, we know that $t < t_0 + \secDelay$ or, equivalently $t - t_0 < \secDelay$. If the round status of the new active contract $\someContract$ is $(r, p) = \roundStatus{\someContract}{\isRun}$ then $r = \left \lfloor \frac{t - t_0}{2\secDelay} \right \rfloor = 0 < |\someContract|$. That is, $\someContract$ is waiting for synchronisation, which is consistent with the status $\isContractState.status = \isStatusChoice$. \\

\paragraph{\underline{Stipulation Right}}

$\isRun = \isRun[0] \xrightarrow{sright(\contractAdv{G}{C}, \someBlockchain)} \isConfig$ and we know by inductive hypothesis that $\xStrategy[A] \xConforms \xRun[0] \land \xRun[0] \xCoherence \isRun[0]$. \\

We are doing an intermediate $sright(\someContract, \someBlockchain)$ move that will change the internal state of a stipulation contract $\isActiveContract{\contract{C}}{\isContractState}{\someBlockchain}{\someStipulation}$ in status $\isContractState.status = \isStatusStipChoice$ to a new state with $\isContractState'.status = \isStatusStipRight$ and $\isContractState'.time = \isContractState.time + \secDelay$.

Intuitively, we are skipping initialization on blockchain $\someBlockchain$. This can be either an asynchronous initialization by the adversary, or part of a regular abort move. \\

For proving Run Conformance, Coherence and Configuration Consistency, we will split by cases on the status of the stipulation of $\someStipulation$ in $\isRun$.

\begin{itemize}

    \item If \underline{$\esStipStatus{\isRun[0]}{\contractAdv{G}{C}} = Advertised $}, then we need to further distinguish the case where this is the last blockchain to move right, which corresponds to aborting the contract, or not.

    \begin{itemize}
        
        \item If \underline{$\forall \someOtherBlockchain \in \activeBlockchains: \isStipStatus{\isRun}{\contractAdv{G}{C}}{\someOtherBlockchain} \in \{ Right, Slashed, Compensated \}$}, then $\someBlockchain$ was the last to move right. We will choose the \bitmlx run $\xRun = \xRun[0] \xrightarrow{abort(\contractAdv{G}{C})}$. \\
    
        \textbf{Run Conformance}
    
        Because we are doing a $sright(\someStipulation, \someBlockchain)$ move, we know that we have a stipulation contract $\isActiveContract{\contract{C}}{\isContractState}{\someBlockchain}{\someStipulation}$ in $\lastConfigOf{\isRun[0]}$. And because $\esStipStatus{\isRun[0]}{\contractAdv{G}{C}} = Advertised $, we know by the Configuration Consistency invariant, that $\contractAdv{G}{C} \in \lastConfigOf{\xRun[0]}$.
        
        Given that this is the only precondition for an $abort(\contractAdv{G}{C})$, $\xRun$ is conformant to $\theHonestUser$'s strategy. \\
    
        \textbf{Coherence}
        
        Given that $\forall \someOtherBlockchain \in \activeBlockchains: \isStipStatus{\isRun}{\contractAdv{G}{C}}{\someOtherBlockchain} \in \{ Right, Slashed, Compensated \}$, by definition we have $\esStipStatus{\isRun}{\contractAdv{G}{C}} = Aborted$. Meanwhile, because we defined our \bitmlx run as $\xRun = \xRun[0] \xrightarrow{abort(\contractAdv{G}{C})}$, the move will consume the contract advertisement and produce no descendants, so we also have $\xStipStatus{\xRun}{\contractAdv{G}{C}} = Aborted$.
    
        Other conditions are trivially held by inductive hypothesis, so we can claim that $\xRun \xCoherence[\theHonestUser] \isRun$. \\
    
        \item If \underline{$\exists \someOtherBlockchain \in \activeBlockchains: \isStipStatus{\isRun}{\contractAdv{G}{C}}{\someOtherBlockchain} \notin \{ Right, Slashed, Compensated \}$}, then we keep $\xRun = \xRun[0]$, which is trivially conformant and coherent by inductive hypothesis.
    \end{itemize}
    
    \item If \underline{$\esStipStatus{\isRun[0]}{\contractAdv{G}{C}} = \{ Aborted, Initialized\}$}, then we also keep $\xRun = \xRun[0]$, which is again trivially conformant and coherent by inductive hypothesis. \\

\end{itemize}

\textbf{Timeout Consistency} \\

Because $\isContractState.status = \isStatusStipChoice$, we know by inductive hypothesis that $\isContractState.time = t_0$ where $advertise(\contractAdv{G}{C}) \in \isRun$ and $(t_0, \someContract^*) = initSettings({\preconditions{G}})$. Then, for the new status $\isContractState'$ after the move, we can reason in the following way:

\[
    \isContractState'.time
    = \isContractState.time \secDelay
    = t_0 + \secDelay
\]

\textbf{Round-Based Execution} \\

Because $\isContractState.status = \isStatusChoice$, we know that $t < t_0 + \secDelay$. But we also know, by the intermediate semantics rule for $sright$ that $t \geq t_0$, so the contract is in compensation phase, which is consistent with the new status $\isContractState'.status = \isStatusStipRight$. Other conditions hold by inductive hypothesis. \\

\paragraph{\underline{Stipulation Abort}}

$\isRun = \isRun[0] \xrightarrow{abort(\contractAdv{G}{C}, \someBlockchain)} \isConfig$ and we know by inductive hypothesis that $\xStrategy[A] \xConforms \xRun[0] \land \xRun[0] \xCoherence \isRun[0]$. \\

We are doing an intermediate $abort(\someContract, \someBlockchain)$ move that will consume the stipulation contract $\isActiveContract{\contract{C}}{\isContractState}{\someBlockchain}{\someStipulation}$ in status $\isContractState.status = \isStatusRight$ and introduce terminal contracts $\someContract_0, \dots, \someContract_n$ with status $\isContractState'.status = \isStatusStipRefunded{\participant{A_i}}$ for each user $\participant{A_i} \in \participants{G}$.

We are aborting the contract in blockchain $\someBlockchain$. This means that we are not finalized at this stage, so there's either some blockchains where the contract was not published, or the dishonest user refused to reveal their init secret. We will split by cases on the status of the stipulation of $\someContract$ in $\isRun$. \\

For proving Run Conformance, Coherence and Configuration Consistency, we will split by cases on the status of the stipulation of $\someStipulation$ in $\isRun$.

\begin{itemize}
    \item We first prove by contradiction that \underline{$\esStipStatus{\isRun[0]}{\contractAdv{G}{C}} \neq Initialized$}.

    Suppose that $\esStipStatus{\isRun[0]}{\contractAdv{G}{C}} = Initialized$. By definition this means that there is some other blockchain $\someOtherBlockchain$ for which $\isStipStatus{\isRun}{\contractAdv{G}{C}}{\someOtherBlockchain} = Initialized$ and consequently $desc(\someStipulation) \neq \emptyset$. But, we also know, because we are doing an $abort(\someStipulation, \someBlockchain)$ move in the intermediate semantics, that $t \ge t_0 + \secDelay$. And according to the round-based execution invariant, this means that $\someStipulation$ is in refund phase, so $desc(\someStipulation) = \emptyset$

    \item If \underline{$\esStipStatus{\isRun[0]}{\contractAdv{G}{C}} = Advertised$}, then this is the first blockchain to abort. We will choose $\xRun = \reductionRule{\xRun[0]}{abort(\someStipulation, \someBlockchain)}{}$.\\

    \textbf{Run Conformance} \\

    Because we are doing an $abort(\someStipulation, \someBlockchain)$ move, we know that we have a stipulation contract $\isActiveContract{\contract{C}}{\isContractState}{\someBlockchain}{\someStipulation}$ in $\lastConfigOf{\isRun[0]}$. We know by the Configuration Consistency invariant (because $\esStipStatus{\isRun[0]}{\contractAdv{G}{C}} = Advertised$), that $\contractAdv{G}{C} \in \lastConfigOf{\xRun[0]}$.
    Given that this is the only precondition for an $abort(\someStipulation)$, $\xRun$ is conformant to $\theHonestUser$'s strategy. \\

    \textbf{Coherence} \\

    This move will instroduce right-descendant contracts, so $\isStipStatus{\isRun}{\contractAdv{G}{C}}{\someOtherBlockchain} = Refunded$. And we know by inductive hypothesis that $\forall \someBlockchain \in \activeBlockchains: \isStipStatus{\isRun}{\contractAdv{G}{C}}{\someBlockchain} \neq Initialized$, so by definition $\esStipStatus{\isRun}{\contractAdv{G}{C}} = Aborted$.

    Meanwhile, the \bitmlx move will consume the contract advertisement $\someAdvertisement$, so 
    $\xStipStatus{\xRun}{\someStipulation} = Aborted$.
    
    Other conditions are trivially held by inductive hypothesis, so we can claim that $\xRun \xCoherence[\theHonestUser] \isRun$. \\

    \textbf{Configuration Consistency} \\

    Holds for $\contractAdv{G}{C}$ because $\esStipStatus{\isRun}{\contractAdv{G}{C}} \neq Advertised$ and for the successor right-descendant contracts, the $abort(\contractAdv{G}{C})$ move will introduce the assigned contracts in the \bitmlx configuration too. \\
    
    \item If \underline{$\esStipStatus{\isRun[0]}{\contractAdv{G}{C}} = Aborted$}, then we have already aborted the contract. We will not extend the \bitmlx run and instead keep $\xRun = \xRun[0]$, which is trivially conformant to the honest user strategy. \\

    \textbf{Coherence}

    We know by inductive hypothesis that $\xStipStatus{\xRun[0]}{\someStipulation} = \esStipStatus{\isRun[0]}{\contractAdv{G}{C}} = Aborted$. On the intermediate semantics side, we are moving the intermediate stipulation status in $\someBlockchain$ from $\isStipStatus{\isRun[0]}{\contractAdv{G}{C}}{\someOtherBlockchain} = \isStatusStipRight$ to $\isStipStatus{\isRun}{\contractAdv{G}{C}}{\someOtherBlockchain} = Refunded$, which will not change the eventual synchronicity stipulation status of $\someAdvertisement$ so $\esStipStatus{\isRun}{\contractAdv{G}{C}} = \esStipStatus{\isRun[0]}{\contractAdv{G}{C}}$. And we also defined $\xRun = \xRun[0]$, so we can conclude that $\xStipStatus{\xRun}{\someAdvertisement} = \esStipStatus{\isRun}{\contractAdv{G}{C}}$.

    Again, other conditions hold trivially by inductive hypothesis, so we can claim that $\xRun \xCoherence[\theHonestUser] \isRun$. \\
\end{itemize}

\textbf{Round-Based Execution} \\

Because $\isContractState.status = \isStatusRight$, we know that $t < t_0 + 2\secDelay$. But we also know, by the intermediate semantics rule for $abort$ that $t \geq t_0 + \secDelay$, so $\someStipulation$ is in refund phase, which is consistent with the new status $\isContractState_i.status = \isStatusStipRefunded{\participant{A_i}}$ for each user $\participant{A_i}$. Other conditions hold by inductive hypothesis. \\

\paragraph{\underline{Stipulation Slash}}
$\isRun = \isRun[0] \xrightarrow{sslash(\someStipulation, \someBlockchain, \participant{B})} \isConfig$ and we know by inductive hypothesis that $\xStrategy[A] \xConforms \xRun[0] \land \xRun[0] \xCoherence[\theHonestUser] \isRun[0]$.

We do not extend the \bitmlx and instead keep $\xRun = \xRun[0]$. Since this intermediate move doesn't change the frontiers or authorizations, Run Conformance and Coherence are trivial. \\

\textbf{Round-Based Execution} \\

Let $\isActiveContract{\contract{C}}{\isContractState}{\someBlockchain}{\someStipulation} \in \isConfig$ be the stipulation contract for $\someStipulation$. We know that according to the intermediate semantics, $\isContractState.status = \isStatusStipRight$ and will change to $\isContractState'.status = \isStatusStipSlashed{\participant{B}}$. By the inductive hypothesis of round-based execution, $\someStipulation$ is either in compensation phase ($t_0 \leq t < t + \secDelay$) or in refund phase ($t_0 + \secDelay \leq t < t + 2\secDelay$). In both cases, we know by inductive hypothesis that $\theHonestUser \notin \userStepSecrets{\someContract}{\isRun}$, so $slash(\someContract, \someBlockchain, \theHonestUser)$ would be an invalid move in the intermediate semantics. Other conditions hold by inductive hypothesis.

\paragraph{\underline{Stipulation Compensation}}
$\isRun = \isRun[0] \xrightarrow{scompensation(\someStipulation, \someBlockchain, \participant{B})} \isConfig$ and we know by inductive hypothesis that $\xStrategy[A] \xConforms \xRun[0] \land \xRun[0] \xCoherence \isRun[0]$.

We are compensating with a revealed step secret on a blockchain that didn't use it to init. This can be either the honest user cleaning up an incomplete initialization or the malicious user hurting themselves. The started initialization, if any, is already in the run, so our only extension will be to add authorizations to future active contracts that we now consider complete when increasing the compensation history sets of those contracts.

We will choose the new \bitmlx run to be $\xRun = \xRun[0]\xRun[a]$, where $\xRun[a]$ consists of authorizations $(\participant{C}: ~\someContract)$ for every user $\participant{C}$ and contract $\someContract$ such that $\someContract \in \maxFrontierFun{\xRun}$ and $\participant{C} \in \finishedContractAuths{A}{\someContract}{\isRun} \setminus \finishedContractAuths{A}{\someContract}{\isRun[0]}$. \\

\textbf{Round-Based Execution} \\

Let $\isActiveContract{\contract{C}}{\isContractState}{\someBlockchain}{\someStipulation} \in \isConfig$ be the stipulation contract for $\someStipulation$. We know that according to the intermediate semantics, $\isContractState.status = \isStatusStipSlashed{\participant{B}}$. By the inductive hypothesis of round-based execution, $\someStipulation$ is either in compensation phase ($t_0 \leq t < t + \secDelay$) or in refund phase ($t_0 + \secDelay \leq t < t + 2\secDelay$). The status after the move will be $\isContractState'.status = \isStatusStipCompensation{\participant{B}}$.  But we know by the inductive hypothesis that $\participant{B} \neq \theHonestUser$. Other conditions hold by inductive hypothesis.\\

\textbf{Run Conformance} \\

We need to prove that $\xRun[a]$ is a valid extension of $\xRun[0]$ and it conforms to the strategy of the honest user. The intuition behind this is that we are only completing already started authorizations, so we know that the honest user intended to authorize. \\

For every authorization $(\participant{C}: ~\someContract) \in \xRun[a]$, we know that the authorization is valid because we are limiting ourselves to contracts $\someContract \in \maxFrontierFun{\xRun}$ and we know by the Frontier Configurations Lemma, that there is an active contract $\someContract$ in the last configuration of $\xRun$.

We are also saying that $\participant{C} \in \finishedContractAuths{A}{\someContract}{\isRun}$, which implies there is at least one partial authorization $(\participant{C}: \someContract, \someBlockchain') \in \isRun$. This means that $\someContract$ has the form of a priority choice, which we know to also be true in $\xRun$ by the Code Symmetry invariant. \\

For the case where $\participant{C} = \theHonestUser$, again, there is at least one partial authorization $(\participant{C}: \someContract, \someBlockchain') \in \isRun$, which implies, by the definition of our honest strategy compilation, that $\theHonestUser: \someContract'$ is conforming to $\theHonestUser$'s strategy. \\

\textbf{Coherence}

Compensations do not change the maximal frontiers. 

For every contract $\someContract \in \maxFrontierFun{\xRun}$, we can reason that:

\[
    \begin{aligned}        
    \contractAuths{\someContract}{\xRun} &=
        \contractAuths{\someContract}{\xRun[0]}
        \cup (\finishedContractAuths{A}{\someContract'}{\isRun} \setminus \finishedContractAuths{A}{\someContract}{\isRun[0]}) \\
        &= \finishedContractAuths{A}{\someContract}{\isRun[0]}
            \cup (\contractAuths{\someContract}{\isRun[0]} \cap \{\theHonestUser\})
            \cup (\finishedContractAuths{A}{\someContract}{\isRun} \setminus \finishedContractAuths{A}{\someContract}{\isRun[0]}) \\
        &= \finishedContractAuths{A}{\someContract}{\isRun} \cup (\contractAuths{\someContract}{\isRun[0]} \cap \{\theHonestUser\})\\
    \end{aligned}
\]

\paragraph{\underline{Top-level Withdraw}}
$\isRun = \isRun[0] \xrightarrow{cwithdraw(\someContract, \someBlockchain)} \isConfig$ and we know by inductive hypothesis that $\xStrategy[A] \xConforms \xRun[0] \land \xRun[0] \xCoherence[\theHonestUser] \isRun[0]$. We start by proving round-based execution of the intermediate run. \\

We are doing a $cwithdraw(\someContract, \someBlockchain)$ in the intermediate semantics, that will consume an active contract $\isActiveContract{\bitmlxWithdraw{\balance}{A}}{\isContractState}{\someBlockchain}{\someContract}$ with $\isContractState.status = \isStatusChoice$ and result in left-descendant terminal contracts $\isActiveContract{\contract{C}}{\isContractState_1}{\someBlockchain}{\someContract_1}, \dots, \isActiveContract{\contract{C}}{\isContractState_n}{\someBlockchain}{\someContract_n}$, where for each withdrawing user $\participant{A_i} \in \vec{\participant{A}}$, $\isContractState_i.status = \isStatusAssigned{\participant{A_i}}$ and $\someContract_i = \someContract| L_i$. \\

\textbf{Timeout Consistency} \\

Trivial, as all successor contracts are terminal.

\textbf{Round-Based Execution} \\

Because $\isContractState.status = \isStatusChoice$, we know by inductive hypothesis that the round status of $\someContract$, $\roundStatus{\someContract}{\isRun} = (r, p)$, can be either waiting to synchronise ($|\someContract| > r$) or in compensation phase ($|\someContract| > r \land p = 0$). We will split by cases on the possible round status of $\someContract$.

\begin{itemize}
    \item If $|\someContract| > r$, then we need to prove that if there is another active contract $\isActiveContract{\bitmlxWithdraw{\balance}{A}}{\isContractState}{\someOtherBlockchain}{\someContract}$ for some other blockchain $\someOtherBlockchain$, then there are no right-descendants. This is the case, because for all descendants, $\someContract_i = \someContract| L_i$.  
    \item If $|\someContract| > r \land p = 0$, then we know by the transition rule that for some user $\participant{B}$, we have a revealed step secret, that is, $\participant{B} \in \userStepSecrets{\someContract}{\isRun}$ and because $\isContractState.status = \isStatusChoice$, we know that $\participant{B} \neq \theHonestUser$.
\end{itemize}

For every successor $\someContract_i$, we know that if they have a round status $\roundStatus{\someContract_i}{\isRun} = (r_i, p_i)$. $|\someContract_i| > |\someContract| \geq r$. That is, the successor contracts are waiting to synchronize, which is consistent with being in an assigned state. \\

For proving Run Conformance and Coherence, we will split by cases on whether the \bitmlx contract $\someContract$ is in the maximal frontier of $\xRun[0]$ or not, i.e. on whether $\someContract \in \maxFrontierFun{\xRun[0]}$.

\begin{itemize}
         \item \underline{If $\someContract \in \maxFrontierFun{\xRun[0]}$} then we are in the case where this is the first partial move for $\someContract$, which will extend the maximal frontier. We will extend the \bitmlx run to mirror this movement, so we choose $\xRun = \reductionRule{\xRun[0]}{cwithdraw(\someContract)}{\xConfig}$ for this case. \\

     \textbf{Run Conformance} \\

     We want to prove that $\xStrategy \xConforms \xRun$. Because we are doing a $cwithdraw(\someContract, \someBlockchain)$ in the intermediate semantics, we know that we have a contract $\someParContract \in \lastConfigOf{\isRun[0]}$ with $\contract{C} = \bitmlxWithdraw{\balance}{A}$. And because $\someContract \in \maxFrontierFun{\xRun[0]}$, by the Frontier Configuration Lemma, we also know that for some balance $\balance$ and some contract $\contract{C'}$, we have $\activeContract{\contract{C'}}{\balance}{\someContract} \in \lastConfigOf{\xRun}$. But we know by the Configuration Consistency invariant that $\contract{C} = \contract{C'}$. \\
     
     Then, according to the \bitmlx semantics, $\reductionRule{\xRun}{cwithdraw(\someContract)}{}$ is a valid move. And because it does not depend on $\theHonestUser$'s strategy, we can claim that $\xStrategy \xConforms \xRun$. \\

    \textbf{Coherence} \\

    We need to prove that after extending both runs, we still have correspondence of maximal frontiers, that is, $\maxFrontierFun{\xRun} = \isJoinFrontier{\isRun}$. Intuitively, this should hold because we are extending the \bitmlx frontier and eventual synchronicity frontier in the same way.
    
    We know that $\someContract \in \maxFrontierFun{\xRun[0]}$ and we are extending the run with a $cwithdraw(\someContract)$ move. If we consider the successors $\vec\someContract$ of $\someContract$ after this move, we can see  that the new maximal frontier of the \bitmlx run will be

    $$\maxFrontierFun{\xRun} = \maxFrontierFun{\xRun[0]} \setminus \{ \someContract \} \cup \{ \someContract' \in \vec{\someContract} \} = \isJoinFrontier{\isRun[0]} \setminus \{ \someContract \} \cup \{ \someContract' \in \vec{\someContract} \}$$

    And, because $\someContract$ is in the eventual synchronicity frontier of $\isRun[0]$, we know by the Eventual Synchronicity Update Lemma that

    \[
        \isJoinFrontier{\isRun} = \isJoinFrontier{\isRun[0]} \setminus \{ \someContract \} \cup \{ \someContract' \in \vec{\someContract} \}
    \]

    In conclusion, we proved that $\maxFrontierFun{\xRun} = \maxFrontierFun{\xRun}$ \\

    \textbf{Configuration Consistency} \\

    The \bitmlx and intermediate semantics rules will introduce the same successor contracts and distribute the funds in the same way, so the invariant holds.

        \item \underline{If $\someContract \notin \maxFrontierFun{\xRun[0]}$} then we are in the case where this is not the first partial move, but we are instead replicating a move started by another blockchain and that we expect to already be present in the \bitmlx run. For this reason, we will not extend the \bitmlx run but instead keep $\xRun = \xRun[0]$. We know by inductive hypothesis that this run is valid and conforming to the user strategy. \\

    \textbf{Coherence} \\

    We didn't change the \bitmlx run, so $\maxFrontierFun{\xRun} = \maxFrontierFun{\xRun[0]}$. And by inductive hypothesis we know that $\maxFrontierFun{\xRun[0]} = \isJoinFrontier{\isRun[0]}$. Additionally, because $\someContract \notin \maxFrontierFun{\xRun[0]}$, we know that $\isJoinFrontier{\isRun} = \isJoinFrontier{\isRun[0]}$. That is, this move didn't change the join of maximal intermediate frontiers. Intuitively, this is because we are replicating a similar move on another blockchain, so the join of maximal frontiers already has descendants of $\someContract$. This means that even if this move expands the maximal frontier of $\someBlockchain$, the join will still be greater.

    In conclusion, we proved that $\maxFrontierFun{\xRun} = \maxFrontierFun{\xRun}$ \\
    
    \textbf{Configuration Consistency} \\
    
    Holds trivially, as $\someContract \notin \maxFrontierFun{\xRun[0]}$.
    
\end{itemize}

\paragraph{\underline{Step secret reveal}}
$\isRun = \isRun[0] \xrightarrow{\participant{B}: \stepSecret{\someContract}{\isRun}} \isConfig$ and we know by inductive hypothesis that $\xStrategy[A] \xConforms \xRun[0] \land \xRun[0] \xCoherence[\theHonestUser] \isRun[0]$.

We do not extend the \bitmlx and instead keep $\xRun = \xRun[0]$. Since this intermediate move doesn't change the frontiers or authorizations, Run Conformance and Coherence are trivial. \\

\textbf{Round-Based Execution} \\

We need to prove that the step secret reveal is not violating the invariants for the honest user.

Let $\someParPrChoice \in \lastConfigOf{\isRun}$ be the active contract for $\someContract$ and $(r, p) = \roundStatus{\someContract}{\isRun}$. If $\participant{B} = \theHonestUser$, we know according to the honest user strategy compilation that $\isContractState.status = \isStatusChoice$ and $t < \isContractState.time = t_0 + 2|\someContract| \secDelay$. Solving for $|\someContract|$, we get that $|\someContract| > \frac{t-t_0}{2\secDelay} \geq \left \lfloor \frac{t - t_0}{2\secDelay} \right \rfloor = r$. That is, $\someContract$ is waiting to synchronise, which means that there is no restriction in round-based execution for revealing the step secret. \\

\paragraph{\underline{Introduction of left}}
$\isRun = \isRun[0] \xrightarrow{left(\someContract, \someBlockchain)} \isConfig$ and we know by inductive hypothesis that $\xStrategy[A] \xConforms \xRun[0] \land \xRun[0] \xCoherence[\theHonestUser] \isRun[0]$.

We are doing a $left(\someContract, \someBlockchain)$ move in the intermediate semantics, that will change the status of an active contract $\isActiveContract{\bitmlxAuthOp{A}{\contract{D}} \prchoice \contract{C}}{\isContractState}{\someBlockchain}{\someContract}$ from $\isContractState.status = \isStatusChoice$ to $\isContractState'.status = \isStatusLeft$ and consume authorizations

We do not extend the \bitmlx and instead keep $\xRun = \xRun[0]$. Since this intermediate move doesn't change the frontiers or authorizations, Run Conformance and Coherence are trivial. \\

\textbf{Round-Based Execution} \\

Let $\someParPrChoice \in \lastConfigOf{\isRun}$ be the active contract for $\someContract$. We know that according to the intermediate semantics, $\isContractState.status = \isStatusChoice$ and by the inductive hypothesis of round-based execution, $\someContract$ is either waiting to synchronize or in compensation phase. In both cases, the change to $\isContractState'.status = \isStatusLeft$ is a valid one.

In the case where $\someContract$ is in compensation phase, we know by the intermediate semantics that there is a step secret $\stepSecret{B}{\someContract}$ revealed for some user $\participant{B}$ and by the inductive hypothesis that $\participant{B} \neq \theHonestUser$. \\

\textbf{Authorized Left} \\

We are introducing a contract with $\isStatusLeft$ status, so we need to prove that if $\theHonestUser$ is one of the users authorizing the contract and we have the active contract in the \bitmlx configuration, then we also have $\theHonestUser$'s authorization for it.

If $\activeContract{\bitmlxAuthOp{A}{\contract{D}} \prchoice \contract{C}}{\balance}{\someContract} \in \lastConfigOf{\xRun}$, then $\someContract \in \maxFrontierFun{\xRun}$. We know that $\theHonestUser \in \contractAuths{\someContract}{\isRun}$, so by Coherence we also know that $\theHonestUser \in \contractAuths{\someContract}{\xRun}$. Then by the Authorizations Persistance lemma, we know that $\userAuthIn{\theHonestUser}{\someContract}{\contract{D}} \in \lastConfigOf{\xRun}$.

\paragraph{\underline{Guarded Withdraw}}
$\isRun = \isRun[0] \xrightarrow{dwithdraw(\someContract, \someBlockchain)} \isConfig$ and we know by inductive hypothesis that $\xStrategy[A] \xConforms \xRun[0] \land \xRun[0] \xCoherence[\theHonestUser] \isRun[0]$. 

We are doing a $dwithdraw(\someContract, \someBlockchain)$ move in the intermediate semantics, that will consume an active contract $\isActiveContract{\contract{C}}{\isContractState}{\someBlockchain}{\someContract}$ with $\contract{C} = \bitmlxAuthOp{B}{\bitmlxWithdraw{\balance}{A}} \prchoice \contract{C'}$ (where $\vec{\participant{B}}$ is possibly empty), $\isContractState.status = \isStatusLeft$ and result in left-descendant terminal contracts $\isActiveContract{\contract{C}}{\isContractState_1}{\someBlockchain}{\someContract_1}, \dots, \isActiveContract{\contract{C}}{\isContractState_n}{\someBlockchain}{\someContract_n}$, where for each withdrawing user $\participant{A_i} \in \vec{\participant{A}}$, $\isContractState_i.status = \isStatusAssigned{\participant{A_i}}$ and $\someContract_i = \someContract| L_i$. \\

The proof for the round-based execution invariant is analogous to the $cwithdraw$ case and for timeout consistency it is also trivial. \\

For Run Conformance, Coherence and Configuration Consistency, we will again start by splitting by cases on whether the \bitmlx contract $\someContract$ is in the maximal frontier of $\xRun[0]$ or not, i.e. on whether $\someContract \in \maxFrontierFun{\xRun[0]}$.

\begin{itemize}
    \item If \underline{$\someContract \in \maxFrontierFun{\xRun[0]}$} then we are in the case where this is the first partial move for $\someContract$, which will extend the maximum frontier. We will extend the \bitmlx run to mirror this movement, like we did on the $cwithdraw$ case. But additionally, we might need to add missing authorizations if needed.

    We will choose $\xRun = \reductionRule{\xRun[0]\xRun[a]}{dwithdraw(\someContract)}{\xConfig}$, where $\xRun[a]$ is a sequence of $\xrightarrow{\participant{B}: \someContract}$ transitions for every $\participant{B} \in \vec{\participant{B}}: \userAuthIn{B}{\someContract}{\bitmlxWithdraw{\balance}{A}} \notin \lastConfigOf{\xRun}$. That is, we first extend with all missing authorizations needed and then with the withdraw move. \\

    \textbf{Run Conformance} \\
    
    We know by the Configuration Consistency invariant that $\activeContract{\contract{C}}{\balance}{\someContract} \in \lastConfigOf{\xRun}$. This makes every individual authorization move $(\participant{B}: \someContract) \in \xRun[a]$ a valid transition, and because their only change to the configuration is adding the authorization object, we can conclude that $\xRun[a]$ is a valid extension. It is also conformant to the honest user strategy because we know by the Honest Left invariant that authorization from $\theHonestUser$ already. That is, $\userAuthIn{\theHonestUser}{\someContract}{\bitmlxWithdraw{\balance}{A}} \in \lastConfigOf{\isRun[0]}$.
    
    And because after $\xRun[a]$ we have all needed authorizations, $dwithdraw(\someContract)$ is a valid move in $\xRun[0]\xRun[a]$. It can be scheduled independently of the honest user, so $\xStrategy \xConforms \xRun$. \\

    \textbf{Coherence} \\

    The reasoning for the successor assigned contracts is analogous to that in the $cwithdraw$ case.

    But because we are possibly adding authorizations, we need to prove that we hold the $\contractAuths{\someContract}{\xRun} = \finishedContractAuths{A}{\someContract}{\isRun} \cup (\contractAuths{\someContract}{\isRun} \cap \{ \participant{A }\})$ invariant. But this is very straight forward, because the $dwithdraw(\someContract)$ move in the \bitmlx run will consume the active contract for $\someContract$, so $ \someContract \notin \maxFrontierFun{\xRun}$. \\
    
    \textbf{Configuration Consistency} \\

    Analogous to the case of $cwithdraw$. \\
         
    \item If \underline{$\someContract \notin \maxFrontierFun{\xRun[0]}$} then we are in the case where this is not the first partial move, but we are instead replicating a move started by another blockchain and that we expect to already be present in the \bitmlx run. For this reason, we will not extend the \bitmlx run but instead keep $\xRun = \xRun[0]$. We know by inductive hypothesis that this run is valid and conforming to the user strategy. The proofs for Coherence and Configuration Consistency are analogous to the case of $cwithdraw$. \\

\end{itemize}

\paragraph{\underline{Split}}
$\isRun = \isRun[0] \xrightarrow{split(\someContract, \someBlockchain)} \isConfig$ and we know by inductive hypothesis that $\xStrategy[A] \xConforms \xRun[0] \land \xRun[0] \xCoherence[\theHonestUser] \isRun[0]$.

We are doing a $split(\someContract, \someBlockchain)$ in the intermediate semantics, that will consume an active contract $\isActiveContract{\bitmlxAuthOp{B}{\bitmlxSplit{\balance}{C}}}{\isContractState}{\someBlockchain}{\someContract}$ with $\isContractState.status = \isStatusLeft$ and result in left-descendant active contracts $\isActiveContract{\contract{C_1}}{\isContractState_1}{\someBlockchain}{\someContract_1}, \dots, \isActiveContract{\contract{C_n}}{\isContractState_n}{\someBlockchain}{\someContract_n}$, where for each contract, $\isContractState_i.status = \isStatusChoice$ and $\someContract_i = \someContract | L_i$. \\

Proofs for Run Conformance Coherence and Configuration Consistency are analogous to the $dwithdraw$ case. \\

\textbf{Timeout Consistency} \\

Because $\isContractState.status = \isStatusLeft$, we know by inductive hypothesis that $\isContractState.time = t_0 + 2|\someContract|\secDelay$ where $t_0$ is the stipulation time of $\someContract$'s root contract. Then, for every successor $\someContract_i$, we can reason in the following way:

\[
    \isContractState_i.time
    = \isContractState.time + 2\secDelay
    = t_0 + 2|\someContract|\secDelay + 2 \secDelay
    = t_0 + 2(|\someContract|+1)\secDelay
    = t_0 + 2(|\someContract_i|)\secDelay
\]

\textbf{Round-Based Execution} \\

Because $\isContractState.status = \isStatusLeft$, we know by inductive hypothesis that the round status $\roundStatus{\someContract}{\isRun} = (r, p)$ of $\someContract$ can be either waiting to synchronise ($|\someContract| > r$) or in compensation phase ($|\someContract| > r \land p = 0$).  We will split by cases on the possible round status of $\someContract$.

\begin{itemize}
    \item If $|\someContract| > r$, then we need to prove that if there is another active contract $\isActiveContract{\bitmlxSplit{\balance}{C}}{\isContractState}{\someOtherBlockchain}{\someContract}$ for some other blockchain $\someOtherBlockchain$, then there are no right-descendants. This is the case, because for all descendants, $\someContract_i = \someContract| L_i$.  
    \item If $|\someContract| > r \land p = 0$, because $\someContract$ has succesor contracts in $\isRun$, we need to prove that $\userStepSecrets{\someStipulation}{\isRun} \setminus \{\theHonestUser\} \neq \emptyset$, but we already know this by inductive hypothesis because $\isContractState.status = \isStatusLeft$.
\end{itemize}

For every successor $\someContract_i$, we know that $|\someContract_i| > |\someContract| \geq r$. That is, the successor contracts are waiting to synchronize, which is consistent with being in a choice state. \\

\paragraph{\underline{Reveal}}
$\isRun = \isRun[0] \xrightarrow{reveal(\someContract, \someBlockchain)} \isConfig$ and we know by inductive hypothesis that $\xStrategy[A] \xConforms \xRun[0] \land \xRun[0] \xCoherence[\theHonestUser] \isRun[0]$.

We are doing a $reveal(\someContract, \someBlockchain)$ move in the intermediate semantics, that will consume an active contract $\isActiveContract{\bitmlxAuthOp{B}{\bitmlxReveal{\secret{s}}{p}{C}} \prchoice \contract{C'}}{\isContractState}{\someBlockchain}{\someContract}$ (where $\vec{\participant{B}}$ and $\vec{\secret{s}}$ are possibly empty), $\isContractState.status = \isStatusChoice$ and result in a left-descendant active contract $\isActiveContract{\contract{C}}{\isContractState'}{\someBlockchain}{\someContract'}$, with $\isContractState'.status = \isStatusChoice$ and $\someContract' = \someContract | L_0$. \\

For Run Conformance and Coherence we again split on whether $\someContract \in \maxFrontierFun{\xRun}$.

\begin{itemize}
    \item If \underline{$\someContract \in \maxFrontierFun{\xRun[0]}$} then we are in the case where this is the first partial move for $\someContract$, which will extend the maximum frontier. We will extend the \bitmlx run to mirror this movement and, like in the $dwithdraw$ case, we will add all of the missing authorizations first.

    We will choose $\xRun = \reductionRule{\xRun[0]\xRun[a]}{reveal(\someContract)}{\xConfig}$, where $\xRun[a]$ is a sequence of $\xrightarrow{\participant{B}: \someContract}$ transitions for every $\participant{B} \in \vec{\participant{B}}: \userAuthIn{B}{\someContract}{\bitmlxReveal{\secret{s}}{p}{C}} \notin \lastConfigOf{\xRun}$. \\ 

    \textbf{Run Conformance} \\
    
    We know by the Configuration Consistency invariant that $\activeContract{\bitmlxAuthOp{B}{\bitmlxReveal{\secret{s}}{p}{C}} \prchoice \contract{C'}}{\balance}{\someContract} \in \lastConfigOf{\xRun}$. This makes every individual authorization move $(\participant{B}: \someContract) \in \xRun[a]$ a valid transition, and because their only change to the configuration is adding the authorization object, we can conclude that $\xRun[a]$ is a valid extension.
    
    We know by the intermediate semantics transition rule for $reveal$ that $\userAuthIn{\theHonestUser}{(\someContract, \someBlockchain)}{\contract{D}} \in \lastConfigOf{\isRun[0]}$. This is only possible if $(\theHonestUser: \someContract, \someBlockchain) \in \isRun[0]$ so by Coherence we also know that $(\theHonestUser: \someContract) \isRun[0]$. Then, by the Authorization Persistance lemma, we know that $\userAuthIn{\theHonestUser}{\someContract}{\bitmlxReveal{\secret{s}}{p}{C}} \in \lastConfigOf{\xRun}$, meaning that we will not be authorizing for $\theHonestUser$ in $\xRun[a]$, so the extension is conformant to the honest user strategy.
    
    We also know by the intermediate semantics transition rule for $reveal$ that $\{\secret{s} \in \vec{\secret{s}}\} \subseteq \userRevealedSecrets{\isRun[0]}$ and Coherence we know that $\userRevealedSecrets{\isRun[0]} = \userRevealedSecrets{\xRun[0]}$. Additionaly we also know by the intermediate semantics that $\llbracket p \rrbracket_{secrets} = true$. And because after $\xRun[a]$ we have all needed authorizations, $reveal(\someContract)$ is a valid move in $\xRun[0]\xRun[a]$. It can be scheduled independently of the honest user, so $\xStrategy \xConforms \xRun$. \\
    
    Proofs for Coherence and Configuration Consistency are analogous to the $dwithdraw$ case. \\
    
    \item If \underline{$\someContract \notin \maxFrontierFun{\xRun[0]}$} then we don't extend the \bitmlx, keeping instead $\xRun = \xRun[0]$. Run conformance is trivial by inductive hypothesis and Coherence is analogous to the $dwithdraw$ case. 
    
\end{itemize}

The proofs for Round-based Execution and Timeout Consistency are analogous to the case of $split$, for the single successor contract. \\

\paragraph{\underline{Right}}
$\isRun = \isRun[0] \xrightarrow{right(\someContract, \someBlockchain)} \isConfig$ and we know by inductive hypothesis that $\xStrategy[A] \xConforms \xRun[0] \land \xRun[0] \xCoherence[\theHonestUser] \isRun[0]$.

We do not extend the \bitmlx and instead keep $\xRun = \xRun[0]$. Since this intermediate move doesn't change the frontiers or authorizations, Run Conformance, Coherence and Configuration Consistency are trivial. \\

\textbf{Timeout Consistency} \\

Because $\isContractState.status = \isStatusChoice$, we know by inductive hypothesis that $\isContractState.time = t_0 + 2|\someContract|\secDelay$ where $t_0$ is the stipulation time of $\someContract$'s root contract. Then, for the new status $\isContractState'$ after the move, we can reason in the following way:

\[
    \isContractState'.time
    = \isContractState.time + \secDelay
    = t_0 + 2|\someContract|\secDelay + \secDelay
    = t_0 + (2|\someContract|+1)\secDelay
\]

\textbf{Round-Based Execution} \\

Let $\someParPrChoice \in \isConfig$ be the active contract for $\someContract$. We know that according to the intermediate semantics, $\isContractState.status = \isStatusChoice$. By the inductive hypothesis of round-based execution, $\someContract$ is either waiting to synchronize ($|\someContract| < r$) or in compensation phase ($|\someContract| = r \land p=0$). But we also know that $t \geq \isContractState.time = t_0 + 2|\someContract| \secDelay$. Solving for $|\someContract|$, we get that $|\someContract| \leq \frac{t-t_0}{2\secDelay}$ which implies that $r = \left \lfloor \frac{t - t_0}{2\secDelay} \right \rfloor \geq |\someContract|$. That means $\someContract$ is in compensation phase. Other conditions hold by inductive hypothesis. \\

\paragraph{\underline{Skip}}
$\isRun = \isRun[0] \xrightarrow{skip(\someContract, \someBlockchain)} \isConfig$ and we know by inductive hypothesis that $\xStrategy[A] \xConforms \xRun[0] \land \xRun[0] \xCoherence[\theHonestUser] \isRun[0]$. 

We are doing a $skip(\someContract, \someBlockchain)$ move in the intermediate semantics, that will consume an active contract $\isActiveContract{\contract{D} \prchoice \contract{C}}{\isContractState}{\someBlockchain}{\someContract}$ with $\isContractState.status = \isStatusRight$ and result in a single right-descendant active contract $\isActiveContract{\contract{C}}{\isContractState^{\downarrow}}{\someBlockchain}{\someDescendant}$, where, $\isContractState^{\downarrow}.status = \isStatusChoice$ and $\someDescendant = \someContract | R \in rdesc(\someContract)$. \\

For proving Run Conformance, Coherence and Configuration Consistency, we will split by cases on whether the \bitmlx contract $\someContract$ is in the maximal frontier of $\xRun[0]$ or not, i.e. on whether $\someContract \in \maxFrontierFun{\xRun[0]}$.

\begin{itemize}
         \item \underline{If $\someContract \in \maxFrontierFun{\xRun[0]}$} then we are in the case where this is the first partial move for $\someContract$, which will extend the maximal frontier. We will extend the \bitmlx run to mirror this movement, so we choose $\xRun = \reductionRule{\xRun[0]}{skip(\someContract, \vec{\someContract})}{\xConfig}$ for this case. \\

     \textbf{Run Conformance} \\

     We want to prove that $\xStrategy \xConforms \xRun$. We know by the Configuration Consistency invariant that we have an active contract $\activeContract{\contract{D} \prchoice \contract{C}}{\balance}{\someContract} \in \lastConfigOf{\isRun}$, but the \bitmlx semantics state that the move needs to be done by consensus. That is, it needs to be in the output of the strategies for all users. In particular, we need to prove that $skip(\someContract) \in \xStrategy(\isRun[0])$. But, because $\isContractState.status = \isStatusRight$, we know this to be true by the honest skip invariant.

    \textbf{Coherence} \\

    Proofs for Coherence and Configuration Consistency are analogous to the $dwithdraw$ case. \\

    \item \underline{If $\someContract \notin \maxFrontierFun{\xRun[0]}$}, then again, we will not extend the \bitmlx run but instead keep $\xRun = \xRun[0]$. We know by inductive hypothesis that this run is valid and conforming to the user strategy. The proofs for Coherence and Configuration Consistency are analogous to the case of $cwithdraw$. \\

\end{itemize}

\textbf{Timeout Consistency} \\

Because $\isContractState.status = \isStatusRight$, we know by inductive hypothesis that $\isContractState.time = t_0 + (2|\someContract|+1)\secDelay$ where $t_0$ is the stipulation time of $\someContract$'s root contract. Then, the successor $\someContract'$, we can reason in the following way:

\[
    \isContractState'.time
    = \isContractState.time + \secDelay
    = t_0 + (2|\someContract|+1)\secDelay + \secDelay
    = t_0 + (2|\someContract|+2)\secDelay
    = t_0 + 2(|\someContract|+1)\secDelay
    = t_0 + 2(|\someContract'|)\secDelay
\]

\textbf{Round-Based Execution} \\

Because $\isContractState.status = \isStatusRight$, we know by inductive hypothesis that the round status $\roundStatus{\someContract}{\isRun} = (r, p)$ of $\someContract$ can be either compensation phase ($|\someContract| > r \land p = 0$) or skipping phase ($|\someContract| > r \land p = 1$). But we also know by the transition rule for $skip$ that $t \geq \isContractState.time = t_0 + (2|\someContract|+1)\secDelay$ which implies that $p=1$.

Because we have a right-descendant for $\someContract$, we need to prove that if there are other contracts $\isActiveContract{\contract{D} \prchoice \contract{C}}{\isContractState}{\someBlockchain}{\someContract}$, then there are no left descendants. We will prove this by contradiction. Suppose the opposite was true and $ldesc(\someContract) \neq \emptyset$. That would imply by inductive hypothesis that $\exists \participant{B} \in \contractUsers{\someContract}{\isRun} \setminus \{\theHonestUser\}: \isContractState.status = \isStatusCompensated{\participant{B}}$. Which is a contradiction because $\isContractState.status = \isStatusRight$.

For the successor $\someDescendant$, we know that $|\someDescendant| > |\someContract| \geq r$. That is, the successor contract is waiting to synchronize, which is consistent with being in a choice state. \\

\paragraph{\underline{Authorizations}}
$\isRun = \isRun[0] \xrightarrow{\participant{B} \colon~ \someContract, \someBlockchain} \isConfig$ and we know by inductive hypothesis that $\xStrategy[A] \xConforms \xRun[0] \land \xRun[0] \xCoherence[\theHonestUser] \isRun[0]$. 

We need to distinguish a few different cases here. Firstly, whether the authorization is for a contract on the maximal frontier. If $\someContract \notin \maxFrontierFun{\xRun[0]}$, then we can ignore this transition and keep $\xRun = \xRun[0]$ which is trivially conforming and (given that the maximal frontier doesn't change) coherent.

If $\someContract \in \maxFrontierFun{\xRun[0]}$, the next distinction  on whether the user authorizing is our honest user $\theHonestUser$ or some other potentially malicious user. That is, whether $\participant{B} = \theHonestUser$.
The general strategy will be that for the case where it's the honest user authorizing we will mirror the authorization in the \bitmlx run when observing the authorization on the first blockchain, while for other users we will take a more conservative approach and only mirror the authorization on the last blockchain.

\begin{itemize}
    \item If \underline{$\participant{B} = \theHonestUser$}, we need to further distinguish whether this is the first blockchain where the honest user is authorizing the move or not.
    \begin{itemize}
        \item If \underline{$\theHonestUser \notin \contractAuths{\someContract}{\isRun[0]}$} then we will mirror this authorization in the \bitmlx run. That is, we will choose $\xRun = \xRun[0] \xrightarrow{\theHonestUser \colon~ \someContract} \isConfig$. \\

        \textbf{Conformance} \\

        We first need to prove that $\theHonestUser \colon~ \someContract$ is a valid move in $\xRun[0]$. For this, the \bitmlx semantics requires that we have an active priority choice contract $\someContract$ in the latest configuration of $\xRun$.
        Because we are doing a $\theHonestUser \colon~ \someContract, \someBlockchain$ transition, the intermediate semantics requires that there is an active contract $\isActiveContract{\contract{D} \prchoice \contract{C}}{\isContractState}{\someBlockchain}{\someContract}$ in $\lastConfigOf{\isRun}$.  By the Frontier Configuration Lemma, we also know that for some balance $\balance$ and some contract $\contract{C'}$, we have $\activeContract{\contract{C'}}{\balance}{\someContract} \in \lastConfigOf{\xRun}$. But we know by the code symmetry invariant that $\contract{C} = \contract{C'}$. \\

        Additionally, we know that $\isStrategy \models \isRun$, so it must be the case that this authorization is on the in intermediate semantics strategy of $\theHonestUser$. Then, by the definition of our honest user strategy compilation, we also know that authorizing is on $\theHonestUser$'s \bitmlx strategy. \\

        \textbf{Coherence} \\

        We need to prove that $\contractAuths{\someContract}{\xRun} = \finishedContractAuths{A}{\someContract}{\isRun} \cup 
        (\contractAuths{\someContract}{\isRun} \cap \{\theHonestUser\}$. If we observe definitions of $\xRun$ and $\isRun$, we can see that $\contractAuths{\someContract}{\xRun} = \contractAuths{\someContract}{\xRun[0]} \cup \{\theHonestUser\}$ and $\contractAuths{\someContract}{\isRun} = \contractAuths{\someContract}{\isRun[0]} \cup \{\theHonestUser\}$. Then, we have:

        \[
            \begin{aligned}        
            \contractAuths{\someContract}{\xRun} &= \contractAuths{\someContract}{\xRun[0]} \cup \{\theHonestUser\} \\
                &= \finishedContractAuths{A}{\someContract}{\isRun[0]}
                    \cup (\contractAuths{\someContract}{\isRun[0]} \cap \{\theHonestUser\}) \cup \{\theHonestUser\}\\
                &= \finishedContractAuths{A}{\someContract}{\isRun[0]}
                    \cup ((\contractAuths{\someContract}{\isRun[0]} \cup \{\theHonestUser\})
                    \cap \{\theHonestUser\}) \\
                &= \finishedContractAuths{A}{\someContract}{\isRun[0]}
                    \cup (\contractAuths{\someContract}{\isRun}
                    \cap \{\theHonestUser\}) \\
            \end{aligned}
        \]

        \item If \underline{$\theHonestUser \in \contractAuths{\someContract}{\isRun[0]}$}, then we don't extend the run and keep $\xRun = \xRun[0]$, which is trivially conforming. To see that it's also coherent, we can do: 
    
        \[
            \begin{aligned}        
            \contractAuths{\someContract}{\xRun} &= \contractAuths{\someContract}{\xRun[0]} \\
                &= \finishedContractAuths{A}{\someContract}{\isRun[0]}
                    \cup (\contractAuths{\someContract}{\isRun[0]} \cap \{\theHonestUser\}) \\
                &= \finishedContractAuths{A}{\someContract}{\isRun[0]}
                    \cup ((\contractAuths{\someContract}{\isRun[0]} \cup \{\theHonestUser\})
                    \cap \{\theHonestUser\}) \\
                &= \finishedContractAuths{A}{\someContract}{\isRun[0]}
                    \cup (\contractAuths{\someContract}{\isRun}
                    \cap \{\theHonestUser\}) \\
            \end{aligned}
        \]
    \end{itemize}

     \item If \underline{$\participant{B} \neq \theHonestUser$}, we need to further distinguish whether this is the last blockchain where the potentially malicious user is authorizing the move or not, up to compensations for $\theHonestUser$. 

     \begin{itemize}
        \item If \underline{$\participant{B} \in \finishedContractAuths{A}{\someContract}{\isRun}$}, then we extend the \bitmlx run with the corresponding authorization. We will choose $\xRun = \xRun[0] \xrightarrow{\participant{B} \colon~ \someContract} \isConfig$. \\

        \textbf{Conformance} \\

        Idem for the case of $\theHonestUser$, except that this time we don't care about conforming to the strategy, because it's not an honest user. \\
        
        \textbf{Coherence} \\

        We need to prove that $\contractAuths{\someContract}{\xRun} = \finishedContractAuths{A}{\someContract}{\isRun} \cup 
        (\contractAuths{\someContract}{\isRun} \cap \{\theHonestUser\}$. We start by observing that $\participant{B}$ is finishing the authorization in this action, so $\finishedContractAuths{A}{\someContract}{\isRun} = \{\participant{B}\} \cup  \finishedContractAuths{A}{\someContract}{\isRun[0]}$. Then, we have:

        \[
            \begin{aligned}        
            \contractAuths{\someContract}{\xRun} &= \{\participant{B}\} \cup \contractAuths{\someContract}{\xRun[0]} \\
                &= \{\participant{B}\} \cup  \finishedContractAuths{A}{\someContract}{\isRun[0]}
                    \cup (\contractAuths{\someContract}{\isRun[0]} \cap \{\theHonestUser\}) \\
                &= \finishedContractAuths{A}{\someContract}{\isRun}
                    \cup (\contractAuths{\someContract}{\isRun}
                    \cap \{\theHonestUser\}) \\
            \end{aligned}
        \]

        \item If \underline{$\participant{B} \notin \finishedContractAuths{A}{\someContract}{\isRun}$}, then we don't extend the run and keep $\xRun = \xRun[0]$, which is trivially conforming. To see that it's also coherent, we can do: 
    
        \[
            \begin{aligned}        
            \contractAuths{\someContract}{\xRun} &=  \contractAuths{\someContract}{\xRun[0]} \\
                &= \finishedContractAuths{A}{\someContract}{\isRun[0]}
                    \cup (\contractAuths{\someContract}{\isRun[0]} \cap \{\theHonestUser\}) \\
                &= \finishedContractAuths{A}{\someContract}{\isRun}
                    \cup (\contractAuths{\someContract}{\isRun}
                    \cap \{\theHonestUser\}) \\
            \end{aligned}
        \]
        
     \end{itemize}
\end{itemize}

\paragraph{\underline{Slashing}}
$\isRun = \isRun[0] \xrightarrow{slash(\someContract, \someBlockchain, \participant{B})} \isConfig$ and we know by inductive hypothesis that $\xStrategy[A] \xConforms \xRun[0] \land \xRun[0] \xCoherence[\theHonestUser] \isRun[0]$.

We do not extend the \bitmlx and instead keep $\xRun = \xRun[0]$. Since this intermediate move doesn't change the frontiers or authorizations, Run Conformance and Coherence are trivial. \\

\textbf{Round-Based Execution} \\

Let $\someParPrChoice \in \isConfig$ be the active contract for $\someContract$. We know that according to the intermediate semantics, $\isContractState.status = \isStatusRight$ and will change to $\isContractState'.status = \isStatusSlashed{\participant{B}}$. By the inductive hypothesis of round-based execution, $\someContract$ is either in compensation phase ($|\someContract| = r \land p=0$) or in skipping phase ($|\someContract| = r \land p=1$). In both cases, we know by inductive hypothesis that $\theHonestUser \notin \userStepSecrets{\someContract}{\isRun}$, so $slash(\someContract, \someBlockchain, \theHonestUser)$ would be an invalid move in the intermediate semantics. Other conditions hold by inductive hypothesis.

\paragraph{\underline{Compensations}}
$\isRun = \isRun[0] \xrightarrow{compensate(\someContract, \someBlockchain, \participant{B})} \isConfig$ and we know by inductive hypothesis that $\xStrategy[A] \xConforms \xRun[0] \land \xRun[0] \xCoherence[\theHonestUser] \isRun[0]$. \\

We are compensating with a revealed step secret on a blockchain that didn't use it to perform a left move. This can be either the honest user cleaning up an incomplete move or the malicious user hurting themselves. The started move, if any, is already in the run, so our only extension will be to add authorizations that we now consider complete when increasing the compensation history sets of those contracts. \\

\textbf{Run Conformance} \\

We will choose the new \bitmlx run to be $\xRun = \xRun[0]\xRun[a]$, where $\xRun[a]$ consists of authorizations $(\participant{C}: ~\someContract')$ for every user $\participant{C}$ and contract $\someContract'$ such that $\someContract' \in \maxFrontierFun{\xRun}$ and $\participant{C} \in \finishedContractAuths{C}{\someContract'}{\isRun} \setminus \finishedContractAuths{A}{\someContract'}{\isRun[0]}$. \\

We need to prove that $\xRun[a]$ is a valid extension of $\xRun[0]$ and it conforms to the strategy of the honest user. The intuition behind this is that we are only completing already started authorizations, so we know that the honest user intended to authorize. \\

For every authorization $(\participant{C}: ~\someContract') \in \xRun[a]$, we know that the authorization is valid because we are limiting ourselves to contracts $\someContract' \in \maxFrontierFun{\xRun}$ and we know by the Frontier Configurations Lemma, that there is an active contract $\someContract'$ in the last configuration of $\xRun$.

We are also saying that $\participant{C} \in \finishedContractAuths{A}{\someContract'}{\isRun}$, which implies there is at least one partial authorization $(\participant{C}: \someContract', \someBlockchain') \in \isRun$. This means that $\someContract'$ has the form of a priority choice, which we know to also be true in $\xRun$ by the Configuration Consistency invariant \\

For the case where $\participant{C} = \theHonestUser$, again, there is at least one partial authorization $(\participant{C}: \someContract', \someBlockchain') \in \isRun$, which implies, by the definition of our honest strategy compilation, that $\theHonestUser: \someContract'$ is conforming to $\theHonestUser$'s strategy. \\

\textbf{Coherence} \\

Compensations do not change the intermediate frontiers, so in particular they don't change the maximal frontier.
If we look at the authorizations for $\someContract'$, we can reason that:

\[
    \begin{aligned}        
    \contractAuths{\someContract}{\xRun} &=
        \contractAuths{\someContract}{\xRun[0]}
        \cup (\finishedContractAuths{A}{\someContract'}{\isRun} \setminus \finishedContractAuths{A}{\someContract'}{\isRun[0]}) \\
        &= \finishedContractAuths{A}{\someContract}{\isRun[0]}
            \cup (\contractAuths{\someContract}{\isRun[0]} \cap \{\theHonestUser\})
            \cup (\finishedContractAuths{A}{\someContract'}{\isRun} \setminus \finishedContractAuths{A}{\someContract'}{\isRun[0]}) \\
        &= \finishedContractAuths{A}{\someContract'}{\isRun} \cup (\contractAuths{\someContract}{\isRun[0]} \cap \{\theHonestUser\})\\
    \end{aligned}
\]

\textbf{Round-Based Execution} \\

We are doing a $compensation(\someContract, \someBlockchain, \participant{B})$ move in the intermediate run. We know by the intermediate semantics that we have an active contract $\someParContract$ with $\isContractState.status = \isStatusSlashed{\participant{B}}$, and by inductive hypothesis that the round status $\roundStatus{\someContract}{\isRun} = (r, p)$ of $\someContract$ can be either compensation phase ($|\someContract| > r \land p = 0$) or skipping phase ($|\someContract| > r \land p = 1$). The status after the move will be $\isContractState'.status = \isStatusCompensated{\participant{B}}$.  But we know by the inductive hypothesis that $\participant{B} \neq \theHonestUser$. \\

\paragraph{\underline{Time Delay}}
$\isRun = \isRun[0] \xrightarrow{\delta} \isConfig$ and we know by inductive hypothesis that $\xStrategy[A] \xConforms \xRun[0] \land \xRun[0] \xCoherence \isRun[0]$. Let $t$ be the time at the latest configuration of $\isRun[0]$. The time for $\isRun$ after the delay move will be $t + \delta$. \\

The intermediate semantics states that a time delay needs to be agreed by all participants. That is, for all participants $\participant{B}$, there should be a delay $\delta_{\participant{B}} \in \isStrategy[\participant{B}](\isRun[0])$ such that $\delta_{\participant{B}} \geq \delta$. In particular, for the honest user $\theHonestUser$, the strategy compilation specifies that they will only include a time delay when no condition for another action is met. Furthermore, the honest user strategy will always wait the minimum time until the next timeout is met for some active contract in $\isRun[0]$, so we know that $\delta \leq \delta_{\theHonestUser} \leq \secDelay$.

We do not extend the \bitmlx run in this case and keep $\xRun = \xRun[0]$. The delay doesn't change frontiers, so this run is trivially conformant, coherent configuration-consistent and timeout-consistent. \\

\textbf{Round-Based Execution} \\

Let $\isContractAdv{G}{C}{\someBlockchain} \in \isConfig$ be a contract advertisement with stipulation time $t_0$. We are interested in the case where $t < t_0 + \secDelay \leq t + \delta$, as for other cases, the invariant holds by inductive hypothesis. For said case, we can prove by contradiction $\isStipStatus{\isRun[0]}{\someStipulation}{\someBlockchain} = DoubleSpent$. Suppose that the opposite is true. That implies in particular that for every user $\participant{A_i} \in \participants{G}$ there is a deposit $\isParticipantDeposit{\participant{A_i}}{v_i}{\someBlockchain}{x_i} \in \lastConfigOf{\isRun[0]}$ fulfilling their deposit precondition $\depositsPre{A_i}{\balance_i}{\vec{x_i}} \in \preconditions{G}$.

But, if $t_0 + \secDelay \leq t + \delta$ and given that we are in the case where $\delta \leq \secDelay$, then even before the delay move, $t \geq t_0$ . In particular for $\theHonestUser$, according to the honest user strategy compilation, this means that $doubleSpend(\isContractAdv{G}{C}{\someBlockchain}, \theHonestUser) \in \isStrategy(\isRun[0])$ and so $\isStrategy \not\vDash \isRun$. \\

We now move to prove the conditions over stipulation contracts. Let $\isActiveContract{\contract{C}}{\isContractState}{\someBlockchain}{\someStipulation} \in \isConfig$ be a stipulation contract with stipulation time $t_0$. We are interested in the cases where the round status of $\someStipulation$ changed with the delay, as for other cases, the conditions will hold by inductive hypothesis. Let's examine those cases:

\begin{itemize}
    \item If $t < t_0 \leq t + \delta$ then $\someStipulation$ is moving from initialization to compensation phase, so we need to prove that the adversary cannot slash the honest user. We know by inductive hypothesis that $\isContractState.status = \isStatusStipChoice$.

    We can prove by contradiction that $\theHonestUser \notin \userStepSecrets{\someStipulation}{\isRun}$. Assume that the opposite is true and $\theHonestUser$ revealed their step secret. Then by inductive hypothesis we know that $\userInitSecrets{\someStipulation}{\isRun} = \contractUsers{\someStipulation}{\isRun}$. According to the honest user strategy compilation rules, this means that $init(\someStipulation, \someBlockchain) \in \isStrategy$ and furthermore, $\delta \notin \isStrategy$ which would imply $\isStrategy \vDash \isRun$.

    \item If $t < t_0 + \secDelay \leq t + \delta$ then $\someStipulation$ is moving from compensation phase to refund phase. We know by inductive hypothesis that $\theHonestUser \notin \userStepSecrets{\someStipulation}{\isRun}$ and $\isContractState.status \notin \{ \isStatusStipSlashed{\theHonestUser}, \isStatusStipCompensation{\theHonestUser}\}$.

    We can prove by contradiction that $\isContractState.status \neq \isStatusStipChoice$ and $\forall \participant{B} \in \contractUsers{\someStipulation}{\isRun} \isContractState.status \neq \isStatusSlashed{\participant{B}}$. If the opposite was true and the contract was in one of those statuses, then another move would be scheduled by the honest user strategy ($skip(\someStipulation, \someBlockchain)$ in the case of $\isContractState.status = \isStatusStipChoice$ or $compensate(\someStipulation, \someBlockchain, \participant{B})$ in the case of $\isContractState.status = \isStatusSlashed{\participant{B}}$. This would again mean that the honest user strategy would not wait and $\isStrategy \not\vDash \isRun$.

    \item If $t < t_0 + 2\secDelay \leq t + \delta$, then $\someStipulation$ is moving from refund phase to finalized. We know by inductive hypothesis that $\isContractState.status = \isStatusStipRight \lor \exists \participant{B} \in \contractUsers{\someContract}{\isRun}: \isContractState.status = \isStatusStipRefunded{\participant{B}} \lor \exists \participant{B} \in \contractUsers{\someContract}{\isRun} \setminus \{\theHonestUser\}: \isContractState.status \in \{ \isStatusStipSlashed{\participant{B}}, \isStatusStipCompensation{\participant{B}} \}$.

    We can prove by contradiction that $\isContractState.status \neq \isStatusStipRight$. The proof is analogous to the previous case. \\
    
\end{itemize}

We now move to proving active contract conditions. Let $\someContract \in \isConfig$ be an active contract with round status $\roundStatus{\someContract}{\isRun[0]} = (r_0, p_0)$ before the delay move and $\roundStatus{\someContract}{\isRun} = (r, p)$ after it. We split by cases on the relation between the status before and after the delay move:

\begin{itemize}

    \item The cases where $r < r_0$ or $r = r_0 \land p < p_0$ are impossible, because they would imply a negative delay, which is not allowed by the intermediate semantics. 
    
    \item If $r_0 \leq r < |\someContract|$ or $|\someContract| < r_0 \leq r$ or $r = r_0 = |\someContract| \land p = p_0$, then the invariant holds trivially by inductive hypothesis.

    \item The cases where $r_0 < |\someContract| \land r = |\someContract| \land p = 1$ or $r_0 < |\someContract| \land r > |\someContract|$ or $r_0 = |\someContract| \land p_0 = 0 \land r > |\someContract|$ are in contradiction with the honest user strategy, which will always wait the minimum time to meet the next timeout.
    
    \item If $r_0 < |\someContract| \land r = |\someContract| \land p=0$, then $\someContract$ is moving from waiting to synchronize into compensation phase. We know by inductive hypothesis that $\isContractState.status \in \{\isStatusChoice, \isStatusLeft\} \lor \exists \participant{B} \in \contractUsers{\someContract}{\isRun}: \isContractState.status = \isStatusAssigned{\participant{B}}$.

    If $\isContractState.status \in \{\isStatusChoice, \isStatusLeft\}$, we can prove by contradiction that $\theHonestUser \notin \userStepSecrets{\someStipulation}{\isRun}$. Assume that the opposite is true and $\theHonestUser$ revealed their step secret. According to the honest user strategy compilation rules, this means that, depending on the form of the contract, $ileft(\someStipulation, \someBlockchain) \in \isStrategy$ or $reveal(\someStipulation, \someBlockchain) \in \isStrategy$ and furthermore, $\delta \notin \isStrategy$ which would imply $\isStrategy \not\vDash \isRun$.

    We know by inductive hypothesis that $desc(\someContract) \subseteq desc(\someContract)$. So if $\isContractState.status = \isStatusLeft \lor desc(\someContract) \neq \emptyset$, there must be an $ileft(\someContract, \someOtherBlockchain)$ or $reveal(\someContract, \someOtherBlockchain)$ move in $\isRun[0]$. According to the intermediate semantics, this implies that for some user $\participant{B}$, we have a step secret $\stepSecret{\participant{B}}{\someContract} \in \lastConfigOf{\isRun[0]}$.

    \item If $r_0 = |\someContract| \land p_0 = 0 \land r = |\someContract| \land p = 1$, then $\someContract$ is moving from compensation to skipping phase.

    We know by inductive hypothesis that $\isContractState.status \notin \{\isStatusSlashed{\theHonestUser}, \isStatusCompensated{\theHonestUser}\}$. We can also prove by contradiction that $\isContractState.status \notin \{\isStatusChoice, \isStatusLeft\}$. Assume that the contrary is true:
    \begin{itemize}
        \item If $\isContractState.status = \isStatusLeft$, then we know by inductive hypothesis that $\userStepSecrets{\someStipulation}{\isRun} \setminus \{\theHonestUser\} \neq \emptyset$ and there is a valid $\alpha(\someContract, \someBlockchain)$ move with $\alpha \in \{cwithdraw, dwithdraw, split\}$. According to the honest user strategy compilation, $\alpha(\someContract, \someBlockchain) \in \isStrategy(\isRun)$.
        \item If $\isContractState.status = \isStatusChoice$, we can observe that by the timeout consistency invariant $\isContractState.time = t_0 + 2|\someContract| \secDelay$, where $t_0$ is the stipulation time of the root contract of $\someContract$. By the definition of round-based status $r_0 = \left \lfloor \frac{t - t_0}{2\secDelay} \right \rfloor$. But we are in the case where $r_0 = |\someContract|$ so we can solve for $t$ as $t \geq t_0 + 2\secDelay|\someContract|$. Combining these conditions, we get $t \geq \isContractState.time$. Then $skip(\someContract, \someBlockchain)$ is a valid move in $\isRun[0]$. According to the honest user strategy compilation, $skip(\someContract, \someBlockchain) \in \isStrategy(\isRun)$.
    \end{itemize}
    In both cases, $\delta \notin \isStrategy(\isRun[0])$ so $\isStrategy \not\vDash \isRun$ which is a contradiction. We conclude then that $\isContractState.status \notin \{\isStatusChoice, \isStatusLeft\}$. \\

    We know by inductive hypothesis that if $(\forall \participant{B} \in \contractUsers{\someContract}{\isRun}.\isContractState.status \neq \isStatusAssigned{\participant{B}}) \implies \theHonestUser \notin \userStepSecrets{\someContract}{\isRun}$. \\

    Now suppose that $ldesc(\someContract) \neq \emptyset$. We know by inductive hypothesis that $rdesc(\someContract) \neq \emptyset$ and also that $\userStepSecrets{\someStipulation}{\isRun} \setminus \{\theHonestUser\} \neq \emptyset$. We can prove by contradiction that $\isContractState.status = \isStatusCompensated{\participant{B}}$. 
    \begin{itemize}
        \item We proved already that $\isContractState.status \notin \{\isStatusSlashed{\theHonestUser}, \isStatusCompensated{\theHonestUser}\}$.
        \item If $\isContractState.status = \isStatusRight$,  then $slash(\someContract, \someBlockchain, \participant{B}) \in \isStrategy(\isRun[0])$.
        \item If $\isContractState.status = \isStatusSlashed{\participant{C}}$ (where $\participant{C}$ is not necessarily $\participant{B}$), then $slash(\someContract, \someBlockchain, \participant{C}) \in \isStrategy(\isRun[0])$.
    \end{itemize}

    For the last two cases, this would imply again that $\delta \notin \isStrategy(\isRun[0])$ so $\isStrategy \not\vDash \isRun$ which is a contradiction. \\

    \item If $r_0 = |\someContract| \land p_0 = 1 \land r > |\someContract|$, then we know by inductive hypothesis that $\isContractState.status \notin \{ \isStatusChoice, \isStatusLeft, \isStatusSlashed{\theHonestUser}, \isStatusCompensated{\theHonestUser}\}$.

    We can prove by contradiction that $\isContractState.status \neq \isStatusRight$ and for every user $\participant{B}$, $\isContractState.status \neq \isStatusSlashed{\participant{B}}$. Again, we will use the tactic of proving that there's another valid move that the honest user strategy would do instead of waiting which would imply that $\delta \notin \isStrategy(\isRun[0])$ and $\isStrategy \not\vDash \isRun$
    \begin{itemize}
        \item If $\isContractState.status = \isStatusRight$, we can observe that by the timeout consistency invariant $\isContractState.time = t_0 + (2|\someContract| + 1) \secDelay$, where $t_0$ is the stipulation time of the root contract of $\someContract$. We also know that $\someContract$ is in skipping phase in $\isRun[0]$, so $r = |\someContract|$ and $(2r+1) \secDelay \leq t - t_0$ or, solving for $t$, $t \geq t_0 + (2|\someContract|+1) \secDelay = \isContractState.time$. Then, $right(\someContract, \someBlockchain)$ is a valid move in $\isRun[0]$ and $right(\someContract, \someBlockchain) \in \isStrategy(\isRun[0])$.
        \item If $\isContractState.status = \isStatusSlashed{\participant{B}}$ for some $\participant{B}$, then $compensate(\someContract, \someBlockchain, \participant{B})$ is a valid move in $\isRun[0]$ and $compensate(\someContract, \someBlockchain, \participant{B}) \in \isStrategy(\isRun[0])$. \\
    \end{itemize}
    
\end{itemize}

\textbf{Honest Skip} \\

Let $\isActiveContract{\contract{D} \prchoice \contract{C}}{\isContractState}{\someBlockchain}{\someContract} \in \lastConfigOf{\isRun}$ with round status $(r_0, p_0) = \roundStatus{\someContract}{\isRun[0]}$ before the $\delta$ move and $(r, p) = \roundStatus{\someContract}{\isRun}$ after it, such that after the wait move $r_0 < |\someContract| = r$.

We will prove by contradiction that $skip(\someContract) \in \xStrategy(\xRun[0])$. Suppose that the opposite is true and $skip(\someContract) \notin \xStrategy(\xRun[0])$. Then because $\xStrategy$ is eager, we know that either $\alpha(\someContract) \in \xStrategy(\xRun[0])$ for $\alpha \in \{dwithdraw, split, reveal\}$, $(\theHonestUser: ~\someContract) \in \xStrategy$ or $\theHonestUser: \secret{s} \in \xStrategy(\xRun[0])$. Then, according to the honest user strategy compilation:

\begin{itemize}
    \item If $\alpha(\someContract) \in \xStrategy(\xRun[0])$, then $(\participant{A}: \stepSecret{A}{\someContract}) \in \isStrategy(\isRun[0])$.
    \item If $(\theHonestUser: ~\someContract) \in \xStrategy(\xRun[0])$, then $(\theHonestUser: ~(\someContract, \someBlockchain)) \in \isStrategy(\isRun[0])$.
    \item If $\theHonestUser: \secret{s} \in \xStrategy(\xRun[0])$, then $\theHonestUser: \secret{s} \in \isStrategy(\isRun[0])$.
\end{itemize}

In all cases, the honest user strategy would not wait, so similarly to previous cases $\isStrategy \not\vDash \isRun$.

\end{proof}

%% file: theory_docs/low_level_proof.tex
\maketitle

\subsection{Batch-Advertisments}

In order to capture cross-chain contracts by compilation to BitML, we need to slightly generalize the BitML language and semantics. 
In particular, we will add the possibility to introduce contracts that share secrets in a batch, so as to give 'global' secrets that can be used by multiple contracts. 
The additions to the BitML syntax and semantics are minor.

We first define the notion of a batch advertisement. 

\begin{definition}[Batch advertisement]
A batch advertisement is a term $\batchAdvertisement{S}{G}{C}$ where $\secret S$ is a possibly empty sequence of shared secret declarations $\participant{A}: \bitmlcode{secret}~s$ and $\vec{\contractAdv{G}{C}}$ is a non-empty sequence of BitML contracts advertisements $\contractAdv{G_i}{C_i}$ such that the following conditions are satisfied:
\begin{enumerate}
    \item Every contract advertisement $\contractAdv{G_i | \secret S}{C_i}$ is a valid BitML advertisments 
        \item The deposit names among all preconditions $\contract{G_i}$ are distinct.
    \item No additional local secrets in preconditions $\contract{G_i}$
    \item No volatile deposits in preconditions $\contract{G_i}$
    \item Every precondition $\contract{G_i}$  states the same set of participants. 
    \end{enumerate}
\end{definition}

Batch advertisements allow for the joint advertisement of contracts between the same set of users and shared secrets. 
We adopt the BitML stipulation rules in the following. 

\newcommand{\runCompStripped}{R^C_*}
\newcommand{\rand}{r}
\newcommand{\batchadvCompV}{\mathcal{C}_{\textit{batch}}}
\newcommand{\advCompV}{\mathcal{C}}
\newcommand{\txoutmap}{{\color{purple}\textbf{txout}}}
\newcommand{\partComp}{{\color{purple}\textbf{part}}}
\newcommand{\valComp}{{\color{purple}\textbf{val}}}
\newcommand{\sechashComp}{{\color{purple}\textbf{sechash}}}
\newcommand{\partGs}{\textsf{Part$\vec{\contract{G}}$}}
\newcommand{\keypair}[2]{K_{#1}(#2)}
\newcommand{\oracle}{{\color{green} O}}
\newcommand{\userA}{\participant{A}}
\newcommand{\userB}{\participant{B}}
\newcommand{\bPrecondV}{\contract{G}}
\newcommand{\NN}{\mathbb{N}}

\[
    \infer[Batch-Advertise]
        {
            \orConfig \xrightarrow[]{advertise(\batchAdvertisement{S}{G}{C})} \orConfig ~|~ 
            \parallelComposition_{\someBlockchain \in \activeBlockchains}
            \isContractAdv{G \cup \secret S }{C}{\someBlockchain}
        }
        {
            \begin{gathered}
                \forall \isContractAdv{G_i}{C_i}{\someBlockchain} \in \batchAdvertisement{S}{G}{C}.~ part(\contract{G_i}) \cap \Hon \neq \emptyset \\
                \forall \isContractAdv{G_i}{C_i}{\someBlockchain} \in \batchAdvertisement{S}{G}{C}.~ ~ \forall
                \big(\depositsPre{A}{v^{\someBlockchain}}{x} \big) \in \contract{G_i}^{\someBlockchain}. ~\isParticipantDeposit{A}{v}{\someBlockchain}{x} \in \orConfig \\
                                                \forall (\participant{A}: \bitmlcode{secret}~\secret a) \in \secret S.~ \secret a ~\fresh
                                            \end{gathered}
        }
\]

\[
    \infer[Batch-AuthCommit]
        {
            \orConfig ~|~ \parallelComposition_{\someBlockchain \in \activeBlockchains}
            \isContractAdv{G}{C}{\someBlockchain}\xrightarrow{\participant{A}: \batchAdvertisement{S}{G}{C}, \Delta} 
             \orConfig ~|~ \parallelComposition_{\someBlockchain \in \activeBlockchains}
            \isContractAdv{G}{C}{\someBlockchain} ~|~ \Delta ~|~ 
            \parallelComposition_{\someBlockchain \in \activeBlockchains} \userAuthIn{A}{\#}{\isContractAdv{G}{C}{\someBlockchain}}
                    }
        {
            \begin{gathered}
                \secret S = secrets(\contract{G^{\someBlockchain}}) \\
                \secret a_1 \dots \secret a_k \text{ secrets of } \participant{A} \textit{ in } \secret S \\
                                                                \forall i \in 1\dots k.~ \not \exists N.~ \secretCommitment{A}{a_i}{N_i} \in \orConfig \\
                \Delta = \parallelComposition_{i=1,\dots,k} \secretCommitment{A}{a_i}{N_i} \\
                \forall i \in [1, k].~ N_i \in 
                    \begin{cases} 
                        \NN & \text{if $\userA \in \Hon$} \\ 
                        \NN \cup \{ \bot \}  & \text{otherwise}
                    \end{cases} 
            \end{gathered}
        }
\]

\[
    \infer[C-AuthInit]
        {
            \orConfig ~|~ \isContractAdv{G}{C}{\someBlockchain} 
            \xrightarrow[]{\participant{A}: \isContractAdv{G}{C}{\someBlockchain}) , x}
            \orConfig ~|~ \isContractAdv{G}{C}{\someBlockchain} ~|~ \userAuthIn{A}{x}{\isContractAdv{G}{C}{\someBlockchain})}
        }
        {
            \begin{gathered}
                \forall \userA \in part(\contract{G}).~ \userAuthIn{A}{\#}{\isContractAdv{G}{C}{\someBlockchain}} \in \orConfig \\
                \big(\depositsPre{A}{v^{\someBlockchain}}{x} \big) \in \contract{G}
            \end{gathered}
        }
\]

\[
    \infer[C-Init]
        {   
            \orConfig ~|~ \isContractAdv{G}{C}{\someBlockchain} ~|~ 
            (\parallelComposition_{i \in I} \isParticipantDeposit{\participant{A_i}}{v_i}{\someBlockchain}{x_i} ) ~|~
            (\parallelComposition_{\participant{A} \in \contract{G}} \userAuthIn{A}{\#}{\isContractAdv{G}{C}{\someBlockchain}}) ~|~
             (\parallelComposition_{i \in I} \userAuthIn{A}{x}{\isContractAdv{G}{C}{\someBlockchain})})
            \xrightarrow{init(\isContractAdv{G}{C}{\someBlockchain})} 
            \orConfig ~|~ \isActiveContract{C}{\sum_{i\in I} v_i}{\someBlockchain}{x}
         }
        {
            \begin{gathered}
                \contract{G} = (\parallelComposition_{i \in I}  \depositsPre{A_i}{v_i^{\someBlockchain}}{x_i}) ~|~ (\parallelComposition_{j} \participant{C_j}: \bitmlcode{secret } \secret a_j) \\ 
                                                                                x \textit{ fresh}
            \end{gathered}
        }
\]

The introduction of batch advertisements induces a small modification in the stipulation protocol.
The following description of the new stipulation protocol heavily relies on notion and definition introduced in the BitML paper \cite{bitml}.

\begin{definition}[Stipulation Protocol]
Let $\participant{A} \in \Hon$, let $\runCompStripped$ be a ($\userA$-stripped) computational run, and let $r_{\userA}$ be a random sequence. 
The stipulation protocol for a computational batch contract advertisement $\batchadvCompV$ is as follows. 
\begin{enumerate}
    \item $\userA$ decodes $\batchadvCompV$, constructing a symbolic batch contract advertisement $\batchAdvertisement{S}{G}{C}$; 
    in doing this; $\userA$ chooses distinct symbolic names for all the transaction outputs in $\batchadvCompV$. 
    The mapping $\txoutmap$ is defined according to the used correspondence between names and transaction outputs.
    \item $\userA$ infers from $\vec{\bPrecondV}$ the parameters $\partComp$, $\partGs$ and $\valComp$ where $\partGs$ refers to the participants in any $\bPrecondV \in \vec{\bPrecondV}$ (not that all preconditions are required to feature the same set of participants). 
    \item $\userA$ uses $\rand_{\userA}$ to obtain the key pairs $K_{\userA}(\rand_{\userA})$ and $\hat{K}_{\userA}(\rand_{\userA})$. 
    The key $K_{\userA}(\rand_{\userA})$ is used by $\userA$ to sign all the protocol messages. 
    Further, $\userA$ reads from the initial prefix of the run $\runCompStripped$, the public keys $K^p_{\userB}(\rand_{\userB})$ and $\hat{K}^p_{\userB}(\rand_{\userB})$ of all the $\userB \in \partGs \backslash \{ \userA \}$.
    The key $K^p_{\userB}(\rand_{\userB})$ is used by $\userA$ to filter out the incoming messages with incorrect signatures. 
    \item $\userA$ generates from $\rand_{\userA}$ a secret nonce of desired length for each ${\userA}:{\secret a_i} \in \secret S$. 
    Then $\userA$ computes the hashes $\vec{h} = h_1, \dots, h_k$ of secret nonces (by querying \oracle), and broadcasts $m^*(\batchadvCompV, \vec{h})$. 
    Dually, $\userA$ receives the hashes $\vec{h'}$ from the other participants. 
    When doing so, $\userA$ starts defining $\sechashComp$ using the first (correctly signed) $m^*(\batchadvCompV, \vec{h'})$ in $\runCompStripped$ which has no duplicate hashes, and has no hashes already occurring (signed) in $\runCompStripped$
    \item After receiving the corresponding hashes of all other participants, for each $\bPrecondV_i \in \vec{\bPrecondV}$, $\userA$ performs steps (4) and (6) to (8) of the original stipulation protocol. 
    Instead of the message $m(\advCompV, \vec{h}, \vec{k})$, messages of the form $m(\batchadvCompV, \advCompV, \vec{h}, \vec{k})$ are used to identify the contract advertisement $\advCompV$ within the batch advertisement $\batchadvCompV$. 
    Step (5) is skipped as local secrets are not supported (and needed) in our model.
\end{enumerate}
Note that as opposed to the original stipulation protocol in~\cite{bitml}, we slightly adapted the order of message exchanges: 
First, the common secrets are committed and exchanged.
At this point, the compiler keys are not generated and exchanged yet.
This, however, should not cause an issue since the compiler keys can also be generated in a later step and broadcasted jointly with the contract-specific secrets.

\end{definition}

The adaptation of the proof follows naturally since we simply add a first authorization phase where the users commit to their secrets.
Intuitively, the broadcast of messages of the form $m^*(\batchadvCompV, \vec{h})$ now reflect the \textit{Batch-AuthCommmit} steps, while broadcasts of messages of the form $m(\batchadvCompV, \advCompV, \vec{h}, \vec{k})$.

\subsection{Low Level Coherence}

Instead of defining a BitML to intermediate level coherence as before, we are defining a function mapping intermediate semantics actions and configurations to corresponding BitML actions and configurations.

\subsubsection{Intermediate Level Configurations}
Individual components of an intermediate configuration $\isConfig$ can be translated to BitML configuration components with the following function $\isconfigcompile$. $\stipulationCompiler$, $\punishCompiler{}$, $\refundCompiler$ and $\topLevelCompiler{}$ are invocation of the \bitmlx compiler.
\begin{itemize}
    \item $ \isconfigcompile{\isContractAdv{G^x}{C^x}{\someBlockchain}} = \advCompiler(\contractAdv{G^x}{C^x}) \downarrow^{\someBlockchain}$ where $\downarrow^{\someBlockchain}$ return the precondition-contract component of blockchain $\someBlockchain$
    \item $ \isconfigcompile{\userAuthIn{A}{\#}{ \batchAdvertisement{S}{G^x}{C^x}}} = \userAuthIn{A}{\#}{\advCompiler(\contractAdv{G^x}{C^x})  \downarrow^{\someBlockchain}}$ 
    \item $\isconfigcompile{\userAuthIn{A}{x}{\isContractAdv{G^x}{C^x}{\someBlockchain}}} =  \userAuthIn{A}{x}{\advCompiler(\contractAdv{G^x}{C^x})  \downarrow^{\someBlockchain}}$ 
    \item Let $\settings{\Omega}$ be the compiler settings of the contract $\kappa$ restored from $\isRun$. 
        Let $\preconditions{G^x}$ be the preconditions of the contract $\kappa$ restored from $\isRun$.
        Let the default balance $v$ be $\isContractState[balance[\someBlockchain]]$ and $n$ be the number of participants in a contract $\kappa$. 
        \[
            \isconfigcompile{\isActiveContract{\contract{C^x}}{\isContractState}{\someBlockchain}{\someContract, x}} = 
            \begin{cases}
                \isActiveContract{\stipulationCompiler(\contractAdv{G^x}{C^x})}{v}{\someBlockchain}{x} &\textit{ if } \isContractState[status] = \isStatusStipChoice \\
                 \isActiveContract{\stipulationCompiler(\contractAdv{G^x}{C^x}) \xrightarrow{\tau}}{v}{\someBlockchain}{x} &\textit{ if } \isContractState[status] = \isStatusStipRight \\
                \isActiveContract{\punishCompiler{\someBlockchain}(\settings{\Omega}) \xrightarrow{\orPut{\emptyset}{\{\stepSecret{A}{\lambda}\}}{x_0}^{\someBlockchain}}}{v}{\someBlockchain}{x} &\textit{ if } \isContractState[status] = \isStatusStipSlashed{\participant{A}} \\
                \isActiveContract{\punishCompiler{\someBlockchain}(\settings{\Omega}) \xrightarrow{\orPut{\emptyset}{\{\stepSecret{A}{\lambda}\}}{x_{-1}}^{\someBlockchain}} \xrightarrow{split(x_0)^{\someBlockchain}}}{v_i}{\someBlockchain}{x_i}  &\textit{ if } \isContractState[status] = \isStatusStipCompensation{\participant{A}} \\
                \isActiveContract{\topLevelCompiler{\someBlockchain}(\refundCompiler(\preconditions{G^x}), \settings{\Omega})}{v}{\someBlockchain}{x} &\textit{ if } \isContractState[status] = \isStatusStipRefunded{\participant{A}} \\
                \isActiveContract{\topLevelCompiler{\someBlockchain}(\contract{C^x}, \settings{\Omega})}{v}{\someBlockchain}{x} &\textit{ if } \isContractState[status] = \isStatusChoice \\
                \isActiveContract{\topLevelCompiler{\someBlockchain}(\contract{C^x}, \settings{\Omega}) \xrightarrow{\orPut{\emptyset}{\{\stepSecret{A}{\lambda}\}}{x_0}^{\someBlockchain}}}{v}{\someBlockchain}{x} &\textit{ if } \isContractState[status] = \isStatusLeft \\
                 \isActiveContract{\topLevelCompiler{\someBlockchain}(\contract{C^x}, \settings{\Omega}) \xrightarrow{\tau}}{v}{\someBlockchain}{x} &\textit{ if } \isContractState[status] = \isStatusRight \\
                \isActiveContract{\punishCompiler{\someBlockchain}(\settings{\Omega}) \xrightarrow{\orPut{\emptyset}{\{\stepSecret{A}{\lambda}\}}{x_0}^{\someBlockchain}}}{v}{\someBlockchain}{x} &\textit{ if } \isContractState[status] = \isStatusSlashed{\participant{A}} \\
                 \isActiveContract{\punishCompiler{\someBlockchain}(\settings{\Omega}) \xrightarrow{\orPut{\emptyset}{\{\stepSecret{A}{\lambda}\}}{x_{-1}}^{\someBlockchain}} \xrightarrow{split(x_0)^{\someBlockchain}}}{v_i}{\someBlockchain}{x} &\textit{ if } \isContractState[status] = \isStatusCompensated{\participant{A}} \\
               \isActiveContract{withdraw \participant{A}}{v}{\someBlockchain}{x} 
                  &\textit{ if } \isContractState[status] = \isStatusAssigned{\participant{A}} \\
            \end{cases}
        \]
        where the notion $ \isActiveContract{\contract{C}\xrightarrow{\alpha}}{v}{\someBlockchain}{x}$ is defined as
        \[
            \isActiveContract{\contract{C}\xrightarrow{\alpha}}{v}{\someBlockchain}{x} ~:= \isActiveContract{\contract{C'}}{v}{\someBlockchain}{x} \textit{ if } (\isActiveContract{\contract{C}}{v}{\someBlockchain}{x_0}\xrightarrow{\alpha}\isActiveContract{\contract{C'}}{v}{\someBlockchain}{x} 
        \]
        with the special case of splits, where
        \[
            \isActiveContract{split~\vec{v} \rightarrow \vec{\contract{C}} \xrightarrow{split(x_0)^{\someBlockchain}}}{\vec{v}}{\someBlockchain}{\vec{x}} ~:= \parallelComposition_{i=1}^{\rightarrow }\isActiveContract{\contract{C_i} }{v_i}{\someBlockchain}{x_i} 
        \]
        
        \item $ \isconfigcompile{  \participant{A_i} [(\someContract, \someBlockchain, x) ~ \triangleright ~ \contract{D^x}]} =  \participant{A_i} [x~ \triangleright ~ \guardedCompiler{\someBlockchain}(\contract{D}, ~ \settings{\Omega(\isConfig, \kappa)})]^{\someBlockchain} $
                                \item $\isconfigcompile{\_}$ maps everything related to secrets and deposits to itself: \\
    $\isconfigcompile{id} = id $ if 
        \begin{align*}
        id \cong \secretReveal{A}{s}{N} \lor
         id \cong \secretCommitment{A}{s}{N}  &\lor 
         id \cong \isCommitedStepSecret{A}{\someContract} \lor 
         id \cong \isRevealedStepSecret{A}{\someContract} \lor  \\
         id \cong \isCommitedInitSecret{A}{\someContract} \lor 
         id \cong \isRevealedInitSecret{A}{\someContract} &\lor 
         id \cong \isParticipantDeposit{\participant{A}}{v}{\someBlockchain}{x} \lor
         id \cong \participant{A}: x,j
        \end{align*}
    \end{itemize}

\subsubsection{Intermediate to BitML Action Translation}

CA on possible moves of $\isStrategy[\participant{A}](\isRun)$:
\begin{itemize}
    \item Advertise: $\isactioncompile{advertise(\contractAdv{G^x}{C^x})} =  \advCompiler(\contractAdv{G^x}{C^x})$ 
    \item Authorize Commit: $\isactioncompile{commit(\participant A, \contractAdv{G^x}{C^x})} = \participant{A}: \advCompiler(\contractAdv{G^x}{C^x}) , \Delta$ 
    \item Authorize Init: $\isactioncompile{authInit(\participant{A}, \isContractAdv{G^x}{C^x}{\someBlockchain})} = \participant{A}:\isContractAdv{G}{C}{\someBlockchain}, x(\participant{A},\contract{G}) $ where $\isContractAdv{G}{C}{\someBlockchain} = \advCompiler(\contractAdv{G^x}{C^x}) \downarrow^{\someBlockchain}$
    \item Publish: $\isactioncompile{publish(\isContractAdv{G^x}{C^x}{\someBlockchain})} = init(\contract{G}, \contract{C})^{\someBlockchain}$ where $\isContractAdv{G}{C}{\someBlockchain} = \advCompiler(\contractAdv{G^x}{C^x}) \downarrow^{\someBlockchain}$
    \item Reveal Init Step Secret: $\isactioncompile{\isRevealedInitSecret{A}{\someContract}} = = \participant{A}: \stepSecret{A}{\uniqueId}$
    \item Init: $\isactioncompile{init(\someContract, \someBlockchain, x)} = \orPut{\emptyset}{\{\stepSecret{A}{\lambda}\} \cup S}{x}^{\someBlockchain}$ 
    \item Stip Skip: $\isactioncompile{from~t:~sskip(\someContract, \someBlockchain)} = \participant{A}: \initSecret{A}{\uniqueId}$
    \item Abort: $\isactioncompile{from~t:~abort(\someContract, \someBlockchain, x)} = \orPut{\emptyset}{\emptyset}{x}^{\someBlockchain}$ 
    \item Intro Stip Compensate: $\isactioncompile{sslash(\someStipulation, \someBlockchain, \participant{A}, x)} = \orPut{\emptyset}{\{\stepSecret{A}{\lambda}\}}{x}^{\someBlockchain}$ 
    \item Elim Stip Compensate: $\isactioncompile{scompensate(\someStipulation, \someBlockchain, \participant{A}, x)} = split(x)$ 
    \item Secret Reveal: $\isactioncompile{\participant A:\secret a} = \participant{A}: \secret{A}{\uniqueId}$
    \item Step Secret Reveal: $\isactioncompile{\isRevealedStepSecret{A}{\someContract}} = \participant{A}: \stepSecret{A}{\uniqueId}$
    \item Intro Left: $\isactioncompile{ileft(\someContract, \someBlockchain, x)} = \orPut{\emptyset}{\{\stepSecret{A}{\lambda}\}}{x}^{\someBlockchain}$
    \item Reveal: $\isactioncompile{reveal(\someContract, \someBlockchain, x)} = \orPut{\emptyset}{\{\stepSecret{A}{\lambda}\} \cup \Vec{s}}{x}^{\someBlockchain}$
    \item Authorize: $\isactioncompile{\participant{A}: (\someContract, \someBlockchain, x)} = \participant{A}: x, \contract{D(\isRun, \kappa)}^{\someBlockchain} $ 
    \item Intro Skip: $\isactioncompile{from ~ t: ~ skip(\someContract, \someBlockchain, x)} =  \orPut{\emptyset}{\emptyset}{x}^{\someBlockchain}$
    \item Elim Skip: $\isactioncompile{from ~ t: ~ right(\someContract, \someBlockchain, x)} = \orPut{\emptyset}{\emptyset}{x}^{\someBlockchain}$
    \item Intro Compensation: $\isactioncompile{slash(\someContract, \someBlockchain, x, \participant{A})} = \orPut{\emptyset}{\{\stepSecret{A}{\lambda}\}}{x}^{\someBlockchain}$
    \item Elim Compensation: $\isactioncompile{compensate(\someContract, \someBlockchain, x, \participant{A})} = split(x_0) $ 
    \item Guarded Withdraw: $\isactioncompile{dwithdraw(\someContract, \someBlockchain, x)} = split(x)^{\someBlockchain}$
    \item Split: $\isactioncompile{split(\someContract, \someBlockchain, x)} = split(x)^{\someBlockchain}$
    \item Withdraw: $\isactioncompile{cwithdraw(\someContract, \someBlockchain, x)} = split(x)^{\someBlockchain}$
\end{itemize}
where 
\begin{itemize}
    \item $x(\participant{A}, \contract{G})$ returns the identifier $x$ for participants $\participant{A}$ deposit in preconditions $\contract{G}$
    \item $\contract{D(\isRun, \kappa)}^{\someBlockchain}$ returns a contract branch of contract $\kappa$ in $\isConfig(\isRun)$ with an authorization (note: there will only be a single branch with authorization in practice)
\end{itemize}

Non contract-related cases:
\begin{itemize}
    \item Time: $\isactioncompile{\delta} = \delta$
    \item Double Spend: $\isactioncompile{doublespend(\isContractAdv{G^x}{C^x}{\someBlockchain}, \participant{A}, x)} = destroy([x])$ 
    \item Auth Double Spend:  $\isactioncompile{\participant{A}: doublespend(\isContractAdv{G^x}{C^x}{\someBlockchain}, \participant{A}, x)} =\participant{A}: x,j$
\end{itemize}

\begin{definition}[BitML Strategy Compilation]
    We define the compiled low-level honest participant strategy for two coherent runs $\isRun \isCoherence \orRun$ with strategy $\orStrategy[\participant{A}]= \isCompiledStrategy$ for participant $\participant{A}$ as 
    \[
        \isactioncompile{\alpha} \in \orStrategy[\participant{A}](\orRun) \iff  \alpha \in \isStrategy[\participant{A}](\isRun)
    \]
\end{definition}

\subsubsection{Translation Correctness}

\begin{lemma}[Translation Correctness I]\label{ll:trans-correct}
    \[
        \forall \isConfig[0] \beta. ~ \isConfig[0] \xrightarrow{\beta} \isConfig ~\Rightarrow~ \isconfigcompile[]{\isConfig[0]} \xrightarrow{\isactioncompile{\beta}} \isconfigcompile[]{\isConfig}
    \]
\end{lemma}

\begin{proof}
    Formal proof by case analysis over all possible intermediate semantics actions $\beta$. 
    The semantics rule of each action $\beta$ implies the existence of certain intermediate semantics configuration elements $\gamma \in \isConfig[0]$ that enabled $\beta$.
    We prove for each case, that the set of BitML configuration elements consisting out of translated intermediate semantics configuration elements (i.e., $\isconfigcompile{\gamma}$), is enough to show the existence of $\isactioncompile{\beta}$.
    Additionally, it is shown in each case that the consumed and added configuration elements coincide.
    \end{proof}

\begin{lemma}[Translation Correctness II]\label{ll:trans-correct-2}
    \[
        \forall \isConfig[0] \alpha. ~ \isconfigcompile[]{\isConfig[0]} \xrightarrow{\alpha} ~\land~ \alpha \not \cong withdraw
                    ~\Rightarrow~ \exists \beta. \isactioncompile{\beta} = \alpha ~\land~
        \isConfig[0] \xrightarrow{\beta} 
                \]
\end{lemma}

\begin{proof}
    Formal proof by case analysis over all possible BitML semantics actions $\alpha$. 
    Every BitML action $\alpha$ implies certain elements $\gamma$ in the configuration, for which another case distinction over $\isconfigcompile[{\isConfig[0]}]{\gamma}$ is made.
    This second case distinction restricts $\alpha$ to contracts from one of the \bitmlx compilers (as defined in $\iscompile_{\isRun}$).
    Lastly, we check that for each case left, there exists an action $\beta \in \isconfigcompile{\alpha}^{-1}$.
        There is a restriction on \textit{withdraws} as they can not (and don't need to) be scheduled in the intermediate semantics.
\end{proof}

\begin{lemma}[BitML Configuration Consistency]\label{ll:bitml-conf-cons}
    \[
        \forall \orConfig[0], \orConfig[0]'.~ \orConfig[0] \xrightarrow{\alpha} \orConfig ~\land~ \orConfig[0] \subseteq \orConfig[0]' ~\Rightarrow~ \orConfig[0]' \xrightarrow{\alpha} \orConfig' ~\land~ \orConfig \subseteq \orConfig'
    \]
\end{lemma}

\begin{proof}
    Trivial by case analysis on BitML semantics.
\end{proof}

\begin{lemma}[Synchronous Actions]\label{ll:static-actions}
Execution of BitML actions in two runs low-level $\orRun[0]$ and $\orRun[0]'$ changes configuration elements synchronously (i.e., adding/deleting the same elements to/from the configuration) as long as contract identifiers and deposit- and secret names have the same meaning.

Let $\orRun[0]$ and $\orRun[0]'$ be two BitML runs where contract identifiers and deposit- and secret names have the same meaning then
    \begin{align*}
            \orRun[0] \xrightarrow{\alpha} \orRun~\land~ \orRun[0]' \xrightarrow{\alpha} \orRun' 
            \Rightarrow \exists \orConfig^-, \orConfig^+.~ &\orConfig(\orRun) = \orConfig[0] \cup \orConfig^+ \setminus \orConfig^- \\
            \land~ &\orConfig(\orRun') = \orRun[0] \cup \orConfig^+ \setminus \orConfig^- 
    \end{align*}
    
\end{lemma}

\begin{proof}
    Induction over bitml semantics action $\alpha$.
\end{proof}

\subsubsection{Low Level Coherence}

Initial configuration consists only out of deposits.
To simplify the presentation of the proof, we are abstracting parts of BitML deposit handling infrastructure, i.e. we only allow the destruction of deposit and disregard joining, dividing and donating deposits.
In the intermediate semantics the destruction of deposits is dubbed under the umbrella term doubleSpend.

\[
\infer[\isCoherence-Initial]
{
    \isRun[0] \isCoherence \orRun[0]
}
{
    \begin{gathered}
    \isConfig(\isRun[0]) = \isConfig[0] ~initial \\
    \orConfig(\orRun[0]) = \orConfig[0]  = \isconfigcompile[]{\isConfig[0]} \\
    \end{gathered}
}
\]

\[
\infer[\isCoherence-Valid]{
    \isRun \isCoherence \orRun \\
}{
    \begin{gathered}
    \isRun[0] \isCoherence \orRun[0] \\
    \orRun = \reductionRule{\orRun[0]}{\alpha}{\orConfig} \\
    \isRun = \reductionRule{\isRun[0]}{\beta}{\isConfig} \\
    \isactioncompile{\beta} = \alpha \\
    \end{gathered}
}
\]

\[
\infer[\isCoherence-Invalid]{
    \isRun[0] \isCoherence \orRun \\
}{
    \begin{gathered}
    \isRun[0] \isCoherence \orRun[0] \\
    \orRun = \reductionRule{\orRun[0]}{\alpha}{\orConfig} \\
    \not \exists \beta. \isactioncompile{\beta} = \alpha \\
    \end{gathered}
}
\]

\subsubsection{Soundness}

\begin{lemma}[Soundness of Intermediate to Low-level Compilation]
    Let $\orStrategy[\participant{A}]$ be a compiled \bitmlx intermediate semantics strategy $\orStrategy[\participant{A}]= \isCompiledStrategy$. The five conditions after strategy conformance and coherence are called coherence invariants.
    \begin{align*}
        \forall \orRun.~ \orStrategy[\participant{A}] \vdash \orRun \Rightarrow&~ \exists \isRun. ~ \isRun \isCoherence \orRun  ~\land~ \isStrategy[\participant{A}] \vdash \isRun \\
                &\land~ \forall \participant{A} \in \Hon.~ \participant{A}:\contractAdv{G}{C}, \Delta_0 \in \orRun  \Rightarrow \forall \participant{A}:\contractAdv{G}{C}, \Delta \in \orRun \\
        & \qquad (\forall \{ s \} \in \Delta: \{ s \} \in \orRun \Rightarrow \{ s \} \in \isconfigcompile[]{\isRun} \\
        & \qquad \phantom{\forall \{ s \} \in \Delta: } s \in \orConfig(\orRun) \Rightarrow s \in \isconfigcompile[]{\isConfig(\isRun)}) \\
        & \qquad \forall \participant{A}:\contractAdv{G}{C}, \Delta \in \orConfig(\orRun) \Rightarrow \participant{A}:\contractAdv{G}{C}, \Delta \in \isconfigcompile[]{\isConfig(\isRun)} \\
        & \qquad \forall \participant{A}:\isContractAdv{G}{C}{\someBlockchain}, x \in \orConfig(\orRun) \Rightarrow \participant{A}:\isContractAdv{G}{C}{\someBlockchain}, x \in \isconfigcompile[]{\isConfig(\isRun)} \\
        &\land~ \forall \isActiveContract{\contract{C}}{v}{\someBlockchain}{x} \in \orConfig(\orRun) \Rightarrow \isActiveContract{\contract{C}}{v}{\someBlockchain}{x} \in \isconfigcompile[]{\isConfig(\isRun)} \\
        &\land~ \forall \participant{A}: x, \contract{D} \in \orConfig(\orRun) \Rightarrow \participant{A}: x, \contract{D} \in \isconfigcompile[]{\isConfig(\isRun)} \\
        &\land~ \forall \isParticipantDeposit{\participant{A}}{v}{\someBlockchain}{x} \in initial(\orConfig(\orRun)).~ \isParticipantDeposit{\participant{A}}{v}{\someBlockchain}{x}  \in \orConfig(\orRun) \Rightarrow \isParticipantDeposit{\participant{A}}{v}{\someBlockchain}{x}  \in \isconfigcompile[]{\isConfig(\isRun)} \\
        &\land~ \forall \isParticipantDeposit{\participant{A}}{v}{\someBlockchain}{x} \in \orConfig.~ \neg initial( \isParticipantDeposit{\participant{A}}{v}{\someBlockchain}{x} )\Rightarrow \isActiveContract{withdraw \participant{A}}{v}{\someBlockchain}{x}  \in \isconfigcompile[]{\isConfig(\isRun)}\\
    \end{align*}
\end{lemma}

\begin{proof}
    Induction over trace $\orRun$ with trivial base case.

    Induction step $\orRun[0] \xrightarrow{\alpha} \orConfig$. 
    Case analysis on $\isCoherence$ rule condition:
    \begin{itemize}
        \item $\exists \beta. \isactioncompile{\beta} = \alpha ~\land~ \isRun[0] \xrightarrow{\beta} \isConfig$:
        Application of $\isCoherence-Valid$ rule.
        Strategy compliance follows directly by definition of strategy compilation.
                                                                        All invariants hold with the synchronous actions lemma (Lemma \ref{ll:static-actions}) and the inductive hypothesis.
        The only special case is the [C-AuthCommit] case where it needs to be reasoned that commitments made prior to the first honest user commitment are present in $\isconfigcompile[]{\isRun}$.
        This is the case, since all valid stipulation commitments are enabled with the only configuration precondition being that the advertisement is present.
        Advertisements are present in $\isconfigcompile[]{\isRun}$ as long the advertisement is a valid \bitmlx contract which is implied by an honest user trying to authorize it (by the definition of strategy compilation).
        
        \item $\not\exists \beta. \isactioncompile{\beta} = \alpha ~\land~ \isRun[0] \xrightarrow{\beta} \isConfig$: 
        Application of $\isCoherence-Invalid$ rule.
        Strategy compliance is correct by induction hypothesis.

        Proof by case analysis on BitML semantics rule emitting $\alpha$. 
                \begin{itemize}
            \item Contract Rules: [C-Split], [C-PutRev], [C-AuthControl] \\
                In each rule, the inductive hypothesis implies that all precondition that were present in $\orConfig(\orRun[0])$ must also be present in $\isconfigcompile[]{\isConfig(\isRun[0])}$ (since every contract has an honest user, all conditions in the hypothesis is met).
                Hence, the step $\isconfigcompile[]{\isConfig(\isRun[0])} \xrightarrow{\alpha}$ is enabled as well.
                The second Translation Correctness lemma (Lemma \ref{ll:trans-correct-2}) of $\iscompile$ guarantees that in that case, there is also at least one enabled action in $\isRun[0]$. This contradicts the case analysis assumption.  
            
            \item Secret [C-AuthRev] \\
                Either the revealed secret was not part of $\isconfigcompile[]{\isRun}$ in the first place or it was part of $\isconfigcompile[]{\isRun}$ in which case the action could have been scheduled as well.
            \item Stipulation: [C-Advertise] \\
                Advertisments can only not be scheduled in the intermediate semantics when they are not a valid \bitmlx contracts or missing non-initial deposits. 
                The relevant parts of the active configuration do not change.
                Hence, the invariants holds by inductive hypothesis.
            \item Stipulation: [C-AuthCommit] \\
                Stipulation commitment authorizations can only not be scheduled in $\isConfig(\isRun[0])$ if the contract advertisement is not present.
                Without contract advertisement in the intermediate run, no honest user can commit to this contract.
                Hence, the invariants holds by inductive hypothesis.
            \item Stipulation: [C-AuthInit] \\
                Stipulation init authorizations can only not be scheduled in $\isConfig(\isRun[0])$ when the contracts advertisement or authorization commitments are missing.
                This case is proven by case distinction on whether the authorized contract was a \bitmlx contract or not. 
                If yes, we show that this move was indeed possible as an honest user has committed to it and the move would have been enabled in $\isconfigcompile[]{\isRun}$ as well.
                If not, no honest user committed, hence, the precondition that at least one honest participant committed is not met in $\orRun$.
            \item Stipulation: [C-Init] \\
                Same reasoning as in the [C-AuthInit] case.
            \item Withdrawel: [C-Withdraw] \\
                Withdraws cannot be scheduled in the intermediate semantics.
                Since the rule adds a non-initial deposit, it needs to be proven that a withdraw contract is part of the intermediate configuraiton.
                This follows from the inductive hypthesis and analysis of $\iscompile^{-1}(\isActiveContract{withdraw \_}{v}{\someBlockchain}{x}$.
        \end{itemize}

    \end{itemize}

\end{proof}

%% file: theory_docs/money_preservation.tex
\maketitle

\begin{definition}[Intermediate User Payouts]
    Let $\isRun$ be an intermediate semantics run. We define the payout of participant $\participant{A}$ in blockchain $\someBlockchain$ (written $\isUserPayout{A}{\someBlockchain}{\isRun}$) as follows
    \begin{align*}
        \sum_{\{
            \isContractState:
                \isActiveContract{\contract{C}}{\isContractState}{\someBlockchain}{\someContract} 
                \in \lastConfigOf{\isRun} \land
                \isContractState [status] = \isStatusAssigned{\participant{A}}
        \}}
            \isContractState.balance
    \end{align*}
\end{definition}

\begin{definition}[\bitmlx User Payouts]
    Let $\xRun$ be a \bitmlx run. We define the payout of participant $\participant{A}$ in $\xRun$ and blockchain $\someBlockchain$ (written $\xUserPayout{A}{\someBlockchain}{\xRun}$) as follows
    \begin{align*}
        \sum_{
            \assignedContract{A}{\balance}{\someContract} \in \lastConfigOf{\xRun}
        }
            \balance[][\someBlockchain]
    \end{align*}
\end{definition}

\begin{definition}[Intermediate Active Contract Funds]
    Let $\isRun$ be an intermediate semantics run.
    We define the contract funds in blockchain $\someBlockchain$ (written $\isContractFunds{\someBlockchain}{\isRun}$) as follows
    \begin{align*}
        \sum_{\{
            \isContractState:
                \isActiveContract{\contract{C}}{\isContractState}{\someBlockchain}{\someContract} \in \lastConfigOf{\isRun}
                \land \isContractState [status] \neq \isStatusAssigned{\participant{A}} \}
                \land \isContractState [status] \neq \isStatusCompensated{\participant{A}} \}
        \}} 
            \isContractState.balance
    \end{align*}
\end{definition}

\begin{definition}[Intermediate Total Contract Funds]
    Let $F$ be the maximal frontier for blockchain $\someBlockchain$ in the intermediate semantics run $\isRun$ ($\isMaxFrontier{F}{\isRun}{\someBlockchain}$).
    We define the contract funds in $F$ (written $\isTotalFunds{\someBlockchain}{\isRun}$ ) as follows
    \begin{align*}
        \sum_{\{
            \isContractState:
                \isActiveContract{\contract{C}}{\isContractState}{\someBlockchain}{\someContract} \in \lastConfigOf{\isRun}
        \}} 
            \isContractState.balance
    \end{align*}
\end{definition}

\begin{definition}[\bitmlx Active Contract Funds]
    Let $F$ be the maximal frontier for \bitmlx run $\xRun$.
    We define the contract funds in $\xConfig$ and blockchain $\someBlockchain$ (written $\xContractFunds{\someBlockchain}{\xRun}$ ) as follows
    \begin{align*}
        \sum_{ \{
            \balance:
                \activeContract{\contract{C}}{\balance}{\someContract} \in \lastConfigOf{\xRun}
        \}}
            \balance[][\someBlockchain]
    \end{align*}
\end{definition}

\begin{definition}[Intermediate User Inputs]
    Let $\isRun$ be an intermediate semantics run. 
    We define the intermediate inputs of $\participant{A}$ in $\isRun$ and blockchain $\someBlockchain$ (written $\isUserInputs{A}{\someBlockchain}{\isRun}$ ) as follows
    \begin{align*}
        \sum_{\{
            v:
                \depositsPre{A}{v \someBlockchain}{x} \in \preconditions{G^{\someBlockchain}}
                \land publish(\contractAdv{G^{\someBlockchain}}{C}) \in \isRun
        \}}    
            v
    \end{align*}
\end{definition}

\begin{definition}[\bitmlx User Inputs]
    Let $\xRun$ be a \bitmlx run. 
    We define the inputs of $\participant{A}$ in $\xRun$ and blockchain $\someBlockchain$ (written $\xUserInputs{A}{\someBlockchain}{\xRun}$ ) as follows
    \begin{align*}
        \sum_{\{
            \balance:
                \depositsPre{A}{\balance}{\vec{x}} \in \preconditions{G}
                \land init(\contractAdv{G}{C}) \in \xRun
        \}}    
            \balance[][\someBlockchain]
    \end{align*}
\end{definition}

\begin{definition}[Intermediate Compensated Funds]
   We define the compensated funds in $\someBlockchain$ of (written $\isCompensatedFunds{\someBlockchain}{\isRun}$) as follows
    \begin{align*}
        \sum_{\{
            \isContractState :
                \isActiveContract{\contract{C}}{\isContractState}{\someBlockchain}{\someContract} \in \lastConfigOf{\isRun}
                \land \isContractState[status] = \isStatusCompensated{\participant{A}} \}
        }
            \isContractState.balance
    \end{align*}
    
\end{definition}

\begin{lemma}[Intermediate Money Preservation]\label{mp:inter-money-preserv}
Let $\isRun$ be an intermediate semantics run. Then, for every blockchain $\someBlockchain$, the following equation holds

$$
            \isTotalFunds{\someBlockchain}{\isRun}
    = \sum_{\participant{A} \in P} \isUserInputs{\participant{A}}{\someBlockchain}{\isRun}
$$
\end{lemma}
\begin{proof}
    By induction over $\isRun$. For the base case of the empty configuration, both sides are $0$.
    
    Let's first consider the case of a move affecting an active contract. Let $\isRun = {\isRun[0]} \xrightarrow{\alpha(\someContract, \someBlockchain)}{\isConfig}$ with the move $\alpha(\someContract, \someBlockchain)$ resulting in successor contracts $\vec\someContract$, and let $F_0, F$ be frontiers such that $\isMaxFrontier{F_0}{\isRun[0]}{\someBlockchain}$ and $\isMaxFrontier{F}{\isRun}{\someBlockchain}$.    

    We know by inductive hypothesis that 
    \[
        \isTotalFunds{\someBlockchain}{\isRun[0]}
        = \sum_{\participant{A} \in P} \isUserInputs{\participant{A}}{\someBlockchain}{\isRun[0]}
    \]

    Given that we are doing an $\alpha(\someContract, \someBlockchain)$ move, we know that $\someContract \in F_0$. Then, by the blockchain frontier update lemma $F = F_0 \setminus \{\someContract\} \cup \{\someDescendant \in \vec\someContract\}$. This move also is also not a $publish$ move so user inputs will remain constant. If we assume that the balance of $\someContract$ is $v_\someContract$ and the successors $\someContract_1, \dots, \someContract_n \in \vec\someContract$ have balances $v_1, \dots, v_n$ respectively, then  we have:
    
    \[
    \begin{aligned}
        &\isTotalFunds{\someBlockchain}{\isRun} \\
        &= \isTotalFunds{\someBlockchain}{\isRun[0]} - v + \sum_{i=1}^n v_i \\
        &= \isUserInputs{\participant{A}}{\someBlockchain}{\isRun[0]} - v + \sum_{i=1}^n v_i \\
        &= \isUserInputs{\participant{A}}{\someBlockchain}{\isRun} - v + \sum_{i=1}^n v_i \\
    \end{aligned}
    \]

    For our goal condition to hold, we only need to prove that $v = \sum_{i=1}^n v_i$, which follow immediately from the transition rules for each case of the intermediate semantics moves.

\end{proof}

\begin{lemma}[\bitmlx Money Preservation]
    Let $F$ be the maximal frontier for \bitmlx run $\xRun$. Then, the following equation holds:
 
$$
    \sum_{\participant{A} \in P} \xUserPayout{A}{\someBlockchain}{\xRun}
    + \xContractFunds{\someBlockchain}{\xRun}
    = \sum_{\participant{A} \in P} \xUserInputs{\participant{A}}{\someBlockchain}{\xRun}
$$
\end{lemma}

%% file: theory_docs/final-statement.tex
\maketitle

\subsection{Eager and deterministic strategies}
We will assume that \bitmlx strategies are \emph{eager} meaning that a user will always schedule a contract action if it is possible. 
This implies in particular that a user will schedule a skip action whenever the priority action is "blocked" by another user.

\begin{definition}[Eager \bitmlx strategies]
A \bitmlx strategy $\xStrategy[\participant{A}]$ is \emph{eager} if the following condition holds: 
\begin{align*}
    \forall \xRun.~  \xStrategy[\participant{A}] \vdash \xRun
    &\Rightarrow \forall \someContract \in contracts(\xRun):
(\exists \alpha \xConfig: \xRun \xrightarrow[]{\alpha(\someContract)} \xConfig) 
\Rightarrow \exists \alpha' \in \xStrategy[\participant{A}](\xRun)\\
&~\land~ (\exists \hat\alpha(\kappa).~ \hat\alpha(\kappa) = \alpha' ~\lor~ \alpha' = \participant{A}: \uniqueId \\
& \qquad ~\lor~ 
    (\activeContract{\bitmlcode{reveal}~S~ \bitmlcode{then } \contract{C} \prchoice\contract{D}}{\balance}{\uniqueId}) \in \xConfig ~\land~ \alpha'=\participant A:\secret a ~\land~ \secret a \in S)\\ 
&~\land~ (\forall \xConfig: \xRun \xrightarrow[]{\alpha'} \xConfig \Rightarrow \xConfig \neq \xConfig(\xRun) )
\end{align*}
\end{definition}

Note that this definition ensures that the honest user strategy needs to eventually schedule $\textit{skip}$ actions (if possible) but at the same time allows for scheduling other actions $\alpha'$, e.g., authorizations or reveal secrets for enabling other (left) moves. 
However, it is required that such $\alpha'$ actions need to change configurations (this, in particular, forbids double authorizations).

In addition to strategies being eager, we will also require them to be \emph{deterministic}.
Intuitively, a strategy is deterministic if will never schedule contradicting actions for the same contract.

\begin{definition}[Deterministic \bitmlx strategies]
    A \bitmlx strategy $\xStrategy[\participant{A}]$ is \emph{deterministic} if the following condition holds: 
    \begin{align*}
    \forall \xRun.~  \xStrategy[\participant{A}] \vdash \xRun
    \Rightarrow \forall \alpha(\someContract), \alpha'(\someContract) \in \xStrategy[\participant{A}](\xRun) 
    \Rightarrow \alpha = \alpha'
\end{align*}
\end{definition}

Note that $\xStrategy[\participant{A}]$ being deterministic and persistent ensures in particular that it will never happen that $\xStrategy[\participant{A}]$ schedules a $skip(\someContract)$ action and then a $left(\someContract)$ action afterwards.

Determinism is from a security perspective a sensible requirement on user strategies, since in cases of non-determinism, it is left to the (adversarial) scheduler to resolve non-determinism. 

Another requirement on x-strategies $\xStrategy$ will be that those strategies will only propose and execute a finite number of contracts $\contractbound$. 
This requirement is needed since, technically it is not prevented that an $\xStrategy$ keeps on proposing (the same) contracts over and over again. 
We will call such strategies to be $\contractbound$-bounded.

\begin{definition}[x-$\contractbound$-Boundedness]
    Let $\participant{A}$ be an honest participant with strategy $\xStrategy[\participant{A}]$ and $\contractbound$ be a set of contracts. 
    We call $\xStrategy[\participant{A}]$ bounded by $\contractbound$ if
    for all runs $\xRun$ with $\xStrategy[\participant{A}] \vdash \xRun$ it holds that if $advertise(\preconditions{G}, \contract{C}) \in \xStrategy[\participant{A}](\xRun)$ and $(t_0, \someStipulation) = initSettings({\preconditions{G}})$ then 
    \begin{enumerate}
        \item $\someStipulation \in \contractbound$
        \item $\someStipulation \not \in \advcontracts{\xRun}$
                    \end{enumerate}
\end{definition}

We define a similar notion for the intermediate level: 

\begin{definition}[Intermediate $\contractbound$-Boundedness]
Let $\participant{A}$ be an honest participant with strategy $\isStrategy[\participant{A}]$ and $\contractbound$ be a set of contracts.   
We call $\isStrategy[\participant{A}]$ bounded by $\contractbound$ if
    for all runs $\isRun$ with $\isStrategy[\participant{A}] \vdash \isRun$ it holds that if $advertise(\contractAdv{G}{C}) \in \isStrategy[\participant{A}](\isRun)$ and $(t_0, \someStipulation) = initSettings({\preconditions{G}})$
    then 
    \begin{enumerate}
        \item $\someStipulation \in \contractbound$
        \item $\someStipulation \not \in \advcontracts{\isRun}$
    \end{enumerate}
\end{definition}

\begin{lemma}
     Let $\participant{A}$ be an honest participant with strategy $\xStrategy[\participant{A}]$ and $\isStrategy[\participant{A}] = \xCompiledStrategy$ its corresponding intermediate strategy. 
        Further let $\contractbound$ be a set of contracts such that $\xStrategy[\participant{A}]$ is $\contractbound$-bounded. 
        Then also $\isStrategy[\participant{A}]$ is $\contractbound$-bounded.
\end{lemma}

We will define Liquidity similar to how it is defined in~\cite{bartoletti2019verifying} and show that all \bitmlx contracts are liquid w.r.t. eager strategies.

Next, we will show how to prove liquidity for compiled (eager) strategies.

\subsection{\bitmlx Liquidity}

We adapt the definition from~\cite{bartoletti2019verifying} to our setting: 
\begin{definition}[Strong \bitmlx Liquidity]
    Let $\participant{A}$ be an honest participant with strategy $\xStrategy[\participant{A}]$.  
    Then $\xStrategy[\participant{A}]$ is strongly liquid if for all runs $\xRun$ with $\xStrategy[\participant{A}] \vdash \xRun$ there exists an extension $\hat{\xRun} = \xRun \xrightarrow[]{\alpha_1} \cdots \xrightarrow[]{\alpha_n}$ of $\xRun$ such that
    \begin{enumerate}
        \item $\forall i \in 1 \dots n : ~ \alpha_i \in \xStrategy[\participant{A}](\xRun \xrightarrow[]{\alpha_1} \cdots \xrightarrow[]{\alpha_{i-1}})$
        \item $\not \exists\activeContract{\contract{C}}{\balance}{\someContract} \in \xConfig(\dot{\xRun})$
    \end{enumerate}
    \end{definition}

In contrast to the BitML-style liquidity notion, this notion takes into account that the strategy $\xStrategy[\participant{A}]$ will only advertise and execute contracts that they can also liquidate.

\begin{lemma}[Non-blocking \bitmlx contracts]
    Given an \bitmlx run $\xRun$, we can proof that every active \bitmlx contract has at least a single active move:
    \[
        \forall \activeContract{\contract{C}}{\balance}{\someContract} \in \xConfig({\xRun}). (\exists \alpha \xConfig.~ \xRun \xrightarrow[]{\alpha(\someContract)} \xConfig) 
    \]
\end{lemma}

\begin{proof}
    Proven by CA on \bitmlx contract semantics using rules [C-Withdraw] and [C-Skip].
\end{proof}

\begin{definition}[\bitmlx Liquidation Distance]
    Let $\participant{A}$ be an honest participant with strategy $\xStrategy[\participant{A}]$ and $\contractbound$ a set of contracts such that $\xStrategy[\participant{A}]$ is $\contractbound$-bounded.
The liquidation distance of an \bitmlx run $\xRun$ w.r.t. to $\xStrategy[\participant{A}]$ (written $\sizeMetric{\xRun, \xStrategy[\participant{A}]}$) is formally defined as follows: 
\begin{align*}
    \sizeMetric{\xRun, \xStrategy[\participant{A}]} := 
    (\ldxTotalMissingContracts, 
    \ldxAdvMissingCommitments,
    \ldxAdvMissingAuths, 
    \ldxAdvContracts,
    \ldxFinDistance, 
    \ldxMissingAuthorizations,
    \ldxMissingRevealSecrets
    )
\end{align*}

where  
\begin{align*}
    \ldxTotalMissingContracts &:= \contractbound / \advcontracts{\xRun} \\
    \ldxAdvMissingCommitments &:= \{ \someStipulation ~|~  \contractAdv{G}{C}
            \in \xConfig(\xRun) 
            ~\land~ (t_0, \someStipulation) = initSettings({\preconditions{G}})
            ~\land~ \participant{A}: 
             [\# \triangleright \contractAdv{G}{C}
             \not \in \xConfig(\xRun) \} \\
                                         \ldxAdvMissingAuths &:= \{ \someStipulation ~|~ 
     \contractAdv{G}{C}
            \in \xConfig(\xRun) 
            ~\land~ (t_0, \someStipulation) = initSettings({\preconditions{G}})
            ~\land~  \depositsPre{A}{\balance}{\vec{x}} \in \preconditions{G} \\
            & ~\land~ \participant{A}: [\vec{x} \triangleright \contractAdv{G}{C}]
            \not \in \xConfig(\xRun) \} \\
    \ldxAdvContracts &:= \{ \someStipulation ~|~ \contractAdv{G}{C}
            \in \xConfig(\xRun) 
            ~\land~ (t_0, \someStipulation) = initSettings({\preconditions{G}})
            ~\land~ \activeContract{\contract{C}}{\balance}{\someStipulation}
            \not \in \xConfig(\xRun) \} \\
\ldxFinDistance &:= \{ 
    (\someContract, n) ~|~ \activeContract{\contract{C}}{\balance}{\someContract}
        \in \xConfig(\xRun) 
    ~\land~ n =  \contractdepth(C) \} \\
\ldxMissingAuthorizations &:= \{ \someContract 
    ~|~ \activeContract{\vec{\participant{A}}: \contract{D} \prchoice \contract{C}}{\balance}{\someContract}
        \in \xConfig(\xRun) 
   ~\land~ \participant{A} \in \vec{\participant{A}}
    ~\land~ \participant{A} [\someContract ~ \triangleright ~ \contract{D}] \not \in \xConfig(\xRun) 
    \} 
    \\
\ldxMissingRevealSecrets &:=  \{ (\someContract, s) 
    ~|~ \activeContract{\vec{\participant{A}}: \contract{D} \prchoice \contract{C}}{\balance}{\someContract}
        \in \xConfig(\xRun) 
    ~\land~ \contract{D} = \bitmlcode{reveal} ~\vec{s} ~\bitmlcode{if} ~p ~\bitmlcode{then} ~\contract{C'}
    ~\land~ s \in \vec{s} \\
    & \qquad ~\land~ \secretReveal{A}{s}{N} \not \in \xConfig(\xRun) \}
\end{align*}
where $\contractdepth$ is defined as
\begin{align*}
    \contractdepth(~\vec{\_}: \bitmlxSplit{\balance}{C} \prchoice \contract{C}) &=  \sum_{\contract{C'}\in\contract{\vec{C}}} \contractdepth(C') +  \contractdepth(C) +1  \\
    \contractdepth(~\_: \contract{D} \prchoice \contract{C}) &=   \contractdepth(D) +  \contractdepth(C) +1 \\
    \contractdepth(~\_ ~ ) &= 1
\end{align*}

\end{definition}

\begin{lemma}[Decreasing \bitmlx finalization distance]\label{fs:xFinDist}
    Let $\participant{A}$ be an honest participant with \textbf{an eager} strategy $\xStrategy[\participant{A}]$ and $\contractbound$ a set of contracts such that $\xStrategy[\participant{A}]$ is $\contractbound$-bounded and let $\xRun$ (with $\xStrategy[\participant{A}] \vdash \xRun$) be a corresponding run.
    Then either 
    \begin{itemize}
        \item $\xRun$ is liquidated i.e. $\not \exists\kappa\in\contractbound. ~\activeContract{\contract{C}}{\balance}{\someContract} \in \xConfig(\dot{\xRun})$
                \item or there is $\alpha \in \xStrategy[\participant{A}](\xRun)$ and $\sizeMetric{\xRun, \xStrategy[\participant{A}]} \gtlex \sizeMetric{\xRun  \xrightarrow[]{\alpha}, \xStrategy[\participant{A}]} $
    \end{itemize}
    where $\gtlex$ denotes the lexicographical ordering on the tuple-output of $\sizeMetric{\cdot}$ with 
    \begin{itemize}
        \item $\ldxTotalMissingContracts \leq \dot{\ldxTotalMissingContracts} :\Leftrightarrow \ldxTotalMissingContracts \subseteq \dot{\ldxTotalMissingContracts}$ 
        \item $\ldxAdvMissingCommitments \leq \dot{\ldxAdvMissingCommitments} :\Leftrightarrow 
        \ldAdvMissingCommitments \subseteq \dot{\ldAdvMissingCommitments}$
        \item $\ldxAdvMissingAuths \leq \dot{\ldxAdvMissingAuths} :\Leftrightarrow 
        \ldxAdvMissingAuths \subseteq \dot{\ldxAdvMissingAuths}$
        \item $\ldxAdvContracts \leq \dot{\ldxAdvContracts} :\Leftrightarrow 
        \ldxAdvContracts \subseteq \dot{\ldxAdvContracts}$
                        \item $\ldxFinDistance \leq \dot{\ldxFinDistance} :\Leftrightarrow 
        \forall (\someContract, n) \in \ldxFinDistance: \exists n' \someContract': (\someContract', n') \in \dot{\ldxFinDistance} ~\land~ n \leq n' ~\land~ \someContract' \in desc(\someContract)$
        \item $\ldxMissingAuthorizations \leq \dot{\ldxMissingAuthorizations} :\Leftrightarrow \ldxMissingAuthorizations \subseteq \dot{\ldxMissingAuthorizations}$ 
        \item $ \ldxMissingRevealSecrets \leq \dot{ \ldxMissingRevealSecrets} :\Leftrightarrow  \ldxMissingRevealSecrets \subseteq \dot{ \ldxMissingRevealSecrets}$ 
    \end{itemize}
\end{lemma}

\begin{proof}
    Assume $\xRun$ is not liquidated, then there exists $\kappa\in\contractbound$ such that $\activeContract{\contract{C}}{\balance}{\someContract} \in \xConfig(\dot{\xRun})$
    By the non-blocking design of \bitmlx contracts and eager strategies we get that 
            \[
            \forall \activeContract{\contract{C}}{\balance}{\someContract} \in \xConfig({\xRun}). \exists \alpha'.~ \alpha'(\someContract) \in \xStrategy[\participant{A}]({\xRun})
        \]
    The decreasing finalization distance of this strategy step $\alpha'$ is proven by CA on possible actions. Note that only contracts in $\contractbound$ can be advertised, committed, signed, etc. This bound is used in the finalization distance as limit. Eagerness is used to ensure progress when revealing secrets and authorizing contracts.

\end{proof}

\begin{lemma}[Liquidity of eager strategies]
    \label{lemma:xliquidty}
    Let $\participant{A}$ be an honest participant and $\xStrategy[\participant{A}]$ be an eager strategy of $\participant{A}$ bounded by $\contractbound$. Then $\xStrategy[\participant{A}]$ is strongly liquid.
\end{lemma} 

\begin{proof}
    Let $\xRun$ be a \bitmlx run such that $\xStrategy[\participant{A}] \vdash \xRun$.

    We are constructing an extension $\hat{\xRun}$ with actions that are decreasing the finalization distance  of $\sizeMetric{\xRun, \xStrategy[\participant{A}]}$ according to Lemma \ref{fs:xFinDist}. 
    This extension is finite since the lower bound of $\sizeMetric{}$ is 0.
    By definition of Lemma \ref{fs:xFinDist}, the last configuration is liquidated.

        \end{proof}

\subsection{Intermediate Liquidity}

We will extend the notion of liquidity to the intermediate semantics:

\begin{definition}[Liquidated runs]
    Let $\contractbound$ be a set of contracts and $\isRun$ an intermediate run. 
    Then $\isRun$ is liquidated 
        if for all $\isActiveContract{\contract{C}}{\isContractState}{\someBlockchain}{\someContract} \in \isConfig(\isRun)$ 
        one of the following holds 
    \begin{itemize}
        \item $\exists \participant{B}: \isContractState [status] = \isStatusAssigned{\participant{B}}$ 
        \item $\exists \participant{B}: \isContractState [status] = \isStatusSlashed{\participant{B}}$ 
        \item $\exists \participant{B}: \isContractState [status] = \isStatusCompensated{\participant{B}}$ 
        \item $\exists \participant{B}: \isContractState [status] = \isStatusStipRefunded{\participant{B}}$
        \item $\exists \participant{B}: \isContractState [status] = \isStatusStipSlashed{\participant{B}}$
        \item $\exists \participant{B}: \isContractState [status] = \isStatusStipCompensation{\participant{B}}$
    \end{itemize}

    We define \textit{Liquidated-Status} to be the set that includes every liquidated status. 
\end{definition}

\begin{definition}[Strong Intermediate Liquidity]
    Let $\participant{A}$ be an honest participant with strategy $\isStrategy[\participant{A}]$. 
    Then $\isStrategy[\participant{A}]$ is strongly liquid if for all runs $\isRun$ there exists an extension $\hat{\isRun} = \isRun \xrightarrow[]{\alpha_1} \cdots \xrightarrow[]{\alpha_n}$ such that 
    
    \begin{enumerate}
        \item $\forall i \in 1 \dots n : ~ \alpha_i \in \isStrategy[\participant{A}](\isRun \xrightarrow[]{\alpha_1} \cdots \xrightarrow[]{\alpha_{i-1}})$
        \item $\isRun$ is liquidated 
    \end{enumerate}
\end{definition}

Note that the second condition accounts for the fact that the intermediate semantics does not delete contracts but changes their status to indicate that they have been executed until finalization. 

We first define the liquidation distance of an intermediate run $\isRun$.
The liquidation distance is a measure of how close an intermediate run is to liquidation. 
Intuitively, an intermediate run will always be liquidated because its contracts (on all chains) will either 
(1) terminate regularly or
(2) will be compensated. 

To ensure liquidation, we need to consider the honest user strategy $\isStrategy$ to be bounded by a set of contracts $\contractbound$, meaning that $\isStrategy$ only proposes contracts from $\contractbound$ and at most once. 
Otherwise, the strategy could continue advertising new contracts without ever agreeing on time to pass. 

This property should be inherited from the underlying $\xStrategy$.
For this reason,

\begin{definition}[Liquidation Distance]
    Let $\participant{A}$ be an honest participant with strategy $\isStrategy[\participant{A}]$ and $\contractbound$ a set of contracts such that $\isStrategy[\participant{A}]$ is $\contractbound$-bounded.
The liquidation distance of an intermediate run $\isRun$ w.r.t. to $\isStrategy[\participant{A}]$ (written $\sizeMetric{\isRun, \isStrategy[\participant{A}]}$) is formally defined as follows: 
\begin{align*}
    \sizeMetric{\isRun, \isStrategy[\participant{A}]} := 
    (\ldTotalMissingContracts, 
    \ldAdvMissingCommitments,
    \ldAdvMissingAuths, 
    \ldAdvContracts,
    \ldMissingStipSecrets,
    \ldStipStatus,
    \ldChainStipDistance,
    \ldESFinDistance,
    \ldESMissingSSecrets, 
    \ldBSynchDistance,
    \ldMissingAuthorizations,
    \ldMissingRevealSecrets,
    \ldChainStatus, 
    \ldChainRoundDistance
    )
\end{align*}

where 
\begin{align*}
    \ldTotalMissingContracts &:= \contractbound / \advcontracts{\isRun} \\
    \ldAdvMissingCommitments &:= \{ \someStipulation ~|~  \isContractAdv{G}{C}{\someBlockchain}
            \in \isConfig(\isRun) 
            ~\land~ (t_0, \someStipulation) = initSettings({\preconditions{G}})
            ~\land~ \participant{A}: 
             [\# \triangleright \isContractAdv{G}{C}{\someBlockchain}
             \not \in \isConfig(\isRun) \} \\
                                         \ldAdvMissingAuths &:= \{ (\someStipulation, \someBlockchain) ~|~ 
     \isContractAdv{G}{C}{\someBlockchain}
            \in \isConfig(\isRun) 
            ~\land~ (t_0, \someStipulation) = initSettings({\preconditions{G}})
            ~\land~  \depositsPre{A}{\balance}{\vec{x}} \in \preconditions{G} \\
            & \qquad ~\land~ \vec{x}[\someBlockchain] = x
             ~\land~ \participant{A}: [x \triangleright \isContractAdv{G}{C}{\someBlockchain}]
            \not \in \isConfig(\isRun) \} \\
    \ldAdvContracts &:= \{ (\someStipulation, \someBlockchain) ~|~ \isContractAdv{G}{C}{\someBlockchain'}
            \in \isConfig(\isRun) 
            ~\land~ (t_0, \someStipulation) = initSettings({\preconditions{G}})
            ~\land~ \isActiveContract{\contract{C}}{\isContractState}{\someBlockchain}{\someStipulation}
            \not \in \isConfig(\isRun) \} \\
    \ldMissingStipSecrets &:=  \{ \someStipulation  
    ~|~ \isActiveContract{\contract{C}}{\isContractState}{\someBlockchain}{\someStipulation}
        \in \isConfig(\isRun)
        ~\land~ \initSecret{A}{\someStipulation} \in \isContractState[initSecrets]
        ~\land~  \isRevealedInitSecret{A}{\someStipulation} \not \in \isConfig(\isRun) \} \\
    \ldStipStatus &:= \{ (\someStipulation, \someBlockchain, \textit{status})
    ~|~ \isActiveContract{\contract{C}}{\isContractState}{\someBlockchain}{\someStipulation}
        \in \isConfig(\isRun) 
        ~\land~ \textit{status} = \isContractState[status] \} \\
\ldChainStipDistance &:= \{ (\someStipulation, \someBlockchain, 1 - p) 
    ~|~ \isActiveContract{\contract{C}}{\isContractState}{\someBlockchain}{\someStipulation}
        \in \isConfig(\isRun) 
        ~\land~ p = ? \}         \\
\ldESFinDistance &:= \{ 
    (\someContract, n) ~|~ \isActiveContract{\contract{C}}{\isContractState}{\someBlockchain}{\someContract}
        \in \isConfig(\isRun) 
    ~\land~ n =  \contractdepth(C) \} \\
\ldESMissingSSecrets &:= \{  \someContract ~|~ 
    \isActiveContract{\contract{C}}{\isContractState}{\someBlockchain}{\someContract}
        \in \isConfig(\isRun) 
    ~\land~ \isRevealedStepSecret{A}{\someContract} \not \in \isConfig(\isRun)
\}\\
\ldBSynchDistance &:= \{ (\someContract, \someBlockchain, n) 
    ~|~  \isActiveContract{\contract{C}}{\isContractState}{\someBlockchain}{\someContract}
        \in \isConfig(\isRun) 
    ~\land~ n = \textit{max}\{ |\someContract^*| :  \isActiveContract{\contract{C'}}{\isContractState}{\someBlockchain'}{\someContract'}
        \in \isConfig(\isRun) ~\land \someContract' = \someContract \cdot \someContract^* \}
\}\\ \ldMissingAuthorizations &:= \{ (\someContract, \someBlockchain) 
    ~|~ \isActiveContract{\vec{\participant{A}}: \contract{D} \prchoice \contract{C}}{\isContractState}{\someBlockchain}{\someContract}
        \in \isConfig(\isRun) 
   ~\land~ \participant{A} \in \vec{\participant{A}}
    ~\land~ \participant{A} [(\someContract, \someBlockchain) ~ \triangleright ~ \contract{D}] \not \in \isConfig(\isRun) 
    \} 
    \\
\ldMissingRevealSecrets &:=  \{ (\someContract, s) 
    ~|~ \isActiveContract{\vec{\participant{A}}: \contract{D} \prchoice \contract{C}}{\isContractState}{\someBlockchain}{\someContract}
        \in \isConfig(\isRun) 
    ~\land~ \contract{D} = \bitmlcode{reveal} ~\vec{s} ~\bitmlcode{if} ~p ~\bitmlcode{then} ~\contract{C'}
    ~\land~ s \in \vec{s} \\
    & \qquad ~\land~ \secretReveal{A}{s}{N} \not \in \isConfig(\isRun) \}
\\
\ldChainStatus &:= \{ (\someContract, \someBlockchain, \textit{status}) 
    ~|~  \isActiveContract{\contract{C}}{\isContractState}{\someBlockchain}{\someContract}
        \in \isConfig(\isRun) 
    ~\land~ \textit{status} = \isContractState[status]
    \}\\
\ldChainRoundDistance &:= \{ (\someContract, \someBlockchain, |\someContract| - r, 1 - p) 
    ~|~  \isActiveContract{\contract{C}}{\isContractState}{\someBlockchain}{\someContract}
        \in \isConfig(\isRun) 
    ~\land~ 
    (r, p) = \roundStatus{\someContract}{\isRun}
    \}
\end{align*}

where $\contractdepth$ is defined as
\begin{align*}
    \contractdepth(~\vec{\_}: \bitmlxSplit{\balance}{C} \prchoice \contract{C}) &=  \sum_{\contract{C'}\in\contract{\vec{C}}} \contractdepth(C') +  \contractdepth(C) +1  \\
    \contractdepth(~\_: \contract{D} \prchoice \contract{C}) &=   \contractdepth(D) +  \contractdepth(C) +1 \\
    \contractdepth(~\_ ~ ) &= 1
\end{align*}
\end{definition}

Intuitively, the different components forming the liquidation distance have the following meaning:
\begin{itemize}
\item $\ldTotalMissingContracts$ denotes the set of contracts $\someStipulation$ from $\contractbound$ which have not been advertised yet.
\item $\ldAdvMissingCommitments$ denotes the set of advertised contracts $\someStipulation$ that are still lacking commitments (from user $\participant{A}$).

\item $\ldAdvMissingAuths$ denotes the set of advertised contracts and blockchains $(\someStipulation, \someBlockchain)$ that are still lacking authorizations (from user $\participant{A}$). Note that authorizations are carried out per blockchain $\someBlockchain$.
\item $\ldAdvContracts$ denotes the set of advertised contracts and blockchains $(\someStipulation, \someBlockchain)$ that have not been published yet. Note that contracts $\someStipulation$ need to be published individually on each blockchain.
\item $\ldMissingStipSecrets$ denotes the set of published contracts $\someStipulation$ that are still lacking the init secrets (from user $\participant{A}$). Note that there is a single init secret per contract $\someStipulation$ and user that honest $\participant{A}$ will only release once the contract $\someStipulation$ has been published on all chains. 
\item $\ldStipStatus$ denotes the set of contracts, blockchain with their corresponding stipulation status $(\someStipulation, \someBlockchain, \textit{status})$. Note that a contract $\someStipulation$ might be in different stages of stipulation (indicated by $\textit{status}$) on the different blockchains
\item $\ldChainStipDistance$ denotes the set of contracts with their corresponding stipulation phase $(\someStipulation, p)$.
\item $\ldESFinDistance$ denotes the set of all running contracts with their finalization distance $(\someContract, n)$, which is given by the depth of the contract. Note that since contracts are non-recursive, the maximal number of execution steps is bounded by the contract structure. 
\item $\ldESMissingSSecrets$ denotes the set of all running contracts $\someContract$ that are missing a step secret from $\participant{A}$.
\item $\ldBSynchDistance$ denotes the set of running contracts, blockchains and their synchronization distance $(\someContract, \someBlockchain, n)$. The synchronization distance $n$ denotes how far the contract $\someContract$ on blockchain $\someBlockchain$ is lacking behind its furthest descendant (on any other chain).
\item $\ldMissingAuthorizations$ denotes the set of running (priority choice) contracts and blockchains $(\someContract, \someBlockchain)$ that are still lacking authorizations from $\participant{A}$.
\item $\ldChainStatus$ denotes the set of running contracts, blockchains and their execution stats $(\someContract, \someBlockchain, \textit{status})$. Note that a contract $\someContract$ may be in different stages of execution on the different blockchains.
\item $\ldMissingRevealSecrets$ denotes the set of running contracts their secrets that need to be revealed $(\someContract, s)$. 
\item $\ldChainRoundDistance$ denotes the set of running contracts, blockchains and the corresponding execution round and phase finalization distances $(\someContract, someBlockchain, r_\delta, p_\delta)$. Here, $r_\delta$ denotes the number of rounds that need to pass until execution of $\someContract$ will be latest enforced and $p_\delta$ denotes the phase within the enforcement round.
\end{itemize}

\begin{lemma}[Non-blocking intermediate contracts]
    Given an intermediate run $\isRun$, we can proof that every non-liquidated intermediate contract has at least a single active move:
    \[
        \forall \isActiveContract{\contract{C}}{\isContractState}{\someBlockchain}{\someContract} \in \isConfig({\isRun}).  status[\isContractState] \not \in \textit{Liquidated-Status} \implies (\exists \delta \alpha \isConfig.~ \isRun \xrightarrow[]{\delta\alpha(\someContract)} \isConfig) 
    \]
\end{lemma}

\begin{proof}
    Proven by CA on intermediate contract semantics using rules [||-ISkip], [||-ESkip], [||-Withdraw], [||-SSkip] and [||-Abort].
\end{proof}

\begin{lemma}[Decreasing finalization distance]\label{fs:isFinDist}
    Let $\participant{A}$ be an honest participant with an eager strategy $\xStrategy[\participant{A}]$ and $\contractbound$ a set of contracts such that $\xStrategy[\participant{A}]$ is $\contractbound$-bounded.
    Let $\isStrategy[\participant{A}] = \xCompiledStrategy$ be the corresponding intermediate strategy and let 
    $\xRun$ (with $\xStrategy[\participant{A}] \vdash \xRun$)
    and $\isRun$ (with $\isStrategy[\participant{A}] \vdash \isRun$) such that $\xRun \xCoherence[\participant{A}]  \isRun$.
    Then either 
    \begin{itemize}
        \item $\isRun$ is liquidated 
                \item or there is $\alpha \in \isStrategy[\participant{A}](\isRun)$ and $\sizeMetric{\isRun, \isStrategy[\participant{A}]} \gtlex \sizeMetric{\isRun  \xrightarrow[]{\alpha}, \isStrategy[\participant{A}]} $
    \end{itemize}
    where $\gtlex$ denotes the lexicographical ordering on the tuple-output of $\sizeMetric{\cdot}$ with 
    \begin{itemize}
        \item $\ldTotalMissingContracts \leq \dot{\ldTotalMissingContracts} :\Leftrightarrow \ldTotalMissingContracts \subseteq \dot{\ldTotalMissingContracts}$ 
        \item $\ldAdvMissingCommitments \leq \dot{\ldAdvMissingCommitments} :\Leftrightarrow 
        \ldAdvMissingCommitments \subseteq \dot{\ldAdvMissingCommitments}$
        \item $\ldAdvMissingAuths \leq \dot{\ldAdvMissingAuths} :\Leftrightarrow 
        \ldAdvMissingAuths \subseteq \dot{\ldAdvMissingAuths}$
        \item $\ldAdvContracts \leq \dot{\ldAdvContracts} :\Leftrightarrow 
        \ldAdvContracts \subseteq \dot{\ldAdvContracts}$
        \item $\ldMissingStipSecrets \leq \dot{\ldMissingStipSecrets} :\Leftrightarrow 
        \ldMissingStipSecrets \subseteq \dot{\ldMissingStipSecrets}$
        \item $\ldStipStatus \leq \dot{\ldStipStatus} :\Leftrightarrow 
        \forall (\someStipulation, \someBlockchain, \textit{status}) \in \ldStipStatus: \exists \textit{status}':(\someStipulation, \someBlockchain, \textit{status}') \in \dot{\ldStipStatus} ~\land~ \textit{status} \leq_{\textit{stip-status}} \textit{status}'$
        \item $\ldChainStipDistance \leq \dot{\ldChainStipDistance} :\Leftrightarrow 
        \forall (\someStipulation, \someBlockchain, n) \in \ldChainStipDistance: \exists n': (\someStipulation, \someBlockchain, n') \in \dot{\ldChainStipDistance} ~\land~ n \leq n'$
                        \item $\ldESFinDistance \leq \dot{\ldESFinDistance} :\Leftrightarrow 
        \forall (\someContract, n) \in \ldESFinDistance: \exists n' \someContract': (\someContract', n') \in \dot{\ldESFinDistance} ~\land~ n \leq n' ~\land~ \someContract' \in desc(\someContract)$
        \item $\ldESMissingSSecrets \leq \dot{\ldESMissingSSecrets} :\Leftrightarrow \ldESMissingSSecrets \subseteq \dot{\ldESMissingSSecrets}$ 
         \item $\ldBSynchDistance \leq \dot{\ldBSynchDistance} :\Leftrightarrow 
        \forall (\someContract, \someBlockchain, n) \in \ldBSynchDistance: \exists  n' \someContract': (\someContract', \someBlockchain, n') \in \dot{\ldBSynchDistance} ~\land~ n \leq n' ~\land~ \someContract' \in desc(\someContract)$
        \item $\ldMissingAuthorizations \leq \dot{\ldMissingAuthorizations} :\Leftrightarrow \ldMissingAuthorizations \subseteq \dot{\ldMissingAuthorizations}$ 
        \item $ \ldMissingRevealSecrets \leq \dot{ \ldMissingRevealSecrets} :\Leftrightarrow  \ldMissingRevealSecrets \subseteq \dot{ \ldMissingRevealSecrets}$ 
        \item $\ldChainStatus \leq \dot{\ldChainStatus} :\Leftrightarrow 
        \forall (\someContract, \someBlockchain, \textit{status}) \in \ldChainStatus: \exists \textit{status}':(\someContract, \someBlockchain, \textit{status}') \in \dot{\ldChainStatus} ~\land~ \textit{status} \leq_{\textit{status}} \textit{status}'$
        \item $\ldChainRoundDistance \leq \dot{\ldChainRoundDistance} :\Leftrightarrow 
        \forall (\someContract, \someBlockchain, r_d, p_d) \in \ldChainRoundDistance: \exists r_d'~  p_d': (\someStipulation, \someBlockchain, r_d', p_d') \in \dot{\ldChainRoundDistance} ~\land~ (r_d, p_d) \leq_{\textit{lex}} (r_d', p_d')$
    \end{itemize}
    and 
    \begin{align*}
        \leq_{\textit{status}} := 
        \{
            &(\isStatusCompensated{\participant{A}},\isStatusSlashed{\participant{A}}), \\
            &(\isStatusSlashed{\participant{A}},
            \isStatusRight), \\
            &(\isStatusRight,\isStatusChoice), \\
            &(\isStatusLeft,\isStatusChoice), \\
            &(\isStatusAssigned{\participant{A}}, \isStatusLeft)
        \}^*
    \end{align*}
    and 
    \begin{align*}
        \leq_{\textit{stip-status}} := 
        \{
            &(\isStatusStipCompensation{\participant{A}},\isStatusStipSlashed{\participant{A}}), \\
            &(\isStatusStipSlashed{\participant{A}},
            \isStatusStipRight), \\
            &(\isStatusStipRight,\isStatusStipChoice), \\
            &(\isStatusChoice,\isStatusStipChoice), \\
            &(\isStatusStipRefunded{\participant{A}},
            \isStatusStipRight)
        \}^*
    \end{align*}
    (With $R^*$ we denote the reflexive and transitive closure of a relation $R$)
\end{lemma}

\begin{proof}

By definition of intermediate strategy compilation, we know that 
\[
    \exists \alpha' \in \isStrategy[\participant{A}](\isRun)
    ~\land~ (\forall \isConfig: \isRun \xrightarrow[]{\alpha'} \isConfig \Rightarrow \isConfig \neq \isConfig(\isRun))
\]

The progress of configuration is natural for timed moves. a
For stipulation commitments and authorizations progress is derived from \bitmlx strategy eagerness and run coherence.

It is left to be shown that each move decreases the finalization distance.
Note that only contracts in $\contractbound$ can be advertised, committed, signed, etc. 
This bound is used in the finalization distance as limit. Eagerness is used to ensure progress when revealing secrets and authorizing contracts.

CA on possible moves of $\isStrategy[\participant{A}]$:
\begin{itemize}
    \item Advertise: $advertise(\contractAdv{G}{C})$: $\ldTotalMissingContracts$ decreases
    \item Authorize Commit: $commit(\participant A, \contractAdv{G}{C})$: only $\ldAdvMissingCommitments$ decreases. $\contractbound$-Eagerness ensures uniqueness of authorization.
    \item Authorize Init: $authInit(\participant{A}, \isContractAdv{G}{C}{\someBlockchain})$: only $\ldAdvMissingAuths$ decreases. 
    \item Publish: $publish(\isContractAdv{G}{C}{\someBlockchain})$: $\ldAdvContracts$ decreases and more significant components stall
    \item Double Spend: $doubleSpend(\participant{A}, \isContractAdv{G}{C}{\someBlockchain})$ 
\end{itemize} 

All of the following moves emitted by $\isStrategy[\participant{A}]$ do not increase/decrease $\ldTotalMissingContracts, \ldAdvMissingCommitments, \ldAdvMissingAuths$ and $\ldAdvContracts$:

\begin{itemize}
    \item Reveal Init Step Secret: $\isRevealedInitSecret{A}{\someContract}$: only decreases $\ldMissingStipSecrets$
    \item Init: $init(\someContract, \someBlockchain)$: only $\ldStipStatus$ decreases from $\isStatusStipChoice$ to $\isStatusChoice$
    \item Stip Skip: $from~t:~sskip(\someContract, \someBlockchain)$:   only $\ldStipStatus$ decreases from $\isStatusStipChoice$ to $\isStatusStipRight$
    \item Abort: $from~t:~abort(\someContract, \someBlockchain)$: only $\ldStipStatus$ decreases from $\isStatusStipRight$ to $\isStatusStipRefunded{\participant{A_i}}$
    \item Intro Stip Compensate: $sslash(\someStipulation, \someBlockchain, \participant{A})$: only $\ldStipStatus$ decreases from $\isStatusStipRight$ to $\isStatusStipSlashed{\participant{A}}$
    \item Elim Stip Compensate: $scompensate(\someStipulation, \someBlockchain, \participant{A})$: only $\ldStipStatus$ decreases from $\isStatusStipSlashed{\participant{A}}$ to $\isStatusStipCompensation{\participant{A}}$
\end{itemize}

In addition, all of the following moves do not increase/decrease $\ldMissingStipSecrets, \ldStipStatus$ and $\ldChainStipDistance$:

\begin{itemize}
    \item Secret Reveal: $\participant A:\secret a$: only $\ldMissingRevealSecrets$ decreases
    \item Step Secret Reveal: $\isRevealedStepSecret{A}{\someContract}$: only $\ldESMissingSSecrets$ decreases
    \item Intro Left: $ileft(\someContract, \someBlockchain, x)$: only $\ldChainStatus$ decreases from $\isStatusChoice$ to $\isStatusLeft$
    \item Reveal: $reveal(\someContract, \someBlockchain, x)$: If ES frontier is moved with this action, $\ldESFinDistance$ is decreased. Otherwise $\ldBSynchDistance$ decreases but $\ldESFinDistance$ and $\ldESMissingSSecrets$ remains the same.
    \item Authorize: $\participant{A}: (\someContract, \someBlockchain, x)$: only $\ldMissingAuthorizations$ decreases. Eagerness ensures that authorization is unique.
    \item Intro Skip: $from ~ t: ~ skip(\someContract, \someBlockchain, x)$: only $\ldChainStatus$ decreases from $\isStatusChoice$ to $\isStatusRight$
    \item Elim Skip: $from ~ t: ~ right(\someContract, \someBlockchain, x)$: If ES frontier is moved with this action, $\ldESFinDistance$ is decreased. Otherwise $\ldBSynchDistance$ decreases but $\ldESFinDistance$ and $\ldESMissingSSecrets$ remains the same.
    \item Intro Compensation: $slash(\someContract, \someBlockchain, x, \participant{A})$: only $\ldChainStatus$ decreases from $\isStatusRight$ to $\isStatusSlashed{\participant{A}}$
    \item Elim Compensation: $compensate(\someContract, \someBlockchain, x, \participant{A})$: only $\ldChainStatus$ decreases from $\isStatusSlashed{\participant{A}}$ to $\isStatusCompensated{\participant{A}}$
    \item Guarded Withdraw: $dwithdraw(\someContract, \someBlockchain, x)$: only $\ldChainStatus$ decreases from $\isStatusLeft$ to $\isStatusAssigned{\participant{A}}$
    \item Split: $split(\someContract, \someBlockchain, x)$: If ES frontier is moved with this action, $\ldESFinDistance$ is decreased. Otherwise $\ldBSynchDistance$ decreases but $\ldESFinDistance$ and $\ldESMissingSSecrets$ remains the same.
    \item Withdraw: $cwithdraw(\someContract, \someBlockchain, x)$: only $\ldChainStatus$ decreases from $\isStatusChoice$ to $\isStatusAssigned{\participant{A}}$ 
\end{itemize}

Time delays $\delta$ do not change the untimed configurations. 
By assumption, the honest user strategy $\isStrategy[\participant{A}]$ proposes only delays that changes the phase/round of at least a single contract.

Let $\hat{\kappa}$ be the identifier of the contract with the next round/phase deadline in $\contractbound$ and $\hat{\delta}$ be the delay to be scheduled by an honest user strategy $\isStrategy[\participant{A}]$ that wants to reach the deadline without scheduling any untimed moves before.
If $\kappa$ is not yet initialised, $\ldChainStipDistance$ is decreased by this $\hat{\delta}$ move by definition of stip-phase. 
Otherwise, $\ldChainRoundDistance$ is decreased by the definition of $\roundStatus{\hat{\kappa}}{\isRun}$.
\end{proof}

\begin{lemma}[Liquidity of intermediate strategies]
    \label{lemma:strong-intermediate-liquidity}
    Let $\participant{A}$ be an honest participant with an eager strategy $\xStrategy[\participant{A}]$ and $\contractbound$ a set of contracts such that $\xStrategy[\participant{A}]$ is $\contractbound$-bounded.
    Let $\isStrategy[\participant{A}] = \xCompiledStrategy$ be the corresponding intermediate strategy.
    Then $\isStrategy[\participant{A}]$ is strongly liquid.
                        \end{lemma}

\begin{proof}
    The definition of strongly liquid requires the construction of a finite trace extension $\hat{\isRun}$ according to $\isStrategy[\participant{A}]$ until $\isRun$ is liquidated.

    We construct the extension $\hat{\isRun}$ with actions that are decreasing the finalization distance of $\sizeMetric{\isRun, \isStrategy[\participant{A}]}$ using Lemma \ref{fs:isFinDist}. 
    This extension is finite since the lower bound of the finalization distance $\sizeMetric{}$ is 0.
    By definition of Lemma \ref{fs:isFinDist}, the last configuration is liquidated.
\end{proof}

\section{\bitmlx Security}

\subsection{Payouts and Contract Funds}

We first introduce some auxiliary infrastructure and lemmas that help us to reason about inputs.

First, we define the intermediate inputs per contract. 
\begin{definition}[Intermediate inputs per contract]
    Let $\isRun$ be an intermediate semantics run and $\someBlockchain$ be a blockchain. 
    We define the intermediate inputs of $\participant{A}$ in $\isRun$ and blockchain $\someBlockchain$ (written $\isContractUserInputs{\someStipulation}{\participant{A}}{\someBlockchain}{\isRun}$) as follows
    \begin{align*}
        \sum_{\{
            v:
                \depositsPre{A}{v \someBlockchain}{x} \in \preconditions{G^{\someBlockchain}}
                \land publish(\isContractAdv{G}{C}{\someBlockchain}) \in \isRun
                \land  (t_0, \someStipulation) = initSettings({\preconditions{G}})
        \}}    
            v
    \end{align*}
\end{definition}

Similarly, we define the total intermediate funds per contracts: 

\begin{definition}[Intermediate total funds per contract]
    Let $\isRun$ be an intermediate semantics run and $\someBlockchain$ be a blockchain.
    We define the contract funds in $\isRun$ (written $\isContractTotalFunds{\someStipulation}{\someBlockchain}{\isRun}$) as follows
    \begin{align*}
        \sum_{\{
            \isContractState:
                \isActiveContract{\contract{C}}{\isContractState}{\someBlockchain}{\someContract} \in \lastConfigOf{\isRun} ~\land~ \someContract \in \descendant(\someStipulation, \isRun)
        \}} 
            \isContractState.balance
    \end{align*}
\end{definition}

Intermediate runs preserve inputs per contract: 

\begin{lemma}[Intermediate money preservation per contract]
    \label{lemma:intermediate-money-preservation-per-contract}
    Let $\isRun$ be an intermediate semantics run and $\someStipulation \in \isRun$. Then, for every blockchain $\someBlockchain$, the following equation holds

    $$
                        \isContractTotalFunds{\someStipulation}{\someBlockchain}{\isRun}
        = \sum_{\participant{A} \in P} \isContractUserInputs{\someStipulation}{\participant{A}}{\someBlockchain}{\isRun}
    $$
\end{lemma}

Further, we can show the following:

\begin{lemma}[Intialized eventual synchronicty implies publication]
    \label{lemma:esInitPublication}
    Let $\isRun$ be an intermediate semantics run and $\someStipulation \in \isRun$ and $\esStipStatus{\isRun}{\someStipulation} = \isStipStatusInit{}$. 
    Then 
    $$
    \forall \someBlockchain: publish(\contractAdv{G^{\someBlockchain}}{C}) \in \isRun
    $$
\end{lemma}

\begin{proof}
    Same proof strategy as in the general intermediate money preservation lemma (Lemma \ref{mp:inter-money-preserv}).
\end{proof}

\begin{lemma}[Liquidated Run Security]
    \label{lemma:liqudated-run-security}
Let $\participant{A}$ be an honest participant with strategy $\xStrategy[\participant{A}]$
and $\isStrategy[\participant{A}] = \xCompiledStrategy$ its corresponding intermediate strategy. 
Let $\xRun$ and $\isRun$ be such that $\xRun \xCoherence[A] \isRun$ and $\isRun$ being liquidated.
Then
$$
    \forall \someBlockchain \in \activeBlockchains: \isUserPayout{A}{\someBlockchain}{\isRun}{} -\isUserInputs{\participant{A}}{\someBlockchain}{\isRun} 
    \geq 
    \xUserPayout{A}{\someBlockchain}{\xRun}{} +  \xContractFunds{\someBlockchain}{\xRun}{} - \xUserInputs{\participant{A}}{\someBlockchain}{\xRun} 
$$
\end{lemma}

Intuitively, this statement says that in a fully liquidated intermediate run $\isRun$, the honest user earned at least as much as in the coherent \bitmlx $\xRun$ plus the money that is still locked in contracts of $\xRun$.
The intuition behind this statement is that contracts in $\xRun$ can only be still present in $\xRun$ because they have been fully compensated.

\begin{proof}
    Let $\someBlockchain \in \activeBlockchains$.
    From the definition of $\xCoherence[A]$ we know that 
    $$\ES = \maxFrontierFun{\xRun} = \bigjoin_{\someBlockchain \in \activeBlockchains}{\isMaxFrontierFun{\isRun}{\someBlockchain}}$$ 
    Consequently, we can conclude that 

    \begin{align}
        \forall \isActiveContract{\contract{C}}{\isContractState}{\someBlockchain}{\someContract}{} \in \lastConfigOf{\isRun} ~\land~ \someContract \in \ES 
        \Rightarrow 
                \exists \activeOrAssigned: \activeOrAssignedContract{\activeOrAssigned}{\balance}{\someContract} \in \lastConfigOf{\xRun}
        ~\land~
        \balance[][\someBlockchain] = \isContractState.balance
        \label{eq:es-coincide}
    \end{align}

    And consequently for all $K \subseteq \ES$ such that $\forall \someContract \in K: \exists \isActiveContract{\contract{C}}{\isContractState}{\someBlockchain}{\someContract}{} \in \lastConfigOf{\isRun}$  we have that 

    \begin{align}
        \sum_{\someContract \in K: \isActiveContract{\contract{C}}{\isContractState}{\someBlockchain}{\someContract}{} \in \lastConfigOf{\isRun}} 
        \isContractState.balance
        =
        \sum_{\someContract \in K: \activeOrAssignedContract{\activeOrAssigned}{\balance}{\someContract} \in \lastConfigOf{\xRun}}
        \balance[][\someBlockchain] 
    \end{align}

    From coherence, we know that
    $$ \forall \someStipulation \in \isRun: \xStipStatus{\xRun}{\someStipulation} = \esStipStatus{\isRun}{\someStipulation} $$

    With that we can show that
    \begin{align*}
        K_0 :=  \{ (\someStipulation, \participant{A}, \someBlockchain, v)& ~|~ init(\contractAdv{G}{C}) \in \xRun ~\land~ (t_0, \someStipulation) = initSettings({\preconditions{G}}) \\
        &~\land~  \depositsPre{A}{\balance}{\vec{x}} \in \preconditions{G} ~\land~ v = \balance[][\someBlockchain] \}
        \\
        = \{ (\someStipulation, \participant{A}, \someBlockchain, v)& ~|~ \xStipStatus{\xRun}{\someStipulation} =  \stipStatusInit{} 
        ~\land~ \contractAdv{G}{C} \in \xRun ~\land~ (t_0, \someStipulation) = initSettings({\preconditions{G}}) \\
        &~\land~  \depositsPre{A}{\balance}{\vec{x}} \in \preconditions{G} ~\land~ v = \balance[][\someBlockchain] \} \\
        = 
        \{ (\someStipulation, \participant{A}, \someBlockchain, v)& ~|~ \esStipStatus{\isRun}{\someStipulation} =  \stipStatusInit{} 
        ~\land~ \contractAdv{G}{C} \in \isRun ~\land~ (t_0, \someStipulation) = initSettings({\preconditions{G}}) \\
        &~\land~  \depositsPre{A}{\balance}{\vec{x}} \in \preconditions{G} ~\land~ v = \balance[][\someBlockchain] \} \\
        = 
        \{ (\someStipulation, \participant{A}, \someBlockchain, v)& ~|~ publish(\contractAdv{G^{\someBlockchain}}{C}) \in \isRun
        ~\land~ \esStipStatus{\isRun}{\someStipulation} =  \stipStatusInit{} \\
        &~\land~ (t_0, \someStipulation) = initSettings({\preconditions{G}})
        ~\land~  \depositsPre{A}{\balance}{\vec{x}} \in \preconditions{G} ~\land~ v = \balance[][\someBlockchain] \}
    \end{align*}

    The last equation holds due to Lemma~\ref{lemma:esInitPublication}.

    As a consequence, we have that 

    \begin{align}
        \sum_{\participant{A} \in P} \xUserInputs{\participant{A}}{\someBlockchain}{\xRun} 
        = \sum_{\participant{A} \in P, \someStipulation \in \{ \someStipulation: \someStipulation \in \isRun ~\land~ \esStipStatus{\isRun}{\someStipulation} =  \stipStatusInit{}  \}} \isContractUserInputs{\someStipulation}{\participant{A}}{\someBlockchain}{\isRun}
        \label{eq:inputs-agreement}
    \end{align}

    Since $\isRun$ is liquidated we know that all contracts are either initialized or aborted, so we get that:

    \begin{align}
        \sum_{\participant{A} \in P} \isUserInputs{\participant{A}}{\someBlockchain}{\isRun}
        &=  
        \sum_{\participant{A} \in P, \someStipulation \in \{ \someStipulation: \someStipulation \in \isRun ~\land~ \esStipStatus{\isRun}{\someStipulation} =  \stipStatusInit{}  \}} \isContractUserInputs{\someStipulation}{\participant{A}}{\someBlockchain}{\isRun} \\
        &+ 
        \sum_{\participant{A} \in P, \someStipulation \in \{ \someStipulation: \someStipulation \in \isRun ~\land~ \esStipStatus{\isRun}{\someStipulation} =  \stipStatusAbort{}  \}} \isContractUserInputs{\someStipulation}{\participant{A}}{\someBlockchain}{\isRun}
        \label{eq:inputs-decomp}
    \end{align}

    Finally, we can show using Lemma~\ref{lemma:intermediate-money-preservation-per-contract} that inputs of the contracts in $\{ \someStipulation: \someStipulation \in \isRun ~\land~ \esStipStatus{\isRun}{\someStipulation} =  \stipStatusAbort{}  \}$ sum up to the corresponding contract funds, which we know must be either refunded or compensated. 
    More precisely, we have that 

    \begin{align*}
        \sum_{\participant{A} \in P, \someStipulation \in \{ \someStipulation: \someStipulation \in \isRun ~\land~ \esStipStatus{\isRun}{\someStipulation} =  \stipStatusAbort{}  \}} \isContractUserInputs{\someStipulation}{\participant{A}}{\someBlockchain}{\isRun} \\
         = \sum_{\{ \someStipulation: \isStipStatus{\isRun}{\someStipulation}{\someBlockchain} = \isStipStatusRefund \}} \isContractTotalFunds{\someStipulation}{\someBlockchain}{\isRun} \\
        + \sum_{\{ \someStipulation: \isStipStatus{\isRun}{\someStipulation}{\someBlockchain} = \isStipStatusComp ~\land~ \esStipStatus{\isRun}{\someStipulation} =  \stipStatusAbort{} \}} \isContractTotalFunds{\someStipulation}{\someBlockchain}{\isRun} \\
    \end{align*}

    Note that if a contract $\someStipulation$ is refunded on one chain, this implies that its eventual synchronicity status is $\stipStatusAbort$.

    For conciseness, we will use the following sets in the remainder of the proof:
    \begin{align*}
        \nES &:= \{ \someContract :  \isActiveContract{\contract{C}}{\isContractState}{\someBlockchain}{\someContract}{} \in \lastConfigOf{\isRun} ~\land~ \someContract \not \in \ES \} \\
        \FB &:=  \{ \someContract : \activeOrAssignedContract{\activeOrAssigned}{\balance}{\someContract} \in \lastConfigOf{\xRun} ~\land~ \exists \isActiveContract{\contract{C}}{\isContractState}{\someBlockchain}{\someContract}{} \in \lastConfigOf{\isRun} \} \\
        \nFB&:=  \{ \someContract : \activeOrAssignedContract{\activeOrAssigned}{\balance}{\someContract} \in \lastConfigOf{\xRun} ~\land~ \not \exists  \isActiveContract{\contract{C}}{\isContractState}{\someBlockchain}{\someContract}{} \in \lastConfigOf{\isRun} \} \\
        \CAB &:= \{ \someStipulation: \isStipStatus{\isRun}{\someStipulation}{\someBlockchain} = \isStipStatusComp ~\land~ \esStipStatus{\isRun}{\someStipulation} =  \stipStatusAbort{} \} \\
        \RE &:= \{ \someStipulation: \isStipStatus{\isRun}{\someStipulation}{\someBlockchain} = \isStipStatusRefund \} \\
        \assignedA &:= \{ \someContract :  \isActiveContract{\contract{C}}{\isContractState}{\someBlockchain}{\someContract}{} \in \lastConfigOf{\isRun} ~\land~ \isContractState[status] =  \isStatusAssigned{\participant{A}} \}\\
        \compensated &:= \{ \someContract :  \isActiveContract{\contract{C}}{\isContractState}{\someBlockchain}{\someContract}{} \in \lastConfigOf{\isRun} ~\land~ \exists \participant{B}: \participant{B} \neq \participant{A} ~\land~ \isContractState [status] = \isStatusCompensated{\participant{B}}\}
    \end{align*}

    We know that for both x and intermediate runs the total funds and inputs sum up. 
    So consequently, it holds that: 
    \begin{align}
        \isTotalFunds{\someBlockchain}{\isRun}{} 
        -  \sum_{\participant{A} \in P}\isUserInputs{\participant{A}}{\someBlockchain}{\isRun} 
        =  \sum_{\participant{A} \in P} \xUserPayout{A}{\someBlockchain}{\xRun}{} 
        +  \xContractFunds{\someBlockchain}{\xRun}{} 
        -  \sum_{\participant{A} \in P} \xUserInputs{\participant{A}}{\someBlockchain}{\xRun} 
        \label{eq:total-sum}
    \end{align}

    Since 
    \begin{align*}
        \isTotalFunds{\someBlockchain}{\isRun}{} 
        =         
        \sum_{\someContract \in \ES: \isActiveContract{\contract{C}}{\isContractState}{\someBlockchain}{\someContract}{} \in \lastConfigOf{\isRun}} 
        \isContractState.balance
        + 
        \sum_{\someContract \in \nES: \isActiveContract{\contract{C}}{\isContractState}{\someBlockchain}{\someContract}{} \in \lastConfigOf{\isRun}} 
        \isContractState.balance
    \end{align*}
    and

    \begin{align*}
        \sum_{\participant{A} \in P} \xUserPayout{A}{\someBlockchain}{\xRun}{} +  \xContractFunds{\someBlockchain}{\xRun}{}
        = 
        \sum_{\someContract \in \FB : \activeOrAssignedContract{\activeOrAssigned}{\balance}{\someContract} \in \lastConfigOf{\xRun}}
        \balance[][\someBlockchain] 
        + 
        \sum_{\someContract \in \nFB : \activeOrAssignedContract{\activeOrAssigned}{\balance}{\someContract} \in \lastConfigOf{\xRun}}
        \balance[][\someBlockchain] 
    \end{align*}

    From ~\ref{eq:es-coincide} we know that 

    \begin{align*}
        \sum_{\someContract \in \ES: \isActiveContract{\contract{C}}{\isContractState}{\someBlockchain}{\someContract}{} \in \lastConfigOf{\isRun}} 
        \isContractState.balance
        = 
        \sum_{\someContract \in \FB : \activeOrAssignedContract{\activeOrAssigned}{\balance}{\someContract} \in \lastConfigOf{\xRun}}
        \balance[][\someBlockchain] 
    \end{align*}
    and consequently, we can conclude from \ref{eq:total-sum} that 

    \begin{align*}
        \sum_{\someContract \in \nFB : \activeOrAssignedContract{\activeOrAssigned}{\balance}{\someContract} \in \lastConfigOf{\xRun}}
        \balance[][\someBlockchain] 
        =
        \sum_{\someContract \in \nES: \isActiveContract{\contract{C}}{\isContractState}{\someBlockchain}{\someContract}{} \in \lastConfigOf{\isRun}} 
        \isContractState.balance
        +
        \sum_{\participant{A} \in P} \xUserInputs{\participant{A}}{\someBlockchain}{\xRun}
        - 
        \sum_{\participant{A} \in P}\isUserInputs{\participant{A}}{\someBlockchain}{\isRun} 
    \end{align*}

    Substituting \ref{eq:inputs-agreement} and \ref{eq:inputs-decomp} we obtain
    \begin{align*}
        \sum_{\someContract \in \nFB : \activeOrAssignedContract{\activeOrAssigned}{\balance}{\someContract} \in \lastConfigOf{\xRun}}
        \balance[][\someBlockchain] 
        = 
        \sum_{\someContract \in \nES: \isActiveContract{\contract{C}}{\isContractState}{\someBlockchain}{\someContract}{} \in \lastConfigOf{\isRun}} 
        \isContractState.balance
        - 
        \sum_{\someStipulation \in \RE} \isContractTotalFunds{\someStipulation}{\someBlockchain}{\isRun} \\
        -
        \sum_{\someStipulation \in \CAB} \isContractTotalFunds{\someStipulation}{\someBlockchain}{\isRun}
    \end{align*}

    In particular, we know that all contracts in $\nES$ have been compensated (since $\isRun$ is liquidated and contracts that are lacking behind the $\xRun$ need to be compensated). 

    So, we know that: 
    \begin{align*}
        \sum_{\someContract \in \nES \cap \compensated: \isActiveContract{\contract{C}}{\isContractState}{\someBlockchain}{\someContract}{} \in \lastConfigOf{\isRun}} 
        \isContractState.balance
        = 
        \sum_{\someContract \in \nFB : \activeOrAssignedContract{\activeOrAssigned}{\balance}{\someContract} \in \lastConfigOf{\xRun}}
        \balance[][\someBlockchain] 
        +
        \sum_{\someStipulation \in \RE} \isContractTotalFunds{\someStipulation}{\someBlockchain}{\isRun} \\
        +
        \sum_{\someStipulation \in \CAB} \isContractTotalFunds{\someStipulation}{\someBlockchain}{\isRun}
    \end{align*}

    Using this, we can show our final statement: 

    \begin{align*}
        & \isUserPayout{A}{\someBlockchain}{\isRun}{} -\isUserInputs{\participant{A}}{\someBlockchain}{\isRun}  \\
        &=  
        \sum_{\someContract \in \assignedA: \isActiveContract{\contract{C}}{\isContractState}{\someBlockchain}{\someContract}{} \in \lastConfigOf{\isRun}} 
        \isContractState.balance
        + 
        \sum_{\someContract \in \compensated: \isActiveContract{\contract{C}}{\isContractState}{\someBlockchain}{\someContract}{} \in \lastConfigOf{\isRun}} 
        \isContractState.balance
        -
        \isUserInputs{\participant{A}}{\someBlockchain}{\isRun} \\
        &=         
        \sum_{\someContract \in \assignedA \cap \ES : \isActiveContract{\contract{C}}{\isContractState}{\someBlockchain}{\someContract}{} \in \lastConfigOf{\isRun}} 
        \isContractState.balance
        +         
        \sum_{\someContract \in \compensated \cap \ES : \isActiveContract{\contract{C}}{\isContractState}{\someBlockchain}{\someContract}{} \in \lastConfigOf{\isRun}} 
        \isContractState.balance
        + 
        \sum_{\someContract \in \compensated \cap \nES : \isActiveContract{\contract{C}}{\isContractState}{\someBlockchain}{\someContract}{} \in \lastConfigOf{\isRun}} 
        \isContractState.balance \\
        & \qquad - 
        \isUserInputs{\participant{A}}{\someBlockchain}{\isRun} \\
        &=         
        \sum_{\someContract \in \assignedA \cap  \FB : \activeOrAssignedContract{\activeOrAssigned}{\balance}{\someContract} \in \lastConfigOf{\xRun}}
        \balance[][\someBlockchain] 
        +         
        \sum_{\someContract \in \compensated \cap \ES : \isActiveContract{\contract{C}}{\isContractState}{\someBlockchain}{\someContract}{} \in \lastConfigOf{\isRun}} 
        \isContractState.balance
        + 
        \sum_{\someContract \in \nFB : \activeOrAssignedContract{\activeOrAssigned}{\balance}{\someContract} \in \lastConfigOf{\xRun}} 
        \balance[][\someBlockchain] \\
        &\qquad +
        \sum_{\someStipulation \in \RE} \isContractTotalFunds{\someStipulation}{\someBlockchain}{\isRun} 
        +
        \sum_{\someStipulation \in \CAB} \isContractTotalFunds{\someStipulation}{\someBlockchain}{\isRun}
        - 
        \isUserInputs{\participant{A}}{\someBlockchain}{\isRun} \\
        & = 
        \xUserPayout{A}{\someBlockchain}{\xRun}{} 
        +  
        \xContractFunds{\someBlockchain}{\xRun}{} 
        +         
        \sum_{\someContract \in \compensated \cap \ES : \isActiveContract{\contract{C}}{\isContractState}{\someBlockchain}{\someContract}{} \in \lastConfigOf{\isRun}} 
        \isContractState.balance \\
        &\qquad +
        \sum_{\someStipulation \in \RE} \isContractTotalFunds{\someStipulation}{\someBlockchain}{\isRun} 
        +
        \sum_{\someStipulation \in \CAB} \isContractTotalFunds{\someStipulation}{\someBlockchain}{\isRun}
        - 
        \isUserInputs{\participant{A}}{\someBlockchain}{\isRun} \\
    \end{align*}

    Consequently, we only need to show that 
    \begin{align*}
        &\xUserInputs{\participant{A}}{\someBlockchain}{\xRun} 
        \geq 
        \isUserInputs{\participant{A}}{\someBlockchain}{\isRun} \\
        & \qquad - \sum_{\someContract \in \compensated \cap \ES : \isActiveContract{\contract{C}}{\isContractState}{\someBlockchain}{\someContract}{} \in \lastConfigOf{\isRun}} 
        \isContractState.balance
        -
        \sum_{\someStipulation \in \RE} \isContractTotalFunds{\someStipulation}{\someBlockchain}{\isRun} 
        -
        \sum_{\someStipulation \in \CAB} \isContractTotalFunds{\someStipulation}{\someBlockchain}{\isRun}
    \end{align*}
This immediately follow from \ref{eq:inputs-agreement} and \ref{eq:inputs-decomp}.

    \end{proof}

\begin{lemma}[Intermediate Security]
    Let $\participant{A}$ be an honest participant and $\contractbound$ be a set of contract identifiers. 
    Let $\xStrategy[\participant{A}]$ be an eager and deterministic strategy of $\participant{A}$ bounded by $\contractbound$ 
    and $\isStrategy[\participant{A}] = \xCompiledStrategy$ its corresponding intermediate strategy. 
    Then for every run $\isRun$ with $\isStrategy[\participant{A}] \vDash \isRun$
    there exists an extension $\hat{\isRun} = \isRun \xrightarrow[]{\alpha_1} \cdots \xrightarrow[]{\alpha_n}$ such that 
    \begin{enumerate}
        \item $\forall i \in 1 \dots n : ~ \alpha_i \in \isStrategy[\participant{A}](\isRun \xrightarrow[]{\alpha_1} \cdots \xrightarrow[]{\alpha_{i-1}})$
        \item $\hat\isRun$ is liquidated 
    \end{enumerate}
    and there exists $\hat{\xRun}$ such that $\xStrategy[\participant{A}] \vdash \hat{\xRun}$, $\hat{\xRun}$ is liquidated and 
    $$
        \forall \someBlockchain \in \activeBlockchains: \isUserPayout{A}{\someBlockchain}{\isRun}{} -\isUserInputs{\participant{A}}{\someBlockchain}{\isRun} 
        \geq 
        \xUserPayout{A}{\someBlockchain}{\hat{\xRun}}{} - \xUserInputs{\participant{A}}{\someBlockchain}{\hat{\xRun}} 
    $$
    \end{lemma}
   \begin{proof}
    From Lemma~\ref{lemma:strong-intermediate-liquidity}, we know that there exists an extension $\hat{\isRun} = \isRun \xrightarrow[]{\alpha_1} \cdots \xrightarrow[]{\alpha_n}$ such that 
    \begin{enumerate}
        \item $\forall i \in 1 \dots n : ~ \alpha_i \in \isStrategy[\participant{A}](\isRun \xrightarrow[]{\alpha_1} \cdots \xrightarrow[]{\alpha_{i-1}})$
        \item $\hat\isRun$ is liquidated 
    \end{enumerate}
    From soundness, we can conclude that there exists also a run $\hat{\xRun}$ such that $\xStrategy[\participant{A}] \vdash \hat{\xRun}$ and $\hat{\xRun} \xCoherence[\participant{A}] \hat{\isRun}$. 

    Consequently, we can use Lemma~\ref{lemma:liqudated-run-security} to conclude that 
    $$
    \forall \someBlockchain \in \activeBlockchains: \isUserPayout{A}{\someBlockchain}{\hat{\isRun}}{} -\isUserInputs{\participant{A}}{\someBlockchain}{\hat{\isRun}} 
    \geq 
    \xUserPayout{A}{\someBlockchain}{\hat{\xRun}}{} +  \xContractFunds{\someBlockchain}{\hat{\xRun}}{} - \xUserInputs{\participant{A}}{\someBlockchain}{\hat{\xRun}} 
$$
    From Lemma~\ref{lemma:xliquidty}, we know that there exists an extension $\dot{\xRun} = \hat{\xRun} \xrightarrow[]{\alpha_1} \cdots \xrightarrow[]{\alpha_n}$ of $\hat{\xRun}$ such that
    \begin{enumerate}
        \item $\forall i \in 1 \dots n : ~ \alpha_i \in \xStrategy[\participant{A}](\hat{\xRun} \xrightarrow[]{\alpha_1} \cdots \xrightarrow[]{\alpha_{i-1}})$
        \item $\not \exists\activeContract{\contract{C}}{\balance}{\someContract} \in \xConfig(\dot{\xRun})$
    \end{enumerate}

    So consequently, 
    \begin{align*}
        \xContractFunds{\someBlockchain}{\dot{\xRun}}{} = 0
    \end{align*}

    The claim hence immediately follows from money preservation.

   \end{proof}

\begin{definition}[Liquidated Low-Level Runs]
    Let $\contractbound$ be a set of contracts and $\orRun$ a low-level BitML run. 
    Then $\orRun$ is considered liquidated 
    w.r.t. $\contractbound$ 
    if for all $\isActiveContract{\contract{C}}{\isContractState}{\someBlockchain}{\someContract} \in \isConfig(\isRun)$ with $\kappa \in \contractbound$, it holds that $\exists \participant{B}.~\contract{C} = \bitmlcode{withdraw}~\participant{B}$.
\end{definition}

\begin{definition}[Strong Low-Level Liquidity]
    Let $\participant{A}$ be an honest participant with strategy $\orStrategy[\participant{A}]$. 
    Then $\orStrategy[\participant{A}]$ is strongly liquid if for all runs $\orRun$ there exists an extension $\hat{\orRun} = \orRun \xrightarrow[]{\alpha_1} \cdots \xrightarrow[]{\alpha_n}$ such that 
    
    \begin{enumerate}
        \item $\forall i \in 1 \dots n : ~ \alpha_i \in \orStrategy[\participant{A}](\orRun \xrightarrow[]{\alpha_1} \cdots \xrightarrow[]{\alpha_{i-1}})$
        \item $\hat\orRun$ is liquidated 
    \end{enumerate}
\end{definition}

\begin{lemma}[Liquidity Connection]\label{fs:liq-connect}
    \[
        \isRun \textit{ is liquidated} \Rightarrow \isconfigcompile[]{\isRun} \textit{ is liquidated}
    \]
\end{lemma}

\begin{proof}
    Holds by definition of translation function $\iscompile$.
\end{proof}

\begin{definition}[Low-level inputs per contract]
    Let $\orRun$ be a low-level BitML semantics run and $\someBlockchain$ be a blockchain. 
    We define the low-level inputs of $\participant{A}$ in $\orRun$ and blockchain $\someBlockchain$ (written $\isContractUserInputs{\vec{x}}{\participant{A}}{\someBlockchain}{\orRun}$) as follows
    \begin{align*}
        \isContractUserInputs{x}{\participant{A}}{\someBlockchain}{\orRun} = \sum_{\{
            v:
                \depositsPre{A}{v^{\someBlockchain}}{x} \in \preconditions{G^{\someBlockchain}}
                \land init(\isContractAdv{G}{C}{\someBlockchain}) \in \orRun
        \}}    
            v
    \end{align*}
\end{definition}

Similarly, we define the total low-level funds per contracts: 

\begin{definition}[Low-level total funds per contract]
    Let $\orRun$ be a low-level BitML semantics run and $\someBlockchain$ be a blockchain.
    We define the contract funds in $\orRun$ (written $\isContractTotalFunds{x_0}{\someBlockchain}{\orRun}$) as follows
    \begin{align*}
        \isContractTotalFunds{x_0}{\someBlockchain}{\orRun} = \sum_{\{
            v:
                \isActiveContract{\contract{C}}{v}{\someBlockchain}{x} \in \lastConfigOf{\orRun} ~\land~ x \in \descendant(x_0, \orRun)
        \}} 
            v
    \end{align*}
\end{definition}

\begin{definition}[Low-level total payout per blockchain]
    Let $\orRun$ be a low-level BitML semantics run and $\someBlockchain$ be a blockchain.
    We define the low-level payout of $\participant{A}$ in $\orRun$ and blockchain $\someBlockchain$ (written $\xUserPayout{\participant{A}}{\someBlockchain}{\orRun}$) as follows
    \begin{align*}
        \xUserPayout{\participant{A}}{\someBlockchain}{\orRun} = \sum_{\{
            v:
                \isActiveContract{\contract{\bitmlcode{withdraw}~\participant{A}}}{v}{\someBlockchain}{x} \in \lastConfigOf{\orRun} 
        \}} 
            v + \sum_{\{
            v:
                \isParticipantDeposit{\participant{A}}{v}{\someBlockchain}{x} \in \lastConfigOf{\orRun} 
                ~\land~ \neg initial(x)
        \}} 
            v
    \end{align*}
\end{definition}

\begin{lemma}[Low-Level Security]
    Let $\participant{A}$ be an honest participant and $\contractbound$ be a set of contract identifiers. 
    Let $\xStrategy[\participant{A}]$ be an eager and deterministic intermediate semantics strategy of $\participant{A}$ bounded by $\contractbound$ 
    and $\isStrategy[\participant{A}] = \xCompiledStrategy$ its corresponding intermediate level strategy and $\orStrategy[\participant{A}] = \isCompiledStrategy$ its corresponding low level strategy. 
    Then for every run $\orRun$ with $\orStrategy[\participant{A}] \vDash \orRun$
    there exists an extension $\hat{\orRun} = \orRun \xrightarrow[]{\alpha_1} \cdots \xrightarrow[]{\alpha_n}$ such that 
    \begin{enumerate}
        \item $\forall i \in 1 \dots n : ~ \alpha_i \in \orStrategy[\participant{A}](\orRun \xrightarrow[]{\alpha_1} \cdots \xrightarrow[]{\alpha_{i-1}})$
        \item $\hat\orRun$ is liquidated 
    \end{enumerate}
    and there exists $\hat{\isRun}$ such that $\isStrategy[\participant{A}] \vdash \hat{\isRun}$, $\hat{\isRun}$ is liquidated and 
    $$
        \forall \someBlockchain \in \activeBlockchains: \isUserPayout{A}{\someBlockchain}{\hat\orRun}{} -\isUserInputs{\participant{A}}{\someBlockchain}{\hat\orRun} 
        \geq 
        \xUserPayout{A}{\someBlockchain}{\hat{\isRun}}{} - \isUserInputs{\participant{A}}{\someBlockchain}{\hat{\isRun}} 
    $$
    \end{lemma}

\begin{proof}
    By low-level soundness we know that there exists an intermediate level run $\isRun$ such that $\isStrategy \vDash \isRun$ and $\isRun \isCoherence \orRun$.
    From Lemma~\ref{lemma:strong-intermediate-liquidity}, we know that there exists an extension $\hat{\isRun} = \isRun \xrightarrow[]{\beta_1} \cdots \xrightarrow[]{\beta_n}$ such that 
    \begin{enumerate}
        \item $\forall i \in 1 \dots n : ~ \beta_i \in \isStrategy[\participant{A}](\isRun \xrightarrow[]{\beta_1} \cdots \xrightarrow[]{\beta_{i-1}})$
        \item $\hat\isRun$ is liquidated 
    \end{enumerate} 
    
    From translation correctness, we know that the extension trace 
    $$\isRun \xrightarrow[]{\beta_1} \cdots \xrightarrow[]{\beta_{n}} \hat{\isRun}$$ 
    where $\hat{\isRun}$ is liquidated can mirrored on $\isconfigcompile[]{\isRun}$:
    $$\isconfigcompile[]{\isRun} \xrightarrow[]{\isactioncompile{\alpha_1}} \cdots \xrightarrow[]{\isactioncompile{\alpha_{n}}} \isconfigcompile[]{\hat{\isRun}}$$
    We will show that $\alpha_1\dots\alpha_n = \isactioncompile{\alpha_1}\dots\isactioncompile{\alpha_{n}}$.
    
    We know from liquidity connection (Lemma \ref{fs:liq-connect}) that $\isconfigcompile[]{\hat{\isRun}}$ is liquidated as well.
    In particular, all contract in $\isconfigcompile[]{\hat{\isRun}}$ have the form $\isActiveContract{withdraw \participant{A}}{v}{\someBlockchain}{x}$.
    The low-level soundness active-contract-invariant states that all contract present in $\orConfig(\orRun)$ were also present in $\isconfigcompile[]{\isRun}$:
    $$\forall \isActiveContract{\contract{C}}{v}{\someBlockchain}{x} \in \orConfig(\orRun) \Rightarrow \isActiveContract{\contract{C}}{v}{\someBlockchain}{x} \in \isconfigcompile[]{\isConfig(\isRun)}$$
    From the synchronous actions lemma (Lemma \ref{ll:static-actions}), we can conclude that all present contract in $\hat\isRun$ are liquidated as well as the non-liquidated contract were removed with the same action in $\isconfigcompile[]{\isConfig(\isRun[i])}$ and $\orRun[i]$.

    \medskip 

    The intermediate to low-level soundness invariants are guaranteeing that all deposits, withdraw contracts and inputs in $\orRun$ are part of $\isconfigcompile{\isRun}$ which can be mapped back to $\isRun$.
    Therefore, we know that for $\orRun$ and $\isRun$ it holds that 
    $$
        \forall \someBlockchain \in \activeBlockchains: \isUserPayout{A}{\someBlockchain}{\orRun}{} = \isUserPayout{A}{\someBlockchain}{\isconfigcompile[]{\isRun}}{}= \xUserPayout{A}{\someBlockchain}{{\isRun}}{}
        ~\land~ \isUserInputs{\participant{A}}{\someBlockchain}{\orRun} 
        = \isUserInputs{\participant{A}}{\someBlockchain}{{\isconfigcompile[]{\isRun}}} =\isUserInputs{\participant{A}}{\someBlockchain}{{\isRun}} 
    $$
    From synchronous actions, we follow that balance is persevered by every action.
    
\end{proof}

%% file: theory_docs/auxiliary.tex
\maketitle

For any contract $\someContract$ in an intermediate or \bitmlx run:
\begin{itemize}
    \item $desc(\someContract)$ is the set of contracts $\someDescendant = \someContract|\dots$
    \item $ldesc(\someContract)$ is the set of contracts $\someDescendant = \someContract|L|\dots$
    \item $ldesc(\someContract)$ is the set of contracts $\someDescendant = \someContract|R|\dots$
\end{itemize}

The following functions make it more easy to analyze the contents of contract preconditions. Suppose a contract advertisement $\contractAdv{G}{C}$ such that we can write $\preconditions{G}$ by extension as:

\[
    \begin{aligned}
    & \preconditions{G} = \parallelComposition_{i=1}^n
        \parallelComposition_{j=1}^k
            \big(\depositsPre{A_i}{v_{i,j}\someBlockchain[j]}{x_{i,j}}\big) \\
    &\quad | \parallelComposition_{i=1}^n
        \parallelComposition_{j=1}^m
            \big(\participant{A_i}: \texttt{secret} ~\secret{s_{i,j}} \big ) \\
    &\quad | {t_0}{\delta}@\uniqueId \\    
    \end{aligned} \\
\]

Then, we can define:

\[
    \chains{G} = [\someBlockchain[1],\dots,\someBlockchain[k]]
\]

\[
    \participants{G} = \{\participant{A_i}, \forall i=1,\dots,n\}
\]

\[
    \deposits{G} = \parallelComposition_{i=1}^n
        \parallelComposition_{j=1}^k
            \big(\depositsPre{A_i}{v_{i,j}\someBlockchain[j]}{x_{i,j}}\big) \\
\]

\[
    \forall \participant{A_i}.~
        \deposits[A_i]{G} = \parallelComposition_{j=1}^k
            \big(\depositsPre{A}{v_{i,j}\someBlockchain[j]}{x_{i,j}}\big) \\
\]

\[
    \secrets{\preconditions{G}} = \{\secret{s_i}, \forall i=1,\dots,m\}
\]

\[
    \forall \participant{A_i}.~
        \userSecrets{A_i}{G} = \{\secret{s_i}, \forall i=1,\dots,m: ~ \participant{A_i} = \participant{A}\}
\]

\[
    \forall \participant{\someBlockchain[j]}.~
        \partialBalance{\someBlockchain[j]}{\preconditions{G}} = \sum_{i=1}^n v_i \someBlockchain[j]
\]

\[
    \totalBalance{G} = [\sum_{i=1}^n v_{i,1}\someBlockchain[1], \dots, \sum_{i=1}^n v_{i,k}\someBlockchain[k]]
\]

The following functions give us information about the history of the moves in an intermediate run $\isRun$

\[
    \contractUsers{\someContract}{\isRun} = \{
        \participant{A} \in \isContractState.participants: 
        \isActiveContract{\contract{C}}{\isContractState}{\someBlockchain}{\someContract} \in \isRun
    \}
\]

\[
    \compensatedBlockchains{A}{\someContract}{\isRun} = \{
        \someBlockchain \in \activeBlockchains:
        \isActiveContract{\contract{C}}{\isContractState}{\someBlockchain}{\someContract} \in \isRun
        \land 
        \isContractState.status = \isStatusCompensated{B}
        \land 
        \participant{A} \in \contractUsers{\someContract}{\isRun} \backslash \{ \participant{B} \}
    \}
\]

\[
    \compensationHistory{A}{\someContract}{\isRun} =
        \bigcup_{\ancestorId \in \ancestor(\someContract)}
            \big( \compensatedBlockchains{A}{\ancestorId}{\isRun}
            \cup \stipCompensations{\ancestorId}{\isRun} \big)
\]

\[
    \fullyCompensatedContracts{A}{\isRun} = \{
        \someContract \in \isRun:
        \compensationHistory{A}{\someContract}{\isRun} = \activeBlockchains
    \}
\]

\[
    \contractAuths{\someContract}{\isRun} = \{
        \participant{A}: \chainAuths{A}{\someContract}{\isRun} \neq \emptyset
    \}
\]

\[
    \contractAuths{\someContract}{\xRun} = \{
        \participant{A}: (\participant{A}: \someContract) \in \xRun
    \}
\]

\[
    \finishedContractAuths{A}{\someContract}{\isRun} = \{
        \participant{B} \in \contractAuths{\someContract}{\isRun}: 
            \chainAuths{B}{\someContract}{\isRun} \cup \compensationHistory{A}{\someContract}{\isRun}
            = \activeBlockchains
    \}
\]

\[
    \userStepSecrets{\theHonestUser}{\isRun} = \{
        \someContract \in \isRun: \isRevealedStepSecret{\theHonestUser}{\someContract} \in \lastConfigOf{\isRun}
    \}
\]

\[
    \userInitSecrets{\theHonestUser}{\isRun} = \{
        \someStipulation \in \isRun: \isRevealedInitSecret{\theHonestUser}{\someStipulation} \in \lastConfigOf{\isRun}
    \}
\]